\documentclass[
12pt,				
openright,			
oneside,
a4paper,			
chapter=TITLE,		
english,			
french,				
spanish,			
brazil				
]{abntex2}

\pdfoutput=1

\usepackage[T1]{fontenc}		
\usepackage[utf8]{inputenc}		
\usepackage{lastpage}			
\usepackage{indentfirst}		
\usepackage{xcolor}				
\usepackage{graphicx}			
\usepackage{pdfpages}          
\usepackage{verbatim}
\usepackage{icomma}
\usepackage{subcaption}

\usepackage{fancyhdr}
\usepackage[Bjornstrup]{fncychap}
\usepackage[fit, breakall]{truncate}

\usepackage[bottom]{footmisc}

\fancypagestyle{plain}{\fancyhead{} }   
\usepackage[sort&compress,numbers,merge]{natbib}
\usepackage{amsthm,mathtools,newpxtext,newpxmath,bm}
\usepackage[thinlines]{easybmat}
\usepackage[cmtip,arrow]{xy} 
\usepackage{tabularx,listings}
\usepackage{multicol}
\usepackage{bm}
\usepackage[detect-none]{siunitx}
\numberwithin{equation}{chapter} 
\numberwithin{figure}{chapter} 

\renewcommand{\ABNTEXchapterfont}{\sffamily\bfseries}
\renewcommand{\ABNTEXsectionfont}{\sffamily}

\makeatletter
\usepackage{hyperref}			
\hypersetup{
	pdfpagemode=UseOutlines, colorlinks=true, breaklinks=true, linkcolor=black, anchorcolor=black, citecolor=[RGB]{138,0,0}, filecolor=black, menucolor=black,  urlcolor=black, bookmarksopen=false, pdfpagelayout=SinglePage, pdfpagetransition=Dissolve}
\makeatother

\theoremstyle{definition}

\DeclareMathAlphabet{\mathbbold}{U}{bbold}{m}{n} 
\DeclareMathOperator{\diag}{diag}
\DeclareMathOperator{\tr}{tr}

\usepackage{titlesec, blindtext, color}
\definecolor{gray75}{gray}{0.75}
\newcommand{\hsp}{\hspace{20pt}}
\titleformat{\chapter}[hang]{\Huge\bfseries}{\thechapter\hsp\textcolor{gray75}{|}\hsp}{0pt}{\Huge\bfseries}

\makeatletter
\renewcommand\tableofcontents{%
	\null\hfill\textbf{\Large\contentsname}\hfill\null\par
	\@mkboth{\MakeUppercase\contentsname}{\MakeUppercase\contentsname}%
	\@starttoc{toc}%
}
\makeatother

\usepackage[labelfont=bf]{caption}
\titulo{{\ABNTEXchapterfont{ ISOMONODROMY METHOD AND BLACK HOLES}}\\
	{\ABNTEXchapterfont{QUASINORMAL MODES: numerical results and \\ extremal limit analysis}}}
\autor{JOÃO PAULO CAVALCANTE}
\local{Brazil}
\data{2023}
\orientador{Prof. Dr. Bruno Geraldo Carneiro da Cunha}
\instituicao{%
	Federal University of Pernambuco -- UFPE
	\par
	Departamento de Física
	\par
	Programa de Pós-Graduação}
\tipotrabalho{Dissertação (Mestrado)}
\makeindex

\pretextual

\begin{document}


\begin{folhadeaprovacao}
        \begin{center}
			\centering
			\includegraphics[scale=0.76]{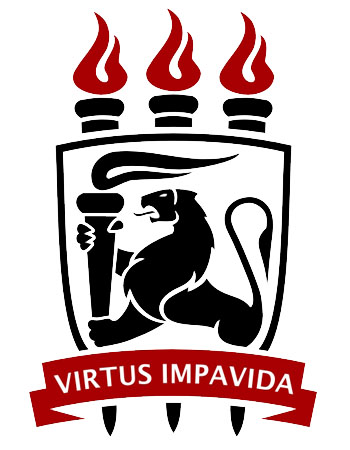}\\
    \begin{center}
            \large{UNIVERSIDADE FEDERAL DE PERNAMBUCO}\\
            \vspace{2mm}
            CENTRO DE CIÊNCIAS EXATAS E DA NATUREZA\\
            \vspace{3mm}
			PROGRAMA DE PÓS-GRADUAÇÃO EM FÍSICA
    \end{center}
            \titulo{{\ABNTEXchapterfont{ ISOMONODROMY METHOD AND BLACK HOLES}}\\
	       {\ABNTEXchapterfont{QUASINORMAL MODES: numerical results and \\ extremal limit analysis}}}
			\vfill
                {\ABNTEXchapterfont\normalfont\large\imprimirautor}
			\vfill
			\ABNTEXchapterfont\Large\imprimirtitulo
			\vfill
			\ABNTEXchapterfont\large\normalfont\imprimirlocal,
                \vspace{0.5cm}
			\ABNTEXchapterfont\large\normalfont\imprimirdata
		\end{center}
\end{folhadeaprovacao}

\begin{folhadeaprovacao}
	\begin{center}
		{\ABNTEXchapterfont{JOÃO PAULO CAVALCANTE}}
	\end{center}
	
	\vspace{1cm}
	\begin{center}
		{\ABNTEXchapterfont{ISOMONODROMY METHOD AND BLACK HOLES}}\\
		{\ABNTEXchapterfont{QUASINORMAL MODES: numerical results and extremal limit analysis}} 
	\end{center}
	\quad
	\vspace{0.05cm}
	
	\begin{flushright}
		\parbox{3in}{{\ABNTEXsectionfont{Thesis presented to the graduation program of the Physics Department of the Federal University of Pernambuco as part of the duties to obtain the degree of Doctor of Philosophy in Physics.}}}
	\end{flushright}
	\normalsize
	
	\vspace{0.05cm}
	
	\begin{flushleft}
		{\ABNTEXsectionfont{Approved on 30/06/2023.}}\\
	\end{flushleft}
	
	\vspace{0.05cm}
	
	\begin{center}
		{\ABNTEXchapterfont{EXAMINING BOARD}}\\
		\vspace{0.7cm}
		{\ABNTEXsectionfont{\underline{Prof. Bruno Geraldo Carneiro da Cunha}}}\\
		{\ABNTEXsectionfont{\textit{Advisor}}}\\
		{\ABNTEXsectionfont{Federal University of Pernambuco}} \\
		\vspace{0.7cm}
		{\ABNTEXsectionfont{\underline{Prof. Shahram Jalalzadeh}}} \\
		{\ABNTEXsectionfont{\textit{Internal Examiner}}}\\
		{\ABNTEXsectionfont{Federal University of Pernambuco}} \\
		\vspace{0.7cm}
		{\ABNTEXsectionfont{\underline{Prof.  Clécio Clemente de Souza Silva}}}\\
		{\ABNTEXsectionfont{\textit{Internal Examiner}}}\\
		{\ABNTEXsectionfont{Federal University of Pernambuco}}\\
		\vspace{0.7cm}
		{\ABNTEXsectionfont{\underline{Prof. Maurício Richartz}}} \\
		{\ABNTEXsectionfont{\textit{External Examiner}}}\\
		{\ABNTEXsectionfont{Federal University of ABC}}\\
		\vspace{0.7cm}
		{\ABNTEXsectionfont{\underline{Prof. Oscar João Campos Dias}}}\\
		{\ABNTEXsectionfont{\textit{External Examiner}}}\\
		{\ABNTEXsectionfont{University of Southampton}}
	\end{center}
\end{folhadeaprovacao}

\begin{folhadeaprovacao}
	\quad
	\vspace{18cm}
	\small{
		\begin{flushright}
			\parbox{2in}{{\ABNTEXsectionfont{
						"Relembro quando criança.\\
						Boneca eu não possuía.\\
						Eu pegava era um sabugo.\\
						Num mulambo eu envolvia.\\
						Numa casinha do mato. \\
						Passava o resto do dia."\\ (EM CANTO E POESIA, 2013)\nocite{encanto}}}}
	\end{flushright}}
	\normalsize
\end{folhadeaprovacao}
\begin{agradecimentos}
	There is no way I can give proper thanks to all the people who have helped me in myriad ways
	during my ten years as a student at UFPE. I offer a partial accounting below and ask for the forgiveness of anyone left out. 
	
	I must first thank my advisor Bruno Carneiro da Cunha that has supported all of my research endeavors with enthusiasm and provided expert advice anytime I came to him for help. His endless curiosity will always be an example to me of how to best approach physics. His technical excellence in a vast array of topics has and will always provide me with something to aspire to.
	
	I would also like to thank Julián, who I was fortunate to work closely with on some of the research presented in this thesis. I am grateful for his generous advice and support during my master's degree and PhD. Of course, I would be nowhere without the wealth of collaborators I have worked with on the projects presented here. In addition to Julián, they are Álvaro Guimarães, Mariana Lima, Matheus Martins, and Gabriel Luz. Everything I have learned, I have learned with the help and encouragement of these people.
	
	I owe a great deal to those people whose friendship and support outside of research helped keep me sane and made the academic trajectory enjoyable and fulfilling. I have shared my days with such luminaries as Thiago José, Jessica Barbosa, Flavia Bezerra, Felipe Nascimento, Bruna Ferreira, Danilo Pontual, Amanda Santos, André Santos, and Wellington Costa; And more important my family, in special my mother Maria Pereira, and my brothers and sisters; Claudia, José, Joelma, Claudio, Verônica, and Clara. And nieces Ana Luisa and Ana Katarina. 
	
	I want to thank my beloved, Helen Ribeiro, for her patience, kindness and encouragement. You make me incredibly happy and a better person. I love you.
	
	Finally, I would like to thank to the agency CNPq for financial support during my PhD.
\end{agradecimentos}

\selectlanguage{english}
\begin{resumo}[Abstract]
	\begin{otherlanguage*}{english}
		
		\hspace{1.25cm}Gravitational waves emitted by different astronomical sources, such as black holes, are dominated by quasinormal modes (QNMs), damped oscillations at unique frequencies that depend explicitly on the parameters that characterize the source of gravitational waves. In the case of black holes, the parameters are charge, mass, angular momentum, and overtones. These quasinormal modes have been studied for a long time, often to describe the time evolution of a given perturbation in a manner very similar to what is done in the analysis of normal modes.
		
		\hspace{1.25cm}More recently, with the first detections of gravitational waves by the Laser Interferometer Gravitational-Wave Observatory (LIGO), the study and analysis of QNMs has become crucial, since they characterize the \textit{ringdown} phase of a given astronomical phenomenon, for example, the coalescence between two black holes or between a black hole and a neutron star. In this phase, there is a superposition of QNMs that, in turn, can be observed by the detectors and analyzed, allowing us to estimate the values for the parameters associated with the astronomical source. Therefore, with the advent of LIGO and other gravitational wave detectors, we have an excellent motivation to study quasinormal modes.
		
		\hspace{1.25cm}From a theoretical point of view, there are in the literature a variety of methods that through different theoretical approaches seek to calculate the quasinormal (QN) frequencies. The best-known methods include; the Wentzel-Kramers-Brillouin (WKB) and Posch-Teller approximations, and the continued fraction method. In this thesis, we present and apply the isomonodromy method (or isomonodromic method) to the study of quasinormal modes, more precisely, we consider the analysis of modes that are associated with linear perturbations in two distinct four-dimensional black holes one with angular momentum (Kerr) and one with charge (Reissner-Nordström). We show, using the method, that the QN frequencies for both black holes can be analyzed with high numerical accuracy and for certain regimes even analytically. We also explore, by means of the equations involved, the regime in which both black holes become extremal. We reveal for this case that through the isomonodromic method, we can calculate with good accuracy the values for the quasinormal frequencies associated with gravitational, scalar, and electromagnetic perturbations in the black hole with angular momentum, as well as spinorial and scalar perturbations in the charged black hole. Extending thus the analysis of QN frequencies in the regime in which the methods used in the literature have generally convergence problems.
		
		\hspace{1.25cm}Through the separation of variables, we show that the equations describing linear perturbations on both black holes can be rewritten in terms of second-order ordinary differential equations (ODEs), where for the cases in which both black holes are non-extremal and extremal, we have that such ODEs are the confluent and double-confluent Heun equations, respectively. In turn, we consider the matrix representation of the solutions of such ODEs and use the method of isomonodromic deformations, which is based on the existence of families of linear matrix systems with the same monodromy parameters that can be deformed isomonodromically. From the method, we derive conditions for the isomonodromic functions $\tau_V$ and $\tau_{III}$, which are strictly connected with isomonodromic deformations in the confluent and double-confluent Heun equations, respectively. By means of these conditions, we are able to perform the numerical analysis of the QN frequencies for both black holes, in the extremal or non-extremal regime.
		
		\hspace{1.25cm}Subsequently, making use of the representation of the two functions $\tau_V$ and $\tau_{III}$ in terms of the Fredholm determinant, we show that it can be possible to reformulate, through the isomonodromic method, the eigenvalue problem of the confluent and double-confluent Heun equations into an initial value problem for both $\tau$-functions. For example, we reveal by means of the $\tau_{V}$-function that it is possible to obtain the values of the QN frequencies for the non-extremal Kerr black hole. The same is observed for the case in which the black hole is extremal, where one has that the frequencies are obtained using the function $\tau_{III}$. For both regimes (non-extremal and extremal), it is considered the analysis of the QN frequencies associated with linear perturbations of gravitational, electromagnetic, and scalar fields in this black hole.

		\hspace{1.25cm}Finally, for the case of the charged Reissner-Nordström black hole, following the same procedure applied to the Kerr black hole, we analyze the values of the QN frequencies for the extremal and non-extremal Reissner-Nordström black hole. For both cases, we present the results for the quasinormal frequencies associated with linear perturbations of charged scalar and spinorial fields. In the analysis of the QN frequencies near extremality, we find that there is a critical value for the coupling between the charge $q$ of the perturbing field and the charge $Q$ of the black hole, at which the quasinormal frequencies become purely real when the black hole becomes extremal, i.e., frequencies associated with normal modes.
		
		\vspace{\onelineskip}
		
	\end{otherlanguage*}
\end{resumo}

\setlength{\absparsep}{18pt} 
\newpage
\begin{resumo}[Resumo]
	
	\hspace{1.25cm}Ondas gravitacionais emitidas por diferentes fontes astronômicas, como buracos negros, são dominadas por modos quase-normais (QNMs), oscilações amortecidas em frequências únicas que dependem explicitamente dos parâmetros que caracterizam a fonte geradora das ondas gravitacionais. No caso dos buracos negros temos que tais parâmetros são a carga, massa, momento angular e sobretons. Esses modos quase-normais têm sido estudados há muito tempo, muitas vezes com o objetivo de descrever a evolução temporal de uma dada perturbação de uma forma muito semelhante ao que é feito na análise de modos normais.
	
	\hspace{1.25cm}Mais recentemente, com as primeiras detecções de ondas gravitacionais pelo LIGO (\textit{Laser Interferometer Gravitational-Wave Observatory}), o estudo e análise dos QNMs passou a ser crucial, dado que eles caracterizam a fase de \textit{ringdown} de um dado fenômeno astronômico, por exemplo, a coalescência entre dois buracos negros ou entre um buraco negro e uma estrela de neutron. Nessa fase tem-se uma superposição dos QNMs que por sua vez podem ser observados pelos detectores e analisados, permitindo assim uma estimativa dos valores dos parâmetros associados com a fonte astronômica. Com o advento do LIGO e outros detectores de ondas gravitacionais, temos portanto uma excelente motivação para estudar modos quase-normais.
	
	\hspace{1.25cm}Do ponto de vista teórico, há na literatura uma variedade de métodos que através de diferentes abordagens teóricas buscam calcular as frequências quase-normais (QN). Os métodos mais conhecidos incluem: as aproximações de Wentzel-Kramers-Brillouin (WKB), e de Posch-Teller, e o método da fração continuada. Nesta tese, nós apresentamos e aplicamos o método de isomonodrômia (ou método isomonodrômico) no estudo dos modos quase-normais, mais precisamente consideramos a análise dos modos que estão associados com perturbações lineares em dois buracos negros quadridimensionais distintos um com momento angular (Kerr) e outro com carga (Reissner-Nordström). Mostramos, por meio do método, que as frequências QN para ambos os buracos negros podem ser analisadas com alta precisão numérica e para certos regimes até mesmo de maneira analítica. Exploramos também, por meio das equações envolvidas o regime no qual ambos os buracos negros tornam-se extremais. Revelamos para esse caso que através do método isomonodrômico conseguimos calcular com boa precisão os valores para as frequências quase-normais associadas com perturbações gravitacionais, escalares e eletromagnéticas no buraco negro com momento angular, bem como perturbações espinoriais e escalares no buraco negro com carga. Estendendo assim a análise das frequências QN no regime no qual os métodos utilizados na literatura apresentam geralmente problemas de convergência.
	
	\hspace{1.25cm}Mostramos, através de separação de variáveis, que as equações que descrevem perturbações lineares em ambos os buracos negros podem ser reescritas em termos de equações diferenciais ordinárias (EDOs) de segunda ordem, onde, para os casos em que ambos os buracos negros são não extremais e extremais, temos que tais EDOs são as equações de Heun confluente e biconfluente, respectivamente. Por sua vez, consideramos a representação matricial das soluções de tais EDOs e utilizamos o método das deformações isomonodrômicas, que fundamenta-se na existência de famílias de sistemas matriciais lineares com os mesmos parâmetros de monodromia e que podem ser deformados isomonodromicamente. A partir do método, derivamos condições para as funções isomonodrômicas $\tau_V$ e $\tau_{III}$, que estão estritamente ligadas com deformações isomonodrômicas nas equações de Heun confluent e biconfluente, respectivamente. Por meio dessas condições conseguimos fazer a análise numérica das frequências QN para ambos os buracos buraco negros, sendo eles extremais ou não.
	
	\hspace{1.25cm}Posteriomente, fazendo uso da representação das duas funções $\tau_{V}$ e $\tau_{III}$ em termos do determinante de Fredholm, mostramos que podemos reformular, através do método isomonodrômico, o problema de autovalores das equacões de Heun confluente e biconfluente em um problema de valor inicial para ambas as funções $\tau$. Por exemplo, revelamos por meio da função $\tau_{V}$ que é possível obter os valores das frequências QN para o buraco negro de Kerr não-extremal. O mesmo é observado para o caso de Kerr extremal, onde temos que as frequências são obtidas usando a função $\tau_{III}$. Para ambos os regimes, não-extremal e extremal, é feita a análise das frequências associadas com perturbações lineares de campos gravitacionais, eletromagnéticas e escalares não massivos nesse buraco negro.
	
	\hspace{1.25cm}Finalmente, para o caso do buraco negro carregado de Reissner-Nordström, seguindo o mesmo procedimento aplicado para o buraco negro de Kerr, analisamos os valores das frequências QN para os casos de Reissner-Nordström extremal e não-extremal. Apresentamos, para ambos os casos, os resultados para as frequências quase-normais associadas com perturbações lineares de campos escalares e espinorias carregados. Na análise das frequências QN próximo da extremalidade verificamos que há um valor crítico para o acoplamento entre a carga $q$ do campo perturbador e a carga $Q$ do buraco negro, no qual as frequências quase-normais tornam-se puramente reais quando o buraco negro se torna extremal, ou seja, frequências associadas com modos normais.

	\textbf{Palavras-chave}: Perturbações lineares. Modos quase-normais. Deformações isomonodrômicas. Funções isomonodrômicas $\tau_V$ e $\tau_{III}$.
\end{resumo}
\thispagestyle{empty}
\pagestyle{fancy} 

\thispagestyle{empty}
\pdfbookmark[0]{\contentsname}{toc}
\thispagestyle{empty}
\tableofcontents
\thispagestyle{empty}
\cleardoublepage
\thispagestyle{empty}
\def\nn{\nonumber}
\thispagestyle{empty}
\def\tbf{\textbf}
\thispagestyle{empty}
\def\ti{\textit}
\thispagestyle{empty}
\def\tn{\texttbf}
\thispagestyle{empty}
\textual
\thispagestyle{empty}
\listoffigures*
\thispagestyle{empty}
\newpage
\thispagestyle{empty}
\listoftables*
\thispagestyle{empty}
\newpage
\thispagestyle{empty}

\chapter{Introduction}
\thispagestyle{myheadings}

\hspace{0.6cm}
We start this thesis with a short historical overview of quasinormal modes (QNMs) and their relevance in the understanding of black holes. In this chapter, we will see that there are, in the literature, a variety of methods that compute the quasinormal modes, such as Wentzel-Kramers-Brillouin (WKB) approximation and continued fraction method. We also give a brief introduction of the isomonodromic deformation theory history, the method presented in the thesis that emerges as a powerful method, extending smoothly the analysis of QNMs for specific values of the parameters, which are challenging to study with other methods. We finish this chapter with a detailed outline of the thesis. 

\section{Historical overview}

\begin{figure}[hbt]
	\centering
	\caption{\small{\small{The berimbau consists of a wooden bow (verga - traditionally made from biribá wood, which grows in Brazil), about 1.2 to 1.5 metres long, with a steel string tightly strung and secured from one end of the verga to the other. A gourd (cabaça), dried, opened and hollowed-out, attached to the lower portion of the verga by a loop of tough string, acts as a resonator.}}}
	\includegraphics[width=5cm]{{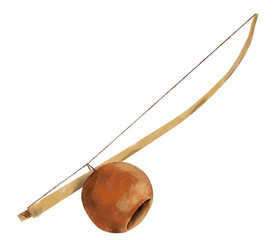}}\\
	Retrieved from Ref.\cite{berimbau}
	\label{fig:berimbau}
\end{figure}

We are all familiar with the fact that the strum of a \textit{berimbau}, single-string percussion instrument originally from Africa commonly used in Brazil, invariably produces a "characteristic sound" -- Fig. \ref{fig:berimbau} provides the reader a visualization of the form of a \textit{berimbau}. Such a system responds to any excitation with a superposition of stationary real frequencies, the normal modes. Black holes have a characteristic sound as well.  These characteristic sounds are called \textit{quasinormal modes} (QNMs) \footnote{Historically, Press was the first person to use the term \textit{quasinormal modes} to describe the black hole oscillations \cite{1971ApJ...170L.105P}.}. The 'normal' part in their names stems from the close analogy to normal modes in the way they are determined. In turn, the 'quasi' part expresses the fact that they are not quite the same: most notably, they are not really stationary in time due to their usually strong damping. 

In general, QNMs have complex frequencies with the imaginary part encoding the decay timescale of the perturbation. Another important feature of the quasinormal modes is that, different of what is observed for normal modes, where the response of the system is given for all times as a superposition of normal modes, the QNMs seem to appear only over a limited time interval and after a very late times QNMs ringing gives way to an exponentially damped. The reason behind the appearance of these complex frequencies is due to the presence of an event horizon, which rules out the  standard normal mode analysis resulting in a non-time symmetric system whose boundary value problem associated is non-Hertmitian. A discussion about these properties can be found in the classical reviews by Nollert \cite{Hans-Peter_Nollert_1999} and Kokkotas and Schmidt \cite{Kokkotas:1999bd}.

\begin{figure}[hbt]
	\centering
	\captionsetup{labelfont=bf}
	\caption{\small{\small{The figure shows the gravitational waveform of the the Zerilli function, $\psi_{\ell=2}$, produced by a test particle of mass $\mu$ falling from rest into a Schwarzschild BH with mass M \cite{PhysRevLett.27.1466}: the shape of the initial precursor depends on the details of the infall, the quasinormal mode ringing dominates the signal after zero.}}}
	\includegraphics[width=8cm]{{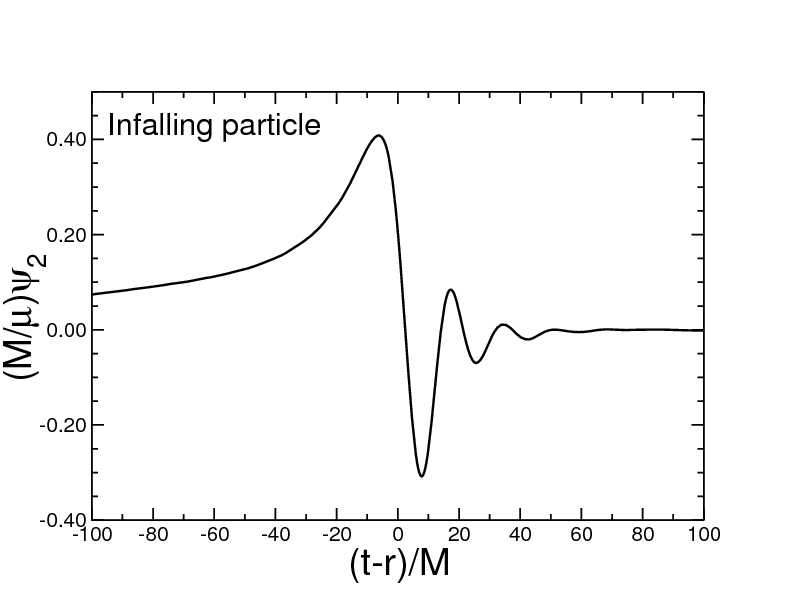}}\\
	Retrieved from Ref.\cite{infallingmass}
	\label{fig:infalling}
\end{figure}

The first investigation of QNMs dated back to the beginning of the 70s, more precisely, in the numerical simulations of the scattering of the Gaussian wavepacket by a static black hole (Schwarzschild black hole) \cite{Vishveshwara:1970zz}. In this paper, Vishveshwara found that during a certain time interval, the evolution of some initial perturbation is dominated by damped frequency oscillations or ringing frequency. These characteristic modes depend only on the parameters of the black hole, where in the Schwarzschild case is its mass, and moreover, they are completely independent of the initial perturbation, responsible for their excitation. The same behavior was observed again in the study of black hole oscillations excited by a test mass falling into the same black hole \cite{PhysRevLett.27.1466}. As a simple illustration, Fig. \ref{fig:infalling} shows how an infalling particle evolves on the Schwarzschild geometry, where we can see the damping behavior of the frequency in the solution plotted.

Decades of experience show that any perturbation in a black hole is likely to end in the same characteristic way: the gravitational wave (or the perturbation) amplitude will die off as exponentially damped sinusoids, whose frequencies and damping times are functions of the parameters of the black holes (i.e. mass, charge, and angular momentum). A great quote from Chandrasekhar's book (p.201) provides insight into the results in Fig. \ref{fig:infalling} and the knowledge developed in the literature: 

\begin{flushright}
	\textit{'[...] we may expect that during the very last stages, the black hole will emit gravitational waves with the frequencies and rates of damping, characteristic of itself, in the manner of a bell sounding its last dying pure notes.'}
\end{flushright}

The results obtained for Schwarzschild black hole made of the study QNMs spectrum (or ringdown spectrum) a relevant topic, where similar to the spectrum of the hydrogen atom in quantum mechanics, the ringdown spectrum acts as a fingerprint of black holes, providing information about the mass, charge, and angular momentum. Based on the Vishveshwara results, new methods for the calculating of the QNMs were created, improving the ringdown analysis and consequently, the understanding of black holes. One of the first methods developed was via the integration of the perturbation equations involved, where Chandrasekhar and Detweiler \cite{1975RSPSA.344..441C} and Detweiler in \cite{Detweiler1980Black} have succeeded in computing numerically some quasinormal mode for perturbations in static (Schwarzschild) and rotating (Kerr) black holes, respectively. They find such results assuming that the QNMs are solutions corresponding to incoming waves on the horizon and outgoing at infinity. Since then, numerous techniques have been developed. Some of them are semianalytical tools like the method presented by Schutz and Will in the paper \cite{osti_6061112}, based on WKB approximation. This approach was applied and extended (high-order terms) to the calculation of the quasinormal modes frequencies in Schwarzschild \cite{PhysRevD.35.3621,PhysRevD.35.3632,Konoplya:2004ip}, Reissner-Nordström \cite{PhysRevD.37.3378}, Kerr \cite{PhysRevD.41.374,KDKokkotas1991} and Kerr-Newman \cite{1993NCimB.108..991K} black holes. Additionally, there is also the Pöschl-Teller potential \cite{1933ZPhy...83..143P} used by Ferrari and Mashhoon to compute the QNMs for Kerr black hole \cite{PhysRevD.30.295}. Finally, the most efficient method was developed by Leaver \cite{Leaver:1985ax}, using the continued fraction form of the equations, which is rather easy to implement numerically. This technique was then applied to Kerr \cite{Leaver:1985ax,PhysRevD.55.3593,PhysRevD.68.124018} and Reissner-Nordström \cite{PhysRevD.41.2986} black holes.

Finally, with the first detection of gravitational waves by the LIGO and Virgo collaborations \cite{Abbott:2016blz}, the theoretical results obtained by a large number methods can be compared with the data now collected. Thus, we have again a relevant motivation to investigate the QNMs.

\section{Isomonodromy Method}

The main method explored in this thesis, the \textit{isomonodromy method} has emerged recently as a powerful method for the computation of the quasinormal modes frequencies. The method is based on the theory of isomonodromic deformations developed by Riemann \cite{Riemann1857} and extended by Schlesinger \cite{Schlesinger1912} and Garnier \cite{Garnier}. The interest in this theory increased enormously in the 1970s, when the connection between isomonodromic deformations in linear matrix systems with simple poles and completely integrable equations of mathematical physics was discovered. The connection is given in detail in the monographs \cite{doi:10.1137/1.9781611970883} and \cite{its2006isomonodromic}. Further in the 1980s more extension for the theory was made in the famous works of M. Jimbo, T. Miwa and K. Ueno \cite{JIMBO198126, JIMBO1981306, JIMBO1981407}, where the concept of isomonodromic deformations for linear matrix systems with poles of arbitrary order was introduced.

In the context of physics, the method was first explored for BHs in \cite{Motl:2003cd,Neitzke:2003mz,PhysRevD.88.044003, Castro2013b} from extensions of the WKB method using the monodromy approach. In turn, using the isomonodromy method, an alternative scheme was proposed for the calculation of quasinormal (QN) frequencies and scattering coefficients in \cite{Novaes:2014lha,daCunha:2015ana}. This method was then used to compute greybody factors in various black hole backgrounds \cite{Novaes:2018fry,Barragan-Amado:2018pxh,Amado:2020zsr}. For the analysis of the QN frequencies, in the series of papers \cite{daCunha:2015ana,CarneirodaCunha:2019tia,Cavalcante:2021scq} properties of the scattering of the fields were related to the monodromy properties of the solutions of the confluent and double-confluent Heun differential equations governing the black hole perturbations. Where these monodromy properties are obtainable from the parameters of the differential equation by means of the \textit{Riemann-Hilbert maps}, and more conveniently expressed in terms of the isomonodromic $\tau_V$ and $\tau_{III}$-functions, whose expansion around their branch points depends on \textit{monodromy parameters}. By using the general expansion of the $\tau_V$ and $\tau_{III}$ given in \cite{Gamayun:2013auu,Lisovyy:2018mnj} and \cite{Cavalcante:2021scq}, respectively, one achieved an analytical solution to the QN frequencies, given implicitly in terms of these functions, which are finally studied numerically. More details will be given in the Chapter \ref{ChapIsoMethod}, where we introduce the method.

\section{Structure of the thesis}

After this bird's eye view of the quasinormal modes calculation literature, we now return to the main matter: quasinormal modes frequencies calculation using the \textit{isomonodromy method}. Inspired by the work of Cunha and Novaes \cite{CarneirodaCunha:2015hzd} on Kerr scattering coefficients via isomonodromy, in this thesis, we present our original works \cite{CarneirodaCunha:2019tia, daCunha:2021jkm, Cavalcante:2021scq, daCunha:2022ewy}, where the isomonodromy method was developed and applied to the equations that describe linear perturbations in the Kerr and Reissner-Nordström black holes.\footnote{In the referencies, we also study the correspondence between confluent and double-confluent Heun equations and semiclassical conformal blocks.} The isomonodromy method approach is much more controllable in the particular case of extremal limit in both BHs. In the method, we reduce the main eigenvalue problem for the QNMs to the study of isomonodromic deformations in the confluent and double-confluent Heun equations.

The second part of the thesis is concerned with extensions to and new results concering the QNM spectrum of perturbed Reissner-Nordström and Kerr black holes. These topics are covered in Chapters \ref{chap:KerrBH}, and \ref{chap:RNBH}, which were originally published in \cite{daCunha:2021jkm, Cavalcante:2021scq}. The thesis is organized as follows.

\begin{itemize}
	\item In Chapter \ref{chap1}, we review the basic concepts of black hole perturbation theory and its relation with quasinormal modes. We further provide the differential equations that describe linear perturbations in the Kerr and Reissner-Nordström black holes, as well as the boundary conditions for the QNMs. We will show that linear perturbation of an $s$-spin field in the Kerr BH is described by a master equation derived from the Newman-Penrose formalism, then, using variables change, we show that such a master equation reduces to a set of differential equations whose boundary conditions allow us to compute the quasinormal frequencies. We also deal with the Reissner-Nordström BH, where, in this case, scalar and spinorial perturbations are governed by another master equation. Then, following the strategy applied to Kerr BH, we use separable variables and impose boundary conditions on the differential equations involved.
	
	The differential equations and boundary conditions presented in this chapter are necessary for a complete understanding of the results presented in the next chapters of this thesis.
	
	\item In Chapter \ref{ChapIsoMethod}, we discuss the main idea behind the isomonodromic deformations theory and its relation with isomonodromic $\tau$-functions. We focus on the deformation theory applied to the confluent and double-confluent Heun equations (CHE and DCHE), which are intrinsically related to the isomonodromic $\tau_V$ and $\tau_{III}$-functions, respectively. We revise the basic theory of linear ODEs in the complex domain and first-order linear systems. It is presented an overview of the solutions of the linear system associated with the CHE, and the monodromy matrices of the system are also defined. We deal with the isomonodromic deformations theory introduced in the chapter, where it is revealed that the application of the theory to the linear matrix system associated with the CHE leads to two conditions for the $\tau_{V}$-function. Following the results obtained for the CHE, it is shown that for the DCHE, we have from the isomonodromic deformations theory two conditions for the $\tau_{III}$-function. We finish by listing the Riemann-Hilbert maps for the $\tau_{V}$ and $\tau_{III}$ functions and deriving the accessory parameter expansion for the CHE and DCHE, this last part allows us to simplify the numerical implementation of the method.
	
	\item In Chapter \ref{chap:KerrBH}, we apply the isomonodromy method introduced in the previous chapter. The basic idea in the method consists in solving a given Riemann-Hilbert (RH) map, where for our case the solutions are related to QN frequencies. All results obtained are numerical and, for specific regimes, analytical, as we will see. It will be shown that, using the isomonodromy method, we can compare the results obtained via the RH map with the frequencies listed in the literature, for generic values of $0\leq a \leq M$. Then we will focus on the near-extremal limit $a\rightarrow M$, relevant to our analysis. We start by presenting the numeric solution for generic rotation parameter $a$. For the extremal $a\rightarrow M$ limit, we found from the studies that some modes display a finite behavior, while the rest display a double confluent limit, being given in the extremal case $a=M$ by the isomonodromic $\tau_{III}$-function, then the QN frequencies and eigenvalues found for some modes in the extremal case are listed. We conclude by discussing the results and showing that overtone modes for $a/M=0$ can be obtained from different RH maps.
	
	\item In Chapter \ref{chap:RNBH}, our main objective is to apply the isomonodromy method to investigate the QN frequencies associated with scalar and spinorial perturbations of the Reissner-Nordström black hole. We will treat the black hole and field charges, respectively $Q$ and $q$, generically, but will focus on the extremal limit. We revise the main equation that encodes the spin-field perturbation in this background, and list the relevant parameters for the confluent Heun equation and present the results of the numerical analysis for the non-extremal case $Q<M$ and extremal limit $Q\rightarrow M$ separately. Additionally, we deal with the extremal case via the Riemann-Hilbert map for the double-confluent Heun equation introduced in Chapter \ref{ChapIsoMethod}. We finish by commenting on the results obtained and comparing the results for the quasinormal frequencies in the subextremal and extremal cases. More results for the overtone frequencies for scalar and spinorial perturbations are also presented.
	
	\item In Chapter 6, we summarize the results of this thesis and present future perspectives.
\end{itemize}


\chapter{Black hole perturbations and\\ Quasinormal modes}
\label{chap1}

In this chapter, we review the basic concepts of black hole perturbation theory and its relation to quasinormal modes. In Secs. \ref{sec:overview} and \ref{sec:NPform}, we discuss the main methods developed in the literature to study perturbations in Black Holes (BHs), in other words, the Regge-Wheeler approach and the Newmann-Penrose formalism. We further provide the differential equations that describe linear perturbations in rotating (Kerr) and charged (Reissner-Nordström) black holes, as well as the boundary conditions for the differential equations involved. In Sec. \ref{sec:TMEeq}, we show that linear perturbation of an $s$-spin field in Kerr BH is described by a master equation derived from the Newman-Penrose formalism, then, using variables change, we reveal that such a master equation simplifies to a set of differential equations whose boundary conditions allow us to compute the quasinormal modes frequencies. In Sec. \ref{sec:RNBH}, we deal with the Reissner-Nordström BH, where, in this case, scalar and spinorial perturbations are governed by another master equation. Then, following the strategy applied to Kerr BH, separable variables are used and boundary conditions are imposed in the differential equations involved. We finish the chapter, in Sec. \ref{DicSection}, with two tables that summarize the differential equations derived and the boundary conditions for the QNMs.

\section{Overview: Black hole perturbations}
\label{sec:overview}

The first investigation of perturbation theory in black holes was made by Regge and Wheeler \cite{PhysRev.108.1063} in the late 1950s. In this seminal paper, they studied whether a small perturbation of the non-rotating Schwarzschild black hole would become unbounded if evolved according to the linearized version of Einstein's equations. If that were the case, Schwarzschild BH could not be considered as astrophysically relevant. For that, they considered perturbations of the spacetime metric directly by taking the following expression
\begin{equation}
g_{\mu\nu} = g^0_{\mu \nu} + \delta g_{\mu \nu},
\label{eq:perturb}
\end{equation}
where $g^{0}_{\mu\nu}$ is the space-time metric of the nonperturbed Schwarzschild black hole when all perturbations have been damped, in other words, the \textit{background} metric. In the lowest linear approximation, the \textit{perturbations} $\delta g_{\mu \nu}$ are supposed to be much less than the background $\delta g_{\mu \nu} \ll g_{\mu \nu}$. Then only terms linear in $\delta g_{\mu \nu}$ are retained in all calculations. Using this procedure they showed that a wave equation with an effective potential governs linear perturbations of a Schwarzschild black hole. 

The Regge-Wheeler approach was later explored by Zerilli \cite{PhysRevLett.24.737,PhysRevD.2.2141} and Vishveshwara \cite{Vishveshwara:1970zz} in the same black hole. With the success of the method, attention naturally turned to the rotating (Kerr) black hole, however, in this case, the description of linear perturbations was more complicated. The main reason was that in the approach the perturbation equations can not be solved via separation of variables\footnote{More precisely, using the symmetries of the spacetime (stationary and axisymmetric) the $t$ and $\phi$ dependence in the perturbation equations are separable, however, for the coordinates $r$ and $\theta$ one obtains an coupled equation \cite{Teukolsky:2014vca}.} and the only separable case were for scalar perturbations, as observed by Brill \textit{et al} \cite{PhysRevD.5.1913}. This problem was solved only in 1972, where, based on the results obtained by Fackerell and Ipser \cite{PhysRevD.5.2455}, Teukolsky solved the problem using the Newman-Penrose (NP) formalism\cite{doi:10.1063/1.1724257}. Moreover, he was able to reduce the perturbation equations for scalar, electromagnetic, and gravitational perturbations into a single master equation - known as the \textit{Teukolsky master equation}- resulting in an elegant description of the perturbations in the Kerr black hole \cite{PhysRevLett.29.1114,Teukolsky:1973ha}. In addition, such a master equation also governs spinorial perturbations, but for a specific case \cite{doi:10.1098/rspa.1976.0090}. We recommend the great review of the Kerr metric written by Teukolsky \cite{Teukolsky:2014vca} and Chandrasekhar's textbook \cite{chandrasekhar1998mathematical}, which summarizes much of the work made by Regge, Wheeler, and Teukolsky. 

In the case of the charged Reissner-Nordström black hole the study of perturbation theory was developed by the work of Zerilli \cite{PhysRevD.9.860} and Moncrief \cite{PhysRevD.10.1057, PhysRevD.9.2707,PhysRevD.12.1526}, which applied the Regge-Wheeler approach to study electromagnetic and gravitational perturbations, as well as linear stability in this spacetime. In turn, Chandrasekhar \cite{doi:10.1098/rspa.1979.0028} treated the equations that govern gravitational and electromagnetic perturbations in the same framework used by Teukolsky, the Newman-Penrose formalism.

The perturbation equations derived in the NP formalism or Regge-Wheeler approach describes waves propagating in the background (e.g. Kerr background). In most cases, these equations, whose symmetries properties are governed by the symmetries of the spacetime, can be separable through appropriate coordinates, reducing the problem to a set of linear ordinary differential equations (ODEs) or a single ODE. The ODEs are supplemented by boundary conditions allowing the study of fundamental excitations of the black hole. These excitations, in turn, represent decaying oscillations of the spacetime itself and are known as the \textit{quasinormal modes} of the black hole. The reduction of the problems into ODEs and the boundary conditions depends on the metric under consideration; some of them are discussed and compared in Chandrasekhar's book \cite{chandrasekhar1998mathematical}. Given the progress in the field and the vast literature on the subject, we will not attempt to describe all these techniques in detail.

In the next sections, we will focus on linear perturbations in the Kerr and Reissner-Nordström (RN) black holes. We will review some definitions of the Newman-Penrose formalism and show that gravitational and electromagnetic perturbations in Kerr BH are two examples of perturbations encoded in the Teukolsky master equation (TME). Regarding Reissner-Nordström black hole, we also reveal in Sec. \ref{sec:RNBH} how linear perturbations can be investigated in this background, more precisely, we are interested in the equations that describe scalar and spinorial perturbations.

\section{The Newman-Penrose formalism} 
\label{sec:NPform}

Introduced by Newman and Penrose (1962), the Newman-Penrose (NP) formalism is a set of notations developed as an effort to treat general relativity in terms of spinor notation. The formalism is a very useful and powerful method for the construction of solutions of the Einstein's equations and for studying physical fields propagating in a curved background. It is especially useful for studying algebraically special spaces and massless fields \cite{stephani_kramer_maccallum_hoenselaers_herlt_2003}. The NP method is itself a special case of the tetrad calculus \cite{Chandrasekhar:1976ap}, where the metric, Einstein field equations, and any additional matter equations are projected onto a complete vector basis defined in the spacetime (e.g. Kerr and Schwarzschild spacetimes). The vector basis chosen is a null tetrad which consists of 2 real null vectors, $\boldsymbol{\ell}$ and $\boldsymbol{n}$, a complex null vector $\boldsymbol{m}$, and its complex conjugate $\boldsymbol{m}^{*}$, that satisfy
\begin{equation}
\boldsymbol{\ell}.\boldsymbol{n}=1, \qquad \boldsymbol{m}.\boldsymbol{m}^{*}=-1,
\label{eq:tetrad}
\end{equation} 
whereas all other self-normalization and orthogonality conditions vanish,
\begin{equation}
\begin{aligned}
\boldsymbol{\ell}.\boldsymbol{\ell} = \boldsymbol{n}.\boldsymbol{n} = \boldsymbol{m}.\boldsymbol{m} = \boldsymbol{m}^{*}.\boldsymbol{m}^{*}=0,\\
\boldsymbol{\ell}.\boldsymbol{m} = \boldsymbol{\ell}.\boldsymbol{m}^{*} = \boldsymbol{n}.\boldsymbol{m} = \boldsymbol{n}.\boldsymbol{m}^{*} =0.
\label{eq:tetradort}
\end{aligned}
\end{equation}

Using these four complex vectors, a variety of primary quantities are introduced: i) Twelve complex \textit{spin coefficients} which describe the change in the tetrad from point to point in the spacetime:$\kappa, \rho, \sigma, \tau, \lambda, \mu, \nu, \pi, \varepsilon, \gamma, \beta, \alpha$. ii) Five complex functions encoding Weyl tensors in the null tetrad basis: $\Psi_0,\Psi_1,\Psi_2,\Psi_3,\Psi_4$, the function are also called by \textit{Weyl scalars}. iii) Ten functions encoding Ricci tensors in the null tetrad basis: $\Phi_{00}, \Phi_{11}, \Phi_{22}, \Lambda; \Phi_{01}, \Phi_{10}, \Phi_{02}, \Phi_{20}, \Phi_{12}, \Phi_{21}$, where the first four are real and the last six are complex. iv) Four covariant derivative operators $D$, $\Delta$, $\delta$, $\delta^{*}$, for each tetrad direction. More relevant quantities are also derived from the formalism, as $\Phi_0,\Phi_1,\Phi_{2}$, which are complex functions associate with the decomposition of the Maxwell tensor in the null tetrad basis. We give in Appendix \ref{sec:NPformApp} the explicit form of these quantities written in terms of the null vector basis, based on the textbooks \cite{frolov1998black, chandrasekhar1998mathematical,stephani_kramer_maccallum_hoenselaers_herlt_2003}.

In many spacetimes, the Newman-Penrose formalism simplifies dramatically and many of the functions listed above are zero. The main examples are the Petrov type D spacetimes (e.g. Kerr spacetime), where some spin coefficients and Weyl scalars vanish \cite{Chandrasekhar:1976ap,stephani_kramer_maccallum_hoenselaers_herlt_2003}. Additionally, more simplifications are also reached when the vector basis is chosen to reflect some symmetry of the spacetime, leading to simplified expressions for physical quantities ($\Psi_{0}$, $\Psi_{4}$, etc).  Given that, let us discuss the main strategy considered by Teukolsky in the derivation of the TME, most part of the procedure will be supplemented by the NP expressions and quantities defined in Appendix \ref{sec:NPformApp}.

\section{Kerr background: The Teukolsky equation}
\label{sec:TMEeq}

In 1972, Teukolsky was able by making use of the Newman-Penrose formalism to arrive at a sufficient set of linear partial differential equations (PDEs) that describe small perturbations of a Kerr black hole. In traditional perturbation theory, one would consider perturbations of the metric directly as in \eqref{eq:perturb}, then find the linear equations governing the perturbations. Taking such an approach in a Kerr spacetime does not lead to a separable set of PDEs. Rather Teukolsky considered the perturbation of all NP functions listed previously in the forms $\ell_{\mu} = \ell_{\mu}^{A}+\ell_{\mu}^{B}+..., $, $\Psi_4= \Psi_4^{A}+ \Psi_4^{B}+...$, $D= D^{A}+ D^{B}+...$, $\sigma = \sigma^{A}+\sigma^{B}+...$, etc., where the $A$ variables denote background quantities, and $B$ the arbitrary functions of the perturbation. The full set of perturbative equations is obtained by first inserting all these expanded quantities into the basic equations of the theory, for instance, Maxwell equations, Ricci and Bianchi identities, etc., and then terms of first order in $B$ are kept. After some algebraic manipulation, one obtains a second-order coupled partial differential equation for relevant physical quantities, for instance, equations for two Weyl scalars, for the gravitational case, or for two electromagnetic components of the Faraday tensor, in the electromagnetic case. In the next pages, both cases will be explored in more details. 

As argued previously, for Petrov type D spacetime, some NP quantities vanish, which simplifies the majority of the equations listed in Appendix \ref{sec:NPformApp}. In this way, for the Kerr metric, we have that four of the five complex Weyl scalars vanish in the background; and only one of them ($\Psi_2^{A}$) does not. Moreover, of the 12 complex spin coefficients, four of them also vanish in the background. To sum up, the vanishing functions are \cite{chandrasekhar1998mathematical}:
\begin{equation}
\begin{aligned}
\Psi_{0}^{A} = \Psi_{1}^{A} = \Psi_{3}^{A} =\Psi_{A}^{A}=0,\\ 
\kappa^{A} = \sigma^{A} = \nu^{A} = \lambda^{A}=0.
\label{eq:backgroundzeros}
\end{aligned}
\end{equation}
For completeness, the conditions above are derived in the study of possible algebraic symmetries of the Weyl tensor, which leads to a classification called, \textit{Petrov classification}. We do not discuss in detail in this thesis such a classification, we will only use the conditions \eqref{eq:backgroundzeros} to show how the Teukolsky master equation is derived. Thus, for a review of the classification we recommend the following textbooks \cite{stephani_kramer_maccallum_hoenselaers_herlt_2003,chandrasekhar1998mathematical}, as well as the Newman and Penrose paper \cite{doi:10.1063/1.1724257}, where it is given a list (see page 571) of how the Petrov classification is made.   

Let us now use the strategy given in the begining of the section and the conditions \eqref{eq:backgroundzeros} to explain schematically how the TME is derived, we will consider only the derivation of the equations that describe gravitational and electromagnetic perturbations in the Kerr black hole. Since the involved computations are quite lengthy, we will provide the main equations and explain the majority of the steps used in Chandrasekhar's book - more details can be verified in the textbook \cite{chandrasekhar1998mathematical} and in Teukolsky's paper \cite{Teukolsky:1973ha}. Additionally, in what follows, we will consider the non-source case, $T_{\mu\nu}=0$, relevant to the analysis of the quasinormal modes in Chapter \ref{chap:KerrBH}. Again, in Appendix \ref{sec:NPformApp} are given the expressions for: spin coefficients, directional derivatives, Weyl and Faraday tensor components derived in the NP formalism and used in the derivation below.

\vspace{0.2cm}
\textbf{Kerr metric}
\vspace{0.2cm}

In four-dimensional asymptotically flat spacetime, the most general vacuum black hole solution of the Einstein's equations is the Kerr metric. In the standard Boyer-Linquist coordinates defined in units such that $c=G=1$, the metric depends on two parameters: the mass $M$ and the spin $a = J/M$, where $J$ is the angular momentum:

\begin{equation}
\begin{aligned}
ds^{2} = -\bigg(1-\frac{2Mr}{\Sigma_{BL}}\bigg) dt^{2}-\bigg(\frac{4Mar \text{sin}^{2}\theta}{\Sigma_{BL}}\bigg)dtd\phi+ \frac{\Sigma_{BL}}{\Delta_{BL}}d^{2}r+\Sigma_{BL} d^{2}\theta \\
-\text{sin}^{2}\theta \bigg(r^{2}+a^{2}+\frac{2Ma^{2}r\text{sin}^{2}\theta}{\Sigma_{BL}}\bigg) d\phi^{2},
\label{eq:kerrmetric}
\end{aligned}
\end{equation}
whereas the functions $\Delta_{BL}$ and $\Sigma_{BL}$ are given by \footnote{Here we are labeling $\Delta$ and $\Sigma$ with "BL" (Boyer-Linquist) because $\Delta$ and $\Sigma$ are used in the NP formalism.}
\begin{equation}
\begin{aligned}
\Delta_{BL} = r^{2}-2Mr+a^{2} =& (r-r_{+})(r-r_{-}), \quad r_{\pm} = M\pm\sqrt{M^2-a^2}\\
\Sigma_{BL} &= r^{2}+a^{2}\cos^{2}\theta.
\end{aligned}
\label{eq:kerrpar}
\end{equation}
For $a^2 < M^2$, one has a \textit{non-extremal} Kerr black hole with the Cauchy horizon at $r=r_{-}$ and an event horizon at $r=r_{+}$, with $r_+ >r_-$. When $a=0$, the metric reduces to the Schwarzschild metric, a nonrotating black hole \cite{Schwarzschild:1916uq}, and if $a^2 > M^2$, $r_{\pm}$ are complex and this corresponds to the unphysical situation of a naked singularity \cite{WALD1974548}. Finally, when both roots are equal $r_{+}=r_{-}$ ($a=M$), one has an \textit{extremal} Kerr black hole.

For Kerr metric, for any Petrov type D metric, simplifications are made by working in a null tetrad basis adapted to the two repeated principal null directions of the corresponding Weyl tensor, where, in this case, the null directions are associated with the real vectors $\boldsymbol{\ell}$ and $\boldsymbol{n}$ \cite{Kinnersley:1968zz,doi:10.1063/1.1724257, Chandrasekhar:1976ap}. Thus, in this background,  it is common to use the Kinnersley tetrad, whose four complex vectors are expressed by
\begin{equation}
\begin{aligned}
\ell^{\mu} &= \frac{1}{\Delta_{BL}}[r^2+a^2,\Delta_{BL},0,a],\\ n^{\mu} &= \frac{1}{2\Sigma_{BL}}[r^2+a^2,-\Delta_{BL},0,a],\\ m^{\mu} &=\frac{1}{\sqrt{2}\bar{\rho}} [ia\sin \theta,0,1,i/\sin \theta],\\ {m^{*}}^{\mu} &=\frac{1}{\sqrt{2}\bar{\rho}^{*}} [-ia\sin \theta,0,1,-i/\sin \theta],
\label{eq:Kinnersleytetrad}
\end{aligned}
\end{equation}
where $\bar{\rho}=r+ia\cos \theta$, $\Sigma_{BL} = \bar{\rho}\bar{\rho}^{*} = \bar{\rho}^2$. The null vectors satisfy the conditions \eqref{eq:tetrad} and \eqref{eq:tetradort} and are labeled by $v^{\mu}=[v^t , v^r , v^{\theta} , v^{\phi}]$.

Using the Kinnersley tetrad, we can compute all the relevant quantities in the NP formalism, as spin coefficients, Weyl and Ricci tensor, Maxwell tensor, etc, (i.e. all expressions listed in Appendix \ref{sec:NPformApp}). In order to start the derivation of the equations that describe gravitational and electromagnetic perturbations in the fixed Kerr background, we first note that only the Weyl scalar component $\Psi_{2}^{A}$ is nonvanishing, while the others satisfy \eqref{eq:backgroundzeros}. On the other hand, for the electromagnetic field case, the Faraday tensor $F_{\mu\nu}$ written in the null tetrad basis is decomposed in three complex functions $\Phi_{0}^{A}$, $\Phi_{1}^{A}$ and $\Phi_{2}^{A}$ which, similar to $\Psi_{2}^{A}$, are nonvanishing in the background -- see equation \eqref{eq:Ftensor}. Said that, let us consider each case:

\vspace{0.2cm}
\textbf{Decoupled Gravitational Equations}
\vspace{0.2cm}

Among the various equations of the NP formalism listed in Appendix \ref{sec:NPformApp} and extracted from the textbooks \cite{Chandrasekhar:1976ap,frolov1998black}, there are six equations - four Bianchi identities (the first four equations in \eqref{eq:BianchiID}) and two Ricci identities (the 2nd and 10th equations in \eqref{eq:RicciId}) - which are linear and homogeneous in the quantities which vanish in the background \eqref{eq:backgroundzeros}. Starting from these equations, we can compute the first-order equations associated with gravitational perturbation in Kerr spacetime, by substituting the following expressions:
\begin{equation}
\begin{aligned}
&\text{Weyl scalars}: \Psi_{i} \rightarrow \Psi_{i}^{A}+\Psi_{i}^{B}, \quad i =0,1,2,3,4\\
&\text{directional derivatives}: \ D \rightarrow D^{A}+D^{B}, \Delta \rightarrow \Delta^{A}+\Delta^{B}, \\ & \qquad \qquad \qquad \qquad \qquad \ \delta\rightarrow \delta^{A}+\delta^{B} , {\delta^{*}}\rightarrow {\delta^{*}}^{A}+{\delta^{*}}^{B}\\
&\text{spin coefficients}: \ \sigma\rightarrow \sigma^{A}+\sigma^{B}, \nu\rightarrow \nu^{A}+\nu^{B}, \kappa\rightarrow \kappa^{A}+\kappa^{B}, \text{etc.}
\end{aligned}
\end{equation}
Then, after some algebraic manipulation, we collect the first order terms in perturbation, resulting in
\begin{subequations}
	\begin{equation}
	(\delta^{*}-4\alpha+\pi)^{A}\Psi_0^{B} - (D-4\rho-2\varepsilon)^{A} \Psi_{1}^{B} - 3\kappa^{B}\Psi_{2}^{A} =0,
	\label{eq:phi0perturba}
	\end{equation}
	\vspace{-0.9cm}
	\begin{equation}
	(\Delta - 4\gamma+\mu)^{A}\Psi_{0}^B - (\delta-4\tau-2\beta)^{A} \Psi_{1}^{B} - 3\sigma^{B}\Psi_{2}^{A} =0,	
	\label{eq:phi0perturbb}
	\end{equation}
	\vspace{-0.7cm}
	\begin{equation}
	(D-\rho-\rho^{*}-3\varepsilon+\varepsilon^{*})^{A}\sigma^{B}-(\delta-\tau+\pi^{*}-\alpha^{*}-3\beta)^{A}\kappa^{B} - \Psi_{0}^{B} =0,
	\label{eq:phi0perturbc}
	\end{equation}
	\label{eq:phi0perturb}
\end{subequations}
and
\begin{subequations}
	\begin{equation}
	(\delta^{*}+4\pi+2\alpha)^{A}\Psi_3^{B} - (D+4\varepsilon-\rho)^{A} \Psi_{4}^{B} - 3\lambda^{B}\Psi_{2}^{A} =0,
	\label{eq:phi4perturba}
	\end{equation}
	\vspace{-0.8cm}
	\begin{equation}
	(\Delta+2\gamma+4\mu)^{A}\Psi_{3}^B - (\delta-\tau+4\beta)^{A} \Psi_{4}^{B} - 3\nu^{B}\Psi_{2}^{A} =0,	
	\label{eq:phi4perturbb}
	\end{equation}
	\vspace{-0.7cm}
	\begin{equation}
	(\Delta+\mu+\mu^{*}+3\gamma-\gamma^{*})^{A}\lambda^{B}-(\delta^{*}-\tau^{*}+\pi+3\alpha+\beta^{*})^{A}\nu^{B} + \Psi_{4}^{B} =0,
	\label{eq:phi4perturbc}
	\end{equation}
	\label{eq:phi4perturb}
\end{subequations}
where the quantities in brackets are computed using the background geometry \eqref{eq:Kinnersleytetrad}, with $T_{\mu\nu}=0$. As anticipated in \eqref{eq:backgroundzeros}, the Weyl scalars $\Psi_0^{A}$ and $\Psi_4^{A}$ are zero for the unperturbed Kerr metric. These functions are related with outgoing and ingoing radiation in
the asymptotic background \cite{doi:10.1063/1.1666203}, thus, to illustrate their contribution to the equations we keep the superscripts A and B in all terms. The differential operators in the above equations are $D = \ell^{\mu}\nabla_{\mu}$, $\Delta = n^{\mu}\nabla_{\mu}$, $\delta = m^{\mu}\nabla_{\mu}$ and $\delta^{*} ={m^{\mu}}\nabla_{\mu}$, which represent the intrinsic derivative along the direction of the null tetrad basis (e.g. for $\ell^{\mu}$ direction one has D, $\Delta$ for $m^{\mu}$, etc.), see Appendix \ref{sec:NPformApp}. The remaining symbols are the spin coefficients of the Kerr background derived using the Kinnersley tetrad \eqref{eq:Kinnersleytetrad}, and are given by
\begin{equation}
\begin{aligned}
\rho &= -1/(r-ia\cos \theta), \qquad \beta = -\rho^{*}\cot \theta / (2\sqrt{2}),\\ \pi &= ia\rho^2 \sin \theta /\sqrt{2},\ \ \quad \qquad \tau = -ia\rho\rho^{*} \sin \theta /\sqrt{2}, \\ \mu &= \rho^2 \rho^{*} \Delta/2, \qquad \ \ \quad \qquad \gamma =\mu+\rho\rho^{*}(r-M)/2, \\ \alpha &= \pi - \beta^{*}, 
\label{eq:spincoef}
\end{aligned}
\end{equation}
whereas the only nonvanishing curvature scalar (or Weyl scalar) is $\Psi_2^{A} = M\rho^{3}$. For the benefit of the reader who is totally unfamiliar with the NP approach, as well as the computation of the spin coefficients \eqref{eq:spincoef} and Weyl scalars, we refer to the \textit{Black Hole Perturbation Toolkit} which provides a \textsc{Mathematica} notebook that computes the spin coefficients \eqref{eq:spincoef} and many others NP functions \cite{BHPToolkit}.

The three equations \eqref{eq:phi0perturba}-\eqref{eq:phi0perturbc} and \eqref{eq:phi4perturba}-\eqref{eq:phi4perturbc} are linearized in the sense that the functions in \eqref{eq:backgroundzeros} are considered as first order quantities, and the next step consists in eliminating $\Psi_{1}^{B}$ and $\Psi_{3}^{B}$ from the first and second systems, respectively. This is most easily done by using the type D background metric relations for $\Psi_{2}^{A}$, derived from the Bianchi identities \eqref{eq:BianchiID} via the background quantities listed in \eqref{eq:backgroundzeros} - see \cite{doi:10.1063/1.1724257, Chandrasekhar:1976ap} for more details about \textit{Goldberg-Sachs Theorem} and its relation with the relations below:
\begin{equation}
D\Psi_{2}^{A} = 3\rho \Psi_{2}^{A}, \quad \delta\Psi_{2}^{A} = 3\tau \Psi_{2}^{A}, \quad \Delta \Psi_{2}^{A} = -3\mu \Psi_{2}^{A}, \quad \delta^{*}\Psi_{2}^{A} = -3\pi \Psi_{2}^{A}.
\label{eq:GSth}
\end{equation}
In what follows, we omit the supercript A for the background quantities to simplify the notation. Thus, multiplying the equations \eqref{eq:phi0perturbc} and \eqref{eq:phi4perturbc} by $\Psi_{2}^{A}$ (or better $\Psi_{2}$) and taking into account the relations
\begin{equation}
\begin{aligned}
\Psi_2 D \sigma^{B} = D(\Psi_{2}\sigma^{B}) - \sigma^{B}D\Psi_{2} = D(\Psi_2 \sigma^{B}) - \sigma^{B}3\rho\Psi_{2},\\
\Psi_2 \delta \kappa^{B} = \delta(\Psi_{2}\kappa^{B}) - \kappa^{B}\delta\Psi_{2}^{B} = \delta(\Psi_2^{B} \kappa^{B}) - \kappa^{B} 3\tau\Psi_{2}^{B},
\end{aligned}
\end{equation}
and
\begin{equation}
\begin{aligned}
\Psi_2 \Delta \lambda^{B} = \Delta(\Psi_{2}\lambda^{B}) - \lambda^{B}\Delta\Psi_{2}^{B} = \Delta(\Psi_2^{B} \lambda^{B}) + \lambda^{B}3\mu\Psi_{2}^{B},\\
\Psi_2 \delta^{*} \nu^{B} = \delta^{*}(\Psi_{2}\nu^{B}) - \nu^{B}\delta^{*}\Psi_{2}^{B} = \delta^{*}(\Psi_2^{B} \nu^{B}) + \nu^{B} 3\pi\Psi_{2}^{B},
\label{eq:GStheorem}
\end{aligned}
\end{equation}
we obtain the expressions
\begin{subequations}
	\begin{align}
		(D-3\varepsilon +\varepsilon^{*}-4\rho-\rho^*)\Psi_2\sigma^B - (\delta+\pi^{*}-\alpha^{*}-3\beta-4\tau)\Psi_{2}\kappa^{B} - \Psi_{0}^{B}\Psi_{2}=0,
		\label{eq:thirdeqPsi0}\\
		(\Delta+3\gamma-\gamma^{*}+4\mu+\mu^*)\Psi_2\lambda^B - (\delta^{*}-\tau^{*}+\beta^{*}+3\alpha+4\pi)\Psi_{2}\nu^{B} + \Psi_{4}^{B}\Psi_{2}=0.
		\label{eq:thirdeqPsi4}
	\end{align}
\end{subequations}
Then, we use the following commutation relations derived by Teukolsky, which are valid for any Petrov type D metric \cite{Teukolsky1974PerturbationsOA}:
\begin{subequations}
	\begin{align}
		[D-3\varepsilon+\varepsilon^{*}-4\rho - \rho^{*}](\delta-2\beta-4\tau) - [\delta-3\beta-\alpha^{*}+\pi^{*}-4\tau](D-2\varepsilon-4\rho)&=0
		\label{eq:commutrel1},\\
		[\Delta + 3\gamma-\gamma^{*}+4\mu+\mu^{*}](\delta^{*}+4\pi+2\alpha) - [\delta^{*}-\tau^{*}+\beta^{*}+3\alpha+4\pi](\Delta+2\gamma+4\mu)&=0.
		\label{eq:commutrel2}	
	\end{align}
\end{subequations}

We remark that, the equations above are related by the interchange $\boldsymbol{\ell} \leftrightarrow \boldsymbol{n}$, $\boldsymbol{m} \leftrightarrow \boldsymbol{m}^{*}$, which means that the second equations \eqref{eq:thirdeqPsi4} and \eqref{eq:commutrel2} are directly derived from the equations \eqref{eq:thirdeqPsi0} and \eqref{eq:commutrel1} via the symmetries of the Newman-Penrose equations listed in Sec. \ref{sec:NPSym}. Additionally, the commutation relations \eqref{eq:commutrel1} and \eqref{eq:commutrel2} are two examples of a general commutation relation defined in Teukolsky's paper \cite{Teukolsky:1973ha}, the same general expression is also written in his thesis \cite{Teukolsky1974PerturbationsOA}.  Finally, using both commutation relations and the expressions \eqref{eq:thirdeqPsi0} and \eqref{eq:thirdeqPsi4}, we eliminate $\Psi_{1}^{B}$ and $\Psi_{3}^{B}$ from the systems to arrive at the following first-order equations for $\Psi_0^{B}$ and $\Psi_{4}^{B}$:
\begin{subequations}
	\begin{equation}
	[(D -3\varepsilon + \varepsilon^{*}-4\rho-\rho^{*})(\Delta-4\gamma+\mu) - (\delta+\pi^{*}-\alpha^{*}-3\beta-4\tau)(\delta^{*}+\pi-4\alpha)-3\Psi_2 ]\Psi_0^{B}=0,
	\label{eq:graveqpsi0}
	\end{equation}
	\setlength{\belowdisplayskip}{0pt} \setlength{\belowdisplayshortskip}{0pt}
	\setlength{\abovedisplayskip}{0pt} \setlength{\abovedisplayshortskip}{0pt}
	\begin{equation}
	[(\Delta + 3\gamma-\gamma^{*}+4\mu+\mu^{*})(D+3\varepsilon-\rho) - (\delta^{*}-\tau^{*}+\beta^{*}+3\alpha+4\pi)(\delta-\tau+4\beta)-3\Psi_{2}]\Psi_{4}^{B}=0.
	\label{eq:graveqpsi4}
	\end{equation}
\end{subequations}\\
where the directional derivatives in the Kinnersley null tetrad \eqref{eq:Kinnersleytetrad} are given by 
\begin{equation}
\begin{aligned}
D = \frac{a^2 +r^2}{\Delta_{BL}}\frac{\partial}{\partial t}+ \frac{\partial}{\partial r}+\frac{a}{\Delta_{BL}}\frac{\partial}{\partial \phi},& \quad \Delta = \frac{a^2 +r^2}{2\Sigma_{BL}}\frac{\partial}{\partial t}- \frac{\Delta_{BL}}{2\Sigma_{BL}} \frac{\partial}{\partial r}+\frac{a}{2\Sigma_{BL}}\frac{\partial}{\partial \phi}, \\
\delta = \frac{ia\sin \theta}{\sqrt{2}\bar{\rho}}\frac{\partial}{\partial t} + \frac{1}{\sqrt{2}\bar{\rho}}\frac{\partial}{\partial \theta}+\frac{i\csc \theta}{\sqrt{2}\bar{\rho}}\frac{\partial}{\partial \phi},& \quad \delta^{*} = -\frac{ia\sin \theta}{\sqrt{2}\bar{\rho}^{*}}\frac{\partial}{\partial t} + \frac{1}{\sqrt{2}\bar{\rho}^{*}}\frac{\partial}{\partial \theta}-\frac{i\csc \theta}{\sqrt{2}\bar{\rho}^{*}}\frac{\partial}{\partial \phi}. 
\end{aligned}
\label{eq:DirecOper}
\end{equation}

Substituting in the equations, the directional derivatives, spin coefficients \eqref{eq:spincoef} and Weyl scalar $\Psi_{2}^{B} = M\rho^{3}$, one obtains a second-order partial differential equations for $\Psi_{0}^{B}$ and $\Psi_{4}^{B}$, which describe gravitational perturbations in the fixed Kerr spacetime. These equations were first derived by Teukolsky in his seminal paper \cite{Teukolsky:1973ha}, he also proved that these equations can be rewritten in a master equation defined in \eqref{eq:TME}, where gravitational perturbations associated with $\Psi_{0}^{B}$ and $\Psi_{4}^{B}$ are related to the s-spin field $\psi_s$ for $s=-2$ and $s=+2$, respectively. Furthermore, Teukolsky showed that these equations can be solved by separation of variables reducing the study of gravitational perturbations to the very manageable task of solving ordinary differential equations.

\vspace{0.2cm}
\textbf{Decoupled Electromagnetic Equations}
\vspace{0.2cm}

Now, for electromagnetic perturbation in Kerr black hole, one decomposes the electromagnetic field strength tensor $F_{\mu\nu}$ onto the null tetrad basis, where the six components of the tensor (three electric and three magnetic field components) are rewritten as three complex Newman-Penrose components: 
\begin{equation}
\Phi_0 = F_{\mu\nu}l^{\mu}m^{\nu},\quad \Phi_{1} = \frac{1}{2}F_{\mu \nu}(\ell^{\mu}n^{\nu}+{m^*}^{\mu}m^{\nu}), \quad \Phi_2 = F_{\mu \nu}{m^*}^{\mu}n^{\mu}.
\end{equation}
Then, the eight homonogeneous Maxwell equations \footnote{The index "$;\nu$" and "$;\lambda$" denote the covariant derivative of the Faraday tensor -- see Frolov's book for review of the notation \cite{frolov1998black}.},
\begin{equation}	
F_{[\mu\nu;\lambda]} =0, \qquad F^{\mu \nu}{}_{;\nu}=0,
\end{equation}
give rise to a first-order system of partial differential equations for the three components $\{\Phi_0$, $\Phi_1$, $\Phi_2\}$ \cite{Teukolsky:1973ha}:
\begin{subequations}
	\begin{align}
		(D - 2\rho)\Phi_1 - (\delta^{*}+\pi-2\alpha)\Phi_{0} =0, \label{eq:Maxeq1}	\\
		(\delta-2\tau)\Phi_1 - (\Delta+\mu-2\gamma)\Phi_0 = 0, \label{eq:Maxeq2} \\
		(D-\rho+2\varepsilon)\Phi_2 - (\delta^{*}+2\pi)\Phi_{1} = 0, \label{eq:Maxeq3}\\
		(\delta-\tau+2\beta)\Phi_{2} - (\Delta+2\mu)\Phi_{1} =0. \label{eq:Maxeq4}
	\end{align}
\end{subequations}

These four equations are already linearized with respect to the components $\Phi_{0}$, $\Phi_{1}$ and $\Phi_{2}$, with the nonvanishing spin coefficients in the Kerr background and the directional derivatives defined in \eqref{eq:spincoef} and \eqref{eq:DirecOper}, respectively. Similar to the gravitational case, where one has two decoupled equations for $\Psi_{0}^{B}$ and $\Psi_{4}^{B}$, we can decouple the equations for the components $\Phi_{0}$ and $\Phi_{2}$ by eliminating the function $\Phi_1$. Moreover, for Kerr metric only the expressions for $\Phi_{0}$ and $\Phi_{2}$ lead to a separable differential equation, while for $\Phi_{1}$ the resulting equation is not a separable equation \cite{PhysRevD.5.2455}. Thus, we remove $\Phi_{1}$ in the system by using the following commutation relation extracted from Teukolsky's thesis \cite{Teukolsky1974PerturbationsOA}:
\begin{subequations}
	\begin{align}
		[D-\varepsilon+\varepsilon^{*}-2\rho - \rho^{*}](\delta-2\tau) - [\delta-\beta-\alpha^{*}+\pi^{*}-2\tau](D -2\rho)&=0,\\
		[\Delta+\gamma-\gamma^{*}+2\mu+\mu^{*}](\delta^{*}+2\pi) - [\delta^{*}+\beta^{*}+\alpha+2\pi-\tau^{*}](\Delta+2\mu)&=0,	
	\end{align}
\end{subequations}
where it can be verified straightforward from the symmetries listed in \eqref{sec:NPSym} that both commutation relations are related by the interchange $\boldsymbol{\ell} \leftrightarrow \boldsymbol{n}$, $\boldsymbol{m} \leftrightarrow \boldsymbol{m}^{*}$. To derive the equation for $\Phi_{0}$ we operate in the equations \eqref{eq:Maxeq1} and \eqref{eq:Maxeq2} with $[D-\varepsilon+\varepsilon^{*}-2\rho - \rho^{*}]$ and $[\delta-\beta-\alpha^{*}+\pi^{*}-2\tau]$, respectively, then we use the first commutation relation. For $\Phi_{2}$, we eliminate $\Phi_{1}$ by applying in equation \eqref{eq:Maxeq3} with $[\delta^{*}+\beta^{*}+\alpha+2\pi-\tau^{*}]$ and \eqref{eq:Maxeq4} with $[\Delta+\gamma-\gamma^{*}+2\mu+\mu^{*}]$. All this procedure leads us to the following equations for $\Phi_{0}$ and $\Phi_{2}$:
\begin{subequations}
	\begin{align}
		[(D - \varepsilon+\varepsilon^{*}-2\rho-\rho^{*})(\Delta+\mu-2\gamma)-(\delta-\beta-\alpha^{*}-2\tau+\pi^{*})(\delta^{*}+\pi-2\alpha)]\Phi_{0}=0,\\
		[(\Delta+\gamma-\gamma^{*}+2\mu+\mu^{*})(D-\rho+2\varepsilon)-(\delta^{*}+\beta^{*}+\alpha+2\pi-\tau^{*})](\delta - \tau + 2\beta)\Phi_{2}=0,
	\end{align}
\end{subequations}
where all functions and operators of the differential equations are defined in \eqref{eq:spincoef} and \eqref{eq:DirecOper}, respectively.

These two equations describe linear perturbations of an electromagnetic field in the fixed Kerr background \cite{Teukolsky:1973ha, PhysRevLett.29.1114}, where, as observed for the components $\Psi_{0}$ and $\Psi_{4}$ in \eqref{eq:graveqpsi0} and \eqref{eq:graveqpsi4}, both equations can be rewritten in the form of the master equation \eqref{eq:TME} with the s-spin field $\psi_{s}$ for the electromagnetic field labeled by $s=+1$ and $s=-1$ for the components $\Phi_0$ and $\Phi_{1}$, respectively.

\vspace{0.2cm}
\textbf{Teukolsky Master Equation and Separability}
\vspace{0.2cm}

After all these steps, one has that the equations for both gravitational and electromagnetic field perturbation in the Kerr background have the same form \eqref{eq:TME}, where the parameter $s$ is the spin-weight of the field. Moreover, the Teukolsky master equation \eqref{eq:TME} do not work only for $s=\pm 2, \pm 1$ \cite{Teukolsky:1973ha,PhysRevLett.29.1114,1973ApJ...185..649P, 1974ApJ...193..443T, Starobinskil:1974nkd, doi:10.1098/rspa.1976.0022}, it appears also for scalar ($s=0$)\footnote{Using the properties of the Kerr spacetime, the study of scalar (linear) perturbations is done by solving directly the Klein-Gordon equation $\Box\Phi =0$ - see reference \cite{PhysRevD.5.1913} for a detailed review.} \cite{Starobinsky:1973aij,PhysRevD.5.1913} and spinoral ($s=\pm1/2, \pm3/2$)\cite{doi:10.1098/rspa.1976.0090,Teukolsky:1973ha,chandrasekhar1998mathematical,PhysRevD.22.2327} perturbations, for the vaccum case $T_{\mu\nu}=0$ or not. We have considered only the electromagnetic and gravitational cases because these two examples have more relevance and provide a notion of how the Newman-Penrose formalism works. One has therefore,
\begin{equation}
\begin{aligned}
\bigg[\frac{(r^{2}+a^{2})}{\Delta_{BL}}-a \text{sin}^{2}\ \theta\bigg]\frac{\partial^{2} \psi_s}{\partial t^{2}}+\frac{4Mar}{\Delta_{BL}}\frac{\partial^{2}\psi_s}{\partial t \partial \phi}+\bigg[\frac{a^{2}}{\Delta_{BL}}-\frac{1}{\text{sin}^{2}\ \theta}\bigg]\frac{\partial^{2}\psi_s}{\partial \phi^{2}}&\\
-{\Delta_{BL}}^{-s}\frac{\partial}{\partial r}\bigg({\Delta_{BL}}^{s+1} \frac{\partial \psi_s}{\partial r}\bigg)-\frac{1}{\text{sin} \ \theta}\frac{\partial}{\partial \theta}\bigg(\text{sin}\ \theta \frac{\partial \psi_s}{\partial \theta}\bigg)-2s\bigg[\frac{a(r-M)}{\Delta_{BL}}+&\frac{i \text{cos}\ \theta}{\text{sin}^{2}\ \theta}\bigg]\frac{\partial \psi_s}{\partial\phi}\\
-2s\bigg[\frac{M(r^{2}-a^{2})}{\Delta_{BL}}-r-ia\text{cos}\ \theta \bigg]\frac{\partial \psi_s}{\partial t}+(s^{2}\text{cot}^{2}\ \theta-s)\psi_s = 0,
\label{eq:TME}
\end{aligned}
\end{equation}
with $\Delta_{BL}$ defined in \eqref{eq:kerrpar}. To summarize, for a given value of $s$, the spin-weight field $\psi_s$ is a function of particular components of the field under the study on the null tetrad \eqref{eq:Kinnersleytetrad} in the Kerr metric \eqref{eq:kerrmetric}: for electromagnetic perturbations, they are components of the Maxwell tensor ($\Phi_{0}$ and $\Phi_{2}$), for gravitational perturbations they are components of the Weyl tensor ($\Psi_{0}^{B}$ and $\Psi_{4}^{B}$), and so on. In order to sum up for all relevant values of $s$, we have the Table \ref{tab:spinTME}, which shows the relation between the spin-weight field $\psi_s$ and the Newman-Penrose components derived in the formalism.\footnote{In this thesis, we do not consider the derivation of the spinorial ($s=\pm\frac{1}{2}$) case, however, the procedure is analogous to the cases treated previously, ($s=\pm 1, \pm 2$) thus, for more details we mention \cite{doi:10.1098/rspa.1976.0090}}
\begin{table}[ht]
	\centering
	\caption{Spin-weight field $\psi_s$ and its relation with each type of perturbation. The spin coefficient $\rho$ defined in \eqref{eq:spincoef} allows us to redefine the NP quantities $\chi_1$, $\Phi_{2}$ and $\Psi_{4}$, then, we write precisely the Teukolsky Master equation in the form \eqref{eq:TME}.}
	\begin{tabular}{l|l|l|l|l|l|l|l|l|l|l|} 
		\cline{2-9}
		& $\quad s\quad  $     &   $\quad   0 \quad  $  &  $\quad  +1/2 \quad $& $ \quad  -1/2 \quad  $ &   $\quad +1 \quad  $ & $ \quad  -1  \quad $   &  $\quad +2  \quad $& $\quad -2 \quad  $    \\ 
		\cline{2-9}
		& $\ \ \psi_s$  &   $\ \ \ \Phi$ &  $\ \ \ \ \ \chi_0$& $ \quad \rho^{-1} \chi_1$  & $\quad \Phi_0$& $\  \rho^{-2} \Phi_2$  & $\ \ \ \Psi_0$& $ \ \rho^{-4} \Psi_4$   \\ 
		\cline{2-9}
		\multicolumn{1}{l}{} & \multicolumn{1}{l}{} & \multicolumn{1}{l}{} & \multicolumn{1}{l}{} & \multicolumn{1}{l}{} & \multicolumn{1}{l}{} \\
		\multicolumn{1}{l}{} & \multicolumn{1}{l}{} & \multicolumn{1}{l}{} & \multicolumn{1}{l}{}  & \multicolumn{1}{l}{} & \multicolumn{1}{l}{}
	\end{tabular}\\
	Adapted from Ref. \cite{Teukolsky:1973ha}
	\label{tab:spinTME}
\end{table}

The Teukolsky Master equation \eqref{eq:TME} has the remarkable property that it can be completely separated into a system of ordinary differential equations (ODEs) for the coordinates $r$ and $\theta$. This is due to the stationarity and axisymmetry properties of the metric \eqref{eq:kerrmetric}, which allows us to separate the $t$- and $\phi$ - dependence of the spin-weight field $\psi_{s}(t,r,\theta,\phi)$ \cite{chandrasekhar1998mathematical}. Thus, we can use the \textit{mode solution} expression \cite{TeixeiradaCosta:2019skg}:
\begin{equation}
\psi_{s}(t,r,\theta,\phi) = e^{-i\omega t}e^{im\phi}{}_{s}S_{\ell m}(\theta){}_{s}R_{\ell m}(r),
\label{eq:wavefunc}
\end{equation}
where $m$ corresponds to the projection of angular momentum onto the axis of symmetry of the black hole (if $s$ is half, so $m$ is also half) and $\omega \in \mathbb{C}$ is the frequency of the field mode. Substituting \eqref{eq:wavefunc} into \eqref{eq:TME}, we arrive at the following ODEs for ${}_{s}R_{\ell m}(r)$ and ${}_{s}S_{\ell m}(\theta)$:
\begin{gather}
	\bigg\{\frac{1}{\sin\theta}\frac{d}{d\theta}\left[\sin \theta
	\frac{d}{d\theta}\right]+
	a^{2}\omega^{2}\cos^{2}\theta-2a\omega s \cos\theta -
	\frac{(m+s\cos\theta)^{2}}{\sin^{2}\theta}+{_sA_{\ell m}}
	\bigg\}{}_{s}S_{\ell m}(\theta)=0,
	\label{eq:Angeq} \\
	\bigg\{{\Delta_{BL}}^{-s}\frac{d}{dr}\left[{\Delta_{BL}}^{s+1}\frac{d}{dr}\right]+
	\frac{K^{2}(r)-2is(r-M)K(r)}{\Delta_{BL}}+4is\omega
	r-\lambda(a\omega)
	\bigg\}{}_{s}R_{\ell m}(r)=0,
	\label{eq:Radeq}
\end{gather}
where ${}_{s}\lambda_{\ell m}$ the separation constant with ${}_{s}A_{\ell m} = s+{}_{s}\lambda_{\ell m}$, $\lambda(a\omega) = {}_{s}{\lambda}_{\ell m}+a^2\omega^2-2ma\omega$, $K(r) = (r^2+a^2)\omega-am$, and $\Delta_{BL} =(r-r_+)(r-r_-)$. 

The solutions of the angular equation \eqref{eq:Angeq} are known in the literature as spin-weighted spheroidal harmonics: ${}_{s}S_{\ell m}(\theta)={}_{s}S_{\ell m}(a\omega,\cos \theta)$ \cite{doi:10.1098/rspa.1977.0187, Casals:2018cgx}. For $a\omega=0$, the eigenfunction reduces to the spin-weighted spherical harmonics ${}_{s}Y_{\ell m}(\theta, \phi)={}_{s}S_{\ell m}(0,\cos\theta)e^{im\phi}$ with the angular separation constant given by ${}_{s}\lambda_{\ell m}=(\ell-s)(\ell+s+1)$ \cite{1967JMP.....8.2155G}.  The equation has two regular points for $\cos \theta =\pm1$ and it is transformed into a Sturm-Liouville problem for ${}_{s}\lambda_{\ell m}$ by requiring a regular behavior of the eigenfunction ${}_{s}S_{\ell m}(\theta)$ at $\theta =0$ and $\theta=\pi$. In this case, for a given set of $a\omega$, $s$, $\ell$ and $m$, with $\ell \geq |s|$ and $-\ell \leq m \leq \ell$, only discrete values of ${}_{s}\lambda_{\ell m}$ are allowed. For a given eigenvalue ${}_{s}\lambda_{\ell m}$ complex if $\omega$ is complex, we can obtain the other value for ${}_{s}\lambda_{\ell m}$ via the following symmetries \cite{Berti:2005gp, PhysRevD.73.024013}:
\begin{equation}
{}_{-s}\lambda_{\ell m} = {}_{s}\lambda_{\ell m}+2s,  \qquad {}_{s}\lambda_{\ell m} = {}_{s}\lambda^{*}_{\ell -m}
\label{eq:lambdasym}
\end{equation}
where the first expression gives the relation between eigenvalues with negative and positive \textit{s}, while the second do the same for \textit{m}. Finally, for $a\neq 0$, the angular eigenvalue ${}_{s}\lambda_{\ell m}$ is a function of $a\omega$, whose expansion for $a\omega \in \mathbb{R}$ and $a\omega \in \mathbb{C}$ can be verified in \cite{Berti:2005gp,Casals:2018cgx, doi:10.1098/rspa.1977.0187, PhysRevD.71.064025}. 

The radial equation, in turn, has more relevance because encodes the response of the black hole to the perturbation and by imposing boundary conditions one obtains an eigenvalue problem for $\omega$. The equation has poles at $r=r_{-}$ (Cauchy horizon) and  $r=r_{+}$ (event horizon), for $r\rightarrow \infty$, the spacetime becomes a Minkoskwi spacetime and ${}_{s}R_{\ell m}(r)$ has an exponential behavior in $r$. It is convenient to look for changes of variables that transform the radial equation \eqref{eq:Radeq} into a Schrödinger-like equation and extend the domain by moving the event horizon to minus infinity. This allows a simple physical interpretation of the quasinormal mode boundary conditions, pertinent to the investigation of linear perturbations in the Kerr BH. Thus, making the transformation ${}_{s}h_{\ell m} = {\Delta_{BL}}^{s/2}(r^2+a^2)^{1/2}{}_{s}R_{\ell m}(r)$ and adopting the \textit{tortoise coordinate} $r_{*}$, defined by $dr_{*}/dr = (r^2+a^2)/\Delta_{BL}$, we arrive at the following expression for \eqref{eq:Radeq} 
\begin{equation}
\frac{d^2 {}_{s}h_{\ell m}}{dr_{*}^2}+{}_{s}V_{\ell m}(\omega,r_{*}){}_{s}h_{\ell m} = 0,
\label{eq:PotentialKerr}
\end{equation}
where the complex function ${}_{s}V_{\ell,m}(\omega,r_{*})$ is given by
\begin{equation}
{}_{s}V_{\ell m}(\omega,r_{*}) =\frac{\Delta_{BL}}{(r^2+a^2)^2}\bigg[\frac{K^2(r)-2is(r-M)K(r)}{\Delta_{BL}}+4is\omega r-{}_{s}\lambda_{\ell m}-\frac{(r^2+a^2)^2}{\Delta_{BL}}\bigg(G^2+\frac{dG}{dr_{*}}\bigg) \bigg]
\end{equation}
with $G = G(r) = s(r-M)/(r^2+a^2)+r\Delta_{BL}/(r^2+a^2)^2$ \cite{10.2307/79115}. 

\begin{figure}[ht]
	\centering
	\caption{\small{\small{The figure illustrates the real part of ${}_{s}V_{\ell m}(\omega,r_{*})$, where given that the potential is a complex function, we just draw the form of the real part based on Chandrasekhar's book. To understand better the function ${}_{s}V_{\ell m}(\omega,r_{*})$, we recommend the book \cite{Chandrasekhar:1976ap} (Chapters 8 and 9) and the reference \cite{10.2307/79115}}}}
	\includegraphics[width=8cm]{{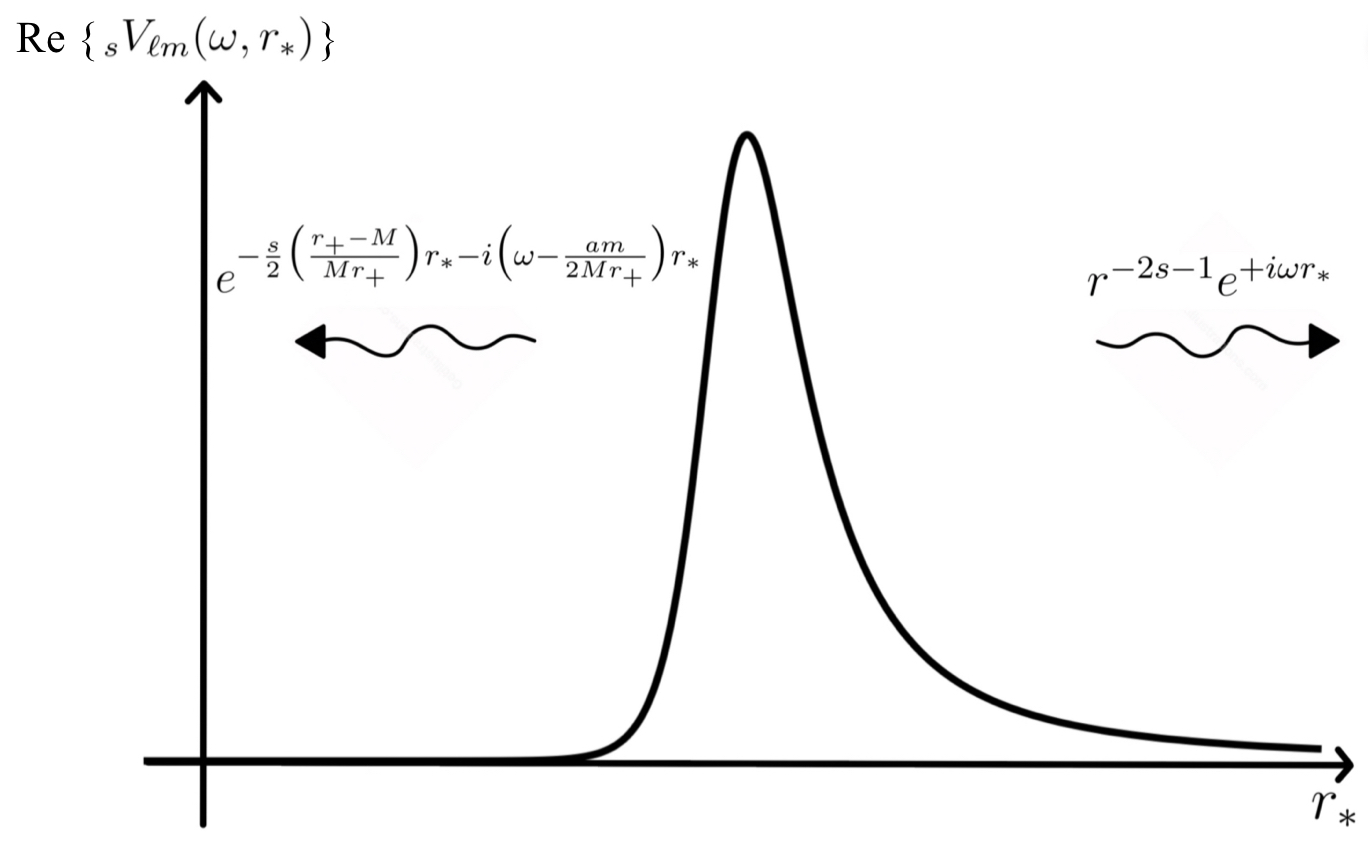}}\\
	Produced by the author.
	\label{fig:potentialform}
\end{figure}

The appropriate boundary conditions for the quasinormal modes are purely ingoing wave at the event horizon and outgoing wave at infinity, as illustrated in Fig. \ref{fig:potentialform}, where in the tortoise coordinate we have $r_{*}=-\infty$ ($r\rightarrow r_{+}$) and $r_{*}=\infty$ ($r\rightarrow \infty$). Thus, ${}_{s}h_{\ell m}$ has the following behavior 
\begin{equation}
{}_{s}h_{\ell m} =\begin{cases}
Z_s^{\text{out}}r^{-2s-1}e^{+i\omega r_{*}},\quad \quad  r_{*} \rightarrow \infty\\Z_s^{\text{in}}e^{-\frac{s}{2}\big(\frac{r_+ -M}{Mr_+}\big)r_*-i\big(\omega-\frac{am}{2Mr_+}\big)r_{*}}, \quad  r_{*} \rightarrow -\infty   
\label{eq:boundKerr}
\end{cases}  
\end{equation}
or for the equation \eqref{eq:Radeq} 
\begin{equation}
{}_{s}R_{\ell m} =\begin{cases}
A_s^{\text{out}}r^{-2s-1}e^{+i\omega r_{*}},\quad \quad \qquad  r \rightarrow \infty\\A_s^{\text{in}}{\Delta_{BL}}^{-s}e^{-i\big(\omega-\frac{am}{2Mr_+}\big)r_{*}}, \qquad  r \rightarrow r_+
\end{cases}
\label{eq:boundKerr1}
\end{equation}
where $Z_s^{\text{out}}$, $Z_s^{\text{in}}$, $A_s^{\text{out}}$ and $A_s^{\text{in}}$ are constants. 

These boundary conditions make sense intuitively, we want to investigate the response of the metric outside the black hole to initial perturbations, therefore, we do not want to take into account perturbations coming from infinity to continue perturbating the Kerr black hole. For the event horizon, one has that classically nothing should leave the black hole at $r=r_+$, consequently, only ingoing waves will be present. Finally, the discrete set of frequencies associated with solutions of the equation \eqref{eq:PotentialKerr} together with the boundary conditions \eqref{eq:boundKerr} are called quasinormal (QN) frequencies, whose solutions constructed from them are the \textit{quasinormal modes}. 
As explained in the Introduction, the 'quasi' part in the word is due to the fact that these frequencies are not really stationary in time since the imaginary part of the frequencies is different of zero. Note that, comparing with the normal-mode analysis, the quasinormal modes do not form a complete set, in other words, the time evolution of any initial perturbation can not be described as a superposition of quasinormal modes. A discussion about this characteristic of the quasinormal modes, as well as their incompleteness can be studied precisely in the reviews written by Nollert \cite{Nollert1999QuasinormalMT} and Kokkotas and Schmidt \cite{Kokkotas:1999bd}.

It was proved by Teukolsky \cite{1973ApJ...185..649P} that for Kerr black hole quasinormal frequencies must have a negative imaginary part; this has also been found for Schwarzschild and Reissner-Nordström BHs by Vishveshwara \cite{Vishveshwara:1970zz} and Moncrief \cite{PhysRevD.10.1057}, respectively. This behavior implies that quasinormal modes decay exponentially in time and the physical significance of this is that the black hole spacetime is losing energy in the form of gravitational waves. Moreover, this also means that Kerr spacetime is perturbatively stable and, in the spectrum of the quasinormal frequencies, there are no frequencies with positive imaginary part\footnote{The two examples of black holes (Kerr and Reissner-Nordström) considered in this thesis have this characteristic and linear perturbation theory can be applied.}. Furthermore, note that, if $\omega$ has a positive imaginary part the exponential $e^{-i\omega t}$ in \eqref{eq:wavefunc} will grow without bound in time and the black hole will never return to its initial state, consequently, the linear perturbation analysis will fail because it would be necessary to take into account the non-linear back reaction of the perturbations upon the metric. For a good review of instabilities in extremal and subextremal Kerr black hole, as well as in Schwarzschild and Reissner-Nordström black holes we recommend \cite{TeixeiradaCosta:2019skg,Casals:2021ugr,1973ApJ...185..649P,doi:10.1063/1.528308, Vishveshwara:1970zz,PhysRevD.10.1057,PhysRevD.9.2707}.

Later, in Chapters \ref{chap:KerrBH} and \ref{chap:RNBH}, we will see that the discrete values for the frequency that satisfy the quasinormal mode conditions \eqref{eq:boundKerr} have negative or, at least, zero imaginary part. Thus, it is not necessary to be concerned about the appearance of growing modes (i.e. modes with positive imaginary part) in the analysis of the quasinormal frequencies for Kerr and Reissner-Nordström black holes.

\subsection{Overtones and numerical methods}
\label{subsec:overtones}

As can be seen in the mode solutions \eqref{eq:wavefunc}, one has that the perturbations are decomposed in spheroidal harmonics ${}_{s}S_{\ell m}(\theta)$, where for a given value of $s$ there is a variety of QN frequencies labeled by the pair of values $(\ell, m)$. Each one of these frequencies is also labeled by an \textit{overtone} index $n$, where the mode solution with $n=1$ (\textit{fundamental mode}) dominates the perturbation, and it is followed by the overtones $n=2$, $n=3$, and so on. The domain of the solution associated with the fundamental mode frequency is due to its longest damping time, i.e. if we express the frequency as $\omega = {}_{s}\omega_{\ell m n}-i/{}_{s}\tau_{\ell mn}$, a small value for the imaginary part implies in a large time decay ${}_{s}\tau_{\ell m n}$, where $s$, $\ell$, $m$ and the overtone $n$ are the parameters of the problem. Additionally, the QN frequencies computed for Kerr black hole are also functions of the parameters M and $a=J/M$, where $J$ is the angular momentum of the black hole.

For a fixed spin-weight $s$, the eigenvalue problem for the radial and angular equations admit two solutions, where for a given ($\ell, m, n$) and $a=J/M$, one has one solution with positive real part of the frequency and other with negative real part of the frequency, and the same damping ${}_{s}\tau_{\ell m n}$, labeled by $(\ell,-m,n)$. Moreover, the positive $m$ frequencies are related to negative $m$ frequencies by the following properties \cite{Berti:2009kk}:
\begin{equation}
{}_{s}\omega_{\ell m n}=-{}_{s}\omega_{\ell -m n}, \ \ \ \ \ 1/{}_{s}\tau_{\ell m n} = 1/{}_{s}\tau_{\ell -m n}, \ \ \ \ \ {}_{s}\lambda_{\ell m n}={}_{s}\lambda^{*}_{\ell -m n}.
\label{eq:pm_m}
\end{equation}
According to this, we have always the superposition of two different solutions for positive and negative $m$ with the same damped factor, but with a different sign in the real part. For a good review of QNMs in black holes, as well as their relevance in black hole perturbation theory we recommend Berti's lecture notes \cite{BertiBangalore}.

\vspace{0.2cm}
\textbf{Numerical Methods}
\vspace{0.2cm}

The first computation of QNMs for the Kerr black hole was carried out by Detweiler, through the integration of the perturbation equations \cite{Detweiler1980Black}. The semianalytical (WKB) method, based on elementary quantum arguments and first developed by Schutz and Will \cite{osti_6061112}, was later applied to Kerr BH by Seidel and Iyer \cite{PhysRevD.41.374} and Kokkotas \cite{Kokkotas1991}. The basic idea of the method is to reduce the quasinormal mode problem into the standard WKB treatment of scattering of waves on the peak of the potential barrier. This method can compute, with great precision, the QN frequencies with a small imaginary part (first overtones), but it fails when the value of the overtone $n$ increases. 

The most successful method which computes, with high precision, the quasinormal frequencies for Kerr BH and many other black holes is the continued fraction method (or Leaver's method) which is numerically stable and has been extensively discussed and improved \cite{Leaver:1985ax,Berti:2003zu,PhysRevD.55.3593,PhysRevD.47.5253}. The method is excellent for the non-extremal case ($M >a$), even at $n\gg1$, however, has a disadvantage, it does not work in the extremal case $a=M$, and results are only obtained via modified version of the method \cite{Onozawa:1995vu,Richartz:2015saa}. The reason lies in the fact that when $a=M$, ($r_+=r_-$), the event horizon of the black hole is an irregular singular point of the perturbation equation \eqref{eq:Radeq}, which implies directly in the convergence of the method. Regarding the non-extremal case, the basic idea of the continued fraction method consists in translate the quasinormal mode boundary conditions for the radial (eq. \eqref{eq:boundKerr}) and angular equations -- the regularity of ${}_{s}S_{\ell m}(\theta)$ at $\theta=0$ and $\theta=\pi$ -- into convergence conditions for the series expansions of the corresponding eigenfunctions ${}_{s}R_{\ell m}(r)$ and ${}_{s}S_{\ell m}(\theta)$. In turn, the convergence conditions associated with each equation is analysed in terms of a recurrence relation that involves continued fractions. Then, the quasinormal frequencies are computed by the following procedure: First, fixing the values of $a$, $s$, $\ell$, $m$ and $\omega$, one uses the continued fraction equation derived from equation \eqref{eq:Angeq} to find the angular separation constant ${}_{s}\lambda_{\ell m}$, then, the value of ${}_{s}\lambda_{\ell m}$ is replaced in the continued fraction equation associated with the radial equation \eqref{eq:Radeq} with the last step consisting of looking for roots of the continued fraction function which are essentially the fundamental and overtone frequencies $\omega$. To provide a good insight of the method, we mention the Leaver's seminal paper \cite{Leaver:1985ax}, as well as the remarks on the continued fraction method \cite{PhysRevD.45.4713}.  

Finally, a new numerical method was presented by the author in \cite{daCunha:2021jkm}, where the isomonodromy method developed in \cite{CarneirodaCunha:2019tia} was applied to the calculation of the quasinormal frequencies for the extremal ($a=M$) and non-extremal ($a<M$) Kerr black hole. The results found agree with the continued fraction method, with the isomonodromy method having an excellent accuracy for the calculation of the QN frequencies near the extremal case ($a=M$). The numerical results will be shown in Chapter \ref{chap:KerrBH}.

\section{Reissner-Nordström black hole}
\label{sec:RNBH}

We now turn our attention to massless charged scalar and spinorial field perturbations in the Reissner-Nordström black hole with focus on the differential equations that describe linear perturbations, and in the boundary conditions for the QNMs. The Reissner-Nordström (RN) metric is a solution to the Einstein-Maxwell field equations, corresponding to a non-rotating charged black hole of mass $M$ and charge $Q$. In static and spherically symmetric coordinates, the metric is given by \cite{Wald:1984}
\begin{equation}
\begin{aligned}
ds^2 =
-\frac{\triangle}{r^2}dt^2+\frac{r^2}{\triangle}dr^{2}+r^2(d\theta^2+\sin^2
\theta d\phi^2), 
\ \ \ \ \  \triangle = r^2-2Mr+Q^2= (r-r_+)(r-r_{-}),
\end{aligned}
\label{eq:RNmetric}
\end{equation}
and in the Coulomb gauge the only non-vanishing component of the electromagnetic potential $A_{\mu}$ is $A_{0} = -\frac{Q}{r}$. The
roots of $\triangle$, given by $r_{\pm} = M\pm\sqrt{M^2-Q^2}$, will be real and distinct if $M^{2}>Q^{2}$. In this case, we have the \textit{non-extremal} RN black hole and $r_+$ and $r_-$ are the event and Cauchy horizons of the black hole, respectively, with $r_+>r_-$. When $Q=M$, the black hole is \textit{extremal} and both roots are equal, $r_+ = r_-$. In turn, for $Q>M$, the black hole spacetime has a naked singularity, with $\triangle$ having complex roots. 

The investigation of scalar and spinorial perturbations in this background was first made in \cite{Wu:2004vb,Chang2007MassiveCQ} (spinorial case) and in \cite{Richartz:2014jla} (scalar and spinorial cases) in the context of quasinormal modes. The author has also studied QNMs in this spacetime in \cite{Cavalcante:2021scq}, in Chapter \ref{chap:RNBH}, we will present the results obtained. Let us start now the derivation of the differential equations that describe scalar and spinorial perturbation in RN BH.

\vspace{0.2cm}
\textbf{Massless charged scalar Field}
\vspace{0.2cm}

The dynamics of a massless charged scalar field $\psi_0$ in the RN spacetime is governed by the usual Klein-Gordon equation \cite{NAKAMURA1976371} 
\begin{equation}
(\nabla^{\mu}-iqA^{\mu})(\nabla_{\mu}-iqA_{\mu})\psi_0=0,
\label{eq:KGRN}
\end{equation}
where $q$ stands for the electric charge of the scalar field and $\nabla_{\mu}-iqA_{\mu}$ includes the minimal coupling between the RN electromagnetic potential and the charge
of the field. One may decompose the field $\psi_0$ as  
\begin{equation}
\psi_0(r,\theta,\phi,t) = {}_{0}R_{\ell m}(r){}_{0}S_{\ell m}(\theta)e^{im\phi}e^{-i\omega t},
\label{eq:decompwave}
\end{equation}
where $\ell$ is the spherical harmonic index, $m$ the azimuthal harmonic index with $-\ell\leq m \leq \ell$, and $\omega$ is the frequency of the mode. With the decomposition \eqref{eq:decompwave}, ${}_{0}R_{\ell m}(r)$ and ${}_{0}S_{\ell m}(\theta)$ obey the following differential equations
\begin{gather}
	\bigg\{\frac{d^{2}}{dr^{2}}+(2r-2M)\frac{d}{dr}+
	\frac{r^4}{\triangle}\bigg(\omega - \frac{qQ}{r}\bigg)^{2}-{}_{0}\lambda_{\ell}\bigg\}{}_{0}R_{\ell m}(r)=0,
	\label{eq:scalarRadeq}\\
	\bigg\{\frac{1}{\sin\theta}\frac{d}{d\theta}\left[\sin \theta
	\frac{d}{d\theta}\right]+
	{_0\lambda_{\ell}}-
	\frac{m^2}{\sin^{2}\theta}\bigg\}{}_{0}S_{\ell m}(\theta)=0,
	\label{eq:scalarAngeq}
\end{gather}
where ${}_{0}\lambda_{\ell}=\ell(\ell+1)$ is the separation constant. Similar to the radial equation \eqref{eq:Radeq}, the radial equation above has three singularities at $r_-$, $r_{+}$ and $\infty$, but now $r_{\pm}$ are given in \eqref{eq:RNmetric}. Additionally, the second equation is the differential equation for the spherical harmonics, $Y_{\ell m}(\theta,\phi) = {}_{0}S_{\ell m}(\theta)e^{im\phi}$. To simplify the notation, the respective radial and angular equations will be given by the forthcoming master equations \eqref{eq:RNRadeq} and \eqref{eq:RNAngeq}, with $s=0$.

\vspace{0.2cm}
\textbf{Massless charged spin-$\frac{1}{2}$ Field}
\vspace{0.2cm}

We now show the form of the radial and angular equations associated with the spinorial (Dirac) field perturbation in RN black hole. Thus, a massless charged Dirac field can be described by a pair of spinors $P^{A}$ and $\bar{Q}^A$, $A=0,1$, which satisfy in this background the following Dirac equations \cite{chandrasekhar1998mathematical},
\begin{equation}
\sigma^{\mu}_{AB'}(\nabla_{\mu}-iqA_{\mu})P^{A}=0, \quad  \sigma^{\mu}_{AB'}(\nabla_{\mu}+iqA_{\mu})\bar{Q}^{A}=0
\label{eq:Diraceq}
\end{equation}
where again $A_{0}=-Q/r$ is the only nonvanishing compoment of the electromagnetic potential $A_{\mu}$. Whereas $\sigma^{\mu}_{AB'}$ stands for the generalizations of the Pauli spin matrices called Infeld-van der Waerden symbol \cite{stewart1993advanced} and is expressed in terms of the null tetrad basis defined in the RN background \cite{chandrasekhar1998mathematical,Wu:2004vb}. Using the following decomposition for the spinors, 
\begin{equation}
\begin{aligned}
P^{0} &= \frac{1}{r}R_{-\frac{1}{2}}(r)S_{-\frac{1}{2}}(\theta)e^{-i\omega t}e^{im \phi}, \quad P^{1} = R_{+\frac{1}{2}}(r)S_{+\frac{1}{2}}(\theta)e^{-i\omega t}e^{im \phi}\\
\bar{Q}^{1'} &= R_{+\frac{1}{2}}(r)S_{-\frac{1}{2}}(\theta)e^{-i\omega t}e^{im \phi}, \quad \bar{Q}^{0'} =- \frac{1}{r}R_{-\frac{1}{2}}(r)S_{+\frac{1}{2}}(\theta)e^{-i\omega t}e^{im \phi}\\
\end{aligned}
\end{equation}
we can separate the two equations in \eqref{eq:Diraceq}. We remark that, the exponential terms is due to the fact that the RN spacetime is stationary and axisymmetric, therefore, one has that the $t$- and $\phi$- dependence are given in terms of $e^{-i\omega t}$ and $ e^{im\phi}$. Thus, the resulting expressions for the radial and angular equations are \cite{Wu:2004vb}
\begin{equation}
\triangle^{\frac{1}{2}}\mathcal{D}_{0}R_{-\frac{1}{2}}(r) = \lambda \sqrt{\frac{\triangle}{2}}R_{\frac{1}{2}}(r), \quad \quad \triangle^{\frac{1}{2}}\mathcal{D}^{\dagger}_0(\triangle^{\frac{1}{2}}R_{\frac{1}{2}}(r)) = \lambda R_{-\frac{1}{2}}(r),
\label{eq:RadDirac}
\end{equation}
and
\begin{equation}
\mathcal{L}_{\frac{1}{2}}S_{\frac{1}{2}}(\theta) =-\lambda S_{-\frac{1}{2}}(\theta), \quad \quad \mathcal{L}^{\dagger}_{\frac{1}{2}}S_{-\frac{1}{2}}(\theta) =\lambda S_{\frac{1}{2}}(\theta),
\label{eq:AngDirac}
\end{equation} 
where $\lambda$ is a separation constant, $\triangle$ is defined in \eqref{eq:RNmetric}, and the operators are given by \cite{chandrasekhar1998mathematical} 
\begin{equation}
\begin{aligned}
\mathcal{D}_{n}=\frac{d}{dr}+i\frac{r^2 \omega}{\triangle}+i\frac{i qQ r^2}{\triangle}+2n\frac{r-M}{\triangle},& \qquad \mathcal{D}_{n}^{\dagger}=\frac{d}{dr}-i\frac{r^2 \omega}{\triangle}-i\frac{ qQ r^2}{\triangle}+2n\frac{r-M}{\triangle},\\
\mathcal{L}_{n}=\frac{d}{d\theta}+\frac{m}{\sin \theta}+n\cot \theta,& \qquad \mathcal{L}_{n}^{\dagger}=\frac{d}{d\theta}-\frac{m}{\sin \theta}+n\cot \theta.
\end{aligned}
\label{eq:operators}
\end{equation}

Since the goal of this chapter is to provide the main differential equations relevant to the next chapters, we only mention that the derivation can be verified in Chandrasekhar's book \cite{chandrasekhar1998mathematical} or in Wu's paper \cite{Wu:2004vb}, where the algebraic manipulation of the operators \eqref{eq:operators}, as well as the explicit form of $\sigma^{\mu}_{AB'}$ are given. Said that, the next step consists in eliminating $R_{+\frac{1}{2}}(r)$ (or $R_{-\frac{1}{2}}(r)$) from \eqref{eq:RadDirac}, in order to obtain an equation for $R_{-\frac{1}{2}}(r)$ (or $R_{+\frac{1}{2}}(r)$) only. For the angular equation \eqref{eq:AngDirac} the same procedure is made, in this way, after some algebraic manipulation, we arrive at
\begin{equation}
\begin{aligned}
\bigg\{\triangle \mathcal{D}_{\frac{1}{2}}^{\dagger}\mathcal{D}_0 - \lambda^2  \bigg\}R_{-\frac{1}{2}}(r)=0, \qquad \qquad 
\bigg\{\mathcal{L}_{\frac{1}{2}}\mathcal{L}_{\frac{1}{2}}^{\dagger} - \lambda^2  \bigg\}S_{-\frac{1}{2}}(\theta)=0.
\end{aligned}
\label{eq:diracAngRadeq}
\end{equation}
Finally, substituting the operators and finding the value of $\lambda$ from the angular equation -- see \cite{doi:10.1063/1.1705135}, we can rewrite the differential equations in the following form:

\begin{gather}
	\bigg\{\triangle^{-s}\frac{d}{dr}\left[\triangle^{s+1}\frac{d}{dr}\right]+
	\frac{K^{2}(r)-2is(r-M)K(r)}{\triangle}+4is\omega r-2isqQ-{}_{s}\lambda_{\ell}\bigg\}{}_{s}R_{\ell}(r)=0,
	\label{eq:RNRadeq}\\
	\bigg\{\frac{1}{\sin\theta}\frac{d}{d\theta}\left[\sin \theta
	\frac{d}{d\theta}\right]+
	s+{_s\lambda_{\ell}}-
	\frac{m^2}{\sin^{2}\theta}-\frac{2ms\cos\theta}{\sin^{2}\theta}-s^2\cot^{2}\theta
	\bigg\}{}_{s}S_{\ell m}(\theta)=0,
	\label{eq:RNAngeq}
\end{gather}
where $K(r) = \omega r^2 - qQr$ and ${}_{s}\lambda_{\ell} = (\ell - s)(\ell + s + 1)$. For $s=0$, equations \eqref{eq:scalarRadeq} and \eqref{eq:scalarAngeq} are recovered, while for $s=\pm1/2$, we have the differential equations \eqref{eq:diracAngRadeq} -- see the references \cite{doi:10.1142/S0218271807009322,PhysRevD.84.104021,Richartz:2014jla} for more details about the differential equations written in \eqref{eq:RNRadeq} and \eqref{eq:RNAngeq}.

Note that the equation \eqref{eq:RNRadeq} is analogous to the radial equation for the Kerr metric \eqref{eq:Radeq}, whereas the second equation is the spin-weighted spherical harmonic \cite{doi:10.1098/rspa.1977.0187}, with the spin assuming the values $s=0$ and $s= \pm\frac{1}{2}$. Finally, the azimuthal value $-\ell \leq m\leq \ell$ can be integer (scalar case) or half-integer (spinorial case). As was made with the radial equation \eqref{eq:Radeq}, we can investigate the boundary conditions for the QNMs by making the transformation ${}_{s}f_{\ell} = \triangle^{s/2}r{}_{s}R_{\ell}(r)$ and adopting the tortoise coordinate $r_{*}$, defined by $dr_{*}/dr = r^2/\triangle$. Then, the master equation \eqref{eq:RNRadeq} is rewritten as
\begin{equation}
\frac{d^2 {}_s f_{\ell}}{dr_{*}^2}+{}_{s}V_{\ell}(\omega,r_{*}){}_{s}f_{\ell} = 0,
\label{eq:Potential}
\end{equation}
where the complex function ${}_{s}V_{\ell}(\omega,r_{*})$ is given by
\begin{equation}
{}_{s}V_{\ell}(\omega,r_{*}) =\frac{\triangle}{r^4}\bigg[\frac{(K(r)-is(r-M))^2}{\triangle}+4is\omega r-2isqQ-l(l+1)+s^2-2\frac{M}{r}+2\frac{Q^2}{r^2} \bigg].
\end{equation}
Again, $r_{*}\rightarrow-\infty$ and $r_{*}\rightarrow\infty$ correspond to the event horizon $r\rightarrow r_{+}$ and the spatial infinity $r\rightarrow \infty$, respectively. Imposing the boundary conditions naturally associated with the QNMs, i.e. purely outgoing waves far away from the black hole and purely ingoing waves near the black hole's event horizon, one has that ${}_{s}f_{\ell}$ satisfies 
\begin{equation}
{}_{s}f_{\ell} =\begin{cases}
T^{\text{out}}r_{*}^{-s-iqQ}e^{+i\omega r_{*}}, \qquad  r_{*} \rightarrow \infty\\T^{\text{in}}e^{-\frac{s}{2}\big(\frac{r_+ -r_-}{r_+^2}\big)r_* -i\big(\omega-\frac{qQ}{r_+}\big)r_{*}}, \qquad  r_{*} \rightarrow -\infty
\end{cases} 
\label{eq:boundRN}
\end{equation}
or for ${}_{s}R_{\ell}$: 
\begin{equation}
{}_{s}R_{\ell} =\begin{cases}
C^{\text{out}}r_{*}^{-s-iqQ}e^{+i\omega r_{*}},\quad \quad \qquad  r \rightarrow \infty\\C^{\text{in}}{\triangle}^{-s}e^{-i\big(\omega-\frac{qQ}{r_+}\big)r_{*}}, \qquad  r \rightarrow r_+
\end{cases}
\end{equation}
where  $T^{\text{in}}$, $T^{\text{out}}$, $C^{\text{in}}$ and $C^{\text{out}}$ are constants. Similar to the radial TME \eqref{eq:Radeq}, the equation \eqref{eq:Potential}, together with the boundary conditions above, becomes an eigenvalue problem for $\omega$, where only discrete values for the frequency are allowed.

\subsection{Quasinormal modes in RN black hole}

For the RN black hole, quasinormal modes are solutions of \eqref{eq:RNRadeq} that correspond to purely ingoing waves at the event
horizon and purely outgoing at infinity, similar to Fig. \ref{fig:potentialform}. The spectrum of these boundary conditions is discrete and labeled by integers $m$, $\ell$, and $n$, with the usual interpretation, the real part represents the oscillation character of the perturbation and the imaginary part the damping factor. Furthermore, the frequencies are functions of the mass $M$ and charge $Q$ of the black hole, and charge $q$ and spin $s$ of the perturbing field. 

The numerical study of QNMs in RN BH was made in \cite{Kokkotas:1988fm}, using the semiclassical (WKB) approach. In \cite{PhysRevD.41.2986}, Leaver considered the same problem using continued fraction method that he introduced in \cite{Leaver:1985ax}. The continued fraction method allowed for improved accuracy when compared with the WKB approximation, but the numerics suffered for the extremal limit $Q\rightarrow M$. The method was later improved by Onozawa \textit{et al} \cite{Onozawa:1995vu}, and used to study the extremal case, obtaining values for the massless scalar, electromagnetic and gravitational cases when $Q=M$. 

The quasinormal modes of spinorial field perturbation in RN black hole have been calculated for small values of the product $qQ$ in \cite{Chang2007MassiveCQ} and \cite{Wu:2004vb} using the WKB approach and the Pöschl-Teller approximation, respectively. In turn, in \cite{Richartz:2014jla} the same analysis was made without assuming a small value for $qQ$. In \cite{Richartz:2014jla}, the continued fraction method was applied to the non-extremal RN black hole to obtain the QN frequencies for spins $0$ and $1/2$, for different values of $qQ$. This analysis was extended for extremal RN black hole in \cite{Richartz:2015saa}, using the same strategy developed in \cite{Onozawa:1995vu}. The study of the extremal limit, where $Q\rightarrow M$ is hindered by the numerical accuracy of the continued fraction method, and the behavior of the QNMs as one approaches the extremal point was not clear. Only in \cite{Cavalcante:2021scq}, the quasinormal frequencies for scalar and spinorial cases were computed via the \textit{isomonodromy method}, where through the method it was possible to obtain the QN frequencies for the extremal and non-extremal RN BH with high precision, complementing the results for the QN frequencies do not obtained by the continued fraction method. The isomonodromy method will be introduced in the next chapter and more results will be discussed in Chapter \ref{chap:RNBH}.

\section{Confluent and Double-Confluent Heun equations}
\label{DicSection}

As presented in the previous sections, quasinormal modes are defined as solutions of the perturbation equations and are found by imposing boundary conditions, as shown in \eqref{eq:boundKerr} and \eqref{eq:boundRN} for Kerr and Reissner-Nordström spacetimes, respectively. In most parts of the methods, one treats with the radial equation in the form \eqref{eq:Radeq} (or \eqref{eq:PotentialKerr}), however, we will transform the radial and angular equations, for both problems, into the canonical form of the confluent Heun equation (CHE), a second-order differential equation in the complex plane. This is done by changing of variables and considering the behavior of the solutions around each singular point (i.e. Cauchy horizon $r=r_{-}$ and event horizon $r=r_+$). Thus, let us start by taking the following transformation in \eqref{eq:Radeq} and \eqref{eq:RNRadeq}:
\begin{equation}
R(r) = (r-r_{-})^{-(\theta_{0}+s)/2}(r-r_{+})^{-(\theta_{z_0}+s)/2}y(z), \quad z = 2i\omega(r-r_{-}),
\label{eq:changVar}
\end{equation}
where, we have removed the subscripts $\ell,m$ and $s$ of ${}_{s}R_{\ell m}(r)$ and ${}_{s}R_{\ell}(r)$. Substituting in both equations and collecting each term, we rewrite the radial equations in the canonical form
\begin{equation}
\frac{d^{2}y}{dz^{2}}+\left[\frac{1-\theta_{0}}{z}+\frac{1-\theta_{z_{0}}}{z-z_{0}}
\right] \frac{dy}{dz}-
\left[\frac{1}{4}+\frac{\theta_{\star}}{2z}+\frac{z_0c_{z_{0}}}{z(z-z_{0})}
\right]y(z)=0.
\label{eq:ConfHeun}
\end{equation}
The $\theta's$ are functions of the parameters of the differential equations \eqref{eq:Radeq} and \eqref{eq:RNRadeq} defined for each problem. In order to simplify the notation, we will label the parameters of the CHE as $\{\theta\} = \{\theta_0,\theta_{z_0},\theta_{\star} \}$\footnote{We will define $\{\theta\}$ to simplify the notation and provide a link with the formalism introduced in the next chapter.}. One has, therefore, from \eqref{eq:changVar} and via Möbius transformation that the singularities at $r_{-}$, $r_{+}$, and $\infty$ are mapped into $0$, $z_0$ and $\infty$, respectively. In turn, $c_{z_0}$ is the accessory parameter of the equation, which does not influence the local behavior of solution $y(z)$. In regard to the behavior of $y(z)$, we have around each singularity of the differential equation the following behaviors:
\begin{equation}
y(z) = \begin{cases}
z^{(\theta_0-1)/2}(1+\mathcal{O}(z)),\qquad \ \text{for} \ \ z=0 \\(z-z_0)^{(\theta_{z_0}-1)/2}(1+\mathcal{O}(z-z_0)),\qquad \ \text{for} \ \ z=z_0\\z^{\theta_{\star}/2}e^{-z/2}(1+\mathcal{O}(z^{-1})),\qquad \ \text{for} \ \ z=\infty
\end{cases}
\end{equation}

Finally, note that in both problems the radial equation has the same canonical form, in this way, we are free to define a dictionary for the radial equation of the Kerr and Reissner-Nordström BHs. Such a dictionary allows us to sum up the parameters and boundary conditions of each problem.

\vspace{0.2cm}
\textbf{Dictionary of the CHE:}
\vspace{0.2cm}

\begin{table}[ht]
	\centering
	\caption{Dictionary for the confluent Heun equation (CHE) with all $\theta's$ expressed in terms of the parameters involved in the problems, as $\omega$, $a$, $s$, $\ell$ and $m$ for Kerr black hole and $\omega$, $s$, $\ell$, $qQ$ for RN black hole.}
	\begin{tabular}{ |c|c|c|c| } 
		\hline
		& Kerr Black hole & RN black hole \\
		\hline
		$ \theta_0$ &\vspace{0.10mm} $s -
		\frac{i}{2\pi T_-}\bigg(\omega-m\Omega_{-}\bigg)$ & $s-\frac{i}{2\pi
			T_{-}}\bigg(\omega-\frac{qQ}{r_{-}}\bigg)$ \\\hline
		$ \theta_{z_0}$ & $  s +
		\frac{i}{2\pi T_+}\bigg(\omega-m\Omega_{+}\bigg)$ & $s+ \frac{i}{2\pi
			T_{+}}\bigg(\omega-\frac{qQ}{r_{+}}\bigg)$ \\ \hline
		$ \theta_{\star}$ & $\small{-2s+ 4iM\omega}$ & $\small{-2s+2i(2M\omega -qQ)}$ \\ \hline
		$z_0$ & $\small{2i\omega(r_+-r_-)}$ & $\small{2i\omega(r_+ - r_-)}$ \\ \hline
		$s$ & $\small{0,\pm1, \pm2}$ & $\small{0,\pm\frac{1}{2}}$ \\ \hline
	\end{tabular}
	\label{tab:CHEDic}	
\end{table}

For the Kerr black hole, in the first column of Table \ref{tab:CHEDic}, we have the black hole temperature and angular velocity for Cauchy horizon ($r_{-}$) and event horizon ($r_{+}$), as well as the roots of $\Delta_{BL}$:
\begin{gather}
	2\pi T_{\pm} = \frac{r_+-r_-}{4Mr_{\pm}}, \quad\quad
	\Omega_{\pm} = \frac{a}{2Mr_{\pm}}, \quad \quad r_{\pm} = M\pm\sqrt{M^2 -a^2},
	\label{eq:temperature}
\end{gather}
where $z_0 c_{z_0} = {_s\lambda}_{\ell,m}+4s-2i(1+2s)M\omega-is
(r_+-r_-)\omega+(M^2a^2-M(r_-+3r_+))\omega^2$. 

On the other hand, for the Reissner-Nordström black hole, the temperature in each horizon and $r_{\pm}$ are defined as
\begin{equation}
\begin{gathered}
2\pi T_{\pm} = \frac{r_{\pm}-r_{\mp}}{2 r^{2}_{\pm}},   \qquad
r_{\pm} = M\pm\sqrt{M^2 -Q^2},
\label{parameters1}
\end{gathered}
\end{equation}
with $z_0c_{z_0} = {}_{s}\lambda_{l}+2s-i(1-2s)qQ+(2qQ+i(1-3s))\omega r_++i(1-s)\omega r_- -2\omega^2 r_+^2$ and ${}_{s}\lambda_{l}=(\ell-s)(\ell+s+1)$.

Note that, the confluent Heun equation has two solutions around each singularity. However, since we want to impose the quasinormal mode boundary conditions for both black holes, we have selected the $\theta$'s that satisfy such conditions. Finally, comparing the first two lines in Table \ref{tab:CHEDic}, it is easy to observe that $\theta_{ 0}$ and $\theta_{z_0}$ for both problems have the same form, with the main difference appearing in $m$ (integer) for Kerr BH and $qQ$ (real) for RN BH. This distinction will be explored in Chapters \ref{chap:KerrBH} and \ref{chap:RNBH}.

\vspace{0.2cm}
\textbf{Angular Differential equations}
\vspace{0.2cm}

Since in both problems, we have angular equations it is also interesting to simplify such equations. For Reissner-Nordström BH the separation constant ${}_{s}\lambda_{\ell m}$, between the equations \eqref{eq:RNRadeq} and \eqref{eq:RNAngeq}, is just a function of $s$ and $\ell$, ${}_{s}\lambda_{\ell}=(\ell-s)(\ell+s+1)$, with $s=0,\pm1/2$ and $\ell = |s|,\ |s|+1,\ ...$. One has therefore that the angular equation for this case is completely solved. On the other hand, for the Kerr BH case, the separation constant is a function of the frequency $\omega$ and the angular differential equation \eqref{eq:Angeq} is a CHE. Thus, we can write such an equation in the canonical form \eqref{eq:ConfHeun} by using the following change of variable 
\begin{equation}
h(z) =
(1+x)^{\theta_{z_{0}}/2}(1-x)^{\theta_{0}/2}
S(x),\quad\quad
z = -2a\omega(1-x),
\label{eq:3.8}
\end{equation}
where $x=\cos \theta$. Then,
\begin{equation}
\frac{d^{2}h}{dz^{2}}+\left[\frac{1-\theta_{0}}{z}+\frac{1-\theta_{z_{0}}}{z-z_{0}}
\right] \frac{dh}{dz}-
\left[\frac{1}{4}+\frac{\theta_{\star}}{2z}+\frac{z_0c_{z_{0}}}{z(z-z_{0})}
\right]h(z)=0,
\label{eq:3.9}
\end{equation}
with 
\begin{equation}
z_{0}= -4a\omega, \quad \quad z_0c_{z_{0}} ={}_{s}\lambda_{\ell, m}+2a\omega,
s+a^2\omega^2,
\label{eq:3.10}
\end{equation}
and the set of parameters $\{\theta\}_{\text{Ang}} = \{\theta_0,\theta_t,\theta_{\star} \}$ \footnote{We will label the angular parameters by $\{\theta\}_{\text{Ang}}$, while for the radial equations we will use $\{\theta\}$, without the label "Ang".} given by
\begin{equation}
\theta_{0}=-m-s, \quad\quad \theta_{z_{0}}=m-s,
\quad\quad \theta_{\star}=2s,
\label{eq:3.11}
\end{equation}
where the $\theta$'s selected agree with the well behavior of the eigenfunction $S(x)$ at $x=1$ and $x=-1$.

\subsection{\texorpdfstring{Extremal limits $a\rightarrow M$ and $Q\rightarrow M$}%
{}}

We are also interested in the extremal limit of the radial equation for both black holes, therefore, let us define a similar dictionary for the extremal cases $a=M$ and $Q=M$. In this situation, we define the following confluent limit, which consists in sending $r_{+}\rightarrow r_{-}$ ($a\rightarrow M$ or $Q\rightarrow M$):
\begin{equation}
\Lambda = \frac{1}{2}(\theta_{z_0}-\theta_0), \ \ \ \theta_{\circ}
=\theta_{z_0}+\theta_0, \ \ \ u_0 = \Lambda z_0 
, \ \ \ \Lambda \rightarrow \infty,
\label{eq:conflim}
\end{equation}
then, if the confluent limit is valid, which happens for specific values of the parameter of the black holes (i.e., $a$,$Q$,$s$,$\ell$,$m$,$qQ$), one has that the CHE \eqref{eq:ConfHeun} transforms into a double-confluent Heun equation (DCHE), 
\begin{equation}
\frac{d^2
	y}{dz^2}+\bigg[\frac{2-\theta_{\circ}}{z}-\frac{u_0}{z^2}\bigg]
\frac{dy}{dz}-\bigg[\frac{1}{4} 
+\frac{\theta_{\star}}{2z}+\frac{u_0k_{u_0}-u_0/2}{z^2}\bigg]y(z)=0, 
\label{eq:DCHE}
\end{equation}
where the two singularities of the equation at $z=0$ and $z=\infty$ are classified as irregular singularities of (Poincaré) rank 1, which is due to the fact that these singularities are double poles in the differential equation -- see \cite{NIST:DLMF}(Chapter 31) for more details about the irregular singularities in this differential equation. Again, $k_{u_0}$ is the accessory parameter and similar to $c_{z_0}$ does not influence the local behavior of the solution $y(z)$. Finally, the dictionary for the DCHE for both problems is given by Table \ref{tab:DCHEDic}.

\vspace{0.2cm}
\textbf{Dictionary of the DCHE:}
\vspace{0.2cm}

\begin{table}[ht]
	\centering
	\caption{Dictionary for the double-confluent Heun equation (DCHE) with all parameters, as frequency $\omega$, $a$, $s$, $\ell$ and $m$ for extremal Kerr black hole and $\omega$, $s$, $\ell$, $qQ$ for extremal RN black hole.}
	\begin{tabular}{ |c|c|c|c| } 
		\hline
		&Extremal Kerr Black hole &Extremal RN black hole \\
		\hline
		$\theta_{\circ}$ &\vspace{0.10mm} $\small{2s+4iM\omega}$ & $\small{2s+2i(2Q\omega-qQ)}$ \\\hline
		$\theta_{\star}$ & $\small{-2s+4iM\omega}$ & $\small{-2s+2i(2Q\omega -qQ)}$ \\ \hline
		$u_0$ & $\small{-4M\omega(2M\omega-m)}$ & $\small{-4M\omega(M\omega-qQ)}$ \\ \hline
		$s$ & $\small{0,\pm1, \pm2}$ & $\small{0,\pm\frac{1}{2}}$ \\ \hline
	\end{tabular}
	\label{tab:DCHEDic}	
\end{table}

The accessory parameters for the first and second column are given by
\begin{equation}
\begin{aligned}
\text{Extremal Kerr BH:}\ \ u_0k_{u_0} =& \ {_s\lambda_{\ell,m}}+4s-2i(1+2s)M\omega-3M^2\omega^2
\\
\text{Extremal RN BH:}\ \ u_0k_{u_0} =&\ (\ell-s)(\ell+s+1)+2s-i(1-2s)qQ -8(M\omega)^2+\\&+2(qQ+i(1-2s))M\omega
\end{aligned}
\end{equation}
where in the case of the Kerr BH, ${}_{s}\lambda_{\ell m}$ is computed from the angular equation \eqref{eq:Angeq} (or \eqref{eq:3.9}) with $a=M$.

\section{Conclusion of the Chapter}

We have presented in this chapter an overview of black hole perturbation theory focusing on linear perturbations in the Kerr and RN black holes. For the first case, we provided the main ingredients relevant to the derivation of the Teukolsky master equation, which describes perturbations of an $s$-spin field in the fixed Kerr spacetime. We also discussed the quasinormal modes boundary conditions for outgoing and ingoing waves at infinity and event horizon, respectively. Then it was shown that there are various methods in the literature that compute the quasinormal frequencies, with the continued fraction method representing the most efficient. For the RN black hole, we have considered linear perturbations of charged scalar and spinorial fields in this spacetime, whose radial differential equation involved has the same form as the radial Teukolsky master equation. Then, we presented the QNMs boundary conditions, and listed the main method used in the literature. We finished the chapter with two dictionaries for the main equations in the chapter, where, since in the non-extremal case both problems are described by a confluent Heun equation, we simplify the notation by putting the two problems on the same table. Moreover, given that we are interested in the behavior of the QN frequencies for $a\rightarrow M$ and $Q\rightarrow M$, a similar dictionary was created for the double-confluent Heun equation involved in the extremal cases, $a=M$ and $Q=M$.

\chapter{Isomonodromy deformation and \\ Riemann-Hilbert maps}
\thispagestyle{myheadings}
\label{ChapIsoMethod}

In this chapter, we discuss the main idea of the isomonodromic deformations theory and how isomonodromic deformations of second-order differential equations are associated with $\tau$-functions. We focus on the deformation theory applied to the confluent and double-confluent Heun equations (CHE and DCHE), which are intrinsically related to the isomonodromic $\tau_V$ and $\tau_{III}$-functions, respectively. This chapter is organized as follows: in Sec. \ref{sec:FnFsystem}, we revise the basic theory of linear ordinary differential equations in the complex domain and linear matrix systems of first-order. In Sec. \ref{sec:solmon}, we present an overview of the solutions to the matrix system associated with the CHE. The monodromy matrices of the system are also defined. In Sec. \ref{subsec:IDCHE}, we deal with the isomonodromic deformations theory introduced in Sec. \ref{sec:isodefthe}, where it is revealed that the application of the theory to the linear matrix system associated with the CHE leads to two conditions for the $\tau_{V}$-function. Based on the results obtained for the CHE, in Sec. \ref{subsec:ID-DCHE}, it is shown that the isomonodromic deformations theory applied to the DCHE leads to two  conditions for the $\tau_{III}$-function. We finish the chapter in Sec. \ref{sec:accpar} by deriving the accessory parameter expansion for the CHE and DCHE, which simplifies the numerical implementation of the conditions for the $\tau_{V}$ and $\tau_{III}$-functions. Finally, in Appendix \ref{chap:AppB}, it is given the relevant derivations used in the chapter.

We present in this part of the thesis results obtained during the PhD and published in \textit{Phys. Rev. D 102, 105013} and, more recently, in \textsc{Arxiv}: \textit{Expansions for semiclassical conformal blocks}\footnote{Arxiv link: \href{https://arxiv.org/abs/2211.03551}{Expansions for semiclassical conformal blocks}}.

\section{Fuchsian and non-Fuchsian matrix systems}
\label{sec:FnFsystem}

We start from the basic theory of linear ordinary differential equations (ODEs) in the complex domain and explore their relations to linear matrix systems of first order. Then, we study the isomonodromic deformations theory applied to linear systems, with focus on those which are related to confluent and double-confluent Heun equations, written in Chapter \ref{chap1}.

In the study of linear matrix systems of first order in the complex domain, the general starting point consists in considering a homogeneous linear ODE given in the scalar form,
\begin{equation}
y^{(N)}+p_{1}(z)y^{(N-1)}+p_{2}(z)y^{(N-2)}+...+p_{N}(z)y(z)=0, \qquad y^{(N)} = \frac{d^Ny}{dz^N}
\label{eq:ODE}
\end{equation}
where $p_1(z), p_2(z),...,p_N(z)$ are holomorphic functions of $z \in \mathbb{C}$. In the most convenient form, the equation can be written as a set of $N$ coupled linear ODEs of first order:
\begin{equation}
\frac{d}{dz}y(z) = A(z)y(z), 
\label{eq:scalarSys}
\end{equation} 
where the $N\times N$ matrix $A(z)$ is a holomorphic matrix function of $z$,
\begin{equation}
\begin{aligned}
y =  \begin{pmatrix}
y_{1} \\
y_{2} \\
\vdots \\
y_{N}
\end{pmatrix}
\end{aligned} ,\qquad A(z) = (a_{ij}(z))_{1\leq i, j\leq N}.
\end{equation}

Both expressions are mathematically equivalent, and the scalar equation \eqref{eq:ODE} can be transformed in the system \eqref{eq:scalarSys} and vise versa -- see Haraoka's book for examples of the system \eqref{eq:scalarSys} \cite{haraoka2020linear}. It is also convenient to deal with the \textit{fundamental matrix} solution $Y(z)\in GL(N,\mathbb{C})$, which is now a $N\times N$ matrix built with a given set of $N$ linear independent vector solutions, with the fundamental matrix satisfying the same equation \cite{somasundaram2001ordinary}
\begin{equation}
\frac{d}{dz}Y(z) = A(z)Y(z).
\label{eq:GenLinearSys}
\end{equation}
Additionally, the $N$ vector solutions are linearly independent if the \textit{Wronskian} of $Y(z)$ does not vanish, $W(Y;z) = \det Y(z)$ and for a given fundamental matrix one has that $A(z) = \big[\frac{d Y(z)}{dz}\big]Y(z)^{-1}$. 

Suppose that $A(z)$ is a holomorphic function on the punctured Riemann sphere $S:\mathbb{C}\mathbb{P}^{1}/\{z_1,z_2,...,z_n\}$ where $z_1, z_2, ..., z_n$ are $n$ distinct points on $\mathbb{C}\mathbb{P}^{1}$. Looking at the set of poles in the matrix $A(z)$, we can classify the linear system as \textit{Fuchsian} or \textit{non-Fuchsian} system, where one has the following definition: if the holomorphic function $A(z)$ has only simple poles, we have that \eqref{eq:GenLinearSys} is a \textit{Fuchsian} system, and its poles are Fuchsian singularities (or regular singularities)\footnote{We can use the term \textit{regular singularity} to describe a Fuchsian singularity in the linear system, whereas multiple poles, i.e., double pole (r=1), triple pole (r=2), and so on, are denoted by \textit{irregular singularities} of Poincaré rank r.}. In turn, if $A(z)$ depends on poles of high order, i.e. double pole, triple pole, etc, the linear system is classified as a \textit{non-Fuchsian} system and its poles are defined in terms of Poincaré rank $r$, a double pole has rank $r=1$, triple pole has rank $r=2$, and so on, \cite{haraoka2020linear}. Note that, Fuchsian singularities have rank $r=0$.

Let us consider now the following expresson for $A(z)$ extracted from Conte's book (Chapter 2, p. 36) \cite{conte}, it is also defined in \cite{Jimbo:1981aa}, where one has a general form for $A(z)$ with the point at infinity may have any rank $r$:
\begin{equation}
A(z) = \sum_{i=1}^{n}\sum_{j=0}^{r_i}\frac{A_{i,j}}{(z - z_i)^{j+1}}+\sum_{j=1}^{r_{\infty}}A_{\infty,j}z^{j-1}
\label{eq:nonfuchsianA}
\end{equation}
where $A_{i,j}$ and $A_{\infty,j}$ are $N\times N$ constant matrices. The nonnegative integers $r_i$ and $r_{\infty}$ are the ranks of the singularity at $z = z_i$, and $z=\infty$, respectively. For a linear Fuchsian system, one has that $A(z)$ has only simple pole in $z_i$,
\begin{equation}
A(z) = \sum_{i=0}^{n}\frac{A_i}{z-z_i}
\label{eq:fuschianA}
\end{equation}
where the $A_i$'s $\in GL(N,\mathbb{C})$ and, for a Fuchsian singularity at $z=\infty$, one has that the following relation is satisfied $\sum_{i=0}^{n}A_i =-A_{\infty}$. For instance, suppose that the system \eqref{eq:GenLinearSys} (or the scalar equation \eqref{eq:ODE}) has four Fuchsian singularities at $z= z_0,z_1,z_2$ and $\infty$, in this situation, the linear Fuchsian system assumes the form
\begin{equation}
\frac{d}{dz}Y(z) = \bigg(\frac{A_0}{z-z_0}+\frac{A_1}{z-z_1}+\frac{A_2}{z-z_2}\bigg)Y(z), \qquad A_\infty = - A_0 - A_1 - A_2, 
\label{eq:FS4points}
\end{equation}
with domain $S:\mathbb{C}\mathbb{P}^{1}/\{z_0,z_1,z_2,\infty\}$. For a non-Fuchsian system, $A(z)$ is given by \eqref{eq:nonfuchsianA} and depends on higher order poles, which results in various examples. To illustrate the richness of the general expression \eqref{eq:nonfuchsianA}, let us list three (aleatory) examples of non-Fuchsian systems:
\begin{itemize}
	\item two regular points ($r_1 = r_2 =0$) + one irregular point ($r_{\infty}= 1$)  \begin{equation}
	\frac{d}{dz}Y(z) = \bigg(\frac{A_{1,0}}{z - z_1}+\frac{A_{2,0}}{z - z_2}+A_{\infty,1}\bigg)Y(z)
	\label{eq:sysofint}
	\end{equation}
	\item one irregular point ($r_1=1$) + one irregular point ($r_{\infty}= 1$)\begin{equation}
	\frac{d}{dz}Y(z) = \bigg(\frac{A_{1,0}}{z - z_1}+\frac{A_{1,1}}{(z - z_1)^2}+A_{\infty,1}\bigg)Y(z)
	\label{eq:nFSDCHE}
	\end{equation}
	\item one regular point ($r_1 =0$) + one irregular point ($r_{\infty}= 2$) \begin{equation}
	\frac{d}{dz}Y(z) = \bigg(\frac{A_{1,0}}{z - z_1}+A_{\infty,1}+A_{\infty,2}\ z\bigg)Y(z)
	\end{equation}
\end{itemize}
where in each case, the solutions $Y(z)$ are defined around each singularity of the systems above. The study of linear matrix systems (or linear ODEs) in the complex domain with regular and/or irregular singularities has been the subject of many textbooks, so, a great review can be made in \cite{hille1997ordinary,haraoka2020linear}. Our aim is to discuss the systems \eqref{eq:sysofint} and \eqref{eq:nFSDCHE}, with a simplification in the size $N$ of the system.

\subsection{Second-order differential equation and non-Fuchsian system}

Let us treat now with the simplification of the general system \eqref{eq:GenLinearSys} (or scalar equation \eqref{eq:ODE}), we are interested in the case $N=2$. More precisely, we will consider a non-Fuchsian system with two regular singularities and one irregular point of Poincaré rank 1 ($r_{\infty}=1$) at infinity. Thus, $A(z)$ is given by
\begin{equation}
\begin{aligned}
\frac{d}{dz}\Phi(z) &= A(z) \Phi(z),\quad\quad
A(z) =\sum^{2}_{i=1}&\frac{A_{i}}{(z-z_{i})} + A_{\infty}, \quad\quad A_i, A_{\infty} \in GL(2,\mathbb{C}),
\end{aligned}
\label{eq:FS}
\end{equation}
where $A_i$ and $A_\infty$ are $2\times2$ matrices associated with the singularities at $z_i$ and $z=\infty$, respectively. Whereas the domain of the system is given by $S:\mathbb{C}\mathbb{P}^{1}/\{z_1,z_2,\infty\}$, with the fundamental matrix $Y(z)$ replaced by $\Phi(z)\in GL(2,\mathbb{C})$. Note that, for $N=2$, the fundamental matrix $\Phi(z)$ is built with two linear independent vectors solutions. For those who are unfamiliar with the system \eqref{eq:FS}, we mention the great book \cite{iwasaki1991gauss}, where more examples of systems in the form \eqref{eq:GenLinearSys} are given and the form of $\Phi(z)$ is shown. 

Regarding the scalar differential equation \eqref{eq:ODE}, we have the second-order differential equation
\begin{equation}
\frac{d^2 y}{d z^2} + p_1(z)\frac{d y}{d z}+p_2(z)y(z)=0,
\label{eq:SODE}
\end{equation}
where $p_1(z)$ and $p_2(z)$ are rational functions of $z\in\mathbb{C}$ with singularities at $z=z_i$ (regular) and $z=\infty$ (irregular). This equation is essentially the confluent Heun equation (CHE), described in the dictionary section, equation \eqref{eq:ConfHeun}.

\section{Solutions and Monodromy Representation}
\label{sec:solmon}

In order to solve the system \eqref{eq:FS}, we need to study how the solutions are given around each singularity, as well as the analytic continuation for each solution via \textit{monodromy matrix}. We will consider first the treatment of the singularities at $z=z_1$ and $z=z_2$, the point at infinity is an irregular singularity of rank 1 and requires a different description.

\subsection{Solutions Around Regular Points}

Around each of the regular singularities, one has that the fundamental solutions are given by the following asymptotic formula \cite{Jimbo:1981aa}
\begin{equation}
\Phi^{(i)}(z) = G_{(i)} \bigg(\mathbb{I}+\sum_{j=1}^{\infty}\Phi^{(i)}_{j}(z-z_{i})^j\bigg)(z-z_{i})^{A^{(i)}_{0}}, \quad \quad i = 1,2
\label{eq:2.4}
\end{equation}
where $G_{(i)}$, $\Phi_{j}^{(i)}$ and $A_0^{(i)}$ are $2\times2$ constant matrices, whereas the term $(z-z_{i})^{A^{(i)}_{0}}$ is responsible for the multivaluedness of \eqref{eq:2.4} with $A^{(i)}_{0}$ controling the branching of the solution. Furthermore, the matrices $A^{(i)}_0$ are diagonal and can be computed from $A_{1}$ and $A_{2}$ by assuming the following expansion for the matrix $A(z)$,
\begin{equation}
A(z) = G_{(i)}\sum_{j=0}^{\infty}A^{(i)}_{j}(z-z_{i})^{j-1}G^{-1}_{(i)}, \quad i =1,2
\label{eq:expA}
\end{equation}
which leads to
\begin{equation}
A^{(i)}_{0} = G^{-1}_{(i)}A_{i}G_{(i)}, \quad i = 1,2, \quad G_{(i)}\in GL(2,\mathbb{C}).
\label{eq:2.9}
\end{equation}

\begin{figure}[htb]
	\centering	
	\caption{\small{\small{Paths on the Riemann sphere, where each loop $\gamma_i$ wraps once around its corresponding singularity. The chosen basepoint at $p$ does not bring any relevance, in other words, any different point is related to the point at $p$ by a conjugate matrix acting on the system and in its monodromy matrices.}}}
	\includegraphics[width=9cm]{{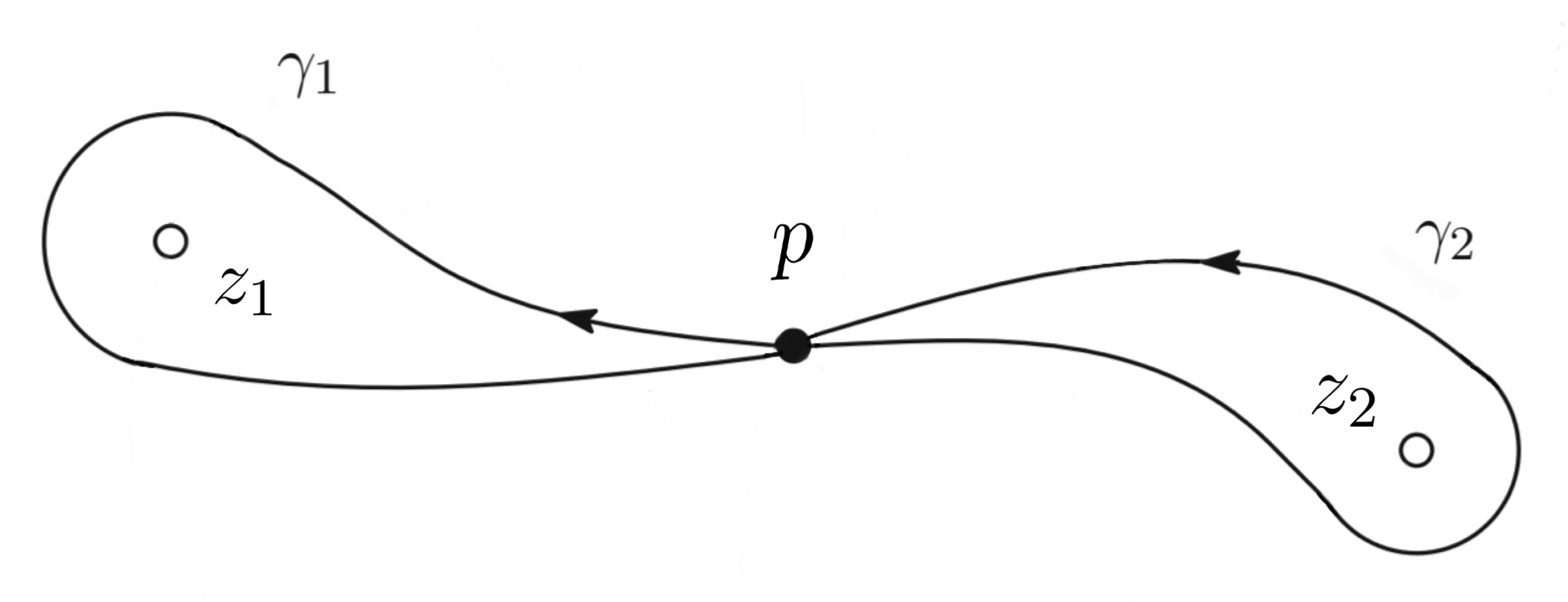}}\\
	Produced by the author, based on the Ref. \cite{haraoka2020linear}.		
	\label{MREV}
\end{figure}

Let us suppose now the analytic continuation of the fundamental matrix solution  $\Phi(z)$, which is multivalued in the branch points $z_1$, $z_2$ and $\infty$. Considering a loop $\gamma$, starting at a fixed point $p\in S$ on the Riemann sphere, one has that when a closed path circles these singular points, the solution $\Phi(z)$ changes to \cite{haraoka2020linear}
\begin{equation}
\Phi(z_{\gamma}) =\Phi(z)\mathcal{M}_{\gamma},
\label{eq:monod}
\end{equation}
where the \textit{monodromy matrix} $\mathcal{M}_{\gamma} \in GL(2,\mathbb{C})$ and only depends on the path $\gamma$. The map $\gamma\rightarrow M_{\gamma}$ defines therefore a monodromy matrix for the solutions around each singularity. In the case of the solutions given by \eqref{eq:2.4}, we have
\begin{equation}
\Phi^{(i)}(z_{\gamma_{i}}) =\Phi^{(i)}(z_{i})\mathcal{M}_{i} \ \ \ \ \ i= 1,2,
\label{eq:2.6}
\end{equation}
as illustrated in the Fig. \ref{MREV}. Furthermore, the solution \eqref{eq:2.4} can be redefined as
\begin{equation}
\tilde{\Phi}^{(i)}(z) =
\Phi^{(i)}(z)C_{i}, \quad\quad \ i =1,2,
\label{eq:2.13}
\end{equation}
where the constant matrix $C_{i}\in GL(2,\mathbb{C})$. Thus, from the expression above, we can derive an explicit formula for the monodromy matrices. This is done by applying the transformation $z \rightarrow e^{2i\pi }z$ around $z_{i}$, resulting in
\begin{equation}
\tilde{\Phi}^{(i)}(e^{2i\pi}(z-z_{i})+z_{i}) = \tilde{\Phi}^{(i)}(z)C_{i}^{-1}e^{2i \pi A^{(i)}_{0}}C_{i},
\label{eq:2.14}
\end{equation}
then, comparing with the equation \eqref{eq:2.6}, we identify the monodromy matrix $\mathcal{M}_{i}$ as
\begin{equation}
\mathcal{M}_{i} = C^{-1}_{i}e^{2i\pi A^{(i)}_{0}}C_{i}, \quad i=1,2.
\label{eq:2.15}
\end{equation}

\subsection{Solution Around The Irregular Point}

Now we want to investigate the fundamental solution around the irregular singularity of Poincaré rank 1 at infinity, whose asymptotic behavior is given by \cite{conte}
\begin{equation}
\Phi^{(\infty)}(z) = \bigg(\mathbb{I} +\sum_{j=1}^{\infty}\Phi^{(\infty)}_{j}z^{-j}\bigg)z^{-A^{(\infty)}_{0}}e^{-A^{(\infty)}_{-1}z},
\label{eq:2.17}
\end{equation}
where $A^{(\infty)}_{0}$ and $A^{(\infty)}_{-1}$ are $2\times2$-diagonal matrices. Similar to the regular singularities, the polynomial term $z^{-A^{(\infty)}_{0}}$ is related to the multivaluedness of the solution $\Phi^{(\infty)}(z)$, on the other hand, the second term, $e^{-A^{(\infty)}_{-1}z}$, is responsible for the growth and decay of the solution at infinity. In this situation, depending of the argument of the $z$ in $e^{-A^{(\infty)}_{-1}z}$, one has that the asymptotic behavior of the solutions differ in different regions of the complex plane, and these differences are described by \textit{Stokes phenomenon} \cite{10.2307/2031404}.

\vspace{0.2cm}
\textbf{Stokes Phenomenon}
\vspace{0.2cm}

First discovered by G. G. Stokes, the Stokes phenomenon is that the asymptotic behavior of functions can differ in different regions (or sectors) of the complex plane, and that these regions are bounded by what are called \textit{Stokes lines} (or \textit{anti-Stokes lines}). This phenomenon has been investigated in a large number of problems in mathematics\footnote{The Airy function, solution of $f''(z) = z f(z)$, is perhaps the main example of solution in which the Stokes phenomenon is observed. For a review of Airy function see Chapter 9 in \cite{NIST:DLMF}} and physics \cite{Berry1988,GILBERT2004247} -- for those who want to know more about this phenomenon we mention Meyer's paper \cite{10.2307/2031404} and Sibuya’s textbook (Chapter 6) \cite{sibuya2008linear}. 

In the case of the system \eqref{eq:FS}, the exponential term in \eqref{eq:2.17} is responsible for the constraint in the solutions at infinity, where depending on the argument of $z \in \mathbb{C}$ the solutions will be restricted in the complex plane by \textit{Stokes lines} $\mathcal{L}$ with the solutions may have different asymptotic expressions in distinct sectors.

\begin{figure}[ht]
	\centering
	\caption{\small{Fig. \ref{fig:sub1}: The solutions around an irregular singular point of rank $r\neq 0$ will be restricted by Stokes lines, commonly labeled by $\mathcal{L}$. The second Fig. \ref{fig:sub2} illustrates how the Stokes sectors $\mathcal{S}_k$ are represented in the literature. The label "$\infty$" represents the sector at $z=\infty$. In addition, the solutions defined in different sectors are related by Stokes matrices \cite{conte}.}}
	\begin{subfigure}{.5\textwidth}
		\centering
		\includegraphics[width=.9\linewidth]{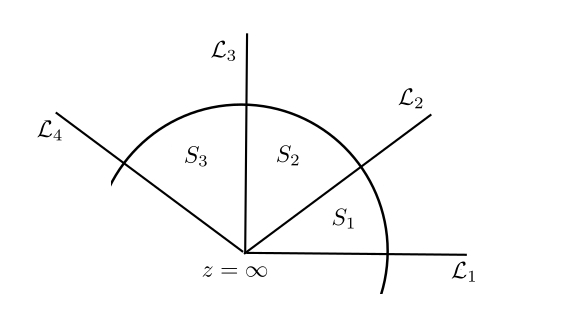}
		\caption{Stokes lines}
		\label{fig:sub1}
	\end{subfigure}%
	\begin{subfigure}{.5\textwidth}
		\centering
		\includegraphics[width=.7\linewidth]{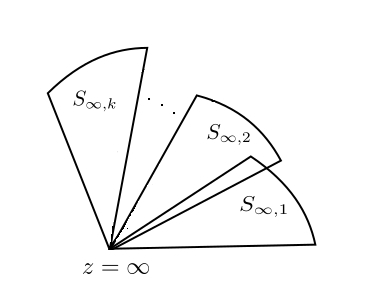}
		\caption{Stokes sectors}
		\label{fig:sub2}
	\end{subfigure}
	Adapted from \cite{conte}.
	\label{fig:test}
\end{figure}

In regard to the solution \eqref{eq:2.17}, it is common to represent the Stokes lines and sectors as shown in Fig. \ref{fig:test}, whose sectors are defined by \cite{Andreev:1995in,Lisovyy:2018mnj}
\begin{equation}
\mathcal{S}_{k} = \bigg\{z\in \mathbb{C}, -\frac{1}{2}\pi+(k-2)\pi < \text{Arg}(z) < \frac{3}{2}\pi+(k-2)\pi \bigg\}, \ \ \ k \in \mathbb{Z}
\label{eq:2.18}
\end{equation}
with $k$ labeling the sectors.

Now, let us focus on the fundamental matrix solution \eqref{eq:2.17}. Suppose that the solution $\Phi^{(\infty)}_{k}(z)$, defined in the $k$-sector, behaves like the general solution $\Phi^{(\infty)}(z)$ defined in \eqref{eq:2.17},
\begin{equation}
\Phi^{(\infty)}(z) \sim \Phi_{k}^{(\infty)}(z),\qquad \Phi_{k}^{(\infty)}(z) \in \mathcal{S}_{k}.
\label{eq:2.19}
\end{equation}

In order to compute $\Phi^{(\infty)}_{k}(z)$ and then the monodromy matrix at $z=\infty$ for the system \eqref{eq:FS}, we assume that the fundamental solutions $\Phi^{(\infty)}_{k+1}(z)$ and $\Phi^{(\infty)}_{k}(z)$ have the same asymptotic expansion in a given sector $\mathcal{S}_{k}$, those two solutions are related by \textit{Stokes matrices} as,
\begin{equation}
\Phi_{k+1}^{(\infty)}(z) =\Phi_{k}^{(\infty)}(z)S_{k},
\label{eq:2.20}
\end{equation}
where for convenience, we consider a basis for the solution $\Phi^{(\infty)}_{k}(z)$ whose Stokes matrices are given by
\begin{equation}
S_{2k} = \begin{pmatrix}
1 & s_{2k}\\
0& 1
\end{pmatrix}, \qquad
S_{2k+1} = \begin{pmatrix}
1 & 0\\
s_{2k+1} & 1
\end{pmatrix},
\label{eq:2.21}
\end{equation}
and the complex parameters $s_{2k}$ and $s_{2k+1}$ are \textit{Stokes multipliers} \cite{Andreev:1995in,oai:repository.dl.itc.u-tokyo.ac.jp:00039422}. A striking consequence of \eqref{eq:2.20}, it is that the equation \eqref{eq:2.19} can be written as
\begin{equation}
\Phi^{(\infty)}(z) \sim \Phi_{0}^{(\infty)}(z)S_{k}S_{k-1}...S_{1} \ \ \ \text{in} \ \ \mathcal{S}_{k},
\label{eq:2.22}
\end{equation} 
where $\Phi_{0}^{(\infty)}(z)$ is the solution at the $k=0$ sector. Furthermore, we have that the map $\gamma \rightarrow \mathcal{M}_{\gamma}$ send solutions from a sector $\mathcal{S}_{k}$ to $\mathcal{S}_{k+2}$ (or in general $\mathcal{S}_{k+n}$) and vice versa, \cite{conte}. Thus, for the solution at $z=\infty$, we have
\begin{equation}
\Phi_{k+2}^{(\infty)}(e^{2i\pi}z) = \Phi_{k}^{(\infty)}(z)e^{2i\pi A^{(\infty)}_0},
\label{eq:2.23}
\end{equation}
then, from \eqref{eq:2.20}, we rewrite the analytic continuation in the $k$-sector as
\begin{equation}
\Phi_{k}^{(\infty)}(e^{-2i\pi}z) = \Phi_{k}^{(\infty)}(z)S_{k}S_{k+1}e^{-2i\pi A^{(\infty)}_0}.
\label{eq:2.24}
\end{equation}
Comparing with \eqref{eq:monod}, we identify the monodromy matrix for $\Phi_{k}^{(\infty)}$ in the k-sector
\begin{equation}
\mathcal{M}_{k}^{(\infty)} = S_{k}S_{k+1}e^{-2i\pi A^{(\infty)}_0}
\label{eq:2.25}
\end{equation}
where, by recurrence, the monodromy associated to the next sector is expressed in terms of Stokes matrices as
\begin{equation}
\mathcal{M}_{k+1}^{(\infty)} = S^{-1}_{k}\mathcal{M}^{(\infty)}_{k}S_{k}.
\label{eq:2.26}
\end{equation}
Finally, replacing \eqref{eq:2.25} in \eqref{eq:2.26}, we arrive at the following expression for the monodromy matrix in terms of the Stokes's matrices and the matrix $A^{(\infty)}_0$,
\begin{equation}
\mathcal{M}_{k+1}^{(\infty)} = S_{k+1}e^{-2i\pi A^{(\infty)}_0}S_{k}.
\label{eq:eqeq}
\end{equation}
Without loss of generality, it is common to choose the first sector $k=1$ to represent the matrix at infinity $\Phi^{(\infty)}(z)\sim \Phi^{(\infty)}_{k=1}(z)$ in \eqref{eq:2.19} -- for a discussion about the derivation above, which only holds for irregular singulariry of rank 1, we recommend the paper written by Andreev and Kitaev \cite{Andreev:1995in}. We have therefore from \eqref{eq:eqeq}:
\begin{equation}
\mathcal{M}_{\infty} = \mathcal{M}^{(\infty)}_{2} = S_2e^{-2i\pi A^{(\infty)}_0}S_{1}.
\label{eq:monoinfty}
\end{equation}

\vspace{0.2cm}
\textbf{Monodromy representation}
\vspace{0.2cm}

Considering the monodromy matrices $\mathcal{M}_{1},\mathcal{M}_{2}$, and $\mathcal{M}_{\infty}$ for each singularity of the system \eqref{eq:FS} and taking a base point $p\in S$, with $S:\mathbb{C}\mathbb{P}^{1}/\{z_1,z_2,\infty\}$, one has that, correspondence between the paths and the monodromy matrices, $\gamma \rightarrow M_{\gamma}$ generates the linear representation
\begin{equation}
\chi : \pi (S,p) \ \rightarrow \ GL(2,\mathbb{C}),
\label{eq:map}
\end{equation}
which is also called by \textit{monodromy representation}. The fundamental group $\pi(S,p)$, which is the domain of the monodromy representation, is expressed by
\begin{equation}
\pi(S,p) = (\gamma_1, \gamma_2, \gamma_{\infty}|\gamma_1 \gamma_2 \gamma_{\infty}=1)
\label{eq:gammacond}
\end{equation}
where, $\gamma_{1,2,\infty}$ are the loops around the singularities. In turn, the three monodromy matrices generate the \textit{monodromy group} of the system with all matrices satisfying the following relation, 
\begin{equation}
\mathcal{M}_1\mathcal{M}_2\mathcal{M}_{\infty}=\mathbb{I}.
\label{eq:Mcond}
\end{equation}

We remark that, it is always possible to choose $\gamma$'s in such a way that the condition $\gamma_1\gamma_2\gamma_\infty=1$ is valid, for a given basis point $b$ on the Riemann sphere. In this case, the monodromy matrices of the system will satisfy the condition above -- see Conte's book for more examples of monodromy representation of linear matrix systems \cite{conte}.

\section{Isomonodromic Deformations Theory}
\label{sec:isodefthe}

Riemann was the first to introduce the idea of monodromy preserving deformation of linear differential equations in the form \eqref{eq:scalarSys} (or \eqref{eq:ODE}) with $A(z)$ expressed by \eqref{eq:fuschianA} \cite{Riemann1857}. He developed the concept idea of constructing a system of functions with Fuchsian singularities (or regular singularities) that has the prescribed monodromy representation, and studied them as function of the singularities when the monodromy matrices of the system is kept invariant. The investigation of deformations in non-Fuchsian systems was made by M. Jimbo, T. Miwa, and K. Ueno in the 1980s, where they extended the classical work developed by Riemann and established a general theory of monodromy preserving deformation for linear systems with singularities of arbitrary Poincaré rank (i.e. regular singularity ($r=0$), irregular singularity of rank 1 ($r=1$), etc.). More precisely, they considered a general $N\times N$ system with $A(z)$ given in the form \eqref{eq:nonfuchsianA}, with the differential equation expressed by \eqref{eq:ODE}. We strongly recommend the references \cite{Jimbo:1981aa, Jimbo:1981ab,Jimbo:1981ac} for a review of the isomonodromic deformations theory.

For linear matrix systems, with $N=2$, it was R. Fuchs \cite{Fuchs1907} who gave first an example of isomonodromic deformation in this type of linear system, where studying a second-order differential equation with four regular singularities, he revealed that the sixth Painlevé equation \footnote{Painlevé equations form six families of non-linear second-order differential equations whose singularities have the Painlevé property: the only movable singularities are poles. Every Painlevé equations has the form $y''=R(y',y,t)$, where $R$ is a rational function -- for a review of the Painlevé equations we mention \cite{iwasaki1991gauss,CLARKSON2003127}} is nothing other than the deformation equation for the linear system with four singularities. Garnier \cite{Garnier} extended the Fuchs resulting by deriving all the six Painlevé equations. In both cases, they supposed the following form for the second-order differential equation \eqref{eq:SODE}, $f''(z) = h(z)f(z)$. In turn, Okamoto \cite{10.3792/pjaa.56.264,okamoto1981isomonodromic} used the fact that the six Painlevé equations can be converted into Hamiltonian systems to show that isomonodromic deformations of equations in the form \eqref{eq:SODE} are governed by Hamiltonian systems.

The aim of isomonodromic deformations theory is to describe a family of linear matrix systems in the form \eqref{eq:FS} (or in general \eqref{eq:GenLinearSys}) that share the same monodromy representation. In the case of the non-Fuchsian system \eqref{eq:FS}, we already know that the monodromy representation is composed by the map \eqref{eq:map}, where the matrices $\mathcal{M}_{1}$, $\mathcal{M}_{2}$, and $\mathcal{M}_{\infty}$ generate the monodromy group, with the monodromy matrix at infinity depending on Stokes' matrices. 

In order to find a family of linear matrix systems in the form \eqref{eq:FS}, we have that the basic idea consists in considering the matrices $\Phi(z)$, $A(z)$ and $A_i$ in \eqref{eq:FS} (or in general \eqref{eq:GenLinearSys}), as functions of the singularities \cite{haraoka2020linear,Mahoux1999}. Thus, for the system \eqref{eq:FS}, we will consider a dependence with respect to one of the singularities. Choosing $z_2 =t$ and, for simplicity $z_1=0$, one has that the system is rewritten as
\begin{equation}
\begin{aligned}
\frac{\partial}{\partial z}\Phi(z,t) = A(z,t) \Phi(z,t),\quad\quad
A(z,t) =\frac{A_{0}(t)}{z}+ \frac{A_{t}(t)}{z-t}+A_{\infty},
\end{aligned}
\label{eq:GenSys1}
\end{equation}
where $A_1(t)$ and $A_2(t)$ were replaced by $A_0(t)$ and $A_t(t)$, respectively. In this case, the monodromy representation \eqref{eq:map}, associated with the non-Fuchsian system, becomes $t$-dependent. However, such a representation will be invariant for any change in $t$ if $\Phi(z,t)$ satisfies the following system\footnote{It is also called by \textit{Lax pair}.},
\begin{equation}
\begin{cases}
\frac{\partial}{\partial z}\Phi(z,t) = A(z,t) \Phi(z,t)\\
\frac{\partial}{\partial t} \Phi(z,t) = B(z,t)\Phi(z,t),
\end{cases}
\end{equation}
where from the compatibility condition $\partial_{z}\partial_{t} \Phi(z,t) =\partial_{t}\partial_{z} \Phi(z,t)$ or the zero-curvature
\begin{equation}
\partial_t A(z,t) - \partial_z B(z,t) +[B(z,t),A(z,t)] =0
\label{eq:flatCond}
\end{equation}
one has that the $2\times2$-matrix function $B(z,t)$ is rational in $z$:
\begin{equation}
B(z,t)=-\frac{A_t(t)}{z-t}.
\label{eq:Bfunction}
\end{equation} 

Essentially, $B(z,t)$ given by the formula \eqref{eq:Bfunction} eliminates the double pole in the compatibility condition, the derivation of $B(z,t)$ for the general case can be verified in \cite{haraoka2020linear, Jimbo:1981aa}. We have, therefore, that with \eqref{eq:flatCond} and \eqref{eq:Bfunction} the non-Fuschian system  \eqref{eq:GenSys1} will be isomonodromic and the monodromy matrices of the fundamental solutions $\Phi(z,t)$ will remain constant for any variation of the deformation parameter $t$. Additionally, from the zero-curvature condition (or flat connection condition), it is possible to derive the following set of equations for $A_0$, $A_t$ and $A_\infty$:  
\begin{equation}
\frac{d A_{0}}{d t} = \frac{[A_{t},A_{0}]}{t},\ \ \ \ \ \ \ \frac{d A_{t}}{d t} =  \frac{[A_{1},A_{t}]}{t} +[A_{\infty},A_t].
\label{eq:2.29}
\end{equation}

These are the \textit{Schlesinger equations} of the system \eqref{eq:GenSys1}\footnote{It can be shown, from the zero-curvature condition, that $A_{\infty}$ is independent of $t$. So, because of that, we have assumed directly that $A_{\infty}$ does not depend on $t$, in \eqref{eq:GenSys1}.}. It means that deformations which preserve the monodromy of the system are governed by an integrable system of partial differential equations (PDEs). For a general system with a large number of Fuchsian (or non-Fuchsian) singularities the Schlesinger equations can be derived directly from the flat connection condition \eqref{eq:flatCond}. For a review of these equations and their relation with isomonodromic deformations theory, we mention the papers written by Jimbo, Miwa and Ueno \cite{Jimbo:1981aa,Jimbo:1981ab,Jimbo:1981ac}.
\vspace{5cm}
\\
\textbf{Isomonodromic $\tau$-function}

It is well-know from \cite{Jimbo:1981aa} that for any solution of the Schlesinger equations \eqref{eq:2.29}, one can associate with a  1-form $\omega$, which is closed when \eqref{eq:2.29} holds. Such a differential form, for the system \eqref{eq:GenSys1}, is defined as
\begin{equation}
\omega=d\ \text{log}\tau= \text{Tr}(A_{0}A_{t})\frac{dt}{t}+\text{Tr}(A_{\infty}A_{t})dt,
\label{eq:2.31}
\end{equation}
where the isomonodromic $\tau$-function is holomorphic everywhere except at the branch points of the system, i.e. $t=0$ and for $t$ in the irregular singular point, $t=\infty$. Since the 1-form is closed and satisfies locally
\begin{equation}
d (d\ \text{log}\tau) =0,
\label{eq:2.32}
\end{equation}
one has that $d\log \tau$ is a constant quantity associated with the isomonodromic flow of the linear matrix system \eqref{eq:GenSys1}. This quantity is essentially the Hamiltonian $H$ of the system, which now assumes a dynamic behavior with respect to the deformation parameter $t$ \cite{Jimbo:1981aa}
\begin{equation}
H =\frac{d}{dt} \text{log}(\tau(t))= \frac{\text{Tr}(A_{0}A_{t})}{t}+\text{Tr}(A_{\infty}A_{t}).
\label{eq:2.33}
\end{equation}

Finally, using the Schlesinger equations \eqref{eq:2.29}, we can write a second condition for the $\tau(t)$-function in term of matrices of the system as
\begin{equation}
\frac{d}{dt}t\frac{d}{dt} \text{log}(\tau(t))-\text{Tr}(A_{\infty}A_{t})=0.
\label{eq:secCondTau}
\end{equation}
With all these definitions, we can start the study of isomonodromic deformations in the differential equation \eqref{eq:SODE}.
For those who is unfamiliar with the isomonodromic deformations theory, we strongly recommend the seminal papers \cite{Jimbo:1981aa,Jimbo:1981ab,Jimbo:1981ac}, which treat with general linear systems with $A(z)$ in the form \eqref{eq:nonfuchsianA}. They also generalize the expression for the isomonodromic $\tau$-function.

\subsection{Isomonodromic Deformations in the CHE}
\label{subsec:IDCHE}

This section is devoted to studying isomonodromic deformation in the scalar differential equation \eqref{eq:SODE}. Thus, we have
\begin{equation}
\frac{d^2y}{dz^2}+p_1(z,t)\frac{dy}{dz}+p_2(z,t)y(z) = 0
\label{eq:SODEdef}
\end{equation}
with the system given by
\begin{equation}
\frac{\partial}{\partial z}\Phi(z,t) = \bigg(\frac{A_0}{z}+\frac{A_t}{z-t}+A_{\infty}\bigg)\Phi(z,t)
\label{eq:CHSys}
\end{equation}
where $A_0$ and $A_t$ are functions of $t$.

As explained previously, the equation \eqref{eq:SODEdef} admits isomonodromic deformations with respect to the parameter $t$, if the monodromy representation of the system does not depend on $t$. If this happens the matrices of the system will satisfy the Schlesinger equations 
\begin{equation}
\begin{aligned}
\frac{d}{d t}A_{0} =-\frac{1}{t}[A_{0},A_{t}],\ \ \ \ \ \
\frac{d}{d t}A_{t} =\frac{1}{t}[A_{0},A_{t}]+[A_{\infty},A_{t}].
\end{aligned}
\label{eq:2.36}
\end{equation}

Given that, let us use a general fundamental matrix $\Phi(z,t)$ in the linear system \eqref{eq:CHSys} to derive a second-order differential equation for $y_i(z)$ in the first line of \eqref{eq:2.37}, $i=1,2$. This procedure will lead us to an explicit form for $p_1(z,t)$ and $p_2(z,t)$ in \eqref{eq:SODEdef}. Thus, we have
\begin{equation}
\Phi(z,t) = \begin{pmatrix}
y_{1}(z) & y_{2}(z)\\
u_{1}(z) & u_{2}(z)
\end{pmatrix},
\label{eq:2.37}
\end{equation}
with the second line obeying an analogue differential equation, where $u_{i}(z)$ is related to $y_{i}(z)$ by
\begin{equation}
u_{i}(z) =
\frac{1}{A_{12}(z,t)}\left(\frac{d}{dz} y_{i}(z)-A_{11}(z,t)y_{i}(z)\right).
\label{eq:2.38}
\end{equation}
Replacing the matrix \eqref{eq:2.37} in \eqref{eq:CHSys}, we arrive at the following differential equation for the first line
\begin{equation}
\begin{aligned}
\frac{d^2 y_{i}}{dz^2}-\bigg[\text{Tr} A + \frac{\partial_{z}A_{12}}{A_{12}}\bigg]\frac{d y_{i}}{dz}+\bigg[\text{det} A -\partial_z A_{11}+A_{11}\frac{\partial_{z}A_{12}}{A_{12}}\bigg]y_{i}(z) =0,
\end{aligned}
\label{eq:2.39}
\end{equation}
where now the holomorphic functions $p_1(z,t)$ and $p_2(z,t)$ are expressed in terms of components of the  matrix $A(z)$, with $\partial_z = \frac{\partial }{\partial z}$. Note that, we have dropped the $"(z,t)"$ to simplify the notation, $A_{12} = A_{12}(z,t)$. It is convenient to assume a basis for the system where the matrix at infinity $A_{\infty}$ is diagonal, this is made by acting on the matrix system with inversible matrices that make $A_{\infty}$ diagonal. In this way, the component $A_{12}$ of $A(z,t)$ in \eqref{eq:CHSys} vanishes in the limit $z \rightarrow \infty$\cite{conte}:
\begin{equation}
A_{12} = \frac{k(z-\lambda)}{z(z-t)}, \qquad k =a^{(0)}_{12}+a^{(t)}_{12}, \qquad \lambda =\frac{t a^{(0)}_{12}}{a^{(0)}_{12}+a^{(t)}_{12}},
\label{eq:2.40}
\end{equation}
where we are assuming 
\begin{equation}
A_0 = \begin{pmatrix}
a^{(0)}_{11} & a^{(0)}_{12}\\
a^{(0)}_{21} & a^{(0)}_{22}
\end{pmatrix}, \qquad A_t = \begin{pmatrix}
a^{(t)}_{11} & a^{(t)}_{12}\\
a^{(t)}_{21} & a^{(t)}_{22}
\end{pmatrix}, \qquad A_{\infty} = \begin{pmatrix}
a^{(\infty)}_{11} & 0\\
0 & a^{(\infty)}_{22}
\end{pmatrix}.
\end{equation}

Then, using the determinant property $\det M = \frac{1}{2}((\text{Tr}M)^2-\text{Tr} M^2)$, we obtain the following expression for the differential equation \eqref{eq:2.39},
\begin{equation}
\begin{aligned}
\frac{d^2y_{i}}{dz^2}-\bigg[\text{Tr}A_{\infty}&+\frac{\text{Tr}A_{0}-1}{z}+ \frac{\text{Tr}A_{t}-1}{z-t}-\frac{1}{z-\lambda}\bigg]\frac{dy_{i}}{dz}+\\+&\bigg[ \text{det}A_{\infty}+\frac{\text{det}A_0}{z^{2}}+\frac{\text{det}A_t}{(z-t)^{2}}+\frac{c_{0}}{z}+\frac{c_t}{z-t}+\frac{\mu}{z-\lambda}
\bigg]y_{i}(z)=0,
\end{aligned}
\label{eq:2.41}
\end{equation}
with
\begin{equation}
\begin{aligned}
c_{0} = \text{Tr}A_{\infty}\text{Tr}A_{0}-\text{Tr}(A_{\infty}A_0)&
+\frac{1}{t}\text{Tr}(A_{0}A_{t})-\frac{1}{t}\text{Tr}A_{0}\text{Tr}A_{t}\\ - a^{(\infty)}_{11}-\frac{1}{\lambda}a^{(0)}_{11}+&\frac{1}{t}(a^{(0)}_{11}+a^{(t)}_{11}),\\
c_{t} = \text{Tr}A_{\infty}\text{Tr}A_{t}+\frac{1}{t}\text{Tr}A_{0}\text{Tr}A_{t}-H& - a^{(\infty)}_{11}-\frac{1}{\lambda-t}a^{(t)}_{11}-\frac{1}{t}(a^{(0)}_{11}+a^{(t)}_{11}),\\
\mu = a^{(\infty)}_{11}+\frac{1}{\lambda}&a^{(0)}_{11}+\frac{1}{\lambda-t}a^{(t)}_{11},\\
c_0+c_t+\mu = \text{Tr}A_{\infty}(\text{Tr}A_{0}+\text{Tr}&A_{t})-\text{Tr}(A_{\infty}(A_{0}+A_{t}))-a^{(\infty)}_{11},
\end{aligned}
\label{eq:2.42}
\end{equation}
where from \eqref{eq:2.33}, we identify the Hamiltonian $H$,
\begin{equation}
\begin{aligned}
H = \text{Tr}(A_{\infty}A_t) & +\frac{1}{t}\text{Tr}(A_{0}A_{t}).
\end{aligned}
\label{eq:hamiltonian}
\end{equation}

Around the singularity $z=t$, we have that the residue of $p_2(z,t)$ is proportional to $H$, whereas $\mu$ is the residue around the new singularity at $z=\lambda$. This extra singularity in the differential equation does not correspond to poles in the system \eqref{eq:CHSys}. A direct calculation shows that, this new singularity has indicial exponents $0$ and $2$, with no logarithmic tails, and hence corresponds to an \textit{apparent singularity}. Consequently, the analytic continuation around this point has a trivial monodromy, $\mathcal{M}_{z=\lambda}=\mathbb{I}$. Seeing that $\lambda$ is an apparent singularity, we can use the absence of logarithmic behavior around this point to derive an expression which relates $c_t$, $\lambda$, $\mu$ and $t$. Thus, considering the residue of $p_1(z,t)$ and $p_2(z,t)$ around the apparent singularity and the Frobenius expansion of $y_{i}(z)$  for $z=\lambda$, we obtain the following indicial equation\footnote{Substituting \eqref{eq:2.44} and \eqref{eq:frobexp} in the differential equation \eqref{eq:2.41}, we obtain the following expressions: $\alpha(\alpha-2)=0$, $(\alpha^2-1)a_1+(\alpha p_1+\mu)a_0=0$, and $\alpha(\alpha+2)a_2+(\mu+(\alpha+1)p_1)a_1+(\alpha p_2+q_1)a_0=0$. The first equation gives the indicial exponents of $z=\lambda$ and assuming $\alpha=0$, one has the indicial equation $(\mu+p_1)\mu+q_1=0$.}  $(\mu+p_1)\mu+q_1=0$, where
\begin{equation}
p_1(z,t) =-\frac{1}{z-\lambda}+p_1 + p_2(z-\lambda)+..., \ \ p_2(z,t)=\frac{\mu}{z-\lambda}+q_1 + q_2(z-\lambda)+...,
\label{eq:2.44}
\end{equation}
\begin{equation}
\begin{aligned}
p_1 = -\text{Tr}A_{\infty}+\frac{1-\text{Tr}A_0}{\lambda}+\frac{1-\text{Tr}A_t}{\lambda-t}&, \qquad
\mu = \frac{a^{(0)}_{11}}{\lambda}+\frac{a^{(t)}_{11}}{\lambda-t},\\
q_1 = \text{det}A_{\infty} +\frac{\text{det}A_0}{\lambda^{2}}+\frac{\text{det}A_t}{(\lambda-t)^{2}}& + \frac{c_0}{\lambda}+\frac{c_t}{\lambda-t},
\end{aligned}
\label{eq:2.45}
\end{equation}
with
\begin{equation}
y_{i}(z)= \sum_{n=0}^{\infty}a_{n}(z-\lambda)^{\alpha+n}.
\label{eq:frobexp}
\end{equation}
Then, substituting in the indicial equation the expressions for $p_1$, $q_1$ and $\mu$, we rewrite the accessory parameter $c_t$ as a polynomial and rational function of $\lambda$, $\mu$ and $t$:
\begin{equation}
\begin{aligned}
c_t(\lambda,\mu;t) = -\frac{\lambda(\lambda-t)}{t}\bigg[\mu^{2}-\bigg(\text{Tr}A_{\infty}+\frac{\text{Tr}A_0}{\lambda}+\frac{\text{Tr}A_t-1}{\lambda-t} \bigg)\mu +\text{det}A_{\infty}&\\+\frac{\text{det}A_0}{\lambda^{2}}+ \frac{\text{det}A_{t}}{(\lambda-t)^{2}}+\frac{ \text{Tr}A_{\infty}(\text{Tr}A_{0}+\text{Tr}A_{t})-\text{Tr}(A_{\infty}(A_{0}+A_{t}))-a^{(\infty)}_{11}}{\lambda}\bigg].
\end{aligned}
\label{eq:2.47}
\end{equation}

Up to here, we did not introduce any parametrization for the matrices $A_0$, $A_t$ and $A_{\infty}$, we have only assumed a basis where  $A_{\infty}$ is diagonal. Let us take the following convenient conditions for the trace and determinant of the matrices
\begin{equation}
\begin{aligned}
\text{Tr}A_{\infty}=0, \ \ \text{Tr}A_{t}=\theta_{t}, \ \ \text{Tr}A_{0}=\theta_{0}&, \ \ \text{Tr}(A_{\infty}(A_0+A_t)) = -\frac{\theta_{\star}}{2},\\
\text{det}A_{0} =\text{det}A_t =0, \ \ \text{det}A_{\infty}=&-\frac{1}{4}, \ \ A_{\infty}=\frac{1}{2}\sigma_3,
\end{aligned}
\label{eq:2.48}
\end{equation}
where, without loss of generality, we assume a basis where $A_{\infty}=\frac{1}{2}\sigma_3$, $\sigma_3 = \diag (1,-1)$. Note that, with this parametrization the double poles in \eqref{eq:2.41}, for $z=0$ and $z=t$, are eliminated, allowing us to recover the form of the confluent Heun equation written in the Chapter \ref{chap1}. In this parametrization the components $a^{(0)}_{11}$ and $a^{(t)}_{11}$ can be written in terms of $\lambda$, $\mu$ and $t$ as
\begin{equation}
\begin{aligned}
a^{(0)}_{11} = -\frac{\lambda(\lambda-t)}{t}\bigg[\mu-\frac{1}{2}- \frac{\theta_{0}+\theta_{t}-\theta_{\star}}{2(\lambda-t)}  \bigg],\\
a^{(t)}_{11} = \frac{\lambda(\lambda-t)}{t}\bigg[\mu-\frac{1}{2}- \frac{\theta_{0}+\theta_{t}-\theta_{\star}}{2\lambda}  \bigg].
\end{aligned}
\label{eq:2.49}
\end{equation}
Finally, replacing the conditions \eqref{eq:2.48} in \eqref{eq:2.47} and using the relation between the accessory parameter $c_t$ and $H$, written in \eqref{eq:2.42}, we arrive at the following expression for the Hamiltonian $H$:
\begin{equation}
H(\lambda,\mu;t)= -\frac{\lambda(\lambda-t)}{t}\bigg[ \mu^{2}-\bigg(\frac{\theta_0}{\lambda}+\frac{\theta_t}{\lambda-t} \bigg)\mu-\frac{1}{4} +\frac{\theta_{\star}}{2\lambda}+\frac{\theta_{0}\theta_t}{\lambda(\lambda-t)} \bigg].
\label{eq:2.50}
\end{equation}

The pair $\{\lambda, \mu\}$ that appears in \eqref{eq:2.41} are interpretated as canonical variables of the matrix system \eqref{eq:CHSys} and are related to the Hamiltonian $H$ by the Poisson bracket\footnote{For a general function f and g, $\{f,g\} = \frac{\partial f}{\partial \lambda}\frac{\partial g}{\partial \mu}-\frac{\partial f}{\partial \mu}\frac{\partial f}{\partial \lambda}$.},
\begin{equation}
\frac{\partial \lambda}{\partial t} = \{\lambda,H\}, \ \ \ \frac{\partial \mu}{\partial t} = \{\mu,H\},
\label{eq:2.51}
\end{equation}
with $H(\lambda,\mu;t)$ describing the evolution of $\lambda$ as a function of the deformation parameter $t$. Furthermore, one has from \eqref{eq:2.51} that such evolution is described by a non-linear second-order differential equation expressed by
\begin{equation}
\begin{aligned}
\frac{d^{2}\lambda}{dt^{2}} = \bigg(\frac{1}{2\lambda} +\frac{1}{\lambda-t}\bigg) \bigg(\frac{d\lambda}{dt}\bigg)^{2} -\frac{1}{t}\frac{d\lambda}{dt}+\frac{(\lambda-1)^{2}}{t^2}\bigg(\frac{\theta^{2}_0 }{2}t-\frac{\theta^{2}_{t}}{2t}\bigg)-(\theta_{\star}+1)\frac{\lambda}{t}-\frac{\lambda(\lambda+1)}{2(\lambda-1)},
\end{aligned}	 
\label{eq:2.52}
\end{equation}
where $\lambda(t)$ is a meromorphic function except in the branch points of the Hamiltonian $H(\lambda,\mu;t)$ at $t=0$ and $ t=\infty$. This equation is known as Painlevé V equation and it is equivalent to the Hamiltonian \eqref{eq:2.50}. More discussions about Hamiltonian systems associated with Painlevé equations can be verified in \cite{cosgrove,CLARKSON2003127,10.3792/pjaa.56.264}.

As presented in \eqref{eq:hamiltonian}, $H(\lambda,\mu;t)$ corresponds to the isomonodromic flow of the matrix system, thus, one has from \eqref{eq:2.33} the following expression for the logarithmic derivative of the $\tau$-function: 
\begin{equation}
\frac{d}{dt}{\text{log}(\tau_{V} (t))} = -\frac{\lambda(\lambda-t)}{t}\bigg[ \mu^{2}-\bigg(\frac{\theta_0}{\lambda}+\frac{\theta_t}{\lambda-t} \bigg)\mu+\frac{\theta_{\star}}{2\lambda}-\frac{1}{4}\bigg]-\frac{\theta_{0}\theta_t}{t},
\label{eq:firstcond}
\end{equation}
the index $V$ labels the $\tau$-function associated with the non-linear second order Painlevé V equation \eqref{eq:2.52}. Based on \eqref{eq:secCondTau}, a second condition for the isomonodromic $\tau_{V}$-function can be written by taking the derivative of the equation above and making use of the Schlesinger equations \eqref{eq:2.36},
\begin{equation}
\frac{d}{d t}t\frac{d}{dt}\log (\tau_V(t)) - \frac{1}{2}\tr\sigma_{3}A_t=0,
\label{eq:seccond}
\end{equation} 
where, in the parametrization \eqref{eq:2.48}, $\tr\sigma_{3}A_t=-\theta_t$. Regarding the second-order differential equation \eqref{eq:2.41}, the parametrization of $A_0$, $A_t$ and $A_\infty$ leads us to the expression
\begin{equation}
\begin{aligned}
\frac{d^2y_{i}}{dz^2}+\bigg(\frac{1-\theta_0}{z}+
\frac{1-\theta_t}{z-t}-\frac{1}{z-\lambda}\bigg)\frac{dy_{i}}{dz}-\bigg(\frac{1}{4}+\frac{\theta_{\star}-1}{2z}+
\frac{tc_t}{z(z-t)}+\frac{\lambda\mu}{z(z-\lambda)}\bigg)y_{i}(z)=0.
\end{aligned}
\label{eq:2.53}
\end{equation}
The equation has four singularities, and given that $\lambda$ is a removable singularity, i.e. apparent singularity, one has that such equation represents the deformed form of the confluent Heun equation, defined in the previous chapter, equation \eqref{eq:ConfHeun}. Again, the apparent singularity $\lambda=\lambda(t)$ is a function of the deformation parameter $t$ and its evolution is described by the Hamiltonian \eqref{eq:2.50}. In what follows, we will denote the three $\theta$'s in the equation as $\{\theta\} = \{\theta_{ 0},\theta_{t},\theta_{\star}\}$.

After a large number of calculations, we have derived from the isomonodromic deformations theory two conditions for the $\tau_{V}$-function and a deformed equation associated with the CHE, defined in \eqref{eq:SODEdef} (or in \eqref{eq:ConfHeun}). The natural procedure consists in recovering the CHE from the equation \eqref{eq:2.53} by imposing conditions in the set of parameters $\{\theta\}$, in the canonical variables $\{\lambda(t), \mu(t)\}$, and consequently in the $\tau_{V}$-function. However, before applying any condition, we need to understand how the $\tau_V$-function is expressed and if any changes in $\{\theta\}$ and $\{\lambda(t),\mu(t)\}$ implies in constraints in such a function.

\subsection{\texorpdfstring{Isomonodromic $\tau_V$-function: an overview}%
{}}

The isomonodromic $\tau_{V}$-function associated with the Hamiltonian $H_{V}$ is a transcendental \footnote{In general, the term transcendental means nonalgebraic, i.e. a function not expressible as a finite combination of the algebraic operations of multiplication, division, addition and subtraction. Such functions are expressible in algebraic terms only as infinite series.}  function whose asymptotic expansion was first obtained by Jimbo in the seminal paper \cite{Jimbo:1982aa}, where using connection and monodromy matrices, he was able to derive the asymptotic form of the function near the branch point $t=0$. Based on Jimbo's result, the asymptotic expansions around $t=0$ and $t=\infty$ were explored in different contexts, such as connection formulae, correlation functions in the Ising model and Bose gas problem \cite{MCCOY198642, MCCOY1986190, JIMBO198080,MCCOY1986187}. In turn, Gamayun, Iorgov, and Lisovyy discovered that such a function can be expressed as a sum of $c=1$ irregular conformal blocks, multiplied by some structure constants written in terms of Barnes functions and an instanton counting function. By using this set of elements they derived the expansion of the $\tau_{V}$-function around the branch points at $t=0$ and $t=\infty$ -- see \cite{Gamayun:2013auu,Nagoya:2015cja,nag2} for more details about the relationship between conformal blocks defined in conformal field theory (CFT) and the isomonodromic $\tau$-functions. 

More recently and based on the formalism introduced in \cite{Gavrylenko:2016zlf,its:hal-01797601}, Lisovyy, Nagoya, and Roussillon derived the Fredholm determinant representation of the transcendental $\tau_{V}$-function \cite{Lisovyy:2018mnj}, leading to a new manner of computing the $\tau_{V}$-function. They arrived at this representation by considering the non-Fuchsian system $\eqref{eq:CHSys}$ with two regular singularities and one irregular singularity of rank 1 on the Riemann sphere, and generic monodromy matrices of the system. Then, via the decomposition of the three-punctured sphere into pair of pants, they were able to obtain an efficient form of computing the expansion for the $\tau_{V}$-function in $t=0$. With respect to the expansion at the irregular point, $t=\infty$, the Fredholm determinant representation for the $\tau_{V}$-function depends on Stokes matrices, and, to the best of our knowledge, there is no representation for such a function around this branch point. The only form of computing the expansion is via $c=1$ irregular conformal blocks, as discussed above. 

We strongly recommend the reference \cite{Lisovyy:2018mnj}, where the Fredholm determinant for the $\tau_{V}$-function was first presented. Based on this reference, we give in Appendix \ref{sec:tools} the main expressions necessary to compute this function. In addition, in order to provide the reader a better knowledge of the form of the expansion around $t=0$, and consequently show the accordance with the expression derived by Jimbo in \cite{Jimbo:1982aa}, we write the first terms of the expansion, whose asymptotic formula is given by
\begin{multline}
	\tau_V(\{\theta\};\sigma,\eta;t)=C_{V}(\{\theta\};\sigma)
	t^{\frac{1}{4}(\sigma^2-\theta_0^2-\theta_t^2)}
	e^{\frac{1}{2}\theta_tt}\bigg[
	1-\left(\frac{\theta_t}{2}-\frac{\theta_\star}{4}
	+\frac{\theta_\star(\theta_0^2-\theta_t^2)}{4\sigma^2}\right)t
	\\ -
	\frac{(\theta_\star+\sigma)((\sigma+\theta_t)^2-
		\theta_0^2)}{8\sigma^2(\sigma-1)^2}\kappa_V^{-1}
	t^{1-\sigma}-
	\frac{(\theta_\star-\sigma)((\sigma-\theta_t)^2-
		\theta_0^2)}{8\sigma^2(\sigma+1)^2}
	\kappa_V\, t^{1+\sigma}+{\mathcal O}(t^2,|t|^{2\pm
		2\Re\sigma})\bigg]
	\label{eq:exptauV}
\end{multline}
with $-1< \Re \sigma<1$ and $\kappa_V$ expressed in terms of ratio of Gamma functions as
\begin{equation}
\kappa_V =
e^{i\pi\eta}\frac{\Gamma(1-\sigma)^2}{\Gamma(1+\sigma)^2}
\frac{\Gamma(1+\tfrac{1}{2}(\theta_\star+\sigma))}{
	\Gamma(1+\tfrac{1}{2}(\theta_\star-\sigma))}
\frac{\Gamma(1+\tfrac{1}{2}(\theta_t+\theta_0+\sigma))
	\Gamma(1+\tfrac{1}{2}(\theta_t-\theta_0+\sigma))}{
	\Gamma(1+\tfrac{1}{2}(\theta_t+\theta_0-\sigma))
	\Gamma(1+\tfrac{1}{2}(\theta_t-\theta_0-\sigma))},
\label{eq:kappV}
\end{equation}
$C_V(\{\theta\};\sigma)$ is a normalization constant for the expansion around $t=0$, and $\{\theta\}$ encodes the parameters of the second-order differential equation \eqref{eq:2.53}. The parameter $\sigma\in\mathbb{C}$ is defined in terms of Stokes multipliers, whereas $\eta\in\mathbb{C}$ is expressed in terms of the (monodromy) parameters $\{\theta\}$ and $\sigma$. In the next pages, we will discuss more about these two terms in the expansion \eqref{eq:exptauV}. Finally, it should be stressed that the isomonodromic $\tau_V$-function is analytic in the whole complex plane except in the branch points at $t=0,\infty$, as mentioned in \eqref{eq:2.31}. The expansion \eqref{eq:exptauV} has thus a infinite radius of convergence. 

As argued previously, the expansion of the $\tau_{V}$-function around the irregular point $t=\infty$ is substantially more complicated. No general expansion exists, not even using Fredholm determinant representation, there are only expansions via $c=1$ irregular conformal blocks for $t= \infty$ along specific rays, such as arg$(t) = 0,\pi/2,\pi,3\pi/2$ \cite{Lisovyy:2018mnj}. Despite the difficulty in computing such expansion, we can still work with the expression for $t = i\infty$ in the analysis of the quasinormal mode' frequencies, which will be discussed in the next chapters. Therefore, considering the expansion for $t = i\infty$, one has the following asymptotic formula from \cite{Lisovyy:2018mnj}
\begin{equation}
\begin{aligned}
\tau_{V}(\{\theta\};\nu,\rho;t) =\mathcal{N}_V(\{ \theta \},\nu) e^{\frac{t}{4} (2+\theta_{0}+\theta_{t}-\nu)} t^{\frac{\theta_\star^2}{8}-\frac{\nu^2}{8}} \bigg[1+t^{-\nu-2}& {U_V}^{-1} e^{-t}+\\+\frac{((\theta_\star+\nu)^2-4\theta_0^2)((\theta_\star-\nu)^2-4 \theta_t^2)}{256}& t^{\nu-2} U_V e^{t}+\mathcal{O}(t^{2(\Re\nu-2)})\bigg],
\end{aligned}
\label{eq:taufirstterm}
\end{equation}
with $-1< \Re \nu <1$ and $U_V$ expressed in terms of Gamma functions as
\begin{equation}
\begin{split}
U_V &=  e^{\pi i \rho}\frac{1}{4 \pi ^2}\Gamma \left(\frac{-2 \theta_0-\theta_\star -\nu}{4} \right) \Gamma\left(\frac{2 \theta_0-\theta_\star -\nu}{4} \right) \Gamma\left( \frac{\theta_\star-2 \theta_t-\nu}{4} \right) \Gamma\left( \frac{\theta_\star+2 \theta_t -\nu}{4} \right),
\label{eq:shat}
\end{split}
\end{equation}
$\mathcal{N}_V(\{ \theta \},\nu)$ is a normalization constant for the expansion at infinity. In contrast to the series around $t=0$, expected to converge in the whole complex plane, the expansion \eqref{eq:taufirstterm} is only asymptotic and more terms are computed from equation (1.12a) in \cite{Lisovyy:2018mnj}. Finally, the complex parameters $\{\nu,\rho\}$ in the expansion above are functions of the parameters $\{\sigma, \eta \}$, for the expansion at $t=0$, \eqref{eq:exptauV}. As conjectured in \cite{Lisovyy:2018mnj}, this dependence is derived from the connection formula between the large and small $t$ expansions for the $\tau_{V}$-function \cite{Andreev:2000aa,its:hal-01797601}, where one obtains, for a convenient parametrization of the linear system \eqref{eq:CHSys}, the following relations
\begin{equation}
e^{-\frac{\pi}{2} i \nu} = X_{-}, \ \ \ \ \ e^{\pi i\rho} =1-X_{-}X_{+}
\label{eq:nurhoparameter}
\end{equation}
with the quantities $X_{-}$ and $X_{-}$ satisfying the generic conditions $X_{+}X_{-}\neq 1$, $X_{-}\neq0$ and given by
\begin{equation}
\begin{aligned}
X_{\pm}\sin^2\pi\sigma=A_{\pm}-A_{\mp}\cos\pi&\sigma\mp i\sum_{\epsilon=\pm}e^{i\pi \epsilon \eta\mp i\frac{\pi}{2}\epsilon \sigma}(\cos\pi(\theta_{t}-\epsilon\sigma)-\cos\pi\theta_{0}) \sin\frac{\pi}{2}(\theta_{\star}-\epsilon\sigma)\\& A_{\pm} = e^{\mp \frac{\pi}{2} i\theta_{\star}}\cos\pi\theta_{0}+e^{\pm \frac{\pi}{2} i\theta_{\star}}\cos\pi \theta_{t}.
\end{aligned}
\end{equation}

\vspace{0.2cm}
\textbf{Monodromy parameters $\{\sigma,\eta\}$}:
\vspace{0.2cm}

The parameters $\{\sigma,\eta\}$\footnote{This pair of parameters are also called by Darboux \textit{coordinates}\cite{Lisovyy:2018mnj}.}, in turn, are computed from the non-Fuchsian system by using monodromy and connection matrices, respectively, with $\eta$ having a direct connection with the boundary conditions for the quasinormal modes discussed in the last chapter. For $\sigma$, we use the constraint \eqref{eq:Mcond}, which is now written as
\begin{equation}
\mathcal{M}_0\mathcal{M}_t = \mathcal{M}_{\infty}^{-1}, \quad\quad \mathcal{M}_i\in GL(2,\mathbb{C}), \quad\quad i=1,2,\infty.
\label{eq:monodromy}
\end{equation}

In the left-hand side, the product of the monodromy matrices $\mathcal{M}_0$ and $\mathcal{M}_t$ is regularly defined as the \textit{composed monodromy} matrix $\mathcal{M}_\sigma$ \cite{Lisovyy:2018mnj}, which is a diagonal matrix associated with the analytic continuation around both Fuchsian singularities in the system \eqref{eq:CHSys}. Taking the trace in \eqref{eq:monodromy} and using the equation \eqref{eq:monoinfty}, we obtain
\begin{equation}
\begin{aligned}
\tr \mathcal{M}_\sigma = \tr \mathcal{M}_{\infty} \quad \rightarrow \quad
2\cos \pi \sigma =2\cos \pi\theta_{\star}+s_1 s_2 e^{-i\pi \theta_{\star}} 
\end{aligned}
\label{eq:sigmastokes}
\end{equation}
where $s_1$ and $s_2$ are Stoke's multipliers, $\mathcal{M}_{\sigma} = e^{i\pi \sigma_3 \sigma}$ and $A^{(\infty)}_{(0)} = \frac{1}{2}\sigma_{3}\theta_{\star}$. In regard to $\eta$, we can derive an expression that relates $\eta$ with $\sigma$ and $\{\theta\}$, this is done by using the connection matrix between local solutions of the confluent Heun equation \eqref{eq:SODE} (or \eqref{eq:ConfHeun}) around $z=t$ and $z=\infty$. Thus, for the equation \eqref{eq:SODE}, we have the following asymptotic behavior for $y(z)$
\begin{equation}
y(z) = \begin{cases}
(z-t)^{(\pm \theta_t-1)/2}(1+\mathcal{O}(z-t)),\quad  z\rightarrow t \\ e^{- z/2}z^{\mp \theta_{\star}/2}(1+\mathcal{O}(z^{-1})), \quad z\rightarrow \infty
\end{cases}
\end{equation}
where the two solutions around t and $\infty$ generate a basis of solutions with $\pm \theta_{t}$ and $\pm\theta_{{\star}}$ encoding the two solutions. By algebraic manipulation of the solutions defined around these points, one obtains the following connection matrix $C_{t,\infty}$ between the local solutions
\begin{equation}
\begin{aligned}
\begin{pmatrix}
\rho_{\infty}y_{\infty,+}(z) \\
\tilde{\rho}_{\infty}y_{\infty,-}(z)
\end{pmatrix}
&=C_{\infty,t}
\begin{pmatrix}
\rho_{t}y_{t,+}(z)\\
\tilde{\rho}_{t}y_{t,-}(z)
\end{pmatrix}
\\&=\begin{pmatrix}
e^{-\tfrac{i\pi}{2}\eta}\zeta'_{t}-e^{\tfrac{i\pi}{2}\eta}\zeta_{t}
& 
-e^{-\tfrac{i\pi}{2}\eta}\zeta_{\infty}\zeta'_{t} +
e^{\tfrac{i\pi}{2}\eta}\zeta'_{\infty}\zeta_{t} \\
e^{-\tfrac{i\pi}{2}\eta}-e^{\tfrac{i\pi}{2}\eta} &
-e^{-\tfrac{i\pi}{2}\eta}\zeta_{\infty}+e^{\tfrac{i\pi}{2}\eta}\zeta'_{\infty}
\end{pmatrix}
\begin{pmatrix}
\rho_{t}y_{t,+}(z)\\
\tilde{\rho}_{t}y_{t,-}(z)
\end{pmatrix},
\end{aligned}
\label{eq:Cm}
\end{equation}
where,
\begin{equation}
\begin{gathered}
\zeta_{\infty} =  e^{-\frac{i\pi}{2}\sigma}\sin\tfrac{\pi}{2}
(\theta_\star+\sigma),
\quad\quad
\zeta'_{\infty}= e^{\frac{i\pi}{2}\sigma}\sin\tfrac{\pi}{2}
(\theta_\star-\sigma),\\
\zeta_{t}=\sin\tfrac{\pi}{2}(\theta_t+\theta_0-\sigma)
\sin\tfrac{\pi}{2}(\theta_t-\theta_0-\sigma),\quad\quad
\zeta'_{t}=\sin\tfrac{\pi}{2}(\theta_t+\theta_0+\sigma)
\sin\tfrac{\pi}{2}(\theta_t-\theta_0+\sigma),
\end{gathered}
\end{equation}
and $\rho_t,\tilde{\rho}_t,\rho_\infty,\tilde{\rho}_\infty$ are normalization constants -- see \cite{Andreev:2000aa,CarneirodaCunha:2015hzd} for more details about the connection matrix $C_{t,\infty}$. The elements of the matrix $C_{\infty,t}$ can be used to relate the scattering coefficients to the monodromy parameters $\{\sigma,\eta\}$. In order to derive such a relation, we first mention that from Chapter \ref{chap1} the quasinormal modes must satisfy the conditions of ingoing and outgoing waves at the event horizon and at infinity, respectively. Such conditions require $C_{\infty,t}$ to be lower triangular. Thus, from \eqref{eq:Cm}, we have that the element $C^{12}_{\infty,t}=0$ allows us to express $\eta$ in terms of $\sigma$ and $\{\theta\}$ as
\begin{equation}
e^{i\pi\eta}=\frac{\zeta_{\infty}\zeta'_{t}}{\zeta'_{\infty}\zeta_{t}}
=e^{-i\pi\sigma}
\frac{\sin\tfrac{\pi}{2}(\theta_\star+\sigma)}{
	\sin\tfrac{\pi}{2}(\theta_\star-\sigma)}
\frac{\sin\tfrac{\pi}{2}(\theta_t+\theta_0+\sigma)
	\sin\tfrac{\pi}{2}(\theta_t-\theta_0+\sigma)}{
	\sin\tfrac{\pi}{2}(\theta_t+\theta_0-\sigma)
	\sin\tfrac{\pi}{2}(\theta_t-\theta_0-\sigma)},
\label{eq:quantV}
\end{equation}
which reduces the number of parameters in the $\tau_{V}$-function \eqref{eq:exptauV}.

\vspace{0.2cm}
\textbf{$\tau_{V}$-function conditions:}
\vspace{0.2cm}

We have described the main relevant features of the isomonodromic $\tau_{V}$-function, as well as its asymptotic expansion for $t=0$ and $t=i\infty$. We now return to the principal goal in this chapter, which consists in finding from the isomonodromic deformations theory conditions for the $\tau_V$-function. These conditions will be relevant to the problems listed in Table \ref{tab:CHEDic}. Therefore, in order to recover the CHE \eqref{eq:ConfHeun} from \eqref{eq:2.53}, it is necessary to take the following choices,
\begin{equation}
\theta_0 \rightarrow\theta_0,\quad\quad
\theta_{t}\rightarrow\theta_{t}-1,\quad\quad
\theta_{\star}\rightarrow\theta_{\star}+1,\quad\quad
\lambda(t)= t,\quad\quad
\mu(t)=-\frac{c_{t}}{\theta_{t}-1}.
\label{eq:fixpar}
\end{equation}

The first three changes are simple shifts in the parameters $\{\theta\}$. In turn, the constraints in $\lambda(t)$ and $\mu(t)$ are depending on the \textit{modulus} $t$ and accessory parameter $c_t$, where $\lambda(t)$ and $\mu(t)$ are related to the isomonodromic family of non-Fuchsian system in the form \eqref{eq:CHSys}. By fixing $t$ and consequently the pair $\{\lambda(t),\mu(t)\}$, we have that the conditions \eqref{eq:firstcond} and \eqref{eq:seccond} simplifies to
\begin{equation}
\frac{d}{d t}\log\tau_V(\{\theta\}^{-},\sigma-1,\eta;t)= c_{t}+\frac{\theta_0
	(\theta_{t}-1)}{2t},\quad\quad
\frac{d}{dt}t\frac{d}{dt}\log\tau_V(\{\theta\}^{-},\sigma-1,\eta;t)+\frac{\theta_{t}-1}{2}=0
\label{eq:3.38}
\end{equation}
where $\{\theta\}^{-}=\{\theta_0,\theta_t-1,\theta_{*}+1\}$ and $e^{i\pi \eta}$ is invariant by the changes \eqref{eq:fixpar}. Finally, the shift in $\sigma$ comes from \eqref{eq:sigmastokes}, where it can be shown straighforward that $\sigma \rightarrow \sigma \pm (2n+1)$, $n \in \mathbb{N}$. To ensure that $\sigma$ satisfy $0<\Re\sigma< 1$, we will restrict ourself to the case $n=0$ resulting in $\sigma \rightarrow \sigma-1$, the choice $\sigma+1$ is also valid, but not relevant for our purpose.

As discussed previously, the second condition in \eqref{eq:3.38} stems from the second derivative of the $\tau_V$ function calculated using the Schlesinger equations. The left-hand side of this equation can be related through the \textit{Toda equation} \cite{198747} to a product of $\tau_V$-functions, the proof of this relation can be verified in Appendix \ref{chap:AppB}. Thus, from \eqref{eq:tauPlusMinus}, we have the expression
\begin{equation}
\frac{d}{dt}t\frac{d}{dt}\log\tau_V(\{\theta\};\sigma,\eta;t)+\frac{\theta_t}{2}
= K_V\frac{\tau_V(\{\theta\}^{+}; \sigma+1, \eta ;t)\tau_V(\{\theta\}^-; \sigma-1,\eta;t)}{\tau_{V}^2(\{\theta\};\sigma,\eta;t)},
\label{eq:3.39}
\end{equation}
where the constant $K_V$ is independent of $t$, and ${\{\theta\}}^\pm$ are related to $\{\theta\}$ by the simple shifts,
\begin{equation}
{\{\theta\}}^\pm =\{\theta_0,\theta_t\pm 1,\theta_{\star}\mp 1\}.
\label{eq:3.40}
\end{equation}
Therefore, if we consider \eqref{eq:fixpar}, the right-hand side of the equation \eqref{eq:3.39} will change and must be zero, then, since the isomonodromic $\tau_V$-function is analytic in $t$ except at the critical
points $t=0$ and $t=\infty$ -- see Miwa's theorem \cite{Miwa:1980yj} --  we have that either $\tau_V^{+}(\{\theta\}^{+};\sigma+1,\eta;t)$ or $\tau_V^{-}(\{\theta\}^{-};\sigma-1,\eta;t)$ has to vanish in \eqref{eq:3.39}, resulting in the conditions:
\begin{equation}
\frac{d}{d t}\log\tau_V(\{\theta\}^{-};\sigma-1,\eta;t)= c_{t}+\frac{\theta_0
	(\theta_{t}-1)}{2t},\quad\quad \tau_V(\{\theta\};\sigma,\eta;t)=0.
\label{eq:tauVcond}
\end{equation}
In this way, instead of treating with the second derivative of the $\tau_{V}$-functions, we reduce the analysis of the second condition in \eqref{eq:3.38} to the equation $\tau_V(\{\theta\};\sigma,\eta;t)=0$. This simplification plays a crucial role in the calculation of quasinormal modes, which will be presented in the following chapters.

For the expansion of the $\tau_{V}$-function around $t = i\infty$, we have also two conditions analogous to \eqref{eq:tauVcond}, where $\sigma$ and $\eta$ are replaced by $\nu$ and $\rho$. Again, $\nu$ and $\rho$ are monodromy parameters that parametrize the expansion around the irregular point and can be calculated from \eqref{eq:nurhoparameter} -- see \cite{Lisovyy:2018mnj,Andreev:1995in} for a review of the parameters $\nu$ and $\rho$, as well as discussions about the expansion for $t=\infty$ with $\arg(t) =\pi$, which is not treated in this thesis \footnote{In this case, the $\tau_V$ expansion is parameterized by different monodromy parameters}. For this case, we have
\begin{equation}
\frac{d}{dt}\text{log}(\tau_{V}(\{\theta\}^{-};\nu-1, \rho;t)) =c_{t}+\frac{\theta_{0}(\theta_{t}-1)}{2t}, \qquad \tau_{V}(\{\theta\};\nu,\rho;t) =0,
\label{eq:tauVcondinf}
\end{equation}
where, from the treatment of isomonodromic deformations in the linear system \eqref{eq:CHSys}, one has the same parameters $\{\theta\}$ and shifts defined in \eqref{eq:3.40}. The shift in $\nu$ is calculated from the relations for the asymptotic parameters $\{\sigma,\eta\}$ and $\{\nu,\rho \}$ expressed in \eqref{eq:nurhoparameter}, where it can be shown from $\{\theta\}^{-}$ and $\sigma-1$ that $A_{\pm}$ and $X_{\pm}$ will change to
\begin{equation}
A_{\pm} \rightarrow \mp i A_{\pm},\ \ \ \ \ X_{\pm} \rightarrow \mp i X_{\pm}.
\end{equation}
Then, substituting $X_{\pm}$ in \eqref{eq:nurhoparameter}, we have that the expression for $e^{\pi i\rho}$ is invariant, whereas the first equation will be invariant if $\nu \rightarrow \nu -(2n+1)$, $n\in\mathbb{N}$, where we consider the first locus $n=0$ where $-1<\Re \nu<1$, resulting in $\nu \rightarrow \nu-1$, as it can be seen in \eqref{eq:tauVcondinf}. Similar to the expansion of the $\tau_{V}$-function for small t \eqref{eq:taufirstterm}, the asymptotic expansion at $t=i\infty$ is defined in the theory of the isomonodromic deformations by the equation \eqref{eq:2.33}, allowing us to define \eqref{eq:tauVcondinf}. Again, in contrast to the expansion at $t=0$, which has an explicit representation in terms of the Fredholm determinant, the expansion for $\tau_{V}(\{\theta\};\nu,\rho;t)$ is only given in terms of $c=1$ irregular conformal block \cite{Lisovyy:2018mnj}.

\begin{figure}[ht]
	\centering
	\caption{\small{The Figure illustrates the map between the parameters of the $\tau_V$-function expanded in the branch points $t=0,i\infty$, and accessory parameter and modulus of the CHE, $\{c_t,t\}$. Where $\{\sigma, \eta\}$ and $\{\nu,\rho\}$, for the expansions at $t=0$ and $t=i\infty$, respectively, are related by the expression \eqref{eq:nurhoparameter}.}}
	\centering
	\includegraphics[width=.5\linewidth]{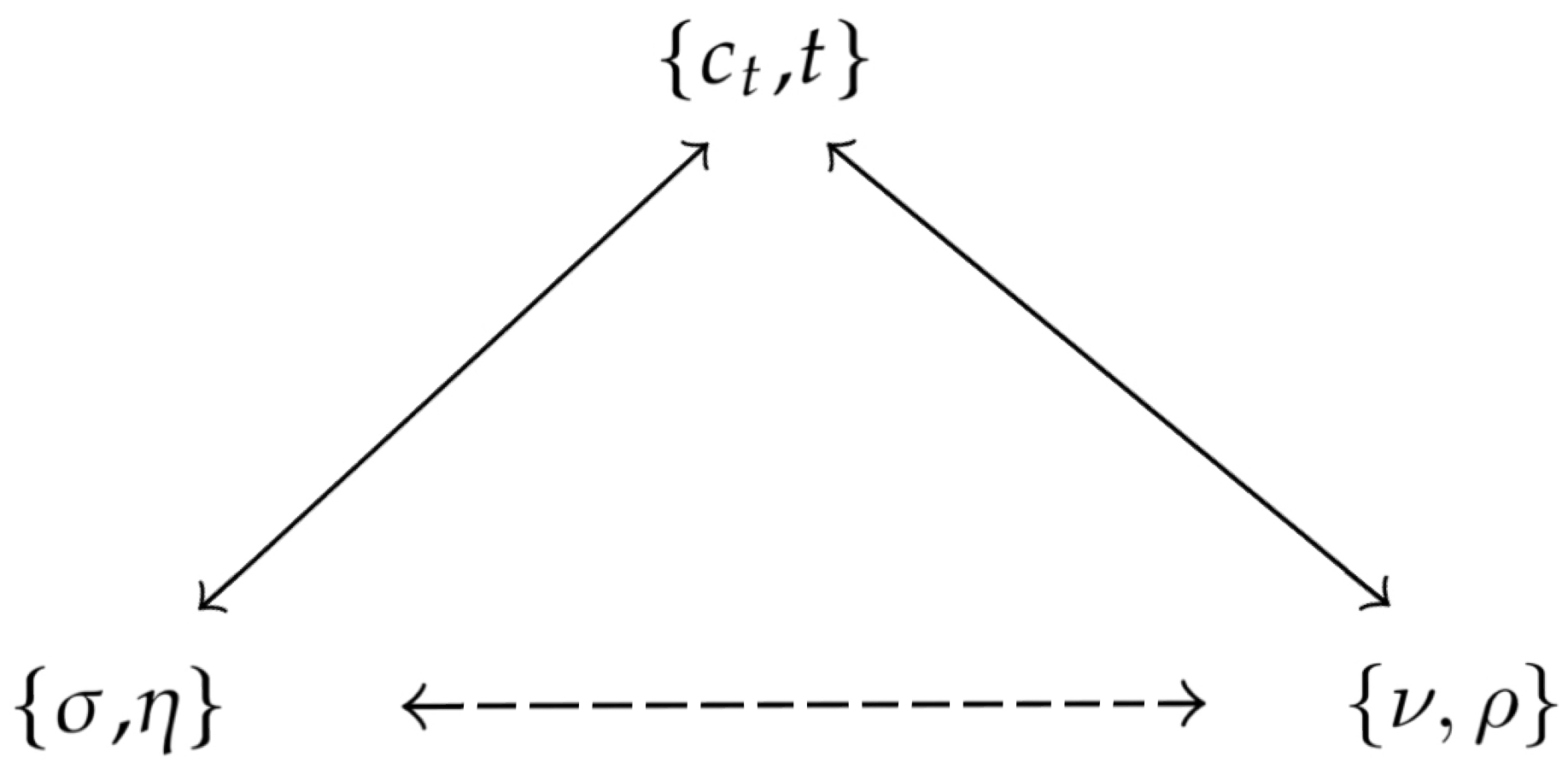}
	\label{fig:maps}
\end{figure}

We remark that the conditions \eqref{eq:tauVcond} and \eqref{eq:tauVcondinf} are also called by \textit{Riemann-Hilbert maps}, the reason is that these maps are related to the Riemann-Hilbert problem, that, in turn, consists in solving a general $N\times N$ linear system through analytic continuation properties (monodromy matrices) of the solutions and the singularities of the linear system -- see references \cite{Bothner_2021,Its2003TheRP, 10.1007/3-540-09996-4_31} for a review of the Riemann-Hilbert problem. Furthermore, in our case, the first \eqref{eq:tauVcond} and second \eqref{eq:tauVcondinf} maps consist in relating the monodromy parameters $\{\sigma, \eta\}$ and $\{\nu, \rho\}$ of the $\tau_{V}$-function with the accessory parameter and modulus of the CHE, $\{c_t,t\}$. To illustrate the two Riemann-Hilbert maps, we summarize in Fig. \ref{fig:maps} the relations between the relevant parameters in the maps, with the dashed line encoding the transformation between $\{\sigma,\eta\}$ and $\{\nu,\rho\}$. 

Riemann-Hilbert (RH) maps in the form \eqref{eq:tauVcond} and \eqref{eq:tauVcondinf} are known in the literature and have been studied in different contexts. For instance, a similar RHm was proposed for the $\tau_{VI}$-function in \cite{Anselmo:2020bmt,Anselmo:2018zre,Amado:2017kao} in the study of conformal mapping and Kerr-AdS/CFT correspondence. For these cases, the isomonodromic deformation theory was applied to the Heun equation, a second-order differential equation in the form \eqref{eq:SODEdef}, with four Fuchsian singularities.

\subsection{Isomonodromic Deformations in the DCHE}
\label{subsec:ID-DCHE}

We are also interested in the isomonodromic deformations theory applied to the double-confluent Heun equation (DCHE), which emerges from the CHE when the two regular singularities at $z=0$ and $z=t$ coalesce in a new singular point of rank $r=1$, as was presented in \eqref{eq:conflim}, whose new differential equation is given in \eqref{eq:DCHE} -- see also \cite{NIST:DLMF}\footnote{For those who are unfamiliar with confluent limits in second-order differential equations, we mention that this type of limit is also observed in the Hypergeometric differential equation, which for a given limit transforms into the confluent Hypergeometric differential equation -- see Chapter 15 in \cite{NIST:DLMF}.}. In this situation, one has a new non-Fuchsian system defined by
\begin{equation}
\frac{d}{dz}\Phi(z) = \bigg(\frac{A_1}{z}+\frac{A_2}{z^2}+A_{\infty}\bigg)\Phi(z),
\label{eq:DCHESys}
\end{equation}
where $A_i\in GL(2,\mathbb{C})$, $i=1,2,\infty$. The linear system has now two irregular singularities at $z=0$ and $z=\infty$ of rank Poincaré $r=1$, and the fundamental matrix solution defined around each singular point is restricted by \textit{Stokes sectors}. Note that, the linear system above has the same form of \eqref{eq:nFSDCHE}, but now with $N=2$. 

Following the same strategy presented in Sec. \ref{subsec:IDCHE}, one has a second-order differential equation in the standard form \eqref{eq:2.39}, where for a basis where $A_{\infty}$ is diagonal the component $A_{12}(z)$ assumes the form
\begin{equation}
A_{12}(z) = \frac{\bar{\kappa}(z-\bar{\lambda})}{z^2},
\end{equation}
$\bar{\kappa}$ and $\bar{\lambda}$ are functions of the elements of matrices $A_1$ and $A_2$. Then, we can derive, from the isomonodromic deformations theory and by imposing the zero-curvature condition \eqref{eq:flatCond}, the Schlesinger equations for the linear system \eqref{eq:DCHESys}. Similar to what was observed for the CHE, the solutions of the given Schlesinger equations are related to another isomonodromic $\tau$-function, where, in this case, the function associated with the DCHE is labeled by the index $III$, which is because the apparent singularity $\lambda(t)$ for this case satisfies the non-linear Painlevé equation $P_{III}$ -- see Jimbo, Miwa, and Ueno paper \cite{Jimbo:1981aa} for a description of the linear system in \eqref{eq:DCHESys}. Finally, one has, from the isomonodromic deformations theory, two conditions for the $\tau_{III}$-function and a deformed double-confluent Heun equation (i.e. similar to the equation \eqref{eq:2.53}), whose apparent singularity is now described by the Hamiltonian $H_{III}$ \cite{Filipuk2019OnTD}. We will not give the form of the equations related to the linear system \eqref{eq:DCHESys} and derived from the isomonodromic deformation theory, as the Painlevé equation $P_{III}$ and the Hamiltonian $H_{III}$. These expressions have been written in a large number of references \cite{Ohyama2006StudiesOT,Okamoto1986StudiesOT,Derezinski2020FromHC}. We want to treat with the Riemann-Hilbert map for the DCHE directly, which can be obtained from the Riemann-Hilbert map \eqref{eq:tauVcond} by applying the confluent limit
\begin{equation}
\Lambda = \frac{1}{2}(\theta_{t}-\theta_0), \quad \theta_{\circ}
=\theta_{t}+\theta_0, \quad u = \Lambda t 
, \quad \Lambda \rightarrow \infty.
\label{eq:conflimit1}
\end{equation}

Note that this limit was defined in Chapter \ref{chap1}. The direct application of the limit above in the RHm \eqref{eq:tauVcond}, is due to the fact that the Painlevé equations and consequently the Hamiltonian systems are related by a process of coalescence, which consists in obtaining by the confluence of singularities a new Hamiltonian system \eqref{eq:2.51} and Painlevé equation. This fact stands also for the isomonodromic $\tau$-functions, where one has that two $\tau$-functions are related by a given confluent limit -- see the series of papers \cite{okamoto1981isomonodromic,10.3792/pjaa.56.264,10.3792/pjaa.56.367,Okamoto1999} for a review of the coalescence process. In this way, in order to obtain the Riemann-Hilbert map for the DCHE, we apply in \eqref{eq:tauVcond} the limit \eqref{eq:conflimit1}, which results in
\begin{equation}
\tau_{III}(\{\tilde{\theta}\};\sigma,\eta;u)=0,\qquad
u\frac{d}{du}\log\tau_{III}(\{\tilde{\theta}\}^-;\sigma-1,\eta;u)
-\frac{(\theta_\circ-1)^2}{8}-\frac{1}{2}
=uk_u,
\label{eq:tauIIIcond}
\end{equation}
where $k_u$ is the confluent limit of $c_{t}$, $\{\tilde{\theta}\} = \{\theta_{\circ},\theta_{\star}\}$, and $\{\tilde{\theta}\}^{-} = \{\theta_{\circ}-1,\theta_{\star}+1\}$\footnote{We will label the parameters for the DCHE as $\{\tilde{\theta}\}$, while for the CHE is $\{\theta\}$, without "$\sim$".}. Again, the shift in $\sigma$ comes from \eqref{eq:sigmastokes}.

The isomonodromic $\tau_{III}$-function is a transcendental function with branch points at $u=0$ and $u=\infty$ and the first terms of the expansion for $u=0$ were derived by Jimbo in \cite{Jimbo:1982aa}. Such a function has been explored in a large number of problems in physics and mathematics, such as random matrix theory and connection problem \cite{CHEN2010270,Its:2014lga}. In turn, all terms in the expansion around $u=0$ can be calculated by using $c=1$ irregular conformal blocks \cite{Gamayun:2013auu} or via the Fredholm determinant \cite{daCunha:2021jkm}. Given the numerical control of the expansion at $u=0$, we will restrict ourselves to the representation of the function in terms of the Fredholm determinant. In addition, to the best of our knowledge, in \cite{daCunha:2021jkm} and now in Appendix \ref{sec:tools}, we give for the first time the representation of the $\tau_{III}$-function in terms of this determinant. 

Therefore, from Appendix \ref{sec:tools}, we have the following asymptotic formula at $u=0$,
\begin{multline}
	\tau_{III}(\{\tilde{\theta}\};\sigma,\eta;u) =
	u^{\frac{1}{4}\sigma^2-\frac{1}{8}\theta_\circ^2}e^{\frac{1}{2}u}\times\\
	\left(
	1-\frac{\sigma-\theta_\circ\theta_\star}{2\sigma^2}u
	-\frac{(\sigma+\theta_\circ)(\sigma+\theta_\star)}{4\sigma^2
		(\sigma-1)^2}\kappa_{III}^{-1}u^{1-\sigma} -
	\frac{(\sigma-\theta_\circ)(\sigma-\theta_\star)}{4\sigma^2
		(\sigma+1)^2}\kappa_{III}u^{1+\sigma}+\mathcal{O}(u^2,u^{2\pm
		2\Re\sigma})\right)
	\label{eq:tauIIIexp}
\end{multline}
with $\kappa_{III}$ derived by applying the limit \eqref{eq:conflimit1} in \eqref{eq:kappV} and collecting the finite term
\begin{equation}
\kappa_{III}=e^{i\pi\eta}\frac{\Gamma(1-\sigma)^2}{
	\Gamma(1+\sigma)^2}
\frac{\Gamma(1+\tfrac{1}{2}(\theta_\star+\sigma))}{
	\Gamma(1+\tfrac{1}{2}(\theta_\star-\sigma))}
\frac{\Gamma(1+\tfrac{1}{2}(\theta_\circ+\sigma))}{
	\Gamma(1+\tfrac{1}{2}(\theta_\circ-\sigma))}
\end{equation}
where $-1<\Re \sigma<1$ -- see Appendix \ref{sec:tools} for a description of the determinant. The first terms for the expansion above agree, unless to the multiplicative term, with \cite{Jimbo:1982aa} and \cite{Gamayun:2013auu}. Finally, for $\eta$, one has the following simplification from the equation \eqref{eq:quantV}, where in the confluent limit the number of sine functions reduces,
\begin{equation}
e^{i\pi\eta}=e^{-2i\pi\sigma}
\frac{\sin\tfrac{\pi}{2}(\theta_\star+\sigma)}{
	\sin\tfrac{\pi}{2}(\theta_\star-\sigma)}
\frac{
	\sin\tfrac{\pi}{2}(\theta_{\circ}+\sigma)}{
	\sin\tfrac{\pi}{2}(\theta_{\circ}-\sigma)}.
\label{eq:quantIII}
\end{equation}

In what follows, we will work only with the two RHm maps for the $\tau_V$-function and the single one in \eqref{eq:tauIIIcond}. The map for the $\tau_{III}$-function expanded for large $u$ exists, however, given the difficulty in obtaining the expansion around the irregular point at $u=\infty$, we will leave the discussion of this map for future work. Furthermore, the RH maps defined in this chapter are already relevant to the analysis of quasinormal mode' frequencies for extremal and subextremal Kerr and Reissner-Nordström black holes, as we will see in the next chapters.

\section{Accessory Parameter Expansions}
\label{sec:accpar}

We finish this chapter by showing how the expansions for the accessory parameters associated with the CHE and DCHE, i.e. $c_t$ in \eqref{eq:tauVcond} and \eqref{eq:tauVcondinf}, and $k_u$ in \eqref{eq:tauIIIcond}, can be computed. Essentially, $c_t$ and $k_u$ can be derived from the second condition for the $\tau_{III}$ (DCHE case) and $\tau_{V}$ (CHE case) or from the series representation of the solution of the equation involved (CHE or DCHE).

Computationally, the series representation is more amenable, since it eliminates the necessity of computing the logarithmic derivative of the $\tau$-functions, defined in the RH maps \eqref{eq:tauVcond}, \eqref{eq:tauVcondinf}, and \eqref{eq:tauIIIcond}. The idea consists in using the series representation for the solution of the equation involved (similar to the Frobenius method) to arrive at a recurrence equation for terms of the series and then compute the expansion for $c_t$ or $k_u$ via continued fraction. By this procedure, one arrives at an efficient form of computing the expansion for the accessory parameters. Therefore, let us show how the series expansion for $c_t$ and $k_u$ are derived. The expansions of this section will be fundamental to the next chapters.

\subsection{Accessory Parameter Expansion for CHE}

Let us consider first the CHE that has two regular singularities at $0$, $t$ and one irregular of rank 1 at $\infty$.
\begin{equation}
\frac{d^{2}y}{dz^{2}}+\left[\frac{1-\theta_{0}}{z}+\frac{1-\theta_{t}}{z-t}
\right] \frac{dy}{dz}-
\left[\frac{1}{4}+\frac{\theta_{\star}}{2z}+\frac{tc_{t}}{z(z-t)}
\right]y(z)=0,
\label{eq:CHeq}
\end{equation}
the accessory parameter $c_t$ can be expanded around the branch points $t=0$ and $t=i\infty$, where, for each case, we have to assume  different series representation for the solution $y(z)$. As can be observed in the logarithmic derivative of the $\tau_{V}$-function in the RH maps \eqref{eq:tauVcond} and \eqref{eq:tauVcondinf}, the expansion for $c_t$ around $t=0$ is parametrized by the complex parameter $\sigma$, whereas for $t= i\infty$, one has a dependence in $\nu \in \mathbb{Z}$, such dependence are also observed for the series representation of the solution $y(z)$, as we will see. We recommend the references \cite{daCunha:2022ewy,Lisovyy:2021bkm} for more details about the derivations below. With this in mind, let us start by considering the expansion for $c_t$ around $t=0$.

\vspace{0.2cm}
\textbf{Expansion for $t=0$ :}
\vspace{0.2cm}

The $\sigma$ parameter in the transcendent $\tau_{V}$ parametrizes the series representation for the solution $y(z)$, which, in turn, is given in terms of \textit{Floquet solutions}\cite{Lisovyy:2021bkm}. In this way, one has that $y(z)$ is written as 
\begin{equation}
y(z) = e^{-\frac{1}{2}z}
z^{\frac{1}{2}(\sigma+\theta_0+\theta_t)-1}\sum_{n=-\infty}^{\infty}
c_nz^n,\quad\quad\text{or}\quad\quad
y(z) = e^{-\frac{1}{2}z}
z^{\frac{1}{2}(-\sigma+\theta_0+\theta_t)}\sum_{n=-\infty}^{\infty}
\tilde{c}_nz^n,
\label{eq:floquetsol}
\end{equation}
where for both cases $y(z)$ converges in an annulus $t<|z|<1$. By using the equations above, we can obtain an expression that relates $c_t$ and $\sigma$. Substituting the first equation into the confluent Heun equation \eqref{eq:CHeq}, one arrives at the following three-term recurrence relation,
\begin{equation}
{ A}_n c_{n-1}-({ B}_n+t{ C}_n)c_n+t{ D}_nc_{n+1}=0,
\label{eq:recurrence}
\end{equation}
where
\begin{gather}
	{ A}_n = 2(\sigma+\theta_\star+2n-2), \\
	{ B}_n = (\sigma+\theta_0+\theta_t+2n-2)(\sigma-\theta_0-\theta_t+2n),\\
	{ C}_n=2(\sigma+\theta_t+\theta_\star+2n-1)-4c_{t}, \\
	{ D}_n=(\sigma+\theta_t+\theta_0+2n)(\sigma+\theta_t-\theta_0+2n).
	\label{eq:abcd}
\end{gather}
The equation \eqref{eq:recurrence} can be solved using continued fractions, where through the relations
\begin{equation}
\frac{c_n}{c_{n+1}}=\frac{t D_n}{tC_n+B_n-
	A_n\frac{c_{n-1}}{c_n}},\qquad
\frac{c_{n}}{c_{n-1}}=\frac{A_n}{t C_n+B_n-
	tD_n\frac{c_{n+1}}{c_n}},
\end{equation}
we rewrite the equation \eqref{eq:recurrence} as
\begin{equation}
\cfrac{t{ A}_0{ D}_{-1}}{{ B}_{-1}+t{ C}_{-1}
	-t\cfrac{{ A}_{-1}{ D}_{-2}}{{ B}_{-2}+t{ C}_{-2}
		-t\cfrac{{ A}_{-2}{ D}_{-3}}{{ B}_{-3}+\ldots}}}
-t{ C}_0
+\cfrac{t{ D}_0{ A}_1}{{ B}_1+t{ C}_1-t
	\cfrac{{ D}_1{ A}_2}{{ B}_2+t{ C}_2-t
		\cfrac{{ D}_2{ A}_3}{{ B}_3+\ldots}}}
={B}_0
\label{eq:contfrac}
\end{equation}
where, after considering some number of terms in the continued fraction, we can truncated the expression by fixing the value for the convergent \footnote{When we truncate a continued fraction after some number of terms, we get what is called a convergent. The convergents in a continued fraction representation of a number are the best rational approximations of that number -- see the textbook \cite{jones1980continued} for a review of continued fraction.} of order $N_c$. Finally, in order to find an expansion for $c_t$ around $t=0$, we replace $c_t = \sum_{i=0}^{n}c_n t^{n}$ and truncate the continued fraction by fixing the value of $N_c$. The final step is to solve the equation and collect each term of the expansion resulting in
\begin{subequations}
	\begin{equation}
	tc_{t}=c_0+c_1t+c_2t^2+c_3t^3+\ldots c_nt^n+\ldots,
	\label{eq:3.58}
	\end{equation}
	with the first terms in the expansion given by
	\begin{equation}  
	c_0=\frac{(\sigma-1)^2-(\theta_{0}+\theta_t-1)^2}{4},\quad\quad
	c_1=\frac{\theta_{\star}(\sigma(\sigma-2)-\theta_0^2+\theta_{t}^2)}{
		4\sigma(\sigma-2)},
	\label{eq:3.59}
	\end{equation}
	\begin{multline}
		c_2=\frac{1}{32}+\frac{\theta_{\star}^2(\theta_0^2-\theta_{t}^2)^2}{64}
		\left(\frac{1}{\sigma^3}-\frac{1}{(\sigma-2)^3}\right)
		+\frac{(1-\theta_\infty^2)(\theta_0^2-\theta_{t}^2)^2+2\theta_{\star}^2
			(\theta_0^2+\theta_{t}^2)}{32\sigma(\sigma-2)}
		\\-
		\frac{(1-\theta_{\star}^2)((\theta_0-1)^2-\theta_{t}^2)((\theta_0+1)^2-
			\theta_{t}^2)}{32(\sigma+1)(\sigma-3)},
		\label{eq:3.60}
	\end{multline}
	\begin{multline}
		c_3 = \frac{\theta^3_{{\star}}(\theta^{2}_0-\theta^{2}_{t})^3}{256}\bigg(\frac{1}{\sigma^5}-\frac{1}{(\sigma-2)^5}\bigg)-\\
		\frac{4(\theta^2_{0}-\theta^2_{t})^3 \theta_{{\star}} -\big(5(\theta^6_{0}-\theta^6_{t})+8 \theta^4_{t}+15\theta^2_0 \theta^4_{t}-\theta^4_{0} (8+15 \theta^2_{t})\big) \theta^3_{{\star}}}{1024}\\
		\bigg(\frac{1}{\sigma^3}-\frac{1}{(\sigma-2)^3}\bigg)-
		\frac{1}{24576}(\theta^2_{t}-\theta^2_{0}) \theta_{{\star}}
		\bigg(64+80 \theta^2_{{\star}}+8 \theta^2_{t} (20-29 \theta^2_{{\star}})+\\
		(\theta^4_{0}+\theta^4_{t})
		(125 \theta^2_{{\star}}-116)+
		\theta^2_{0} \big(160-232 \theta^2_{{\star}}+\theta^2_{t} (232-250 \theta^2_{{\star}})\big)\bigg) \bigg(\frac{1}{\sigma}-\frac{1}{\sigma-2
		}\bigg)\\
		+\frac{((-1+\theta_0)^2-\theta^2_{t}) (\theta_{0}^2-\theta^2_{t}) ((1+\theta_{0})^2-\theta^2_{t}) \theta_{{\star}} (1-\theta^2_{{\star}})}{96 (3-\sigma) (1+\sigma)}\\
		-\frac{((-2+\theta_{0})^2-\theta^2_{t}) (\theta^2_{0}-\theta^2_{t})((2+\theta_{0})^2-\theta^2_{t}) \theta_{{\star}}(4-\theta^2_{{\star}} )}{4096 (4-\sigma ) (2+\sigma)}.
		\label{eq:3.61}
	\end{multline}
\end{subequations}

The expansion for $c_t$ above agrees to order $t^n$ with the logarithmic derivative of $\tau_V$ expansion written in the RHm \eqref{eq:tauVcond}. Furthermore, we have that through the Floquet solution \eqref{eq:floquetsol}, the accessory parameter or, more precisely, the expression \eqref{eq:contfrac} allows us to implement the equation \eqref{eq:contfrac} instead of consider the logarithmic derivative of $\tau_{V}$-function \eqref{eq:exptauV}. Therefore, in the next chapters, we will focus on the implementation of the equation \eqref{eq:contfrac}. 

In addition, the three-term recurrence equation \eqref{eq:recurrence} and its solution in terms of continued fractions \eqref{eq:contfrac} are the basis for the  so-called continued fraction (or Leaver's method \cite{Leaver:1985ax}) to compute angular eigenvalues and quasi-normal modes for the Teukolsky Master Equation \eqref{eq:TME}.

\vspace{0.2cm}
\textbf{Expansion for large $t$:}
\vspace{0.2cm}

For large $t$, with $\arg(t)=\pi/2$, the series basis for $y(z)$ is given in terms of confluent Hypergeometric functions, as observed in \cite{Casals:2018cgx}. In this case, the complex parameter $\nu$ in the $\tau_{V}$ expansion \eqref{eq:taufirstterm} is related to the solution $y(z)$ by the expression
\begin{equation}
y(z)=\sum_{n\in\mathbb{Z}}a_n e^{-z/2}z^{\theta_0}\,F_n(z),\qquad
\,F_n(z)=\left\{\begin{matrix} M \\ U \end{matrix}\right\}
(\tfrac{1}{2}(1+\theta_0+\theta_\star)+\tfrac{1}{4}\nu-\tfrac{1}{4}\theta_\star-n;1+\theta_0;z),
\label{eq:serrepre}
\end{equation}
where the convergence of the series depends on the \textit{Stokes sectors}. $M(a,b;z)$ and $U(a,b;z)$ are Kummer and Tricomi functions, respectively, and satisfy the confluent Hypergeometric differential equation in the following form,
\begin{equation}
z\frac{d^{2}w}{dz^2}+(b-z)\frac{dw}{dz}-aw =0,
\end{equation}
with $a, b \in \mathbb{C}$ -- see \cite{NIST:DLMF} (Chapter 13) for more details about the functions $M(a,b;z)$ and $U(a,b;z)$, as well as their relations with Stokes phenomenon. For both functions, the three term recursion relation is given by
\begin{equation}
\begin{gathered}
\bar{A}_n a_{n-1} - (\bar{B}_n+t\bar{C}_n) a_n + \bar{D}_n a_{n+1} = 0,\\
\bar{A}_n=(n+(-\tfrac{1}{4}\nu+\tfrac{1}{4}\theta_\star)-\tfrac{1}{2}(1+\theta_t))(n+(-\tfrac{1}{4}\nu+\tfrac{1}{4}\theta_\star)-\tfrac{1}{2}(1-\theta_0+\theta_\star)),
\\
\bar{B}_n=2(n+(-\tfrac{1}{4}\nu+\tfrac{1}{4}\theta_\star))(n+(-\tfrac{1}{4}\nu+\tfrac{1}{4}\theta_\star)-\tfrac{1}{2}\theta_\star)+\tfrac{1}{2}(1-\theta_0)(1-\theta_t),\\
\bar{D}_n =
(n+(-\tfrac{1}{4}\nu+\tfrac{1}{4}\theta_\star)+\tfrac{1}{2}(1+\theta_t))(n+(-\tfrac{1}{4}\nu+\tfrac{1}{4}\theta_\star)+\tfrac{1}{2}(1-\theta_0-\theta_\star))\\
\bar{C}_n = c_t-n-(-\tfrac{1}{4}\nu+\tfrac{1}{4}\theta_\star).
\label{eq:recursionM}
\end{gathered}
\end{equation}

The relation between $c_t$ and $\nu$ that arises from this recurrence equation will be the same for both solutions in \eqref{eq:serrepre}. Such a relation is solved by the same method considered for the case $t=0$, where in order to solve the equation in \eqref{eq:recursionM}, we use
\begin{equation}
\frac{a_n}{a_{n+1}}=\frac{t^{-1}\bar{D}_n}{\bar{C}_n+t^{-1}\bar{B}_n-
	t^{-1}\bar{A}_n\frac{a_{n-1}}{a_n}},\qquad
\frac{a_{n}}{a_{n-1}}=\frac{t^{-1}\bar{A}_n}{\bar{C}_n+t^{-1}\bar{B}_n-
	t^{-1}\bar{D}_n\frac{a_{n+1}}{a_n}}.
\end{equation}
Then, substituting the relations above, we arrive at
\begin{equation}
\cfrac{t^{-2}{\bar{A}}_0{\bar{D}}_{-1}}{{\bar{C}}_{-1}+t^{-1}{\bar{B}}_{-1}
	-t^{-2}\cfrac{{\bar{A}}_{-1}{\bar{D}}_{-2}}{{\bar{C}}_{-2}+t^{-1}{\bar{B}}_{-2}
		-\ldots}}
-t^{-1}{\bar{B}}_0 
+\cfrac{t^{-2}{\bar{A}}_1{\bar{D}}_0}{{\bar{C}}_1+t^{-1}{\bar{B}}_1-t^{-2}
	\cfrac{{\bar{A}}_2{\bar{D}}_1}{{\bar{C}}_2+t^{-1}{\bar{B}}_2-\ldots}}
={\bar{C}}_0.
\label{eq:contfracINF}
\end{equation}
Following the same strategy, we obtain the $c_t$ expansion for large $t$, by truncating the convergent of order $N_c$ and substituting $c_t=\sum_{i=1}^{n}\bar{c}_n t^{-n}$ in the equation above, leading to
\begin{equation}
\begin{gathered}
c_t = \bar{c}_{0} + \bar{c}_{1}t^{-1}+\bar{c}_{2}t^{-2}+\bar{c}_{3}t^{-3}+\ldots+\bar{c}_{n}t^{-n}+\ldots,\\
\bar{c}_{0}=-\frac{\nu}{4}+\frac{\theta_\star}{4},\qquad
\bar{c}_{1}=-\frac{\nu^2}{8}-\frac{(1-\theta_0)(1-\theta_t)}{2}+
\frac{\theta_\star^2}{8},\\
\bar{c}_{2}=\frac{\nu^3}{16}+
\frac{(4-2\theta_0^2-2\theta_t^2-\theta_\star^2)\nu}{16}+
\frac{(\theta_0^2-\theta_t^2)\theta_\star}{8},\\
\bar{c}_{3}=-\frac{5\nu^4}{64}-
\frac{(20-6\theta_0^2-6\theta_t^2-3\theta_\star^2)\nu^2}{32}-
\frac{(\theta_0^2-\theta_t^2)\theta_\star\nu}{4}\\ \qquad \qquad +
\frac{(8+4\theta_0^2+4\theta_t^2-\theta_\star^2)\theta_\star^2}{64}
-\frac{(1-\theta_0^2)(1-\theta_t^2)}{4},
\end{gathered}
\label{eq:accparINF}
\end{equation}
which agrees to order $t^n$ with the logarithmic derivative of $\tau_V$ expansion for $t= i\infty$ defined in \eqref{eq:tauVcondinf} -- see \cite{daCunha:2022ewy} for a review of the expansion above and its connection with semiclassical conformal blocks.

\subsection{Accessory Parameter Expansion for DCHE}
\label{subsec:accparDCHE}

Finally, for the RHm \eqref{eq:tauIIIcond}, one has that the confluent limit \eqref{eq:conflimit1} applied to the expression \eqref{eq:CHeq} leads to the double-confluent Heun equation (DCHE):
\begin{equation}
\frac{d^2
	y}{dz^2}+\bigg[\frac{2-\theta_{\circ}}{z}-\frac{u}{z^2}\bigg]
\frac{dy}{dz}-\bigg[\frac{1}{4} 
+\frac{\theta_{\star}}{2z}+\frac{uk_{u}-u/2}{z^2}\bigg]y(z)=0.
\label{DCHEeq}
\end{equation}

In turn, the expansion for the accessory parameter $k_u$ follows the same lines and has a structure parallel to \eqref{eq:3.58}. Thus, taking the confluent limit \eqref{eq:conflimit1} in the equations \eqref{eq:recurrence}-\eqref{eq:abcd}, we arrive at the following three-term recurrence relation for the DCHE:
\begin{equation}
u(\tilde{A}_n a_{n-1}+\tilde{C}_n a_{n+1})+a_n(4k_u+\tilde{B}_n)=0
\end{equation}
with
\begin{gather}
	\tilde{A}_n = 2(\sigma+\theta_\star+2n-4)+4\bigg(1-\frac{\theta_\star-\theta_{\circ}}{2}\bigg), \\
	\tilde{B}_n = (\sigma+\theta_{\star}+2n-2)(\sigma-\theta_{\star}+2n),\\
	\tilde{C}_n=2(\sigma+\theta_{\star}+2n).
	\label{eq:abcdnew}
\end{gather}
Then, considering the same strategy utilized for the CHE case, we use continued fraction to write the three-term recurrence relation as
\begin{equation}
\cfrac{u\tilde{A}_0\tilde{D}_{-1}}{\tilde{B}_{-1}-4k_u
	-u\cfrac{\tilde{A}_{-1}\tilde{D}_{-2}}{\tilde{B}_{-2}-4k_u
		-u\cfrac{\tilde{A}_{-2}\tilde{D}_{-3}}{\tilde{B}_{-3}-\ldots}}}
+4{k}_u
+\cfrac{u\tilde{D}_0\tilde{A}_1}{\tilde{B}_1+u\tilde{C}_1-4k_u
	\cfrac{\tilde{D}_1\tilde{A}_2}{\tilde{B}_2-4k_u-u
		\cfrac{\tilde{D}_2\tilde{A}_3}{\tilde{B}_3-\ldots}}}
=\tilde{B}_0.
\label{eq:contfracDCHE}
\end{equation}
Assuming $k_u = \sum_{i=0}^{n}k_n u^{n}$ and replacing in the equation above, we arrive at the expansion for $k_u$: 
\begin{equation}
uk_{u} = k_0 + k_1 u + k_2 u^2 + k_3 u^3 + ... + k_n u^{n}+... 
\label{eq:accparinf}
\end{equation}
with the first four terms given by
\begin{equation}
\begin{aligned}
k_{0} &=\frac{(\sigma-1)^2-(\theta_{\circ}-1)^2}{4} \\
k_{1} &=-\frac{1}{2}-\frac{\theta_{{\circ}} \theta_{\star}}{4(\sigma-2)}+\frac{\theta_{{\circ}} \theta_{\star}}{4 \sigma}\\
k_{2} &=\frac{\theta_{{\circ}}^{2} \theta_{\star}^{2}}{16}\left(\frac{1}{\sigma^{3}}-\frac{1}{(\sigma-2)^{3}}\right)+\frac{\theta_{\star}^{2}-\theta_{{\circ}}^{2}\left(\theta_{\star}^{2}-1\right)}{8(\sigma-2) \sigma}+\frac{\left(-1+\theta_{{\circ}}^{2}\right)\left(-1+\theta_{\star}^{2}\right)}{8(\sigma-3)(1+\sigma)}
\end{aligned}
\end{equation}

\begin{equation}
\begin{aligned}
k_{3} &=-\frac{\theta_{{\circ}}^{3} \theta_{\star}^{3}}{32}\left(\frac{1}{(\sigma-2)^{5}}-\frac{1}{\sigma^{5}}\right)-\frac{4 \theta_{{\circ}} \theta_{\star}^{3}+\theta_{{\circ}}^{3} \theta_{\star}\left(4-5 \theta_{\star}^{2}\right)}{128}\left(\frac{1}{\sigma^{3}}-\frac{1}{(-2+\sigma)^{3}}\right)-\\&
\frac{\theta_{{\circ}} \theta_{\star}\left(80-116 \theta_{\star}^{2}+\theta_{{\circ}}^{2}\left(-116+125 \theta_{\star}^{2}\right)\right)}{3072}\left(\frac{1}{\sigma-2}-\frac{1}{\sigma}\right)+\frac{\theta_{{\circ}}\left(\theta_{{\circ}}^{2}-1\right) \theta_{\star}\left(\theta_{\star}^{2}-1\right)}{12(\sigma-3)(1+\sigma)} \\
&-\frac{\theta_{{\circ}}\left(\theta_{{\circ}}^{2}-4\right) \theta_{\star}\left(\theta_{\star}^{2}-4\right)}{512(\sigma-4)(2+\sigma)}.
\end{aligned}
\label{eq:c3exp}
\end{equation}

Again, it can be verified, after a large number of calculations, that the expansion for $k_u$ is exactly the expansion derived from the second equation in \eqref{eq:tauIIIcond}. The procedure consists in implementing the $\tau_{III}$-function and then computing the logarithmic derivative, the terms will be the same as listed above. We remark that the  $k_u$ expansion for large $u$ can be calculated from the equation for $c_t$ written in \eqref{eq:accparINF}, however, such an expansion is not relevant to the next chapters, and it seems to be non-trivial of obtaining. 

\section{Conclusion of the Chapter}

We have shown in this chapter that the study of isomonodromic deformation theory, for both confluent and double-confluent Heun equations, is given in terms of monodromy parameters and isomonodromic $\tau$-functions. From the approach, one has that the conditions for the $\tau$-functions are related to Riemann-Hilbert maps, which in turn relate the accessory parameter and modulus of the differential equation involved (CHE or DCHE) to monodromy parameters that parametrize the expansions for the $\tau$-functions. 

We started the chapter by explaining, in Sec. \ref{sec:FnFsystem}, the relation between linear systems of first-order and linear ODEs, then, in Sec. \ref{sec:solmon}, we presented the solutions and monodromy matrices of the linear system associated with the confluent Heun equation. In Sec. \ref{sec:isodefthe}, we discussed the main feature of the isomonodromic deformations theory, then, based on the theory introduced, we deal in Sec. \ref{subsec:IDCHE} with the isomonodromic deformations approach in the CHE, which led us to two conditions for the transcendental $\tau_{V}$-function. Given the branch points structure, we revealed that such a function is parametrized by two monodromy parameters, where for small and large $t$, the expansion is parametrized by $\{\sigma,\eta\}$ and $\{\nu,\rho\}$, respectively. 

In Sec. \ref{subsec:ID-DCHE}, it was shown that via the confluent limit \eqref{eq:conflimit1}, we can define a new Reimann-Hilbert map associated with the double-confluent Heun equation. In this case, the map is written in terms of the $\tau_{III}$-function, whose monodromy parameters for small $u$ are $\{\sigma, \eta\}$. We concluded the chapter in Sec. \ref{sec:accpar}, where it was shown that the accessory parameters for the CHE and DCHE can be computed using series
representation for the solution of the second-order differential equations, which allows us to implement numerically the accessory parameter expansion using continued fraction rather than computing using the logarithmic derivative of the $\tau$-functions involved in the three RH maps listed in the chapter.


\newpage
\thispagestyle{empty}
\begin{center}\centering
	\vspace*{\fill}
	\resizebox{!}{0,40cm}{\textbf{ISOMONODROMY METHOD AND QNMs:}}\\
	\resizebox{!}{0,40cm}{Kerr and Reissner-Nordström black holes}
	\vspace*{\fill}
\end{center}
\newpage

\chapter{Kerr black hole}
\label{chap:KerrBH}

In this chapter, we make use of the isomonodromy method introduced in the previous chapter, based on \cite{CarneirodaCunha:2019tia}. The basic idea in the method consists in solving a given RH map, where for our case the solutions are related to the QN frequencies. All results obtained are numerical and, for specific regimes, analytical, as we will see. It will also be shown that using the isomonodromy method, we can compare the results obtained with the literature for generic values of $0\leq a\leq M$, then we will focus on the near-extremal limit $a\rightarrow M$, relevant to our analysis. 

In Sec. \ref{sec:kerrequations}, we give the main differential equations governing perturbations of the Kerr black hole. In Sec. \ref{sec:angradialsys}, the angular eigenvalue expansion for spheroidal harmonics \eqref{eq:angulareq} is derived, using monodromy matrices and the accessory parameter expansion derived in the previous chapter. In Sec. \ref{sec:secradeq}, we deal with the radial equation \eqref{eq:radialeq}, where the relevant parameters of the differential equation are defined, and the description of the RHm associated with the radial equation is made. In Sec. \ref{sec:generica}, the numerical solution for generic rotation parameter $0\leq a\leq M$ is given, and in Sec. \ref{sec:extremallimit} we discuss the extremal $a\rightarrow M$ limit. In this case, we found in the numerical and analytical studies that some QN frequencies display a finite behavior, analyzed in Sec. \ref{sec:smallnu}, whereas the rest display a double confluent limit, studied in Sec. \ref{sec:painleveIII}, being given in the extremal case $a=M$ by the isomonodromic $\tau_{III}$-function, then the frequencies and eigenvalues found for some modes in the extremal case are shown. We close in Sec. \ref{sec:RHmapslst} with a discussion of the results and showing that overtone modes for $a/M=0$ can be obtained from different RH maps. Finally, we include in Appendix \ref{sec:tools} the main equations used in the chapter. 

The results presented in this chapter are published in \textit{Physical Review D} \textbf{104}, \textit{084051 (2021)} and in \textsc{Arxiv:} \textit{Expansions for semiclassical conformal blocks}\footnote{Arxiv link: \href{https://arxiv.org/abs/2211.03551}{Expansions for semiclassical conformal blocks}}.

\section{Kerr background: the Teukolsky equation}
\label{sec:kerrequations}

As discussed in Chapter \ref{chap1}, the Teukolsky Master Equation \eqref{eq:TME} governs linear perturbations in the Kerr metric. For vacuum perturbations $(T_{\mu\nu}=0)$, its solutions $\psi_s$, in \eqref{eq:wavefunc}, can be written as product of the solutions of two ordinary differential equations 
\begin{gather}
	\bigg\{\frac{1}{\sin\theta}\frac{d}{d\theta}\left[\sin \theta
	\frac{d}{d\theta}\right]+
	a^{2}\omega^{2}\cos^{2}\theta-2a\omega s \cos\theta -
	\frac{(m+s\cos\theta)^{2}}{\sin^{2}\theta}+{}_{s}A_{\ell m}
	\bigg\}{}_{s}S_{\ell m}(\theta)=0,
	\label{eq:angulareq} \\
	\bigg\{{\Delta_{BL}}^{-s}\frac{d}{dr}\left[{\Delta_{BL}}^{s+1}\frac{d}{dr}\right]+
	\frac{K^{2}(r)-2is(r-M)K(r)}{\Delta_{BL}}+4is\omega
	r-\lambda(a\omega)
	\bigg\}{}_{s}R_{\ell m}(r)=0,
	\label{eq:radialeq}
\end{gather}
where ${}_{s}A_{\ell m} = s +{_s\lambda_{\ell,m}}$ and $\lambda(a\omega)= {_s\lambda_{\ell m}}+a^2\omega^2-2ma\omega$, with
\begin{equation}
K(r)=(r^2+a^2)\omega-am,\quad\quad \Delta_{BL} =
r^2-2Mr+a^2=(r-r_+)(r-r_-).
\end{equation}
$M$ and $a=J/M$ are the mass and angular momentum parameter of the black hole, whereas $\omega$, $m$ and $s$ are the frequency, azimuthal angular momentum parameter and spin of the perturbation.

We start our study of the equations \eqref{eq:angulareq} and \eqref{eq:radialeq} with the angular equation, where the main goal consists in finding the expansion for ${}_{s}\lambda_{\ell m}$ in terms of $a\omega$, this is done by imposing the regularity of the solution ${}_{s}S_{\ell m}(\theta)$ at $\theta=0$ and $\theta=\pi$.

\section{Angular Teukolsky Master Equation}
\label{sec:angradialsys}

From Chapter \ref{chap1}, we know that the angular TME \eqref{eq:angulareq} is a spin-weighted spheroidal harmonics equation, where using change of variables 
\begin{equation}
y_{\mathrm{Ang}}(z) =
(1+\cos\theta)^{\theta_{\mathrm{Ang},z_0}/2}
(1-\cos\theta)^{\theta_{\mathrm{Ang},0}/2}
{}_{s}S_{\ell m}(\theta),\quad\quad
z_{\mathrm{0}} = -2a\omega(1-\cos\theta),
\label{eq:shomotopic}
\end{equation}
it can be brought to the standard confluent Heun form, as discussed in Sec. \ref{DicSection},
\begin{equation}
\frac{d^2 y}{dz^2}+\bigg[\frac{1-\theta_{{\mathrm{Ang}},0}}{z}+\frac{1-\theta_{{\mathrm{Ang}},z_0}}{z-z_0} \bigg]\frac{d y}{dz}-\bigg[\frac{1}{4}+\frac{\theta_{\mathrm{Ang},\star}}{2z}+\frac{z_0 c_{z_0}}{z(z-z_0)}\bigg]y(z)=0
\label{eq:angCHE}
\end{equation}
where the parameters $\{\theta\}_{\mathrm{Ang}}$ are
\begin{equation}
\theta_{\mathrm{Ang},0}=-m-s, \quad\quad \theta_{\mathrm{Ang},z_0}=m-s,
\quad\quad \theta_{\mathrm{Ang},\star}=-2s,
\label{eq:singlemonoangular}
\end{equation}
with the modulus $z_0$ and accessory parameter $c_{z_0}$ given by
\begin{equation}
z_{0}= -4a\omega, \quad\quad
{z}_{0}c_{z_0} ={}_{\ell}\lambda_{s,m}+2a\omega
s+a^2\omega^2.
\label{eq:accessoryangular}
\end{equation}

As discussed in the Chapter \ref{chap1}, we can compute the angular eigenvalue ${}_{s}\lambda_{\ell, m}$ by imposing the regularity of the solution $y(z)$ around the singular regular points $0$ and $z_0$, i.e. between $\theta=0$ and $\theta=\pi$, in \eqref{eq:angulareq}. Thus, in other to find ${}_{s}\lambda_{\ell m}$, we assume the following behavior for $y(z)$
\begin{equation}
y(z)=\begin{cases}
z^{\theta_{{\mathrm{Ang}},0}}(1+{\mathcal O}(z)) & z\rightarrow 0, \\
(z-z_0)^{\theta_{{\mathrm{Ang}},z_0}}(1+{\mathcal O}(z-z_0)) & z\rightarrow z_0,
\end{cases}
\label{eq:3.12}
\end{equation}
which will place a restriction on the value of the angular eigenvalue ${}_{s}\lambda_{\ell,m}$ and eliminates divergent solutions.

In turn, from Chapter \ref{ChapIsoMethod}, the confluent Heun equation \eqref{eq:angCHE} can be cast in the matricial form, 
\begin{equation}
\frac{d}{dz}\Phi(z) = \bigg(\frac{A_0}{z}+ \frac{A_{z_0}}{z-z_0} +A_{\infty} \bigg)\Phi(z).
\end{equation}
Therefore, rather than treating with the explicit form of the solutions $y(z)$ that satisfy \eqref{eq:3.12}, we can investigate the  conditions \eqref{eq:3.12} in terms of the fundamental matrix solutions of the system around $z=0$ and $z=z_0$, in other words, we can work directly with the expression \eqref{eq:2.4}. Furthermore, since the constraint is over the regular singularities, we can use the composed monodromy $\mathcal{M}_{\sigma}$ to derive an expression for the monodromy parameter $\sigma$. This procedure allows us to compute the angular eigenvalue ${}_{s}\lambda_{\ell, m}$ from the accessory parameter expansion \eqref{eq:3.58}. 

Following the definitions in \eqref{eq:2.4}, one has that the two fundamental matrix solutions around $z=0$ and $z=z_0$ are expressed by
\begin{equation}
\Phi^{(0)}(z) = G_{(0)}\big[\mathbb{I}+\sum_{j=1}^{\infty}\Phi^{(0)}_{j}z^{j}\big]z^{A^{(0)}_{0}},
\label{eq:3.14}
\end{equation}
\begin{equation}
\Phi^{(z_0)}(z) = G_{(z_0)}\big[\mathbb{I}+\sum_{j=1}^{\infty}\Phi^{(z_0)}_{j}(z-z_0)^{j}\big](z-z_0)^{A^{(z_0)}_{0}},
\label{eq:3.15}
\end{equation}
with $G_{(0)}$, $G_{(z_0)}$, $\Phi^{(0)}_{j}$, and $\Phi^{(z_0)}_{j}$ constant matrices of $GL(2,\mathbb{C})$. In turn, $A_{0}^{(0)}$ and $A_{0}^{(z_0)}$ are defined in \eqref{eq:2.9} as,
\begin{equation}
A_{0}^{(0)} = G^{-1}_{(0)}A_0G_{(0)}, \quad\quad
A_{0}^{(z_0)} = G^{-1}_{(z_0)}A_{z_0}G_{(z_0)}.
\label{eq:3.16}
\end{equation}
Without loss of generally, we can consider a parametrization for $A_0$ and $A_{z_0}$ that makes $A_{0}^{(0)}$ and $A_{z_0}^{(0)}$ diagonal and dependent of $\theta_{{\mathrm{Ang}},0}$ and $\theta_{{\mathrm{Ang}},z_0}$, respectively,
\begin{equation}
A_{0}^{(0)} = \frac{1}{2}\sigma_3\theta_{{\mathrm{Ang}},0}, \qquad A^{(z_0)}_{0} = \frac{1}{2}\sigma_{3}\theta_{{\mathrm{Ang}},z_0}, \qquad  \sigma_3= \diag(1,-1).
\end{equation}
Then, one has from the definition \eqref{eq:2.15} that the monodromy matrices for $\Phi^{(0)}(z)$ and $\Phi^{(z_0)}(z)$ are given by
\begin{equation}
\mathcal{M}_0 = C_0^{-1}e^{i\pi\sigma_3\theta_{{\mathrm{Ang}},0}}C_0, \qquad \mathcal{M}_{z_0} = C_{z_0}^{-1}e^{i\pi\sigma_3\theta_{{\mathrm{Ang}},z_0}}C_{z_0}.
\end{equation}

The next step consists in using the monodromy matrices to derive an expression for $\sigma$. From the linear system, we have defined the composed monodromy as the analytic continuation around both Fuchsian singularities as $\mathcal{M}_{\sigma} = \mathcal{M}_{0}\mathcal{M}_{t}$. Thus, the trace of $\mathcal{M}_{0}\mathcal{M}_{z_0}$ results in
\begin{equation}
\begin{aligned}
\tr \mathcal{M}_0 \mathcal{M}_{z_0} &= \tr (C_0^{-1}e^{i\pi\sigma_3\theta_{{\mathrm{Ang}},0}}C_0 C_{z_0}^{-1}e^{i\pi\sigma_3\theta_{{\mathrm{Ang}},z_0}}C_{z_0})\\&=\tr (C_{z_0}C_0^{-1}e^{i\pi\sigma_3\theta_{{\mathrm{Ang}},0}}C_0 C_{z_0}^{-1}e^{i\pi\sigma_3\theta_{{\mathrm{Ang}},z_0}}),
\end{aligned}
\end{equation}
where we have used the cyclic property $\tr(AB) = \tr(BA)$, with $A$ and $B$ general matrices. The product  $C_{z_0}C_0^{-1}$ is essentially the connection matrix between the fundamental solutions $\Phi^{(0)}(z)$ and $\Phi^{(z_0)}(z)$. Then, defining $C_{z_0,0} = C_{z_0}C_0^{-1}$ and assuming a generic form for $C_{z_0,0}$:
\begin{equation}
C_{z_0,0} = \begin{pmatrix}
c_{11}&c_{12} \\
c_{21}& c_{22}  \\
\end{pmatrix}, \qquad \det C_{z_0,0} \neq 0,
\end{equation}
one obtains
\begin{equation}
\tr (\mathcal{M}_0\mathcal{M}_{z_0}) = \frac{2c_{11}c_{22}\cos\pi(\theta_{{\mathrm{Ang}},0}+\theta_{{\mathrm{Ang}},z_0})+2c_{12}c_{21}\cos\pi(\theta_{{\mathrm{Ang}},0}-\theta_{{\mathrm{Ang}},z_0})}{c_{11}c_{22}-c_{12}c_{21}}. 
\end{equation}

In terms of connection matrix, the conditions \eqref{eq:3.12} leads to an upper triangular matrix form for $C_{z_0,0}$, which implies in $c_{12}$ zero. The constraint \eqref{eq:3.12} is then translated to $\sigma$ as
\begin{equation}
\sigma = \theta_{{\mathrm{Ang}},0}+\theta_{{\mathrm{Ang}},z_0}+2j, \quad j\in \mathbb{Z}
\label{eq:angularquant}
\end{equation}
where we have used that $\tr \mathcal{M}_{\sigma}= \tr (e^{i\pi\sigma_{3} \sigma}) = 2\cos \pi \sigma$, with $\sigma_{3}=\diag(1,-1)$. Furthermore, in order to recover the asymptotic behavior for ${}_{s}\lambda_{\ell, m}$ computed in the literature, we choose $j=\ell+s+1$ -- see \cite{daCunha:2021jkm}. Where the minimum eigenvalue of $\ell$ is $|s|$ and the azimuthal momenta are constrained by  $|m|\leq\ell$ \cite{Teukolsky:1973ha}.  

Finally, the accessory parameter expansion for the CHE, written in \eqref{eq:3.58}, can be used, together with the condition \eqref{eq:angularquant} to derive an expression for the angular eigenvalue ${_s\lambda_{\ell,m}}$ defined in \eqref{eq:accessoryangular}. Substituting hence the parameters $\{\theta\}_{\mathrm{Ang}}$ and $\sigma$ in \eqref{eq:3.58}, we have from \eqref{eq:accessoryangular} that each term of ${}_{s}\lambda_{\ell,m}(a\omega)$ can be computed. Thus, assuming
\begin{equation}
{_s\lambda}_{\ell,m}(a\omega)= \sum^{\infty}_{n=0}f_n (a\omega)^{n},
\label{eq:3.65}
\end{equation}
where in order to simplify the expressions for the $f_n$'s we define
\begin{equation}
h(\ell) = \frac{2(\ell^{2}-m^{2})(\ell^{2}-s^{2})^{2}}{(2\ell-1)\ell^{3}(2\ell+1)}
\end{equation}
one has that the first seven coefficients are given by
\begin{equation*}
	\begin{aligned}
		f_0 &= (\ell-s)(\ell+s+1),\\ 
		f_1 &= -\frac{2ms^2}{\ell(\ell+1)},\\
		f_2 &= h(l+1)-h(l)-1,\\
		f_3 &=\frac{2h(l)ms^2}{(l-1)l^2(l+1)}-\frac{2h(l+1)ms^2}{l(l + 1)^2(l+2)},\\
		f_4 &= m^2s^4\bigg(
		\frac{4h(\ell+1)}{
			\ell^2(\ell+1)^4(\ell+2)^2}-
		\frac{4h(\ell)}{(\ell-1)^2\ell^4(\ell+1)^2}\bigg)
		-\frac{(\ell+2)h(\ell+1)h(\ell+2)}{
			2(\ell+1)(2\ell+3)}\\& +\frac{h^2(\ell + 1)}{2(\ell+1)}+\frac{h(\ell)h(\ell+1)}{2\ell^2 + 2\ell}
		-\frac{h^2(\ell)}{2\ell}+\frac{(\ell-1)h(\ell-1)h(\ell)}{4\ell^2-2},\\
		f_5 &= m^3s^6\bigg(\frac{8h(\ell)}{\ell^6(\ell + 1)^3(\ell - 1)^3}-\frac{8h(\ell + 1)}{\ell^3(\ell+1)^6(\ell+2)^3}\bigg)+
		\\&ms^2h(\ell)\bigg(-\frac{h(\ell+1)(7\ell^2 + 7\ell + 4)}{\ell^3(\ell+2)(\ell+1)^3(\ell-1)}-
		\frac{h(\ell-1)(3\ell-4)
		}{\ell^3(\ell+1)(2\ell-1)(\ell-2)}\bigg)+\\&ms^2\bigg(
		\frac{(3\ell+7)h(\ell+1)h(\ell+2)}{\ell(\ell + 1)^3(\ell + 3)(2\ell+3)} -\frac{3h^2(\ell+1)}{\ell(\ell+1)^3(\ell+2)}+\frac{3h^2(\ell)}{\ell^3(\ell-1)(\ell+1)}\bigg),
	\end{aligned}
\end{equation*}
\begin{equation}
\begin{aligned}
f_6&= \frac{16m^4s^8}{\ell^4(\ell + 1)^4}\bigg(
\frac{h(\ell+1)}{(\ell+1)^4(\ell+2)^4}-\frac{h(\ell)}{\ell^4(\ell-1)^4}\bigg)-\\&
\frac{4m^2
	s^4}{\ell^2(\ell+1)^2}\bigg(\frac{(3\ell^2+14\ell+17)h(\ell+1)h(\ell+2)}{(\ell+1)^3(\ell+2)(\ell+3)^2(2\ell+3)}-\frac{3h^2(\ell+1)}{(\ell+1)^3(\ell+2)^2}+\frac{3h^2(\ell)}{\ell^3(\ell-1)^2}\bigg)\\&
+\frac{4m^2s^4}{\ell^2(\ell+1)^2}
\bigg(\frac{(11\ell^4+22\ell^3+31\ell^2+20\ell+6)h(\ell)h(\ell+1)}{\ell^3(\ell-1)^2(\ell+1)^3(\ell+2)^2}
\\&+\frac{(3\ell^2-8\ell+6)h(\ell-1)h(\ell)
}{\ell^3(\ell-2)^2(\ell-1)(2\ell-1)}\bigg)
+\frac{h(\ell+1)h(\ell+2)}{4(\ell+1)(2\ell+3)^2}
\bigg(\frac{(\ell+3)h(\ell+3)}{3}\\&+\frac{\ell+2}{\ell+1}\bigg((\ell+2)h(\ell+2)-(7\ell+10)h(\ell+1)+ \frac{(3\ell^2+2\ell-3)h(\ell)}{\ell}\bigg)\bigg)\\&+\frac{h(\ell)h(\ell+1)}{4\ell^2(\ell+1)^2}
\bigg((2\ell^2+4\ell+3)h(\ell)-(2\ell^2+1)h(\ell+1)\\&-
\frac{(\ell^2-1)(3\ell^2+4\ell-2)h(\ell-1)}{(2\ell-1)^2}\bigg)+\frac{h^3(\ell + 1)}{2(\ell+1)^2}
-\frac{h^3(\ell)}
{2\ell^2}\\&+\frac{h(\ell-1)h(\ell)
}{4\ell^2(2\ell-1)^2}
\bigg((\ell-1)(7\ell-3)h(\ell)-(\ell-1)^2h(\ell-1)-\frac{\ell(\ell-2)h(\ell-2)}{3}\bigg).
\end{aligned}
\end{equation}

These terms are in agreement with \cite{Seidel:1988ue} - see \cite{Berti:2005gp} for a thorough review. Again, to calculate the next terms in \eqref{eq:3.65}, we need to find more terms in the accessory parameter expansion \eqref{eq:3.58}. In other words, we have to consider more terms in the continued fraction expansion. This is done by increasing the value of the convergent of order $N_c$. Finally, the expansion above has various symmetry properties \cite{PhysRevD.73.024013, Berti:2005gp} as presented in \eqref{eq:lambdasym}, namely:
\begin{enumerate}
	\item The eigenvalues for negative and positive \textit{m} are related by ${_s\lambda}_{\ell,m}(a\omega) = {_s\lambda}^{*}_{\ell,-m}(a\omega)$.
	\item Eigenvalues for negative and positive \textit{s} are related by ${_{-s}\lambda}_{\ell,m}(a\omega)={_s\lambda}_{\ell,m}(a\omega)+2s$.
	\item If $\omega$ and ${_{-s}\lambda}_{\ell,m}(a\omega)$ correspond to a solution for given $(s,\ell,m)$, another solution can be obtained by the following replacements: $m\rightarrow -m$, $\omega \rightarrow -\omega^{*}$ and ${_{-s}\lambda}_{\ell,m}(a\omega) \rightarrow {_{-s}\lambda}^{*}_{\ell,-m}(a\omega)$.
\end{enumerate}

\section{Radial Teukolsky Master Equation}
\label{sec:secradeq}

As discussed in Chapter \ref{chap1}, the radial equation \eqref{eq:radialeq} can be brought to the canonical form of the CHE by using change of variables:
\begin{equation}
{}_{s}R_{\ell m}(r)=R(r)
=(r-r_{-})^{-(\theta_{{\mathrm{Rad}},0}+s)/2}(r-r_{+})^{-(\theta_{{\mathrm{Rad}},z_0}+s)/2}
y(z),
\quad\quad z=2i\omega(r-r_{-}),
\end{equation}
with
\begin{equation}
\frac{d^2 y}{dz^2}+\bigg[\frac{1-\theta_{{\mathrm{Rad}},0}}{z}+\frac{1-\theta_{{\mathrm{Rad}},z_0}}{z-z_0} \bigg]\frac{d y}{dz}-\bigg[\frac{1}{4}+\frac{\theta_{\mathrm{Rad},\star}}{2z}+\frac{z_0 c_{\mathrm{Rad},z_0}}{z(z-z_0)}\bigg]y(z)=0
\label{eq:radCHE}
\end{equation}
where the set of parameters $\{\theta\}_{\mathrm{Rad}}$, written in the dictionary of the Table \ref{tab:CHEDic} and labeled by "Rad", are given by
\begin{gather}
	\theta_{{\mathrm{Rad}},0}= s -
	i \frac{\omega-m\Omega_{-}}{2\pi T_-}, \quad
	\theta_{\mathrm{Rad},z_0}= s +
	i \frac{\omega-m\Omega_{+}}{2\pi T_+},\quad
	\theta_{\mathrm{Rad},\star}=2s-4iM\omega,\\
	2\pi T_{\pm} = \frac{r_+-r_-}{4Mr_{\pm}}, \quad\quad
	\Omega_{\pm} = \frac{a}{2Mr_{\pm}}, \quad \quad r_{\pm}=M\pm\sqrt{M^2-a^2},
	\label{eq:singlemonosr}
\end{gather}
with the accessory parameter and modulus for the radial equation expressed by
\begin{equation}
\begin{gathered}
z_{\mathrm{Rad}}=z_0=2i(r_+-r_-)\omega,\\
{z}_{0}c_{\mathrm{Rad},z_{0}} =
{_s\lambda}_{\ell,m}+4s-2i(1+2s)M\omega-is
(r_+-r_-)\omega+(M^2a^2-M(r_-+3r_+))\omega^2,
\label{eq:accessoryradialr}
\end{gathered}
\end{equation}
where ${}_{s}\lambda_{\ell m}$ is a function of $a\omega$, as expressed in \eqref{eq:3.65}. We remark that, the $\theta_{\mathrm{Rad}}$'s chosen for $\{\theta\}_{\mathrm{Rad}}$ are in according with the QNMs boundary conditions \eqref{eq:boundKerr1}, i.e. the condition of outgoing and ingoing waves at infinite and event horizon of the black hole, respectively.

Now, a suitable parametrization to treat the extremal limit $a\rightarrow M$ in Kerr BH is
\begin{equation}
\sin \nu = \frac{r_+-r_-}{r_++r_-}=\frac{r_+-r_-}{2M},\qquad
a= \sqrt{r_+r_-}=M\cos\nu,\qquad
\nu \in [0,\pi/2],
\label{eq:nuparameter}
\end{equation}
with the parameters \eqref{eq:singlemonosr} and \eqref{eq:accessoryradialr} rewritten as
\begin{equation}
\begin{gathered}
\label{eq:singlemonos}
\theta_{\mathrm{Rad},0}=s-i\frac{2(1-\sin\nu)M\omega-m\cos\nu}{\sin\nu},\quad
\theta_{{\mathrm{Rad}},z_0}=s+i\frac{2(1+\sin\nu)M\omega-m\cos\nu}{\sin\nu},\\
\theta_{\mathrm{Rad},\star} = 2s-4iM\omega,
\end{gathered}
\end{equation}
and
\begin{equation}
\begin{aligned}
z_0c_{\mathrm{Rad},z_0}={_s\lambda_{\ell,m}}+4s-2i(1+(2+\sin\nu)s)&M\omega-
(3+4\sin\nu+\sin^2\nu)(M\omega)^2 \\ z_0 = z_{\mathrm{Rad}}=-4iM&\omega\sin\nu,
\end{aligned}
\label{eq:accessoryradial}
\end{equation}
where, $\theta_{\mathrm{Rad},z_0}+\theta_{\mathrm{Rad},0}=2s+4iM\omega$ has no explicit dependence on $\nu$, and the $\nu$ parameter defined by \eqref{eq:nuparameter} correlates with the black hole temperature as $\nu\rightarrow 0$:
\begin{equation}
2\pi T_+ = \frac{r_+-r_-}{4Mr_+}=\frac{\sin\nu}{2M(1+\sin\nu)},
\end{equation}
or, for $r_{+}\simeq r_{-}$
\begin{equation}
2\pi T_+ =\frac{1}{2M}\nu+\mathcal{O}(\nu^{2}).
\end{equation}

With all parameters of the radial TME defined, we can bring the numerical results for the QN frequencies obtained from the RH map \eqref{eq:tauVcond}. Initially, we will focus on the frequencies related to gravitational perturbations, as well as their behavior for $a\rightarrow M$. It will be also shown that, for the scalar case, the Riemann-Hilbert map can be used to study the behavior of the overtones ($n>1$) in the limit $a\rightarrow M$. In the final part, the accessory parameter expansion for large $t$ (i.e. large $z_0$), derived in \eqref{eq:contfracINF}, is used to plot a contour map for the first overtones, for the case $a/M=0$, i.e. Schwarzschild black hole, providing a better insight of the maps in Fig. \ref{fig:maps}.

\section{\texorpdfstring{Numerical Results for $0\leq a \leq M$}%
{}}
\label{sec:generica}

In contrast to the angular TME, where the eigenvalue problem is solved by imposing the well behavior of the solution $y(z)$ around the regular points at $z=0$ and $z_0 =-4a\omega$. One has, for the radial TME, that the quasinormal modes are solutions that satisfy the boundary conditions between the regular singularity at $z_{0}(=2i\omega(r_+-r_-))$ and the irregular point of rank 1 at infinity, as defined in \eqref{eq:boundKerr1}. In this case, the eigenvalue problem for the frequencies can be solved in terms of the isomonodromic $\tau_{V}$-function through of the Riemann-Hilbert map presented in the Chapter \ref{ChapIsoMethod}. Thus, the map \eqref{eq:tauVcond} for the radial TME is written as
\begin{equation}
\begin{aligned}
z_{0}\frac{d}{dz_0}\log\tau_V(\{\theta\}_{\mathrm{Rad}}^-;\ 
\sigma-1,\ \eta;\ z_{0})&-
\frac{\theta_{\mathrm{Rad},0}(\theta_{\mathrm{Rad},z_0}-1)}{2}=
z_{0}c_{\mathrm{Rad},z_0}\\
\tau_V(\{\theta\}_{\mathrm{Rad}};\ \sigma,\ &\eta;\ z_{0})=0,
\end{aligned}
\label{eq:radialsyst}
\end{equation}
where $\{\theta\}_{\mathrm{Rad}}^{-} = \{\theta_{\mathrm{Rad},0},\ \theta_{\mathrm{Rad},z_0}-1,\ \theta_{\mathrm{Rad},\star}+1\}$ with $\eta$ giving in terms of $\{\theta\}_{\mathrm{Rad}}$ and $\sigma$ by the expression \eqref{eq:quantV}:
\begin{equation}
e^{i\pi\eta}=e^{-i\pi\sigma}
\frac{\sin\tfrac{\pi}{2}(\theta_\star+\sigma)}{
	\sin\tfrac{\pi}{2}(\theta_\star-\sigma)}
\frac{\sin\tfrac{\pi}{2}(\theta_t+\theta_0+\sigma)
	\sin\tfrac{\pi}{2}(\theta_t-\theta_0+\sigma)}{
	\sin\tfrac{\pi}{2}(\theta_t+\theta_0-\sigma)
	\sin\tfrac{\pi}{2}(\theta_t-\theta_0-\sigma)},
\label{eq:quantVrad}
\end{equation}
where, in order to simplify the notation, we dropped the index "Rad" in the $\theta$'s of the equation for $e^{i\pi\eta}$ and in \eqref{eq:c5expker}. Again, the equation for $e^{i\pi\eta}$ is derived by imposing the QNMs boundary conditions in the connection matrix defined between the fundamental solutions, for $z=z_0= 2i\omega(r_+-r_-)$ and $z=\infty$. In turn, the accessory parameter expansion derived in Chapter \ref{ChapIsoMethod} is given by
\begin{multline}
	z_0c_{\mathrm{Rad},z_0}=\frac{(\sigma-1)^2-(\theta_t+\theta_0-1)^2}{4}+
	\frac{\theta_\star(\sigma(\sigma-2)+\theta_t^2-\theta_0^2)}{
		4\sigma(\sigma-2)}z_0 \ +\\
	+\left[\frac{1}{32}+\frac{\theta_\star^2(\theta_t^2-\theta_{0}^2)^2}{64}
	\left(\frac{1}{\sigma^3}-\frac{1}{(\sigma-2)^3}\right)
	+\frac{(1-\theta_\star^2)(\theta_0^2-\theta_{t}^2)^2+2\theta_\star^2
		(\theta_0^2+\theta_{t}^2)}{32\sigma(\sigma-2)}\right.
	\\ \left.-
	\frac{(1-\theta_\star^2)((\theta_0-1)^2-\theta_{t}^2)((\theta_0+1)^2-
		\theta_{t}^2)}{32(\sigma+1)(\sigma-3)}\right]z_0^2+
	{ O}(z_0^3),
	\label{eq:c5expker}
\end{multline}
where again the expression above is obtained by assuming an expansion for $c_{z_0}$ at $z_0=0$ in the equation \eqref{eq:contfrac}. 

We have that the system \eqref{eq:radialsyst} can be seen as transcendental equations determining $\omega$ and $\sigma$. Thus, these equations can be solved numerically by the following procedure: We first substitute $z_0$ and the parameters $\{\theta \}_{\mathrm{Rad}}= \{\theta_{{\mathrm{Rad}},0},\theta_{{\mathrm{Rad}},z_0},\theta_{\mathrm{Rad},{\star}}\}$ into the equations for the $\tau_{V}$-function (second equation in \eqref{eq:radialsyst}) and into the expression \eqref{eq:contfrac}. Then, the values for $s$, $\ell$, $m$, and $a/M$ are fixed, which reduces the number of parameters in the equations to four: $M\omega$, ${}_{s}\lambda_{\ell m}$, $\eta$, and $\sigma$. Note that, the $\eta$-dependence is eliminated by \eqref{eq:quantVrad}. Using as guess the values for ${}_{s}\lambda_{\ell m}$ and $M\omega$ listed in  Berti's \href{https://pages.jh.edu/eberti2/ringdown/}{website}\footnote{Berti's website provides a large number of archives, where the values for the QN frequencies associated with gravitational, scalar and electromagnetic perturbations can be verified. In the archives only values for $a/M$ from 0 to $0.9999$ are listed -- see \href{https://pages.jh.edu/eberti2/ringdown/}{Ringdown}.}, we search for values of $\sigma \in \mathbb{C}$, in the first locus $-1<\Re \sigma < 1$. This is done by replacing \eqref{eq:accessoryradialr} in \eqref{eq:contfrac}, then a root finding algorithm (Muller's method) is used to compute the roots (i.e.$\sigma$) of the function \eqref{eq:contfrac}. The following step consists in replacing $\sigma$ and $\eta$ into the expression $\tau_V(\{\theta\}_{\mathrm{Rad}};\ \sigma,\ \eta;\ z_{0})=0$. This substitution allows us to arrive at an equation that depends on $M\omega$. Finally, the Muller's method is used to calculate the roots of the $\tau_V$-function, resulting in the QN frequencies.\footnote{Note that, $a/M$ is defined in terms of the mass, the same can be done with $\omega$, simplifying the analysis of the QN Frequencies.}

\begin{figure}[thb]
	\caption{Real (left) and imaginary (right) parts of the fundamental quasi-normal frequency for $s=-2$, $\ell=2$ and $m=2$ (top), $m=1$ (middle) and $m=0$ (bottom). The continuous line shows the numerical results from \eqref{eq:radialsystemeqn} and the dashed
		refers to the results obtained with the continuous fraction method.}
	\begin{center}
		\includegraphics[width=0.95\textwidth]{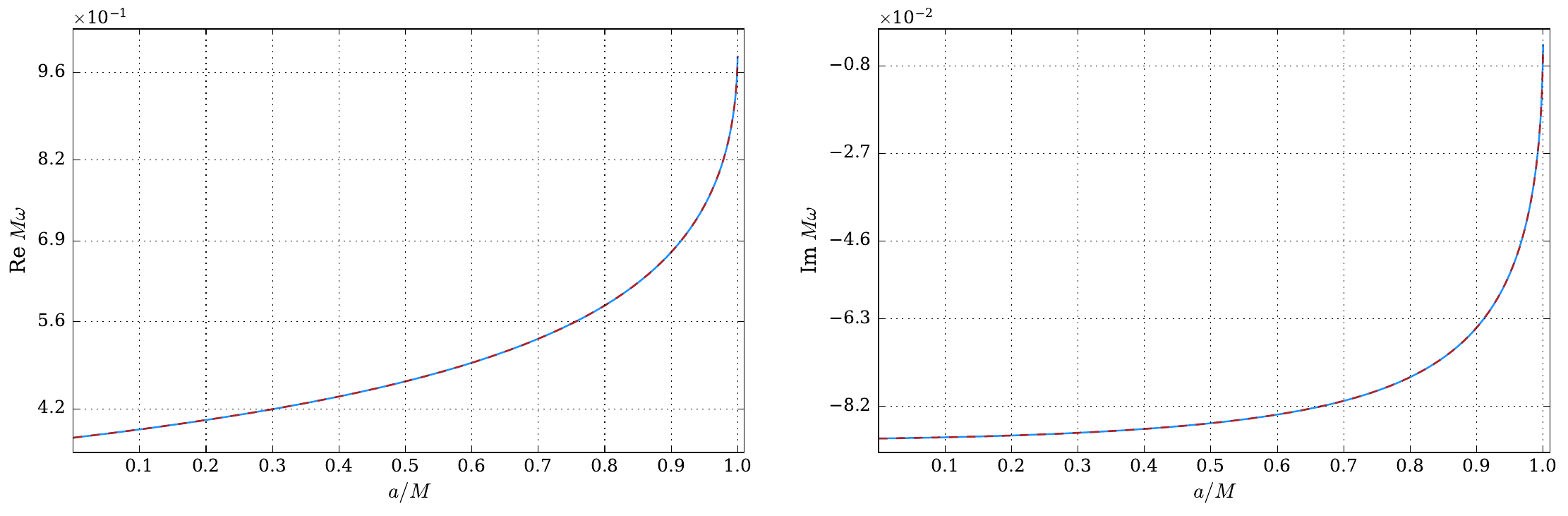}
		\includegraphics[width=0.95\textwidth]{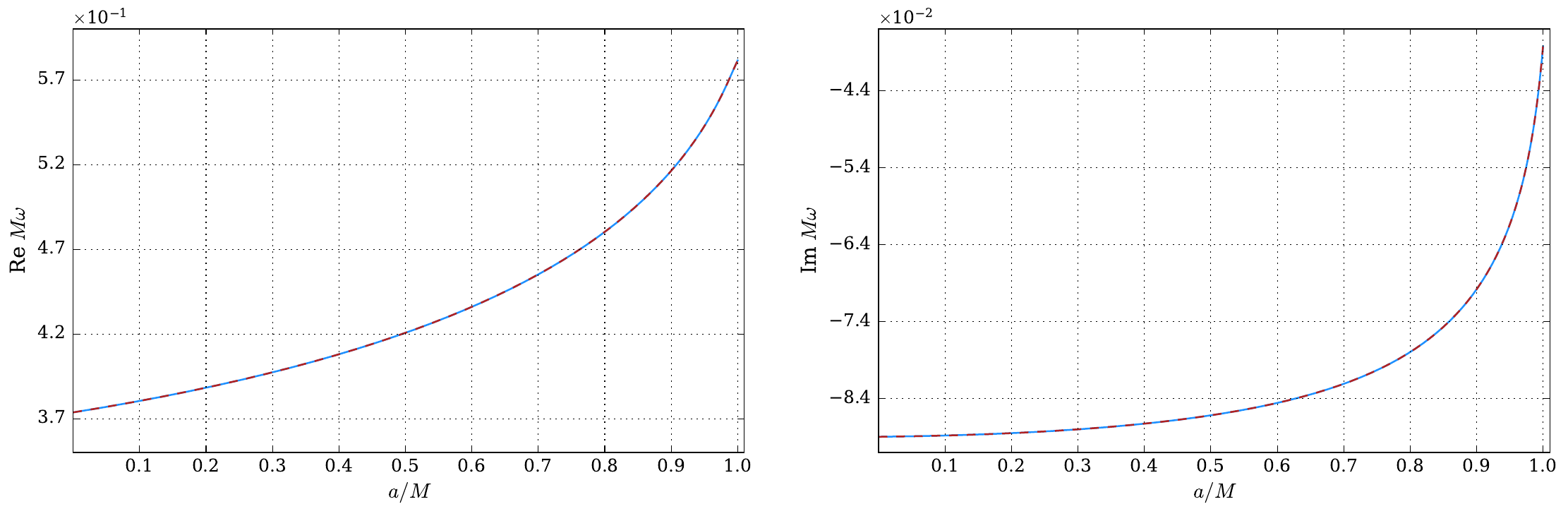}
		\includegraphics[width=0.95\textwidth]{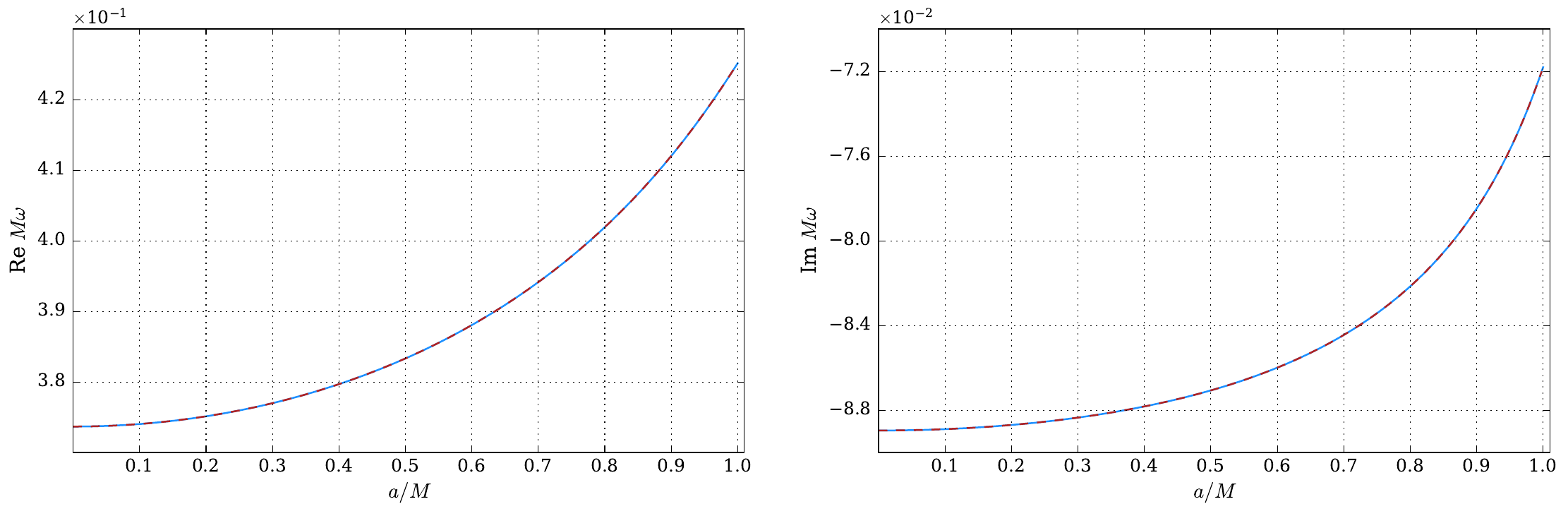}
	\end{center}
	\label{fig:s-2l2m0}
\end{figure}

\begin{figure}[htb]
	\centering
	\caption{The real (left) and imaginary (right) parts of the relative difference between the isomonodromic method based and the continued fraction method for the $s=-2$, $\ell=2$ and $m=2$. We note that the agreement worsens as $a\rightarrow M$, where the expansion parameter $t$ of $\tau_V$ is smaller.}
	\begin{center}
		\includegraphics[width=0.99\textwidth]{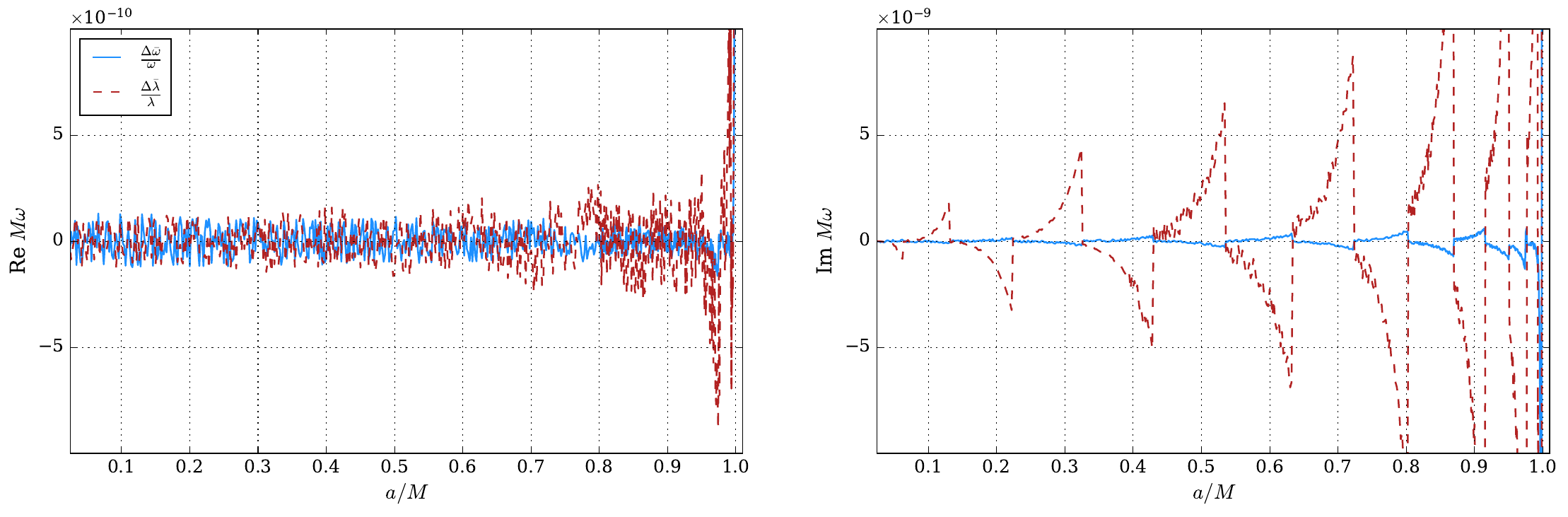}
	\end{center} 
	\label{fig:difference}
\end{figure}

Regarding the Fredholm determinant of the $\tau_{V}$-function, the computational implementation was done in \href{http://julialang.org}{Julia} language using \href{https://github.com/JeffreySarnoff/ArbNumerics.jl}{ArbNumerics}, a port of the \href{https://arblib.org/index.html}{Arb} C language library for arbitrary-precision, with $192$-digit accuracy. The determinant in \eqref{eq:fredholmV} was truncated at $N_f=64$ Fourier decomposition, whereas ${}_{s}\lambda_{\ell m}$ for the angular TME and the accessory parameter for the radial TME are computed from \eqref{eq:contfrac} by fixing the level of convergence $N_c$ of the  continued fraction in 128, $N_c = 128$. In Fig. \ref{fig:s-2l2m0}, we compare the results with those using the continued fraction method \cite{Berti:2005ys,Berti:2009kk} for
$s=-2$ and $\ell=2$. For all fundamental and overtone modes studied, there was good agreement with the modes studied and
listed on Berti's website, except for $a/M>0.9999$, where the continued fraction method (Leaver's method) has a slow convergent behavior as $a\rightarrow M$. For the plots above, one has that in order to make sure we are following the right root, an initial guess for $a/M=0$ is made based on the results in the \href{https://pages.jh.edu/eberti2/ringdown/}{literature}, then we calculate the next frequency for $a/M\neq0$ using the previous value for $M\omega$. We have considered $10^{-3}$ as increment for $a/M$, with $a/M \in [0,1]$. The script used to compute the quasinormal frequencies is available in \cite{GithubIM}. In Appendix \ref{sec:tools}, we give the relevant equations written in the scripts. The difference between the frequencies computed using the isomonodromy method (i.e. the map \eqref{eq:radialsyst}) and Leaver's method is shown in Fig. \ref{fig:difference}, with the trend that the difference becomes smaller as one approaches the extremal limit. We will see below in more detail the dependence of frequencies as $a \rightarrow M$ ($r_-\rightarrow r_+$), as well as the results for the QNMs in the extremal case ($a=M$).

\vspace{-1.0cm}
\hspace{0.0cm}\textbf{Data:} $\omega$ and $\lambda$ are values extracted from the Ringdown's website -- see \href{https://pages.jh.edu/eberti2/ringdown/}{Ringdown} 
\begin{equation}
\frac{\Delta \bar{\omega}}{\omega} = \frac{\omega-\omega_{RHm}}{\omega}, \qquad \frac{\Delta \bar{\lambda}}{\lambda} = \frac{\lambda-\lambda_{RHm}}{\lambda}
\end{equation}

In addition, the theorem on the Painlevé property of the isomonodromic $\tau$-functions assures that the roots sought for $\tau_V(\{\theta\}_{\mathrm{Rad}};\ \sigma,\ \eta;\ z_{0})=0$ are isolated \cite{Miwa:1981aa} away from the essential singularity at $z_0=0$. Near that point, there is an accumulation of roots, which complicates the numerical analysis in the extremal limit. Moreover, for a fixed $a/M$, with $\sigma$ calculated from equation \eqref{eq:contfrac}, one has that the zeros of the $\tau_{V}$-function are associated with the fundamental and overtone modes, resulting in an efficient form of computing the QN frequencies for any value of the overtone $n$.

\subsection{Overtones and comments about QNMs}

\begin{figure}[htb]
	\centering
	\caption{Fig. \ref{fig:berti1}: Fundamental gravitational mode with $\ell=2$. We mark by dots the points corresponding to $a/M = 0, 0.1,0.2,...,1.0$, the meeting point represents the case $a/M =0$. Note that, as discussed in Sec. \eqref{subsec:overtones} the imaginary part for positive and negative $m$ has the same value, corroborating with equation \eqref{eq:pm_m}. Fig. \ref{fig:berti}: It is shown the first seven overtones $n=1,...,7$, for $s=-2$, $\ell=2$ and $m=2$. }
	\begin{subfigure}{.5\textwidth}
		\centering
		\includegraphics[width=0.99\textwidth]{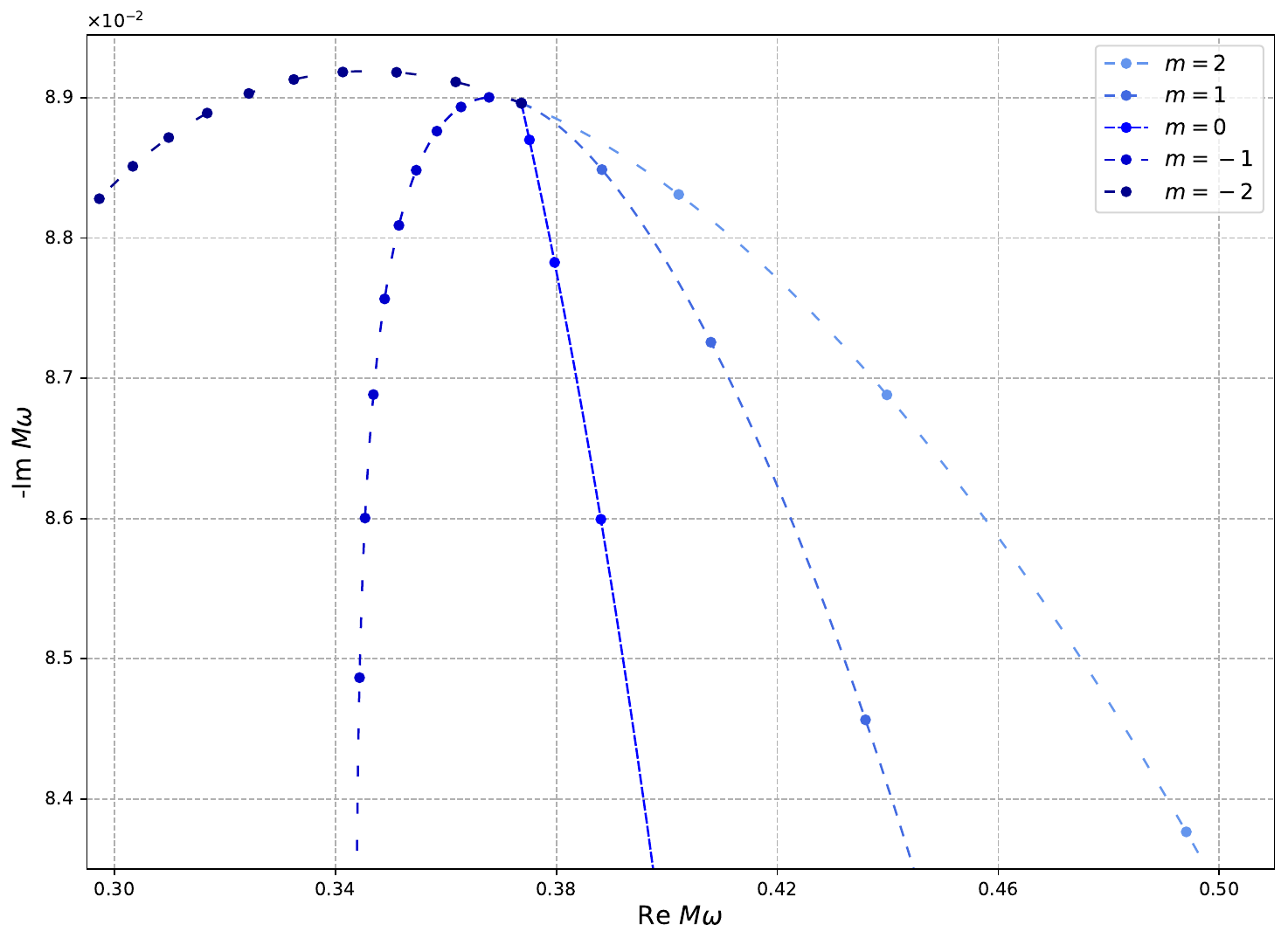}
		\caption{Splitting of the fundamental mode.}
		\label{fig:berti1}
	\end{subfigure}%
	\begin{subfigure}{.5\textwidth}
		\centering
		\includegraphics[width=0.99\textwidth]{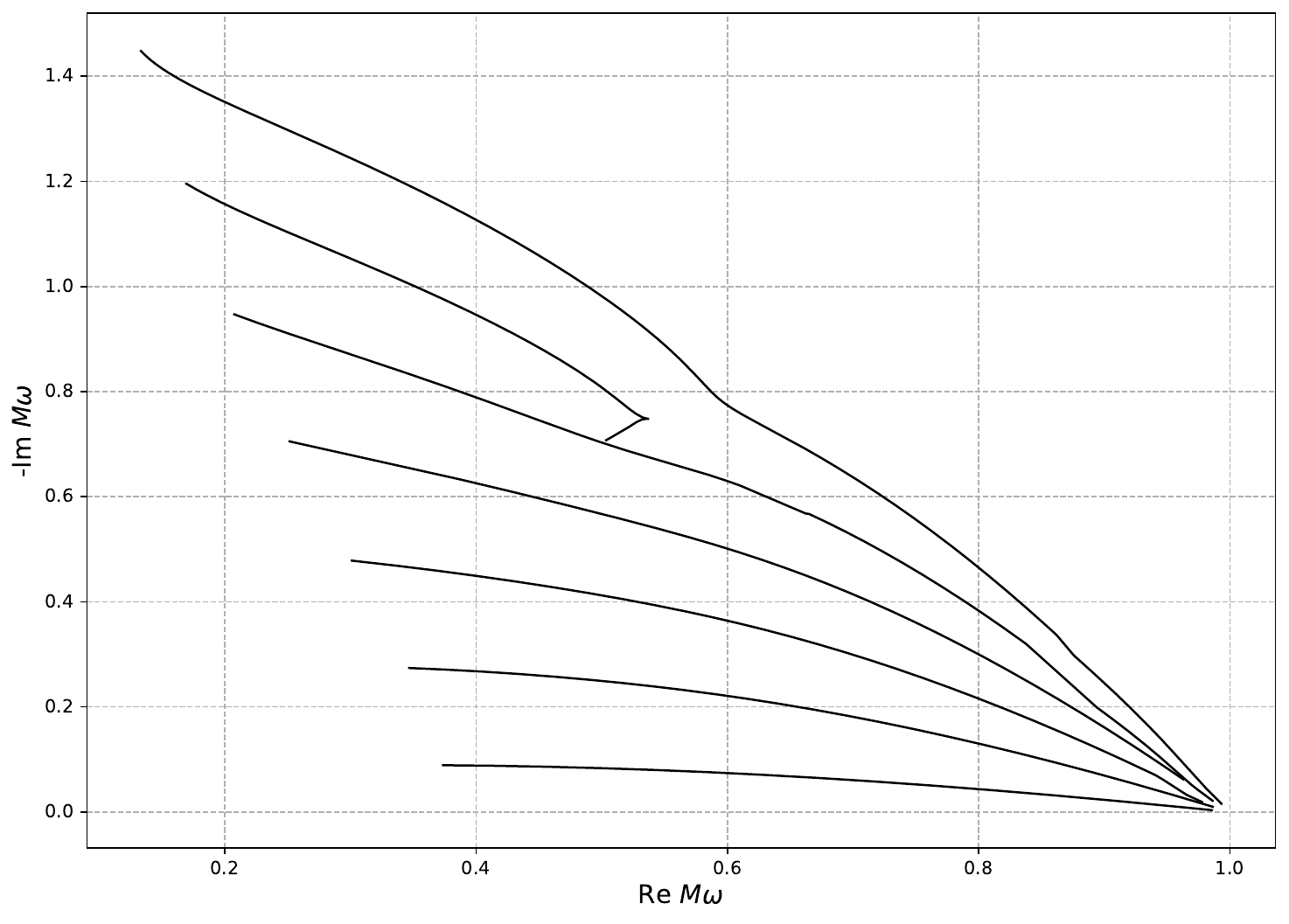}
		\caption{First overtones for $s =-2,\ \ell= m = 2$.}
		\label{fig:berti}
	\end{subfigure}\\
	Reproduced based on \cite{Berti:2009kk}.
\end{figure}

It is customary to plot only frequencies for positive values of the azimuthal number $m$ \cite{Leaver:1985ax}. However, in order to illustrate the relations between the QN frequencies for both positive and negative $m$, and follows the discussion made in Sec. \ref{subsec:overtones}, we show in Fig. \ref{fig:berti1} the modes for $s=-2$, $\ell=2$ with $m=0,\pm1,\pm2$. Note that, as the rotation parameter $a/M$ increases, the branches with $m=2$ and $m=-2$ move in opposite directions, being tangent to the branches with $m=1$ and $m=-1$ in the limit $a/M \rightarrow 0$. In this way, the most relevant feature of the Kerr spectrum is that rotation acts very much like an external magnetic field on the energy levels of an atom, causing a "Zeeman splitting"\footnote{We recomend to the reader the good review of Zeeman effect \cite{cohen2019quantum} } of the QN frequencies, as it is shown in Fig. \ref{fig:berti1}. In other words, when $a/M\neq0$ the values for $m=0,\pm1,\pm2$ splits. In Fig. \ref{fig:berti}, the first seven overtones are plotted, with the characteristic behavior for the overtone $n=6$ appearing naturally, corroborating with \cite{Berti:2009kk}.

\begin{figure}[ht]
	\centering
	\caption{First overtones for scalar perturbations, where we are assuming $s=\ell=m=0$. One has, therefore, that when the overtone $n$ increases and $a\rightarrow M$ the frequency has a spiral behavior. Similar behavior is observed in \cite{Cardoso:2003pj}, but now with more precision for $a \rightarrow M$, as observed for $n=2$.}
	\begin{subfigure}{.5\textwidth}
		\centering
		\includegraphics[width=.99\linewidth]{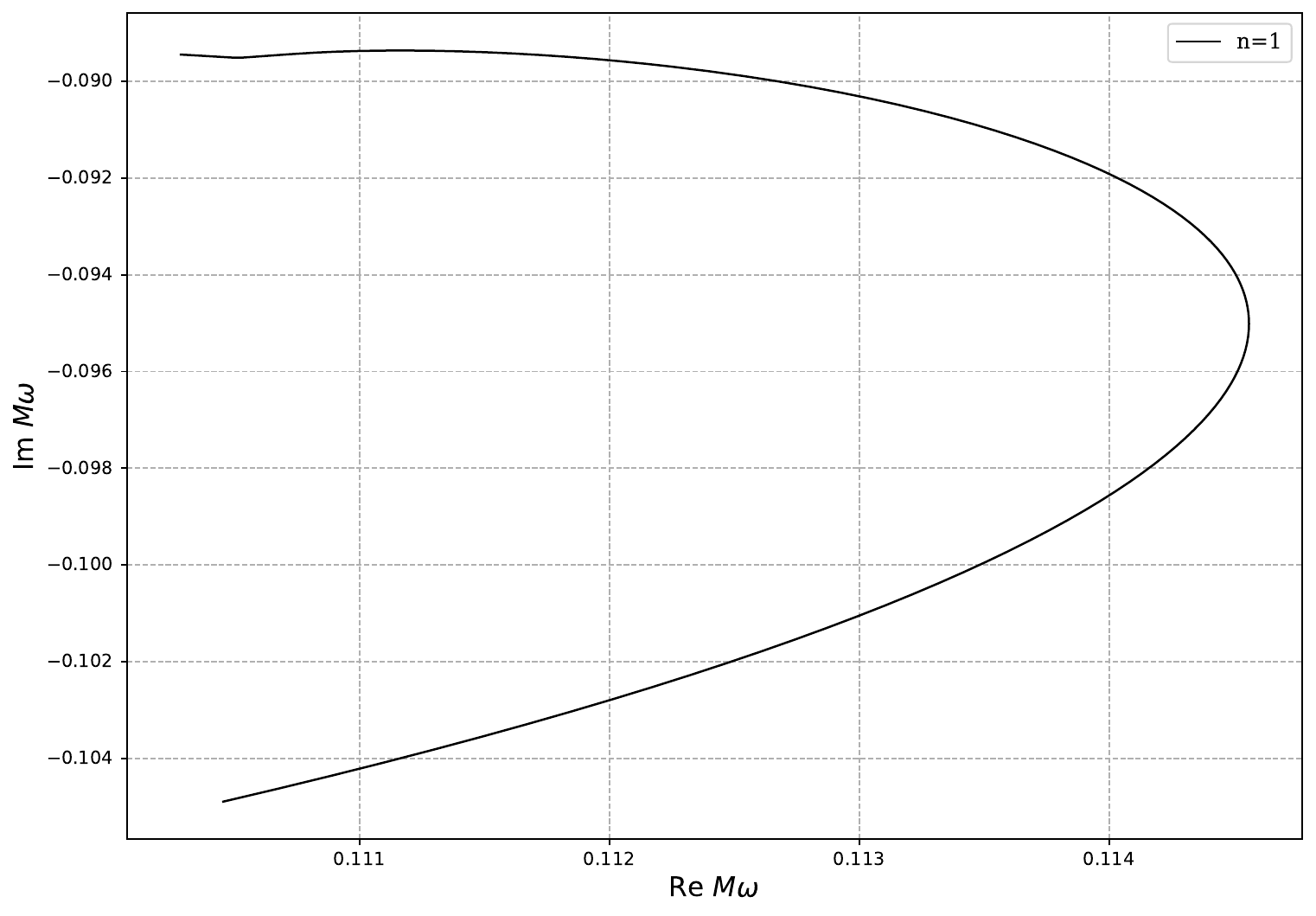}
	\end{subfigure}%
	\begin{subfigure}{.5\textwidth}
		\centering
		\includegraphics[width=.99\linewidth]{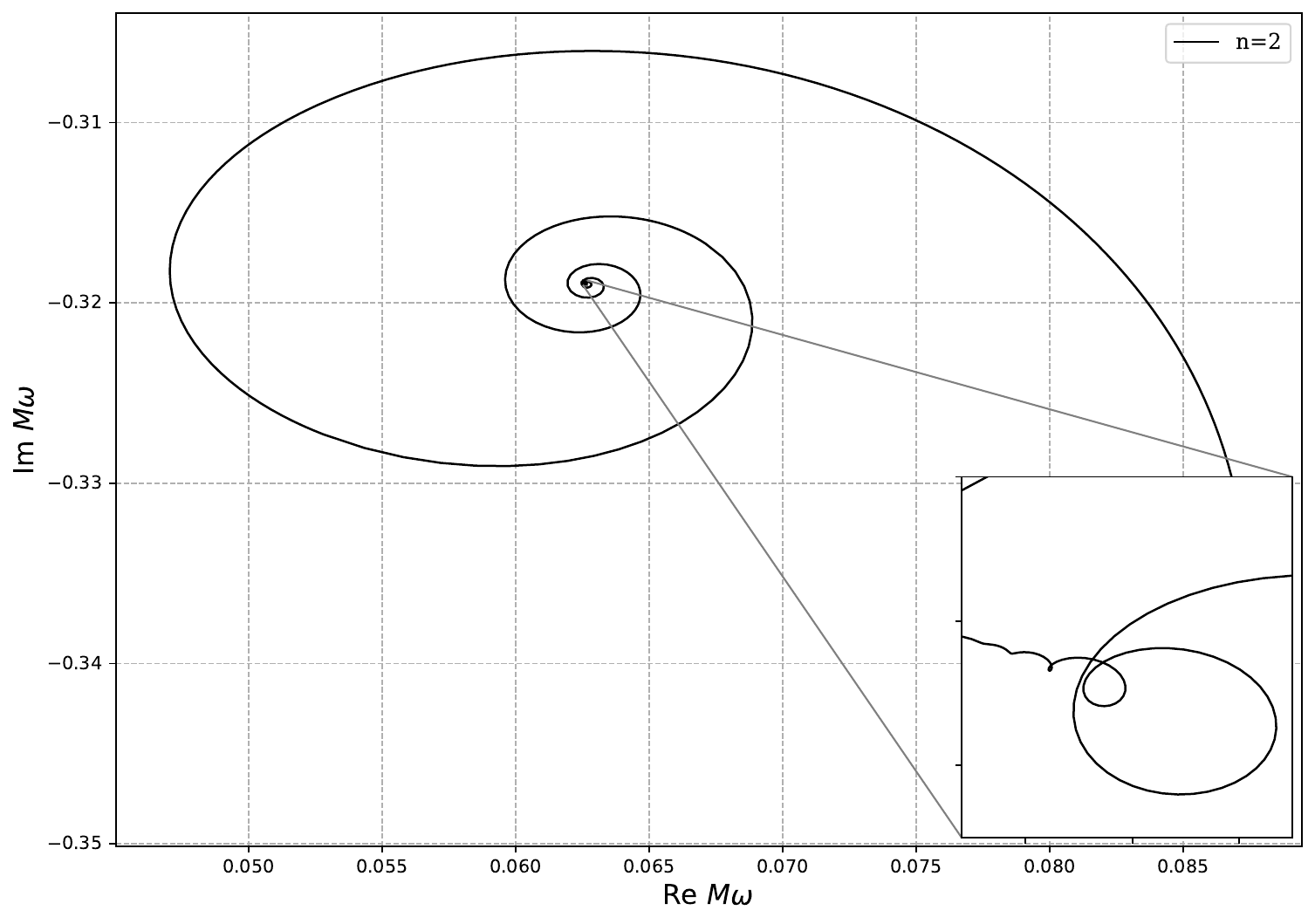}
	\end{subfigure}
	\label{fig:s0l0m0_n2}
\end{figure}

From Figs. \eqref{fig:s0l0m0_n2} and \eqref{fig:berti}, one has that the RH map can be applied not only to fundamental modes, but also for different values of the overtone $n$. Furthermore, given the numerical control of the isomonodromic $\tau_{V}$-function via the Fredholm determinant, it is possible to analyse, with great precision, how the overtones behave in the extremal limit. As it can be seen in Fig. \eqref{fig:s0l0m0_n2} for the overtone $n=2$.

\section{\texorpdfstring{The $a\rightarrow M$ Limit}%
{}}
\label{sec:extremallimit}

The extremal limit, which consists in the process of coalescence of the singularities $r_{\pm}$ (i.e. $r_- \rightarrow r_+$) can be studied using the $\tau_V$-function by taking the appropriate limit of the conditions \eqref{eq:radialsyst}. Moreover, the limit provides analytic tools to study the extremal case.

Let us first note that the angular equation has a smooth limit of the parameters \eqref{eq:singlemonoangular} and \eqref{eq:accessoryangular}, which means that when $a\rightarrow M$, we just need to replace $a$ by $M$ in the parameters of the angular equation \eqref{eq:angulareq}, and the expansion for ${}_{s}\lambda_{\ell m}$ will be the same as \eqref{eq:3.65}, with $a=M$.

The radial equation, on the other hand, will have a more complex limit depending on the particular mode. Given the $\nu$ parameter defined in \eqref{eq:nuparameter},  we now define in terms of $\nu$ the confluence parameter
\begin{equation}
\Lambda = \frac{\theta_{z_0}-\theta_0}{2} =
i\frac{2M\omega-m\cos\nu}{\sin\nu},
\label{eq:lambdaparameter}
\end{equation}
which, for $a\rightarrow M$, behaves as
\begin{equation}
\Lambda=\frac{i(2M\omega-m)}{\nu}+\frac{1}{3}(m+M\omega)\nu + \mathcal{O}(\nu^2),
\label{eq:4.36}
\end{equation}
and the new modulus of the equation in this limit is given by
\begin{equation}
\begin{aligned}
u_0 = \Lambda z_0
&=4M\omega(2M\omega-m)+4m\,M\omega \left(1-\frac{a}{M}\right)\\
&=4M\omega (2M\omega - m)+4m\,M\omega (1-\cos\nu).
\end{aligned}
\end{equation}
We observed two distinct behaviors for $M\omega$ as we approach the
extremal limit $\nu\rightarrow 0$:
\begin{enumerate}
	\item[\textbf{A.}] $M\omega$ converges to $m/2$ with $\nu$ or higher order as $\nu\rightarrow 0$. In this case the confluence parameter has a	finite limit, and the system is actually well described by \eqref{eq:radialsyst};
	\item[\textbf{B.}] $M\omega$ does not go to $m/2$ as $\nu\rightarrow 0$. In this case, the confluence parameter $\Lambda$ will diverge and one has	to consider the confluent limit of the equations for the RHm \eqref{eq:radialsyst}.
\end{enumerate}
Let us now analyze each case separately.

\subsection{\texorpdfstring{The Finite $\Lambda$ Limit}%
{}}
\label{sec:smallnu}

We observed numerically that for the modes $\ell=m$, with $m\neq 0$, the eigenfrequencies tend to $m/(2M)$ in the extremal limit corroborating with the literature \cite{Berti:2009kk} and as observed for the overtones in Fig. \ref{fig:berti}. To describe the behavior of the solutions of \eqref{eq:radialsyst} in this limit, we propose the expansion
\begin{equation}
M\omega = \frac{m}{2}+\beta_1\nu+\beta_2\nu^2+\ldots,\qquad
\sigma = 1+\alpha,\quad
\alpha=\alpha_0+\alpha_1\nu+\alpha_2\nu^2+\ldots,
\label{eq:sigmazexpansion}
\end{equation}
where $\nu \simeq4M\pi T_+$, and the coefficients $\beta_i$ and $\alpha_i$ can be computed recursively from the equations \eqref{eq:c5expker} and \eqref{eq:zerochi5}, in Appendix \ref{sec:tools}. The consideration of the series is simplified from the fact that the expansion parameter $z_0$ of the expressions \eqref{eq:c5expker} and \eqref{eq:chi5} is small near the extremality,
\begin{equation}
z_0 = -4iM\omega\sin\nu = -2im\nu+{ O}(\nu^2),
\end{equation}
and $\Lambda$ defined through \eqref{eq:lambdaparameter} is finite and goes as $\Lambda \simeq \frac{1}{3}(m-M\omega)\nu+\mathcal{O}(\nu^2)$. We then have that each term of the expansions \eqref{eq:c5expker} and \eqref{eq:chi5} is finite and the term of order $z_0^n$ is of order $\nu^n$.

In terms of $\nu$, the accessory parameter for the radial equation $c_{\mathrm{Rad},z_0}$ is
\begin{equation}
z_0c_{\mathrm{Rad},z_0} = {_s\lambda_{\ell,m}}+4s-2i(1+2s)M\omega -3M^2\omega^2-
2(isM\omega+2M^2\omega^2)\nu-M^2\omega^2\nu^2+{O}(\nu^3).
\label{eq:radC0}
\end{equation}
Finally, since the angular eigenvalue ${}_{s}\lambda_{\ell m}$ depends on $a=M\cos\nu$, we will also expand \eqref{eq:3.65} in terms of $\nu$,
\begin{equation}
{_s\lambda_{\ell,m}}=\lambda_0+\lambda_1\nu+\lambda_2\nu^2+\ldots.
\end{equation}
where we note that, despite being an expansion in $a\omega=M\omega\cos\nu$, it can have odd terms in $\nu$ through its
dependence with $M\omega$. For $\nu \rightarrow 0$, the first term is derived from \eqref{eq:3.65}, resulting in $\lambda_0 = {_s\lambda_{\ell,m}}(m/2)$.

\begin{figure}[htb]
	\centering
	\caption{The near-extremal behavior for the fundamental quasi-normal frequency for $s=0$, $\ell=m=1$ (top) and $s=-2$, $\ell=m=2$
		(bottom). Both refer to the extremal behavior described in Sec. \ref{sec:smallnu}. In the top case, the $\alpha_0$ parameter
		is real, and near-horizon corrections to the real part of the frequency are of higher order in $\nu$. In the bottom, $\alpha_0$ is imaginary	and corrections to both real and imaginary parts are linear.}
	\begin{center}
		\includegraphics[width=0.95\textwidth]{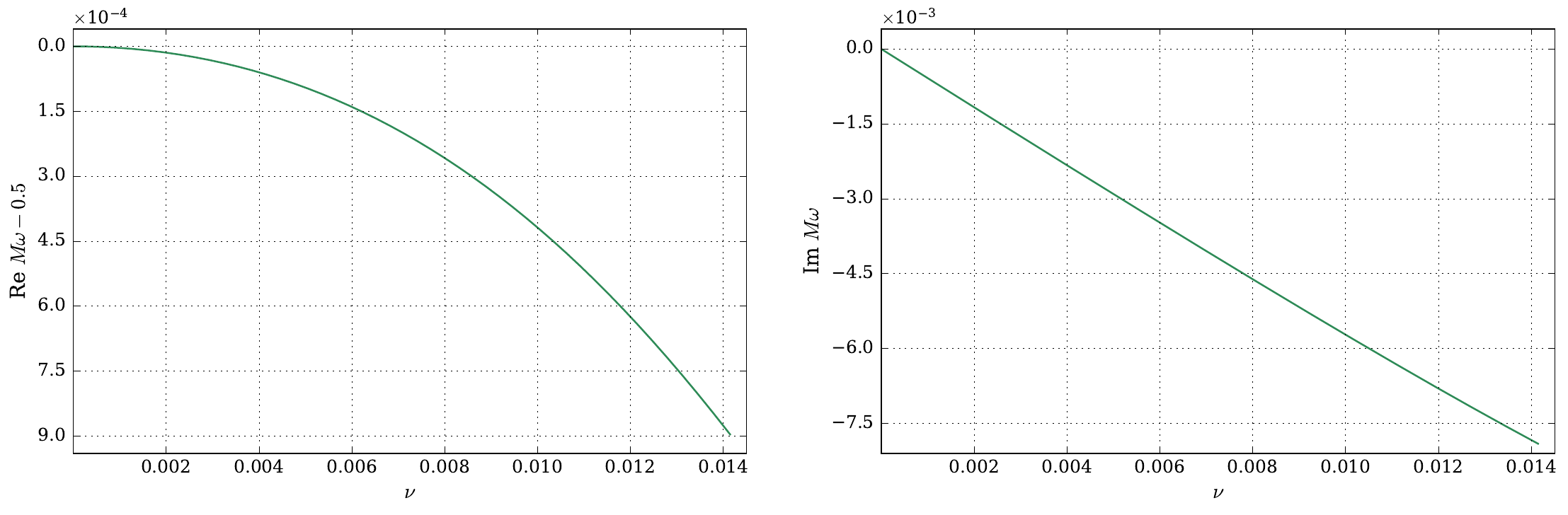}
		\includegraphics[width=0.95\textwidth]{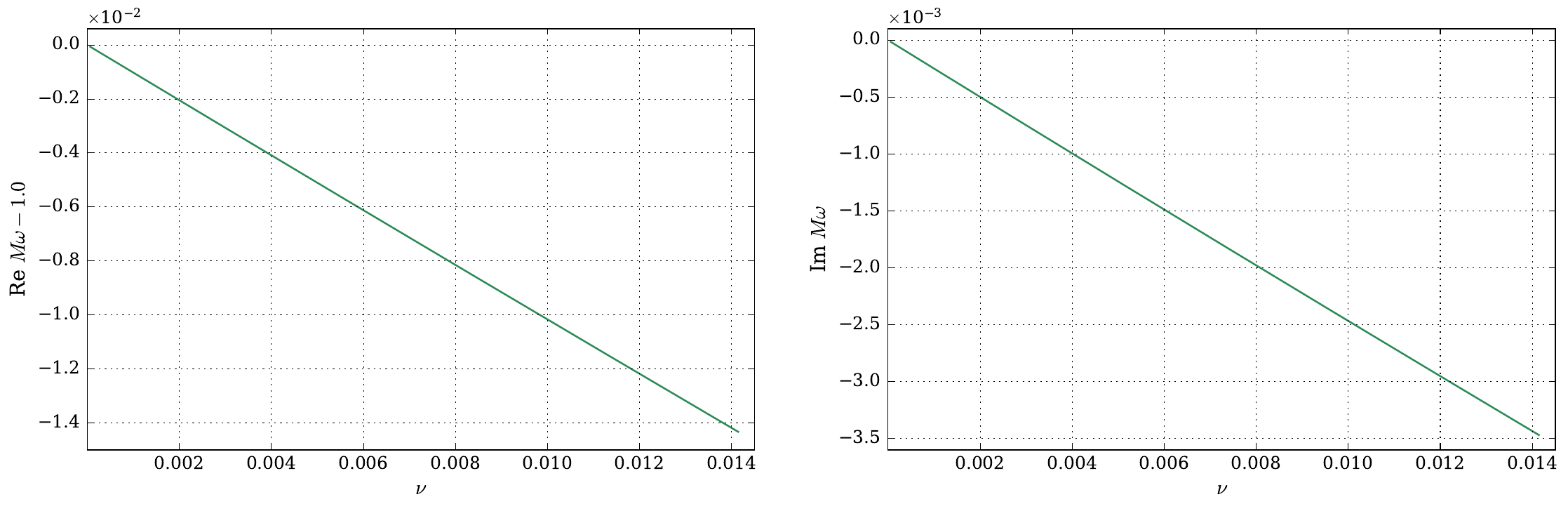}
	\end{center}
	\label{fig:smallnu}
\end{figure}

Substituting the series for $M\omega$ and $\sigma$ into the expansion for the accessory parameter \eqref{eq:c5expker} and taking \eqref{eq:radC0}, we can relate the coefficients $\alpha_i$ with $\beta_i$ and $\lambda_i$, and compute them recursively. The first terms are:
\begin{equation}
\alpha_0 = \pm\sqrt{1+4\lambda_0+4s(s+1)-7m^2},
\qquad
\alpha_1=\frac{2m
	(28\lambda_0+36s^2+28s-41m^2)}{
	\alpha_0(1-\alpha_0^2)}\beta_1+\frac{2\lambda_1}{\alpha_0}.
\label{eq:alphaexpansion5}
\end{equation}
The sign of $\alpha_0$ chosen will actually depend on the mode. We can now substitute the expansions \eqref{eq:sigmazexpansion} into \eqref{eq:zerochi5} and find 
a transcendental equation for the $\beta_i$. Supposing $\sigma=1+\alpha_0$, with $-1<\Re \alpha_0<1$, the first non-trivial term from \eqref{eq:zerotau5p} is 
\begin{equation}
\begin{aligned}
i^{\alpha_0}\frac{\Xi(\alpha_0,\beta_1,s,m)}{\Xi(-\alpha_0,\beta_1,s,m)}
(2m\nu)^{\alpha_0}=
1+{ O}({\nu},\nu\log\nu),
\label{eq:quantbeta}
\end{aligned}
\end{equation}
where $\Xi(\phi,\beta_1,s,m)$ is given by the product of Gamma functions, 
\begin{equation}
\Xi(\phi,\beta_1,s,m)= 
\Gamma(1-\phi)^2\Gamma(\tfrac{1}{2}(1+\phi)-2i\beta_1)\Gamma(\tfrac{1}{2}(1+\phi)-s-im)\Gamma(\tfrac{1}{2}(1+\phi)-s+im).
\end{equation}

The expansion of $\chi_V$ in \eqref{eq:zerotau5p} is analytic in $\nu$, whereas the expansion of the $t^{\sigma-1}$ and
$\Theta_V$ will include non-analytic terms like $\nu\log\nu$. As one takes $\nu$ to zero, the term $\nu^{\alpha_0}$ above will
go to zero if $\Re \alpha_0>0$, and the only way to satisfy the equation will be if one of the gamma functions' arguments in the
numerator becomes very close to zero. In this way, we have that
\begin{equation}
\beta_1=-\frac{i}{4}(1+\alpha_0)
+\frac{i}{2} \frac{e^{i\frac{\pi}{2} \alpha_0}}{
	\Gamma(-\alpha_0)}\frac{\Upsilon(\alpha_0,m,s)}{\Upsilon(-\alpha_0,m,s)}
(2m\nu)^{\alpha_0}
+\ldots
\label{eq:solbeta1}
\end{equation}
with $\Upsilon(\pm\alpha_0,m,s)$,
\begin{equation}	
\Upsilon(\pm\alpha_0,m,s) =	\Gamma(1\mp\alpha_0)^2 
\Gamma(\tfrac{1}{2}(1\pm\alpha_0)-s-im)
\Gamma(\tfrac{1}{2}(1\pm\alpha_0)-s+im).
\end{equation}

So, if $\alpha_0$ is real, which should be expected if $m$ is small, then one picks the sign so that $\Re\alpha_0>0$. This behavior seems to be restricted to the mode $s=0$, $\ell=m=1$ and it is shown in the top part of Fig. \ref{fig:smallnu}. Note that in this case the correction to the real part of the frequency is of higher order in $\nu$. When $m$ is large enough, one expects $\alpha_0$ to be purely imaginary, and then the $\nu$-dependent term in \eqref{eq:solbeta1} will oscillate logarithmically. This correction term will still be small if $\Im \alpha_0<0$, which selects the negative root in \eqref{eq:alphaexpansion5}. The behavior is much more common and it is represented in the bottom part of Fig. \ref{fig:smallnu}, where both real and imaginary parts of the frequency display the linear behavior with $\nu$. Furthermore, in the case where $\alpha_0$ is purely imaginary, the imaginary slope is approximately $-\nu/4$. These modes are stable, but their decay time diverges as one approaches the extremal limit. 

Finally, we note that the equation \eqref{eq:quantbeta} admits infinite roots of the sort $\beta_1 \rightarrow \beta_1-in/2$,
corresponding to the poles of the gamma function at negative integral values of the argument. This facts leads to the accumulation of zeros of the $\tau_V$-function. Similar remarks in the same context were made from the continuous fraction expansion in the papers \cite{Richartz:2017qep} and \cite{Casals:2019vdb}, albeit the last one with a slightly different value for $\alpha_0$.

\section{Confluent Limit and Extremal Kerr Black Hole}
\label{sec:painleveIII}

All modes with $m\neq \ell$, including those with negative $m$, will \textit{not} tend to $M\omega=m/2$ in the extremal limit. In this case, the parameter $\Lambda$ \eqref{eq:lambdaparameter} goes off to infinity, and the equation \eqref{eq:radCHE} will undergo a confluent limit in the form
\begin{equation}
\Lambda = \frac{1}{2}(\theta_{\mathrm{Rad},z_0}-\theta_{\mathrm{Rad},0}), \ \ \ \theta_{\mathrm{Rad},\circ}
=\theta_{\mathrm{Rad},z_0}+\theta_{\mathrm{Rad},0}, \ \ \ u_0 = \Lambda z_0 
, \ \ \ \Lambda \rightarrow \infty,
\label{eq:conflimi}
\end{equation}
where for $M\omega\neq m/2$, one has, from \eqref{eq:4.36} that $\Lambda \simeq i(2M\omega-m)/\nu$ will diverge. Thus, when $a=M$ we have that the limit leads to the double-confluent Heun equation (DCHE)
\begin{equation}
\frac{d^2
	y}{dz^2}+\bigg[\frac{2-\theta_{\mathrm{Rad},\circ}}{z}-\frac{u_0}{z^2}\bigg]
\frac{dy}{dz}-\bigg[\frac{1}{4} 
+\frac{\theta_{\mathrm{Rad},\star}}{2z}+\frac{u_0k_{u_0}-u_0/2}{z^2}\bigg]y(z)=0
\label{radDCHE}
\end{equation}
where from the first column in Table \ref{tab:DCHEDic}, we have the following parameters for the equation
\begin{equation}
\begin{aligned}
\theta_{\mathrm{Rad},\star} = -2s+4iM\omega,&\qquad \theta_{\mathrm{Rad},\circ} = 2s+4iM\omega,\\
u_0 = 4M\omega(m-2M\omega),\qquad
u_0k_{u_0} =& {_s\lambda_{\ell,m}}+2s-2i(1+2s)M\omega-3M^2\omega^2,
\end{aligned}
\label{eq:parameterDCHE}
\end{equation}
with ${}_{s}\lambda_{\ell, m}$ given in \eqref{eq:3.65} with $a=M$.

As discussed in Chapter \ref{ChapIsoMethod}, we have that the RHm associated with the DCHE is given in terms of the isomonodromic $\tau_{III}$-function. Thus, from \eqref{eq:tauIIIcond}, we have that the eigenvalue problem for the radial equation can be cast as
\begin{equation}
\tau_{III}(\{\tilde{\theta}\}_{\mathrm{Rad}};\sigma,\eta;u_0)=0,\quad
u_0\frac{d}{du_0}\log\tau_{III}(\{\tilde{\theta}\}_{\mathrm{Rad}}^-;\sigma-1,\eta;u_0)
-\frac{(\theta_\circ-1)^2}{8}-\frac{1}{2}
=u_0k_{u_0},
\label{eq:tauIIIconditions}
\end{equation}
where $\{\tilde{\theta}\}_{\mathrm{Rad}} = \{\theta_{{\mathrm{Rad}},\circ},\theta_{{\mathrm{Rad}},\star}\}$, $\{\tilde{\theta}\}_{\mathrm{Rad}}^{-} = \{\theta_{{\mathrm{Rad}},\circ}-1,\theta_{{\mathrm{Rad}},\star}+1\}$. We solve these conditions by using the same procedure as the one used with the isomonodromic $\tau_{V}$-function defined in the RHm for the non-extremal case \eqref{eq:radialsyst}. In turn, the expansion of the accessory parameter $k_{u_0}$ is given from the confluent limit of $c_{z_0}$,  as discussed in \eqref{subsec:accparDCHE}. Thus, from \eqref{eq:accparinf}:
\begin{multline}
	u_0k_{u_0}=\frac{(\sigma-1)^2-(\theta_\circ-1)^2}{4}
	+\frac{\theta_\circ\theta_\star}{2\sigma(\sigma-2)}u_0
	-\left[\frac{\theta_\circ^2\theta_\star^2}{2\sigma^3(\sigma-2)^3}
	\right. \\ \left.
	+\frac{3\theta_\circ^2\theta_\star^2}{8\sigma^2(\sigma-2)^3}
	-\frac{\theta_\circ^2+\theta_\star^2-\theta_\circ^2\theta_\star^2}{
		8\sigma(\sigma-2)}
	-\frac{(\theta_\circ^2-1)(\theta_\star^2-1)}{8(\sigma+1)(\sigma-3)}      
	\right]u_0^2+\mathcal{O}(u_0^3).
	\label{eq:C3expansion}
\end{multline}

In order to simplify the notation, we have dropped the label "$\mathrm{Rad}$" in the equation above, the same is made with equation \eqref{eq:quanIII}. Finally, the expression for $\eta$ is derived from \eqref{eq:quantVrad} by applying the confluent limit \eqref{eq:conflimi}, where if $\Lambda$ goes to $\infty$ in a ray with argument sufficiently close to $\pi/2$, we have
\begin{equation}
e^{i\pi\eta}=e^{-2\pi i\sigma}
\frac{\sin\tfrac{\pi}{2}(\theta_\star+\sigma)}{
	\sin\tfrac{\pi}{2}(\theta_\star-\sigma)}
\frac{\sin\tfrac{\pi}{2}(\theta_\circ+\sigma)}{
	\sin\tfrac{\pi}{2}(\theta_\circ-\sigma)}
\label{eq:quanIII}
+ \mathcal{O}(e^{2i\Lambda}).
\end{equation}
For our application, this limit should hold if $\Re M\omega > m/2$ in the extremal limit, as it can be seen in \eqref{eq:4.36}. If $\Re M\omega < m/2$, then the argument of the exponential factor should be replaced by $+2\pi i\sigma$. 

\begin{figure}[htb]
	\centering
	\caption{The near-extremal behavior for the fundamental quasi-normal frequency for $s=-1$, $\ell=2$ and $m=1$. We note the roughly quadratic dependence for small $\nu$, which is consequence of the first near-confluent correction to the $\tau_{III}$-function involved.}
	\begin{center}
		\includegraphics[width=0.95\textwidth]{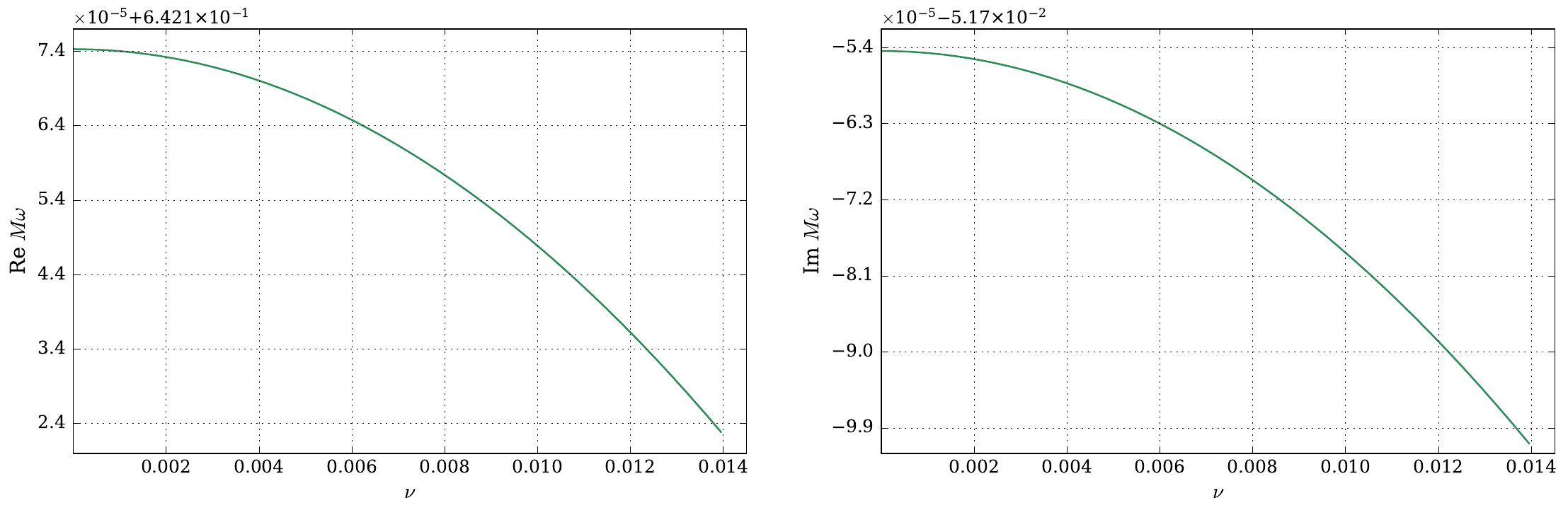}
	\end{center}
	\label{fig:nearextremalqnm}
\end{figure}

Following the same strategy applied to the non-extremal case, we start by fixing the values for $s$, $\ell$, and $m$, and replacing ${}_{s}\lambda_{\ell m}$ in \eqref{eq:parameterDCHE}. Then, the expression \eqref{eq:contfracDCHE}, with $N_c =128$, and the equation \eqref{eq:quanIII} are used to transform the expression $\tau_{III}(\{\tilde{\theta}\}_{\mathrm{Rad}};\sigma,\eta;u_0)=0$ into a single function that depends on the frequency $M\omega$. The final step consists in using Muller's method to compute the roots of the function. Regarding the expansion for the isomonodromic $\tau_{III}$-function, we have implemented the Fredholm determinant \eqref{eq:fredholmIII} using the Julia language port of the Arb C library for arbitrary precision arithmetic, set at $192$-digit accuracy \cite{Github}. In \cite{GithubIM}, the reader can access the script that computes the extremal modes. We also provided a tutorial of how the code works, as well as many archives with results for extremal and non-extremal modes. We present below the values obtained for the fundamental QN frequencies for low values of $\ell$ and $m$, with $10$-digit accuracy.

\setlength\tabcolsep{1.0mm}
{\renewcommand{\arraystretch}{1.2}
	\begin{table}[htb]
		\centering
		\caption{The fundamental mode for scalar $s=0$ perturbations of the
			extremal black hole obtained from solving \eqref{eq:c3expansion}
			and \eqref{eq:tauIIIconditions}. The $\ell=m=1$, $\ell=m=2$ and
			$\ell=m=3$ modes fall into the analysis outlined in Sec. \ref{sec:smallnu}.}
		\scalebox{0.84}{
			\centering
			\begin{tabular}{|c|c|c|}  
				\hline
				$\ell,\,m$ & $M\omega$ & ${}_{0}\lambda_{\ell,m}$ \\  \hline
				$ \ell=0, m=0$ & $0.1102454759 - 0.0894331855i$ & $-0.0013797497 + 0.0065754944i$ \\ 
				\hline $ \ell=1,\, m=0$ & $0.3149861271 - 0.0817137749i$ &
				$1.9444359816 + 0.0309517044i$ \\
				\hline $\ell=1,\,m=-1$ & $0.2394237989 - 0.0938214466i$ &
				$1.9902942278 + 0.0090051928i$ \\
				\hline $ \ell=2,\, m=1$ & $0.6643114515 - 0.0560535566i$ &
				$5.8114879904 + 0.0321717438i$
				\\
				\hline $ \ell=2,\, m=0$ & $0.5241220471 - 0.0813229203i$ &
				$5.8602337552 + 0.0441805059i$ \\
				\hline $\ell=2,\, m=-1$ & $0.4391457612 - 0.0902770953i$ & $5.9207350548 + 0.0340952252i$
				\\
				\hline $\ell=2,\, m=-2$ & $0.3803109539 - 0.0932780508i$ &
				$5.9805544767 + 0.0101732386i$ \\
				\hline $ \ell=3,\, m=2$ & $1.0715947258 - 0.0322379809i$ &
				$11.6146118676 + 0.0233900184i$\\
				\hline $ \ell=3,\, m=1$ & $0.8617579225 - 0.0660049906i$ &
				$11.6562448960 + 0.0528368609i$ \\  
				\hline $ \ell=3,\, m=0$ & $0.7333028611 - 0.0811680413i$ &
				$11.7294008968 + 0.0604329503i$ \\
				\hline $ \ell=3,\, m=-1$ & $0.6433808795 - 0.0883909414i$ &
				$11.8106880688 + 0.0529451496i$\\
				\hline $ \ell=3,\, m=-2$ & $0.5757561619 - 0.0917564731i$ &
				$11.8920985029 + 0.0353732242i$\\
				\hline $ \ell=3,\, m=-3$ & $0.5225142116 - 0.0930632653i$ &
				$11.9705661993 + 0.0108575306i$ \\
				\hline
			\end{tabular}
		}
		\label{tab:scalarextremal}
	\end{table}
	
	\begin{table}[htb]
		\centering
		\caption{The fundamental mode for vector $s=-1$
			perturbations of the extremal black hole obtained from
			\eqref{eq:c3expansion} and \eqref{eq:tauIIIconditions}. Again, the
			$\ell=m=1$, $\ell=m=2$ and $\ell=m=3$ modes don't involve
			confluent limits.}
		\scalebox{0.84}{
			\begin{tabular}{|c|c|c|} 
				\hline $\ell,\,m$ & $M\omega$  &
				$\,{}_{-1}\lambda_{\ell,m}$ \\
				\hline $\ell=1,\,m=0$ & $0.2748281298 - 0.0752324478i$ &
				$1.9719892377 + 0.0166548111i$
				\\
				\hline $\ell=1,\,m=-1$ & $0.2043492138 - 0.0913479776i$ &
				$2.1862313294 - 0.0715890886i$ \\
				\hline $\ell=2,\,m=1$ & $0.6421742977 - 0.0517543959i$
				& $5.6473843696 + 0.0388039369i$ \\
				\hline $\ell=2,\,m=0$ & $0.5010131351 - 0.0793652981i$ & $5.9073835684 + 0.0298519612i$ \\
				\hline $\ell=2,\,m=-1$ & $0.4175669677 - 0.0890692358i$ &
				$6.0754297445 - 0.0004305330i$ \\
				\hline $\ell=2,\,m=-2$ & $0.3604214984 - 0.0924215081i$ & $ 6.2024454710 - 0.0413134591i$
				\\
				\hline $\ell=3,\,m=2$ & $1.0594453891 - 0.0288707600i$ &
				$11.2889266226 + 0.0284814980i$\\
				\hline $\ell=3,\,m=1$ & $0.8454254991 - 0.0643820591i$ &
				$11.5786857314 + 0.0527811581i$\\
				\hline $\ell=3,\,m=0$ & $0.7169356912 - 0.0801969267i$ & $11.7804802031 + 0.0496300959i$\\
				\hline $\ell=3,\,m=-1$ & $0.6277439795 - 0.0877125609i$ &
				$11.9429953032 + 0.0319152006i$ \\
				\hline $\ell=3,\,m=-2$ & $0.5609879632 - 0.0912326502i$ & $12.0811096271 + 0.0053853626i$ \\
				\hline $\ell=3,\,m=-3$ & $0.5085923435 - 0.0926316762i$ & $12.2016953171 - 0.0269126158i$ \\
				\hline
			\end{tabular}
		} 
		\label{tab:vectorextremal}
	\end{table}
	\begin{table}[htb]
		\centering
		\caption{The tensor $s=-2$ quasi-normal frequencies for the extremal
			case. Again, the $\ell=m=2$, $\ell=m=3$ and $\ell=m=4$ modes are
			special in that they do not undergo the confluent limit and are best described
			by the method in Sec. \ref{sec:smallnu}.}
		\scalebox{0.84}{
			\begin{tabular}{|c|c|c| } 
				\hline
				$\ell,\,m$ & $M\omega$ & ${}_{-2}\lambda_{\ell,m}$ \\ 
				\hline $\ell=2,\, m=1$ & $0.5814332024 - 0.0382554552i$ &
				$3.0207354411 + 0.0786829388i$
				\\
				\hline $\ell=2,\, m=0$ & $0.4251451091 - 0.0718061840i$ &
				$3.9076436236 + 0.0322811895i$
				\\
				\hline $\ell=2,\, m=-1$ & $0.3438615570 - 0.0833840937i$ &
				$4.3957973918 - 0.0793575127i$
				\\
				\hline $\ell=2,\, m=-2$ & $0.2915534644 - 0.0880258373i$ &
				$4.7217273455 - 0.1982855830i$
				\\
				\hline $\ell=3,\, m=2$ & $1.0285533392 - 0.0185716489i$ &
				$8.1886516502 + 0.0410275646i$
				\\
				\hline $\ell=3,\, m=1$ & $0.7952833561 - 0.0589650419i$ &
				$9.2539870727 + 0.0711300827i$ \\
				\hline $\ell=3,\, m=0$ & $0.6494427364 - 0.2316683583i$ &
				$9.8776142720 + 0.0988361573i$ \\
				\hline $\ell=3,\, m=-1$ & $0.5777578894 - 0.0851643733i$ &
				$10.2684370456 - 0.0213456601i$ \\
				\hline $\ell=3,\, m=-2$ & $0.5135859425 - 0.0891982485i$ &
				$10.5798962415 - 0.0811223415i$ \\
				\hline $\ell=3,\, m=-3$ & $0.4638245318 - 0.0909282625i$ &
				$10.8285024487 - 0.1424496701i$ \\
				\hline $\ell=4,\, m=3$ & $1.5032224673 - 0.0043707042i$ &
				$15.3854120716 + 0.0100091174i$
				\\
				\hline $\ell=4,\, m=2$ & $1.1924746738 - 0.0446945449i$ &
				$16.5634175634 + 0.0709127271i$ \\
				\hline $\ell=4,\, m=1$ & $1.0132544572 - 0.0667186221i$ &
				$17.2408483261 + 0.0723498037i $ \\
				\hline $\ell=4,\, m=0$ & $0.8905096648 - 0.0785749845i$ &
				$17.7172910887 + 0.0500242267i$ \\
				\hline $\ell=4,\, m=-1$ & $0.7990890146 - 0.0852100050i$ &
				$18.0861558459 + 0.0168416633i$
				\\
				\hline $\ell=4,\, m=-2$ & $0.7233591194 - 0.0604378092i$ &
				$18.3852012389 - 0.0152698249i$ 
				\\
				\hline $\ell=4,\, m=-3$ & $0.6715929718 - 0.0570714631i$ &
				$18.6433459145 - 0.0405034950i$
				\\
				\hline $\ell=4,\, m=-4$ & $0.5477717327 - 0.0321354863i$ &
				$18.7744842479 - 0.0397101220i$
				\\
				\hline
			\end{tabular}
		}
		\label{tab:tensorextremal}
	\end{table}
}

\markboth{}{}

The results for fundamental QN frequencies for the scalar, electromagnetic, and gravitational modes are shown in Tables \ref{tab:scalarextremal}, \ref{tab:vectorextremal}, and \ref{tab:tensorextremal}, respectively. Finally, in Fig. \ref{fig:nearextremalqnm}, we show how the electromagnetic modes converges to the extremal mode, with the first correction for $M\omega$ near $a=M$ given in order of $T_+^{2}$, which is zero in $a=M$.

\section{\texorpdfstring{Overtone Results: $\{\sigma,\eta\}$ and $\{\nu,\rho\}$ maps}%
{}}
\label{sec:RHmapslst}

We finish this chapter with a remarkable feature observed recently in \cite{daCunha:2022ewy}. We have verified that from the expression for $e^{i\pi\eta}$ and $X_{\pm}$ defined in \eqref{eq:nurhoparameter}, we can use the accessory parameter expansion for large $t$ \eqref{eq:contfracINF} to investigate the overtone modes through the RHm that depends of the monodromy parameters $\{\nu,\rho\}$, as defined in \eqref{eq:tauVcondinf}. It is important to point out that, in this subsection, $\nu$ is the parameter of the $\tau_{V}$-function expanded for large $t$, as written in \eqref{eq:taufirstterm}. The previous "$\nu$" was defined only in the study of the extremal limit $a\rightarrow M$.

In Fig. \ref{fig:maps}, we have illustrated the maps for the CHE, where $\{\sigma,\eta\}$ and $\{\nu,\rho\}$ are related by the expression $X_{\pm}$ in \eqref{eq:nurhoparameter}. From this relation, we can show that the overtone modes obtained from the RHm \eqref{eq:radialsyst} are exactly the same computed from the RHm for large $t$ \eqref{eq:tauVcondinf}, where in terms of the radial equation parameters it is given by
\begin{equation}
\frac{d}{dz_0}\text{log}(\tau_{V}(\{\theta\}_{\mathrm{Rad}}^{-};\nu-1, \rho;z_0)) =c_{z_0}+\frac{\theta_{\mathrm{Rad},0}(\theta_{\mathrm{Rad},z_0}-1)}{2z_0}, \quad \tau_{V}(\{\theta\}_{\mathrm{Rad}};\nu,\rho;z_0) =0,
\label{eq:tauVcondinfkerr}
\end{equation}

To show the equivalence between the maps, we use the expression for $e^{i\pi\eta}$ in $X_{\pm}$ and then compute a condition for $\nu$ from the equation $e^{-i\frac{\pi}{2} \nu} = X_-$. The first step is to use the QNM boundary conditions, which relates $\eta$ with $\{\theta\}(=\{\theta\}_{\mathrm{Rad}})$ and $\sigma$,
\begin{equation}
e^{2\pi i\eta}=e^{-\pi  i\sigma}
\frac{\sin\frac{\pi}{2}(\sigma+\theta_t+\theta_0)
	\sin\frac{\pi}{2}(\sigma+\theta_t-\theta_0)
	\sin\frac{\pi}{2}(\sigma+\theta_{\star})}{
	\sin\frac{\pi}{2}(\sigma-\theta_t+\theta_0) 
	\sin\frac{\pi}{2}(\sigma-\theta_t-\theta_0)
	\sin\frac{\pi}{2}(\sigma-\theta_{\star})},
\label{eq:quantcond}
\end{equation}
in the equation \eqref{eq:nurhoparameter} to show that the value of $\nu$ is determined by
\begin{equation}
\nu_k = -\tfrac{1}{2}\theta_{\mathrm{Rad},z_0}-\tfrac{1}{4}(\theta_{\mathrm{Rad},\star}+1)+
k,\qquad k\in\mathbb{Z},
\label{eq:nuradquantcond}
\end{equation}
where, for high values of the overtone modes, one has $k>0$.

Using the condition for $\nu_{k}$, we can compute the overtone modes for the case $a/M=0$. Thus, the parameter of the equation \eqref{eq:radCHE} simplifies to
\begin{gather}
	\theta_{{\mathrm{Rad}},0}= s , \quad
	\theta_{\mathrm{Rad},z_0}= s +
	4iM\omega,\quad
	\theta_{\mathrm{Rad},\star}=2s-4iM\omega,
	\label{eq:schpar}
\end{gather}
with the accessory parameter and modulus of the radial equation expressed by
\begin{equation}
\begin{gathered}
z_{\mathrm{Rad}}=z_0=4iM\omega,\\
{z}_{0}c_{\mathrm{Rad},z_{0}} =
{_s\lambda}_{\ell,m}+4s-2i(1+2s)M\omega-2Mis
\omega-6M^2\omega^2,
\label{eq:accessoryradialrschw}
\end{gathered}
\end{equation}
where, for $a/M=0$, the angular eigenvalue simplifies to ${}_{s}\lambda_{\ell m}=(\ell-s)(\ell+s+1)$. Fixing the values of $s$ and $\ell$ and substituting the parameters above, and the conditition \eqref{eq:nuradquantcond} into equation \eqref{eq:contfracINF}, one obtains a function that depends only on $M\omega$. The final procedure consists in fixing the value of $N_c$ and computes the overtone frequencies. For the computation, we have assumed a large value for $N_c$, around $10^{4}$ and used the Muller's method. 

We remark that the expression above for $\nu_k$ simplifies the RHm \eqref{eq:tauVcondinfkerr}, where, with $\nu_{k}$ computed, we arrive at the values for the overtone' frequencies just searching for roots of the accessory parameter $c_{\mathrm{Rad},z_0}$ expanded for large $z_0$, as derived in \eqref{eq:contfracINF}.

\begin{table}[htb]
	\centering
	\caption{First overtones for $s=-2$ $\ell=2$ in the Schwarzschild black	hole. To the left, we list the values for $M{}_s\omega_{\ell,m}$ found from \eqref{eq:contfracINF}, using the condition \eqref{eq:nuradquantcond}. To the right, we present the values obtained from the RHm \eqref{eq:tauVcond} for the $\tau_{V}$-function expanded around $t=0$ with the  condition \eqref{eq:quantcond}.}
	\scalebox{0.84}{
		\begin{tabular}{|c|c|c|}
			\hline
			$n$ &  $M{}_{-2}\omega_{20}$ -- large $t$
			& $M{}_{-2}\omega_{20}$ -- small $t$ \\ \hline
			1 & $0.37367168441804 - 0.08896231568894i$ 
			& $0.37367168441804 - 0.08896231568894i$ \\ \hline
			2 & $0.34671099687916 - 0.27391487529123i$
			& $0.34671099687916 - 0.27391487529123i$ \\ \hline
			3 & $0.30105345461236 - 0.47827698322307i$
			& $0.30105345461236 - 0.47827698322307i$ \\ \hline
			4 & $0.25150496218559 - 0.70514820243349i$
			& $0.25150496218559 - 0.70514820243349i$ \\ \hline
			5 & $0.20751457981306 - 0.94684489086635i$
			& $0.20751457981306 - 0.94684489086635i$ \\ \hline
			6 & $0.16929940309304 - 1.19560805413585i$
			& $0.16929940309304 - 1.19560805413585i$ \\ \hline
			7 & $0.13325234024519 - 1.44791062616204i$
			& $0.13325234024519 - 1.44791062616204i$ \\ \hline
			8 & $0.09282233367020 - 1.70384117220614i$
			& $0.09282233367020 - 1.70384117220614i$ \\ \hline
			9 & $0.06326350512560 - 2.30264476515854i$
			& $0.06326350512560 - 2.30264476515854i$ \\ \hline
			10& $0.07655346288598 - 2.56082661738151i$
			& $0.07655346288598 - 2.56082661738151i$ \\ \hline
			11& $0.08259814464162 - 2.81544237260799i $
			& $0.08259814464162 - 2.81544237260799i$ \\ \hline
			12& $0.08572792206332 - 3.06869474393119i $
			& $0.08572792206332 - 3.06869474393119i $ \\ \hline
			13& $0.08739426237360 - 3.32123023843256i $
			& $0.08739426237360 - 3.32123023843256i $ \\ \hline
			14& $0.08823894076059 - 3.57332069830792i $
			& $0.08823894076059 - 3.57332069830792i $ \\ \hline
			15& $0.08859038712017 - 3.8251052081451i $
			& $0.08859038712017 - 3.82510520814541i $ \\ \hline
		\end{tabular} 
		\label{tab:gravrad}
	}
\end{table}

\begin{figure}[htb]
	\begin{center}
		\caption{Contour plot for equation \eqref{eq:contfracINF}, for $\nu_{k=0}$, with $s=-2$, $\ell=2$ and $m=0$ for Schwarzschild case $a/M=0$. Note the first three zeros corresponding to the first three overtones, listed in Table \ref{tab:gravrad}.}
		\includegraphics[width=0.50\textwidth]{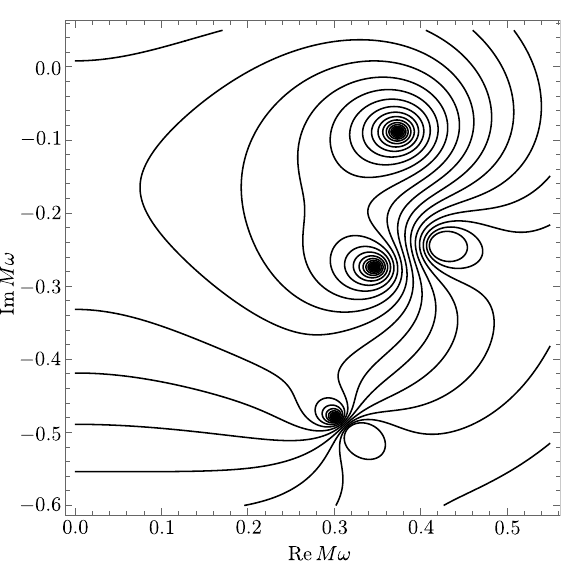}
		\label{fig:contour_schw}
	\end{center}
\end{figure}

The numerical results for the first fifteen overtone modes can be checked in Table \ref{tab:gravrad}, and match those obtained from the $\tau_{V}$-function in the map \eqref{eq:radialsyst} and the values accepted in the literature from \cite{Berti:2009kk}. Additionally, we present the contour plot for \eqref{eq:contfracINF} in Fig. \ref{fig:contour_schw}, where we can see that the zeros are the overtone frequencies. 

\section{Conclusion of the Chapter}

In this chapter, we applied the isomonodromy method outlined in \cite{daCunha:2015ana,CarneirodaCunha:2019tia}, and revised in Chapter \ref{ChapIsoMethod} to study the QN frequencies of generic rotating Kerr black holes. The three Riemann-Hilbert maps defined in Chapter \ref{ChapIsoMethod} provide us with a numerical procedure as an alternative to the continued fraction method which, despite having a slower convergence, has a more controlled behavior at the extremal limit $r_-\rightarrow r_+$. In the method, the eigenvalue problem for the radial equation is reduced to solving two transcendental equations for the non-extremal \eqref{eq:radialsyst} and extremal \eqref{eq:tauIIIconditions} Kerr BH, which are numerically amenable given the analytical properties of the functions involved. In Sec. \ref{sec:angradialsys}, we treated the eigenvalue problem of the angular equation in terms of fundamental matrix solutions, and monodromy matrices. The derivation of the angular eigenvalue ${}_{s}\lambda_{\ell m}$ was also shown. In Sec. \ref{sec:secradeq}, we listed the parameters of the radial equation, and the eigenvalue problem for such an equation is translated to conditions for the $\tau_{V}$-function. In Sec. \ref{sec:generica}, the RH map is solved showing excellent accordance with the continued fraction method for the generic value of the rotating parameter. 

In Sec. \ref{sec:extremallimit} we turned to the extremal limit $a\rightarrow M$. Based on the results of the previous section,
we found two distinct behaviors as $r_-\rightarrow r_+$. For the $\ell=m$ modes, we observed in Sec. \ref{sec:smallnu} that the
eigenfrequencies approached $\omega=m/(2M)$, in a manner linear with the temperature. In the rest of the
modes, the limit was seen in Sec. \ref{sec:painleveIII} to be described by the isomonodromic $\tau_{III}$-function, whose Fredholm determinant formulation for the $\tau$-function derived. By keeping the first near-extremal correction, we could assert that the extremal value for the frequencies approached the extremal value in a manner quadratic with the temperature.  

We finished by showing in Sec. \ref{sec:RHmapslst} that the QN frequencies for $a/M=0$ can be obtained using two different maps defined in \eqref{eq:radialsyst} and \eqref{eq:tauVcondinfkerr}. In this way, one has two forms of computing the overtones for the Schwarzschild black hole. In addition, Table \ref{tab:gravrad} provides a proof of the relation between the RH map described in Fig. \ref{fig:maps}. 

\chapter{Reissner-Nordström black hole}
\label{chap:RNBH}

Our main objective in this chapter is to apply the isomonodromy method to investigate the QN frequencies associated with scalar and spinorial perturbations in the Reissner-Nordström black hole. We will treat the black hole and field charges, respectively $Q$ and $q$, generically, but will focus on the extremal limit. This chapter is organized as follows: in Sec. \ref{sec:RNback}, we revise the main equation that encodes the spin-field perturbation in this background, and in Sec. \ref{sec:Qgeneric}, we list the relevant parameters for the confluent Heun equation and present the results of the numerical analysis for the non-extremal case $Q<M$ and extremal limit $Q\rightarrow M$, separately. In Sec. \ref{sec:Conflimit_and_RNextremal}, we deal with the extremal case via the Riemann-Hilbert map for the double-confluent Heun equation. We finish by commenting on the results in Sec. \ref{sec:Conflimit_and_RNextremal} and comparing the results for the quasinormal frequencies in the subextremal and extremal cases. For completeness, Appendix \ref{sec:tools} supplies relevant technical details on the $\tau$ functions.

The results presented in the chapter are published in \textit{Physical Review D} \textbf{104}, 124040. (2021). In the final part, we also bring more results, where the analysis of the overtone frequencies is made via the RHm for large $t$ defined in Chapter \ref{ChapIsoMethod}.

\section{Reissner-Nordström Background}
\label{sec:RNback}

As presented in Chapter \ref{chap1}, the Reissner-Nordström (RN) metric is a solution to the Einstein-Maxwell field equations, corresponding to a non-rotating charged black hole of mass $M$ and charge $Q$. In static and spherically symmetric coordinates, one has the metric \eqref{eq:RNmetric} and in the Coulomb gauge the electromagnetic potential $A_{\mu}$ is $A_{\mu} = (-Q/r,0,0,0)$. 

In turn, linear perturbations of massless charged scalar ($s=0$) and spinor ($s=\frac{1}{2}$) fields in this background can be effectively unified in a scheme similar to the radial Teukolsky master equation (TME) for the Kerr black hole \eqref{eq:radialeq}. The master equation can be derived by separating the Klein-Gordon \eqref{eq:KGRN} and the Dirac \eqref{eq:Diraceq} equations in the RN background, arriving at 
\begin{equation}
\triangle^{-s}\frac{d}{dr}\bigg(\triangle^{1+s}\frac{d{}_{s}R_{\ell}(r)}{dr}
\bigg)+\bigg(\frac{K(r)^{2}-2is(r-M)K(r)}{\triangle}+4is\omega
r-2isqQ-{}_{s}\lambda_{\ell}\bigg){}_{s}R_{\ell}(r)=0 
\label{eq:radeq}
\end{equation}
where, $s=0,\pm 1/2$, $K(r) = \omega r^2-qQ r$ and
${}_{s}\lambda_{\ell}=(\ell-s)(\ell+s+1)$ is the separation constant. The angular dependence of the solutions is given by the usual
spin-weighted spherical harmonics $Y^{s}_{\ell m}(\theta,\phi)$, where the parameters $\ell$ and $m$ satisfy the usual constraints: $\ell$ is a non-negative integer and a non-negative half-integer for bosonic and fermionic case, respectively,  with $\ell \geq |s|$. For $m$, one has essentially $-\ell \leq m \leq \ell$ \cite{doi:10.1063/1.1705135,Berti:2005gp}. Furthermore, one has in \eqref{eq:radeq} the roots of $\triangle = (r-r_+)(r-r_-)$ with $r_{\pm} = M\pm\sqrt{M^{2}-Q^2}$. For $Q<M$, we have the non-extremal RN black hole and $r_+$ and $r_-$ are the event and Cauchy horizons of the black hole, respectively. When $Q=M$, the black hole is extremal, where the roots are equal, $r_+ = r_-$. Finally, for $Q>M$, the black hole space-time has a naked singularity, with $\triangle$ having no real root. 

\section{\texorpdfstring{Numerical Results for $0\leq Q \leq M$}%
{}}
\label{sec:Qgeneric}

After the comments and definitions above, let us consider the RH map \eqref{eq:tauVcond} for the radial equation \eqref{eq:radeq}. As presented in Chapter \ref{chap1}, Sec. \ref{DicSection}, the radial equation \eqref{eq:radeq} can be brought to the standard form \eqref{heuneq1} by the change of variables 
\begin{equation}
{}_{s}R_{\ell}(r)=(r-r_{-})^{-(s+\theta_{\mathrm{Rad},0})/2}(r-r_{+})^{-(s+\theta_{\mathrm{Rad},z_0})/2}y(z),
\qquad z=2i\omega(r-r_{-})
\end{equation}
with
\begin{equation}
\frac{d^2
	y}{dz^2}+\bigg[\frac{1-\theta_{\mathrm{Rad},0}}{z}+\frac{1-\theta_{\mathrm{Rad},z_0}}{z-z_0}
\bigg]\frac{dy}{dz}-\bigg[\frac{1}{4}
+\frac{\theta_{\mathrm{Rad},\star}}{2z}+\frac{z_0c_{z_0}}{z(z-z_0)}\bigg]y(z)=0. 
\label{heuneq1}
\end{equation}

The equation is defined in the complex domain $z\in \mathbb{C}$, with regular singular behavior at $0$ and $z_0$ and an irregular singularity of rank 1 at $z=\infty$. The set $\{\theta\}_{{\mathrm{Rad}}}$ in the equation are functions of the parameters of the radial equation \eqref{eq:radeq}, as listed in Table \ref{tab:CHEDic},
\begin{equation}
\begin{gathered}
\theta_{\mathrm{Rad},0}=  s+\frac{i}{2\pi
	T_{-}}\bigg(\omega-\frac{qQ}{r_{-}}\bigg), \quad \theta_{\mathrm{Rad},z_0}=
s+ \frac{i}{2\pi
	T_{+}}\bigg(\omega-\frac{qQ}{r_{+}}\bigg), \\
\theta_{\mathrm{Rad},\star}=2i(2M\omega -qQ)-2s, \qquad
2\pi T_{\pm} = \frac{r_{\pm}-r_{\mp}}{2 r^{2}_{\pm}},   \qquad
r_{\pm} = M\pm\sqrt{M^2 -Q^2},
\label{parameters}
\end{gathered}
\end{equation}
with modulus $z_0= 2i\omega(r_+ - r_-)$ and accessory parameter:
\begin{equation}
z_0c_{z_0} =   {}_{s}\lambda_{l,m}
+2s-i(1-2s)qQ+(2qQ+i(1-3s))\omega r_++i(1-s)\omega r_-
-2\omega^2 r_+^2.
\label{modacce}
\end{equation}

Note that, now $r_{\pm}$ are functions of $Q$, while for Kerr BH we have the momentum per mass units, $\textit{a=J/M}$. Moreover, there is a similarity between the $\theta_{ 0}$'s and $\theta_{ z_0}$'s defined in the problem for Kerr and RN BHs, where it can observed that the azimuthal number $m$ was 'replaced' by the decoupled term $qQ$ -- see Table \ref{tab:CHEDic}. This distinction allows us to variety continually the value of $qQ$, different of the case with dependence in $m$. The results for $qQ$ varying will be shown in the next pages. Following the same strategy used for Kerr BH, we parametrize $Q/M = \cos\nu$, with $\nu \in [0,\pi/2]$, where again the extremal limit $r_-\rightarrow r_+ (Q \rightarrow M) $ corresponds to $\nu\rightarrow 0$. In terms of $\nu$ , the monodromy parameters \eqref{parameters} are then rewritten as
\begin{equation}
\begin{gathered}
\theta_{\mathrm{Rad},0} = s -\frac{i}{2\pi
	T_{-}}\bigg(\omega-\frac{qQ}{M(1-\sin\nu)}\bigg),
\qquad \theta_{\mathrm{Rad},z_0} = s + \frac{i}{2\pi
	T_{+}}\bigg(\omega-\frac{qQ}{M(1+\sin\nu)}\bigg)\\
\theta_{\mathrm{Rad},\star} = -2s+2i(2M\omega-qQ),
\label{eq:nuparametrization} \quad 2\pi T_{\pm}
=\pm\frac{\sin\nu}{M(1\pm\sin\nu)^2}, \qquad
r_\pm =M(1\pm \sin\nu),
\end{gathered}
\end{equation}
with $z_0$ and $z_0c_{z_0}$ in \eqref{modacce} given by
\begin{equation}
\begin{aligned}
z_0= 4iM\omega\sin\nu, \quad \ \qquad  z_0c_{z_0}={}_{s}\lambda_{l,m}&+2s-i(1-2s)qQ +
2(qQ+i(1-2s)+\\&+(qQ-is)\sin\nu) M\omega
-2(1+\sin\nu)^2(M\omega)^2.
\end{aligned}
\label{modaccASnufunction}
\end{equation}
Note that, for $\nu\rightarrow 0$, $T_+$ behaves as
\begin{equation}
T_+ = \frac{\nu}{2M\pi}-\frac{\nu^{2}}{M\pi}+\mathcal{O}(\nu^3).
\end{equation} 

We know from Chapter \ref{ChapIsoMethod} that the Riemann-Hilbert map in \eqref{eq:tauVcond} relates the accessory parameter $c_{z_0}$ and modulus $z_0$ of \eqref{heuneq1} to the monodromy parameters $\{\sigma, \eta \}$ for the $\tau_{V}$ function expanded around $z_0=0$. For the radial equation \eqref{eq:radeq} the map is given by 
\begin{equation}
\tau_V(\{\theta\}_{\mathrm{Rad}};\sigma,\eta;z_{0})=0,
\quad
z_{0}\frac{d}{dz_0}\log\tau_V(\{\theta\}_{\mathrm{Rad}}^{-};
\sigma-1,\eta;z_{0})-
\frac{\theta_{\mathrm{Rad},0}(\theta_{\mathrm{Rad},z_0}-1)}{2}=
z_{0}c_{z_0},
\label{eq:radialsystemeqn}
\end{equation}
with
$\{\theta\}_{\mathrm{Rad}}=\{\theta_{\mathrm{Rad},0},\theta_{\mathrm{Rad},z_0},\theta_{\mathrm{Rad},\star}
\}$ and
$\{\theta\}_{\mathrm{Rad}}^{-}=\{\theta_{\mathrm{Rad},0},\theta_{\mathrm{Rad},z_0}-1,\theta_{\mathrm{Rad},\star}+1
\}$. As observed for the Kerr BH, one has that the numerical analysis is more convenient made directly with the accessory parameter expansion derived in the final part of the Chapter \ref{ChapIsoMethod}, rather than considering the logarithm derivative of the $\tau_{V}$-function, i.e. second equation in \eqref{eq:radialsystemeqn}. Therefore, in our analysis, we consider only the numerical implementation of the $\tau_{V}$-function and equation \eqref{eq:contfrac}. 

From Chapter \ref{ChapIsoMethod}, we have noted that $\sigma$ parametrizes the so-called Floquet solutions of \eqref{heuneq1}, and a direct relation between $\sigma$ and $c_{z_0}$ was derived from the three recurrence relation written in terms of continued fraction in \eqref{eq:contfrac}, where for small $z_0$, $c_{z_0}$ is given by
\begin{multline}
	z_0c_{z_0}=\frac{(\sigma-1)^2-(\theta_t+\theta_0-1)^2}{4}+
	\frac{\theta_\star(\sigma(\sigma-2)+\theta_t^2-\theta_0^2)}{
		4\sigma(\sigma-2)}z_0 \ +\\
	+\left[\frac{1}{32}+\frac{\theta_\star^2(\theta_t^2-\theta_{0}^2)^2}{64}
	\left(\frac{1}{\sigma^3}-\frac{1}{(\sigma-2)^3}\right)
	+\frac{(1-\theta_\star^2)(\theta_0^2-\theta_{t}^2)^2+2\theta_\star^2
		(\theta_0^2+\theta_{t}^2)}{32\sigma(\sigma-2)}\right.
	\\ \left.-
	\frac{(1-\theta_\star^2)((\theta_0-1)^2-\theta_{t}^2)((\theta_0+1)^2-
		\theta_{t}^2)}{32(\sigma+1)(\sigma-3)}\right]z_0^2+
	{ O}(z_0^3).
	\label{eq:c5expansion}
\end{multline}

In order to simplify the notation, the label "Rad" was removed. For the calculation of quasinormal modes' frequencies, we must impose the boundary conditions defined in \eqref{eq:boundRN}, which corresponds to outgoing and ingoing waves at infinity and event horizon in the RN black hole, respectively. As shown in Chapter \ref{ChapIsoMethod}, these boundary conditions can be cast in terms of $\{\theta\}$ and $\sigma$ as,
\begin{equation}
e^{i\pi\eta}=e^{-i\pi\sigma}
\frac{\sin\tfrac{\pi}{2}(\theta_\star+\sigma)}{
	\sin\tfrac{\pi}{2}(\theta_\star-\sigma)}
\frac{\sin\tfrac{\pi}{2}(\theta_t+\theta_0+\sigma)
	\sin\tfrac{\pi}{2}(\theta_t-\theta_0+\sigma)}{
	\sin\tfrac{\pi}{2}(\theta_t+\theta_0-\sigma)
	\sin\tfrac{\pi}{2}(\theta_t-\theta_0-\sigma)},
\label{eq:quantizationV}
\end{equation}
which is equivalent to assume that the connection matrix between $z=\infty$ and $z=z_0$, $C_{z_0,\infty}$, is lower triangular.

\subsection{\texorpdfstring{Numerical Results for $\ell >|s|$}%
{}}

We are now ready to solve the eigenvalue problem \eqref{eq:radialsystemeqn} to determine the fundamental modes frequencies for spin-$0$ and
spin-$\frac{1}{2}$ perturbations of the non-extremal RN black hole. The initial analysis is concentrated on the case $\ell > |s|$, which means $\ell=1,2,3,\ldots$ for the scalar case and $\ell=3/2,5/2,\ldots$ for the spinorial case.  As explained in
\cite{Chang2007MassiveCQ}, the potential terms for spin-$1/2$ produce the same QNMs spectrum for both values $s = \pm1/2$ and there is no loss of generality in considering only the $s = -1/2$ case. Additionally, since the RN black hole is symmetric with respect to transformations $q\rightarrow -q$ and $\omega \rightarrow -\omega^{*}$, we consider only positive $q$ and, consequently, $\Re(M\omega)>0$.
\begin{figure}[htb]
	\caption{Fundamental modes calculated for scalar and spinor perturbations in non-extremal RN black hole. Results for  $s=0,
		l=1$ (top) and $s=-\frac{1}{2}, \ell=\frac{3}{2}$ (bottom) as a function of $Q/M$ for $qQ= 0.0$ (lightest), $0.2,0.4,0.6,0.8$ and
		$1.0$ (darkest).}
	\begin{center}
		\includegraphics[width=0.95\textwidth]{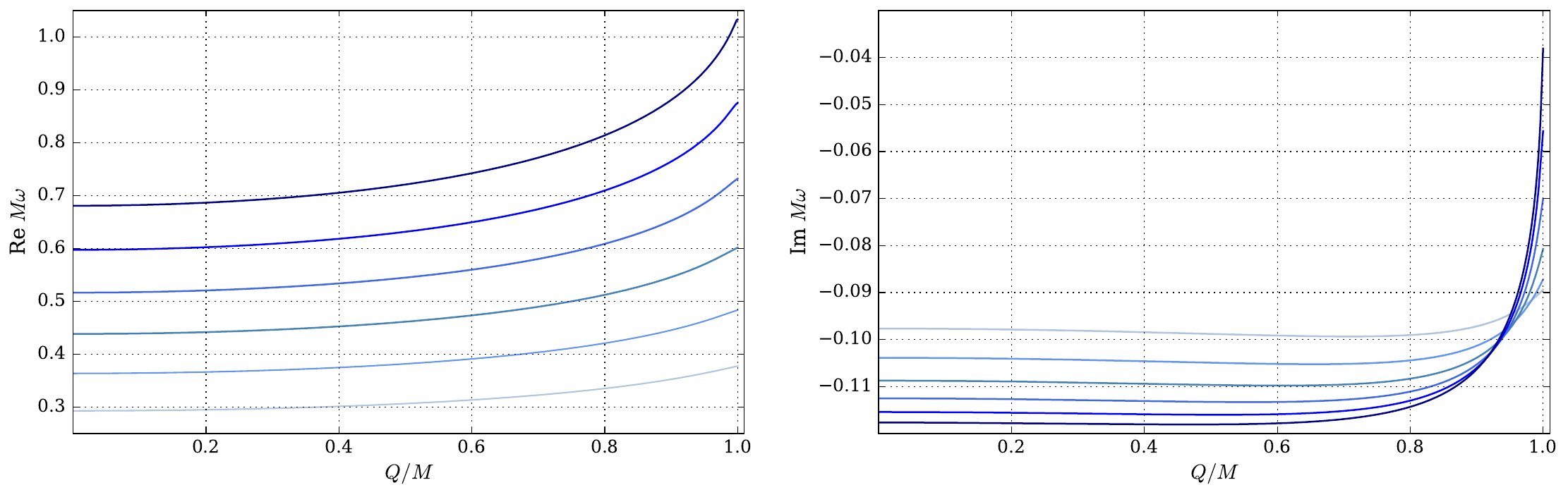}
		\includegraphics[width=0.95\textwidth]{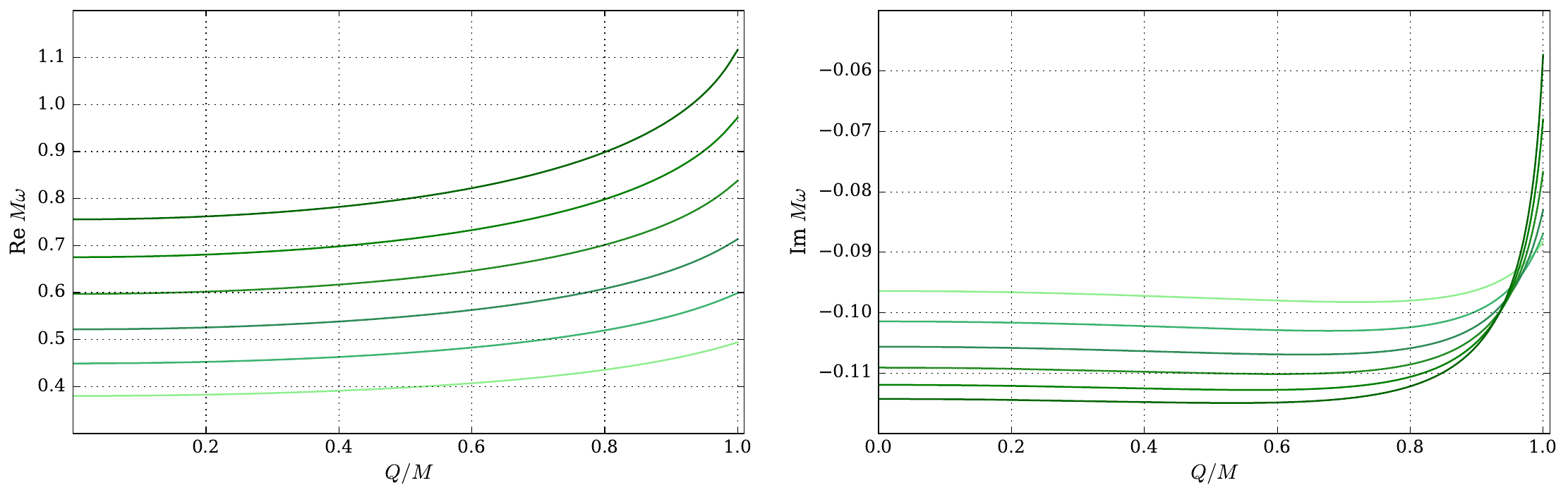}
	\end{center} 
	\label{fig:s0l1}
\end{figure}

In Fig. \ref{fig:s0l1}, we show the fundamental modes frequencies for $s=0$, $\ell=1$ and $s=-\frac{1}{2}$, $\ell= \frac{3}{2}$ as a function of $qQ$. These values were obtained by directly solving \eqref{eq:radialsystemeqn}, where following the same procedure applied to Kerr BH, we use the numeric implementation of the $\tau_{V}$-function \eqref{eq:expansiontauV} and the equation \eqref{eq:contfrac} in \href{http://julialang.org}{Julia} language, then, the roots of the $\tau_V$ are calculated using the Muller's method. The Fredholm determinant involved in the definition of $\tau_V$-function was truncated at $N_f=64$ Fourier components and \eqref{eq:contfrac} at $N_c=128$ convergents. The method converges fastly due to Miwa's
theorem \cite{Miwa:1981aa}, which asserts that $\tau_V$ only has isolated and simple zeros away from its critical values $t=0,\infty$. However, the zero locus structure can become very intricate, particularly close to extremality, so when varying $\frac{Q}{M}$ we use the QN frequency value for a particular point as the guess for the following point. The implementation of the isomonodromic $\tau_{V}$ and $\tau_{III}$ based on the Fredholm determinants is available in \cite{Github}. In \cite{GithubIM} the script that computes the results for this chapter and Chapter \ref{chap:KerrBH} is also available.

When solving the set of equations \eqref{eq:radialsystemeqn}, we have verified that the fundamental modes for $qQ=0$, $Q/M =0$, in the cases $s=0$, $l=1,2,3,...$, and $s=-\frac{1}{2}$, $l=\frac{3}{2},\frac{5}{2},\frac{7}{2},...$, are in agreement with the
modes for Schwarzschild black hole calculated using the WKB approximation \cite{Konoplya:2004ip,Cho:2003qe} and continued fraction (CF) method in
the \href{https://pages.jh.edu/eberti2/ringdown/}{literature}, see Table \ref{fig:s0l1}. In turn,  in the case of the
non-extremal RN black hole with $qQ\neq 0$ and $Q/M=0.999$, we recovered the fundamental modes for $s=0$, $l=1,2,3$ and
$s=-\frac{1}{2}$, $\ell=\frac{3}{2},\frac{5}{2},\frac{7}{2}$ calculated using the modified version of the continued fraction method
and listed in \cite{Richartz:2015saa}. Given that the CF method has convergence problems near the extremality, we have extended the
analysis by  solving the system \eqref{eq:radialsystemeqn} for $Q\rightarrow M$, as we can seen in Fig. \ref{fig:s0l1}. By employing
the similar procedure explained for Kerr BH, we can compute the QN frequencies values up to $Q/M = 1-10^{-11}$ in reasonable time and accuracy. One notes that the method infers a smooth extremal limit for these modes, a point we will return to below. 

\begin{table}[htb]
	\centering
	\caption{To the left, the fundamental modes for Schwarzschild black	hole $q=Q=0$ recovered with $\ell>|s|$. For comparison, we show to the right the values in the	\href{https://pages.jh.edu/eberti2/ringdown/}{literature} \cite{Berti:2005gp}.}
	\scalebox{0.96}{
		\begin{tabular}{|c|c|c|c|c|}
			\hline $s$& $\ell$ & 
			${}_{s}\omega_{\ell}$\, ($\tau_V
			\text{ function}$)
			&  ${}_{s}\omega_{\ell}$\,
			($\text{CF method}$)  \\
			\hline
			0 &1  & $0.292936133267 - 0.097659988914i$   &   $ 0.292936133267
			- 0.097659988913i $
			\\ \hline
			0 &2  & $0.483643872211 - 0.096758775978i$   &  $ 0.483643872211 -
			0.096758775978i $
			\\ \hline
			0 &3  & $0.675366232537 - 0.096499627734i$   &  $ 0.675366232537 -
			0.096499627734i $
			\\ \hline
			$-\frac{1}{2}$ & $\frac{3}{2}$ &$0.380036764833 - 0.096405208085i$
			&  $0.380036764833 - 0.096405208085i $  \\
			\hline
			$-\frac{1}{2}$ & $\frac{5}{2}$  & $0.574093974298 -
			0.096304784939i$  &
			$0.574093974298 - 0.096304784939i$   \\ \hline
			$-\frac{1}{2}$ & $\frac{7}{2}$  & $ 0.767354592773 -
			0.096269878994i$  &
			$0.767354592773 - 0.096269878994i$   \\ \hline
		\end{tabular}
	} 
	\label{tab:schwarzschild_s01over2}
\end{table}

\subsection{\texorpdfstring{Numerical Results for $\ell =|s|$ }%
{ }}

We now focus our attention on the case $\ell =|s|$ and investigate the behavior of the fundamental modes frequencies as $Q/M$ and $qQ$ vary. In our
analysis, we have obtained QN frequencies for $Q\in[0,M]$ as a function of $qQ$ and observed two different types of behavior in the
limite $Q\rightarrow M$, as follows:
\begin{enumerate}
	\item For $qQ$ below a spin-dependent critical value $qQ_c(s)$ modes calculated from \eqref{eq:radialsystemeqn} behave similarly to $\ell > |s|$, with finite limits for the real and imaginary parts of the QN frequencies as  $Q\rightarrow M$. We will call these ``damping modes'', as defined in \cite{Richartz:2014jla}. 
	\item For $qQ$ above the spin-dependent critical value $qQ_c(s)$, the
	frequency converges to $q$ as $Q\rightarrow M$ ($\nu\rightarrow
	0$). In this case, the imaginary part of the QNM frequency tends to
	zero, in a non-damping behavior. Also, the existence of a
	natural small parameter allows for the analytical treatment of these
	modes.
	\label{item:case}
\end{enumerate}

\begin{figure}[htb]
	\centering
	\caption{Fundamental modes for  $s=0, l=0$ (top) and
		$s=\frac{1}{2},\ell = \frac{1}{2}$ (bottom) with $qQ$ varying from
		$0$ (lightest) to $0.3$ (darkest) with $0.1$ of increment.}
	\begin{center}
		\includegraphics[width=0.95\textwidth]{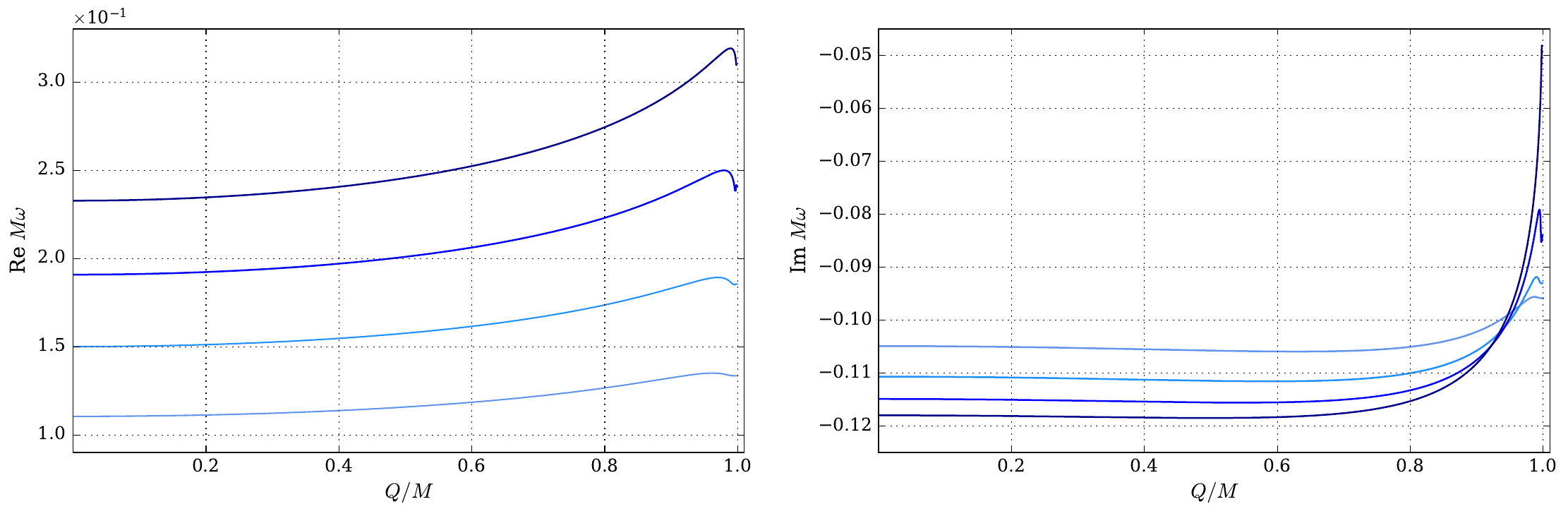}
		\includegraphics[width=0.95\textwidth]{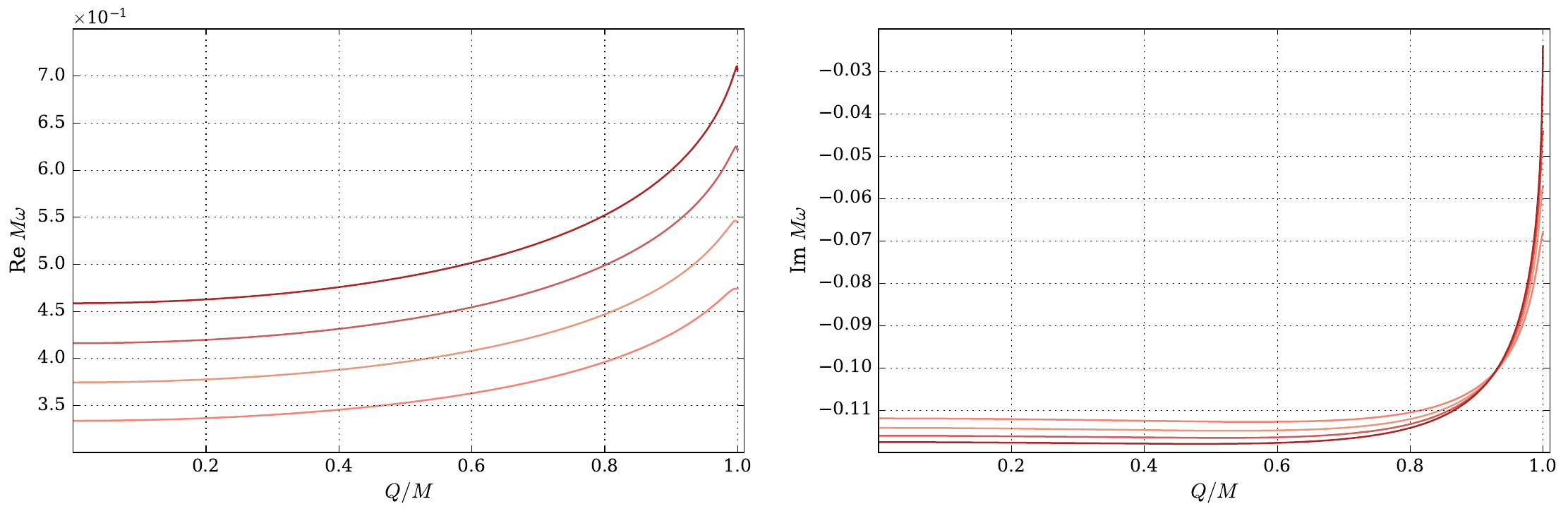}
	\end{center}
	\label{fig:s01over2l01over2}
\end{figure}

Before checking each case separately, we show in Fig. \ref{fig:s01over2l01over2} the qualitative difference between
cases I and II in the usual graphs where $M\omega$ is plotted against $Q/M$. For $s=\ell=0$, one observes the appearance of non-trivial behavior for $Q\rightarrow M$ as one increases $qQ$. In this regime, the parametrization using $\nu$ \eqref{eq:nuparametrization} is more adequate, so we will switch to it. Also, one can appreciate the difficulty in studying these different behaviors using the usual CF method, given that it does not converge easily (if at all) as one approaches the extremal limit $\nu\rightarrow 0$.

The numerical analysis for these modes follows the same procedure described in the previous chapter, with $qQ$ developing a crucial role in the analysis. For $\ell=|s|$: we simply solve the RHm \eqref{eq:radialsystemeqn}. As a way of checking the procedure, we have recovered the fundamental mode for $qQ=0$, $s=\ell=0$, and $Q/M=0$ obtained for
scalar perturbation in the Schwarzschild black hole and listed in the \href{https://pages.jh.edu/eberti2/ringdown/}{literature}. Also the fundamental mode for Dirac perturbation in this background was also retrieved and compared with our implementation of the CF method. This frequency was obtained in \cite{Konoplya:2004ip}, using the WKB approximation which does not permit accurate comparisons. These values are presented in Table \ref{tab:schwarzschilds=ell}. We have also checked our results for $qQ\neq 0$ and $Q/M\lesssim 0.999$ against \cite{Richartz:2015saa} and found excellent agreement.

\begin{table}[htb]
	\centering
	\caption{Comparison of the fundamental QNMs frequencies for
		Schwarzschild $q=Q=0$ black hole in the case $\ell=|s|$.}
	\begin{tabular}{|c|c|c|c|c|}
		\hline
		\ $s$& $\ell$ & ${}_{s}\omega_{\ell}$\,
		($\tau_V\text{ function}$) &
		${}_{s}\omega_{\ell}$\, 
		($\text{CF  method}$)  \\
		\hline
		\ $0$ & $0$ &  $0.110454939080 - 0.104895717087i$   &  $
		0.110454939080
		-
		0.104895717087i
		$  \\
		\hline
		$-\frac{1}{2}$ & $\frac{1}{2}$  & $0.182962870255 -
		0.096982392762i$  & $
		0.182962870255
		-
		0.096982392762i
		$   \\
		\hline
	\end{tabular}
	\label{tab:schwarzschilds=ell}
\end{table}

\begin{figure}[htb]
	\caption{\textit{Blue lines}: Fundamental modes for  $s=0, l=0$, with $qQ$ varying from
		$0.0$ (lightest) to $0.2$ (darkest) with $0.05$ increment in the left-hand side graphic and from $0.25$ (darkest) to $0.45$ (lightest) with $0.05$ in the right-hand side. \textit{Red lines}: Fundamental modes for  $s=-1/2, l=1/2$, with $qQ$ varying from $0.6$ (lightest) to $0.64$ (darkest) with $0.01$ increment in the left-hand side graphic and from $0.65$ (darkest) to $0.69$ (lightest) with $0.01$ in the right-hand side. For both cases, it is clear that before the critical values, the quasinormal modes' frequencies spiral with the imaginary part assuming a finite value. After the critical value, this characteristic is not observed.}
	\begin{center}
		\includegraphics[width=0.96\textwidth]{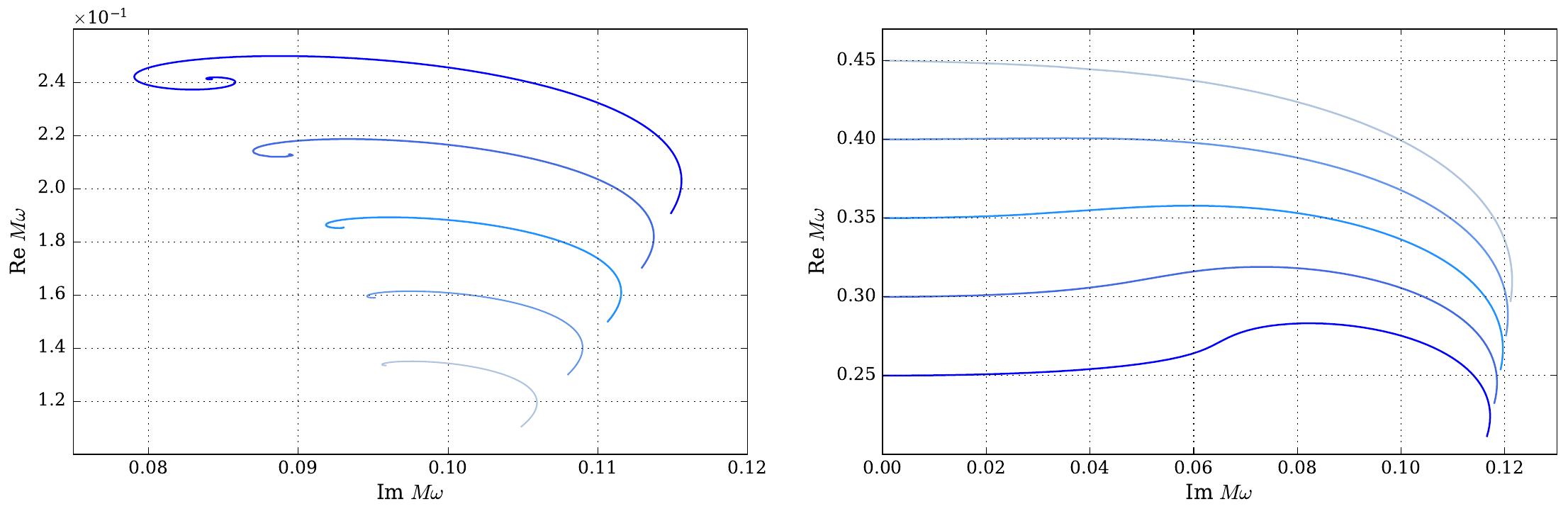}
		\includegraphics[width=0.97\textwidth]{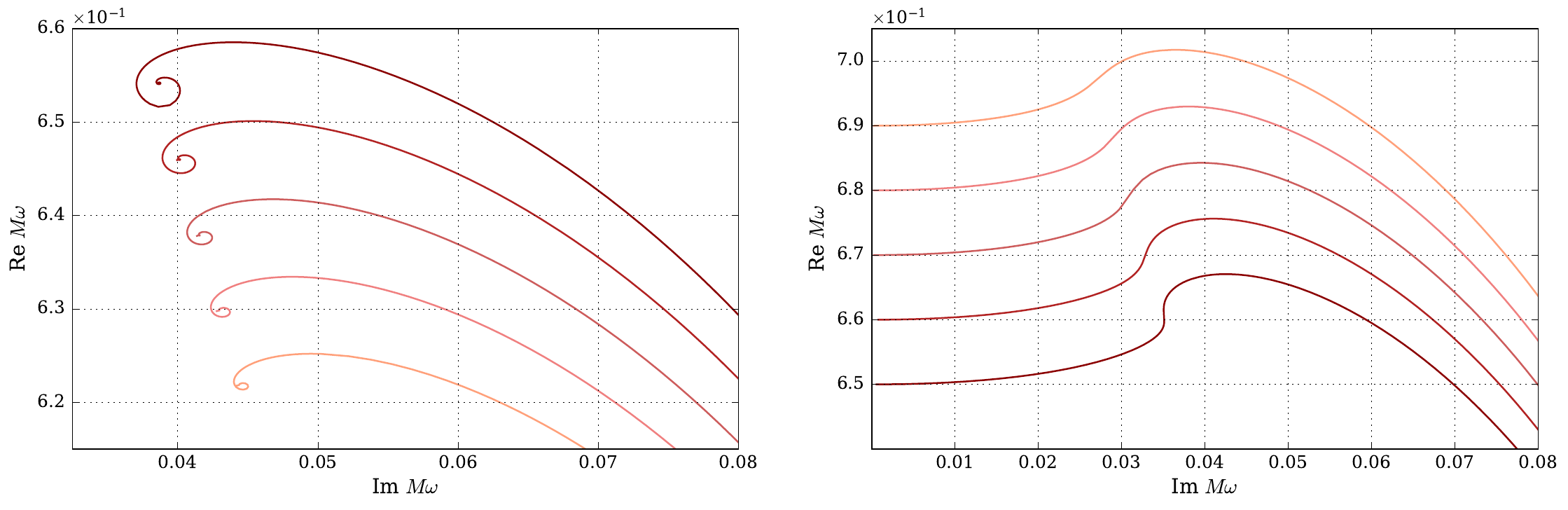}
	\end{center}
	\label{fig:s01over2l01over21}
\end{figure}

As observed in Fig. \ref{fig:s01over2l01over2}, in the extremal limit $Q/M\rightarrow 1$ a non-trivial structure appears when the value for $qQ$ increases. In order to investigate how the real and the imaginary part of the frequency behave in this limit, we plotted the Fig. \ref{fig:s01over2l01over21}, where it is shown that for different values of $qQ$ the real and imaginary parts go to a finite value following a spiral behavior. For $qQ\geq 0.25$ (scalar case) and $qQ\geq0.65$ (spinorial case) the real part converges to $qQ$ and the imaginary part becomes zero in the extremal limit $Q/M \rightarrow 1$. Therefore, it seems that there is a critical value for $qQ$ such that the QN frequencies become normal mode frequencies in $Q=M$.

\subsection*{Non-damping modes for $\nu \rightarrow 0$}
\label{sec:lambfinite}

\begin{figure}[htb]
	\caption{Damping and non-damping modes for $s=0, l=0$ (top) and	$s=-\frac{1}{2}$, $\ell =\frac{3}{2}$ (bottom) in near-extremal RN
		black hole. The dark color line identifies the critical values $qQ_c\approx 0.216$  ($s=0$, $\ell=0$) and $qQ_c\approx 0.643$
		($s=-\frac{1}{2}$, $\ell=\frac{1}{2}$).}
	\begin{center}
		\includegraphics[width=0.95\textwidth]{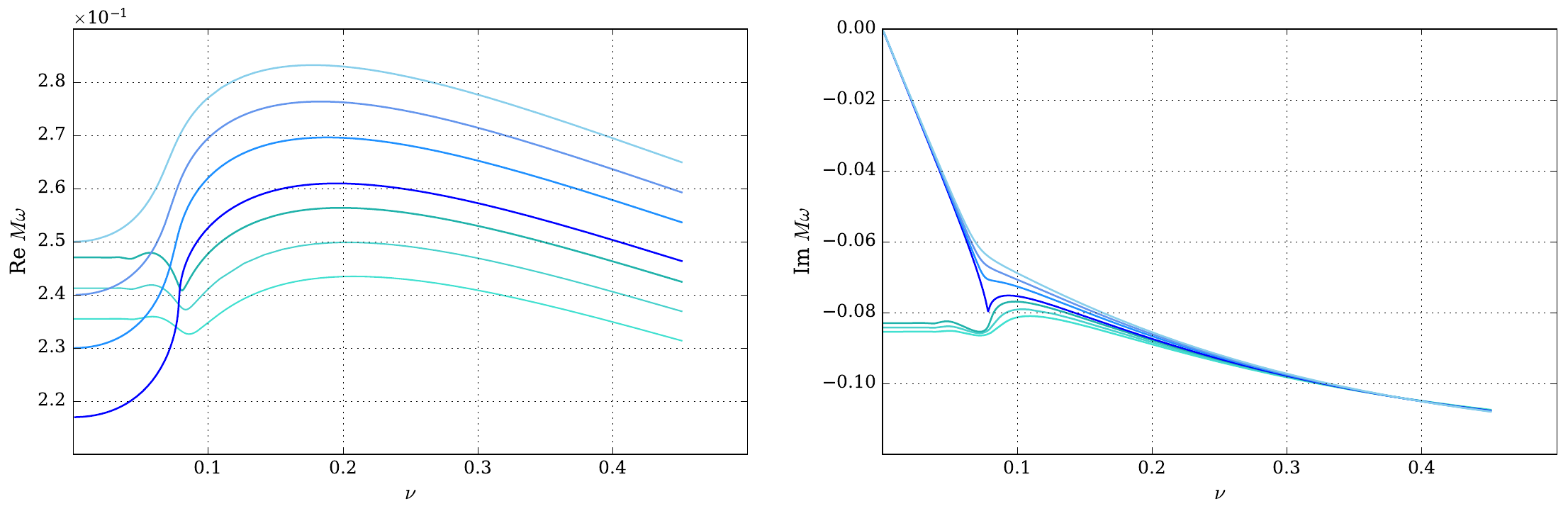}
		\includegraphics[width=0.95\textwidth]{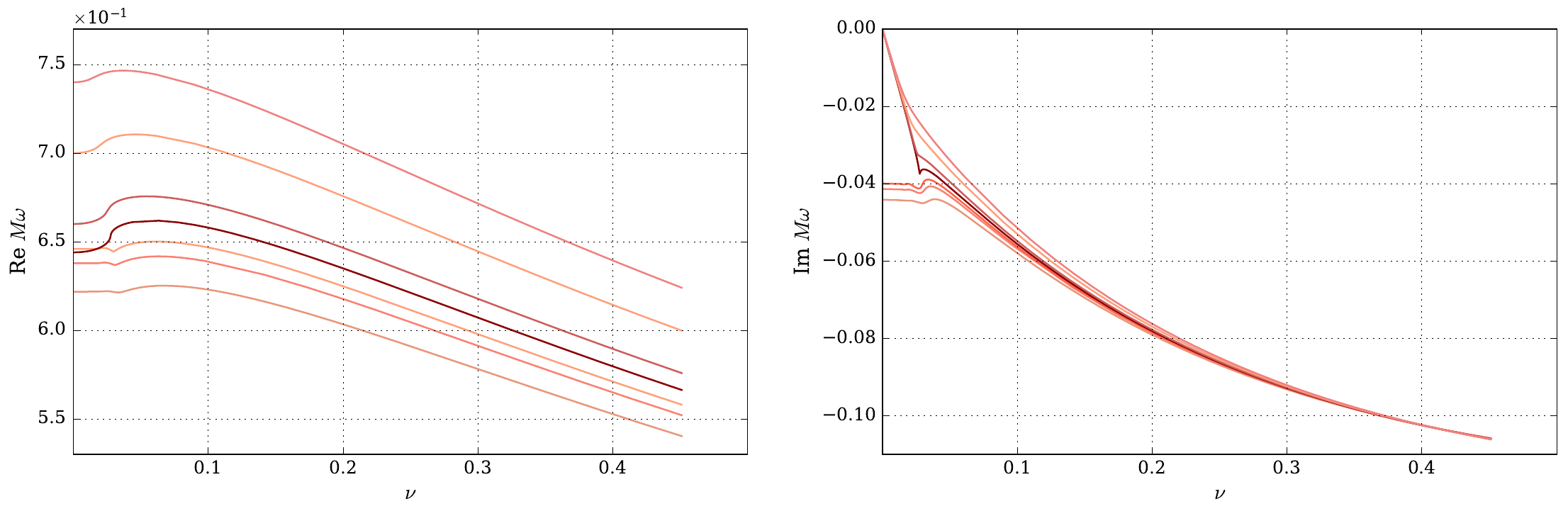}
	\end{center}
	\label{fig:s0l0_nu}
\end{figure}

As the black hole approaches extremality, and $qQ$ is large enough, the fundamental mode frequencies approach the purely real value $M\omega \rightarrow qQ$. Using the isomonodromy method, we can solve the RHm \eqref{eq:radialsystemeqn} numerically and study the behavior of the spin-$0$ and spin-$\frac{1}{2}$ modes as $\nu\rightarrow 0$. The result is displayed in Fig. \ref{fig:s0l0_nu}, which can be understood as a zoomed version of Fig. \ref{fig:s01over2l01over2} at the near-extremal region, covering the transition part observed in Fig. \ref{fig:s01over2l01over21}. One notes a drastic bifurcation at a critical value $qQ_c(s)$, above which the modes become non-damping. We note that the extremal values for $qQ<qQ_c$ also have a distinct ``almost constant'' behavior for $\nu\rightarrow 0$ in contrast with the $\ell>|s|$ case.

\begin{figure}[htb]
	\caption{Non-damping transitions for $\ell = |s|$ scalar (top) and	spinor (bottom) modes for the RN black hole. The modes represented	with darker color have coupling slightly above the critical value, whereas the lighter color are slightly below. Even with $qQ$ differing only in the sixth decimal plane, they display radically different extremal limits in the region $\nu\leq \nu_c$.}
	\begin{center}
		\includegraphics[width=0.95\textwidth]{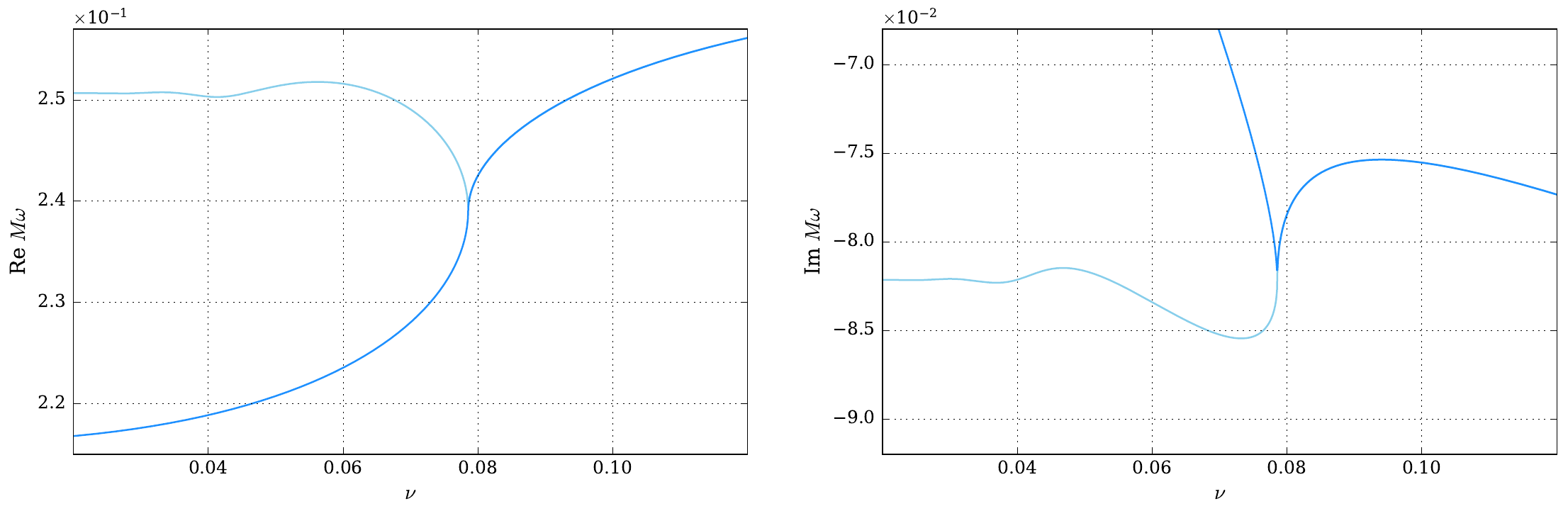}
		\includegraphics[width=0.95\textwidth]{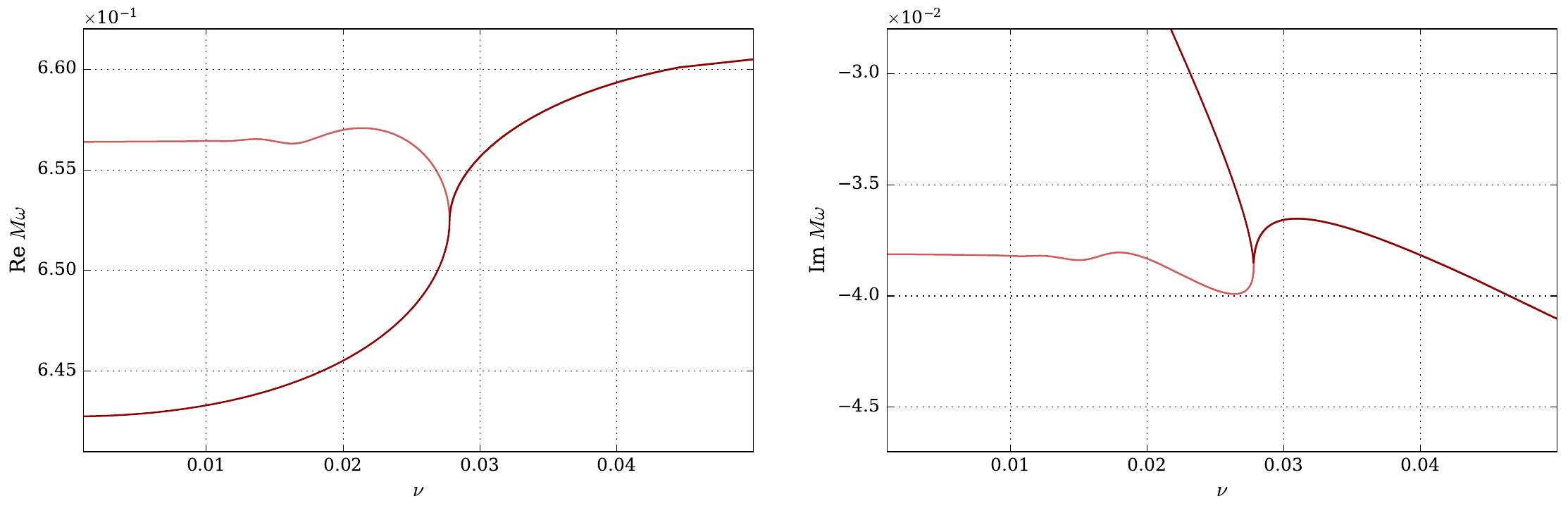}
	\end{center} 
	\label{fig:criticalqQ}
\end{figure}

In Fig. \ref{fig:criticalqQ}, we zoom into the transition for $s=0$ and
$s=-\frac{1}{2}$. A more careful numerical analysis yields critical values:
\begin{equation}
qQ_c(s=0) \simeq 0.216228,\qquad
\text{and }\qquad qQ_c(s=-1/2) \simeq 0.642745,
\label{eq:criticalqQ}
\end{equation}
with the split at $\nu=\nu_c\simeq 0.078547$ for $s=0$ and
$\nu_c\simeq 0.027797$ for $s=-1/2$, corresponding to $Q/M\simeq
0.996917$ and $Q/M\simeq 0.999614$, respectively.

The non-damping modes also have the property that the associated single monodromy parameters $\theta_{\mathrm{Rad},0}$ and $\theta_{{\mathrm{Rad}},0}$ in
\eqref{eq:nuparametrization} have finite limits as $\nu\rightarrow 0$. Given that the expansion parameter $z_0$ is
proportional to $\sin\nu$, it is also small in this limit and one can actually solve the equations \eqref{eq:radialsystemeqn} approximately by considering the first terms of the expansions involved. For the calculation of the QN frequencies, one can tackle the expansions for $c_{z_0}$ in \eqref{eq:c5expansion} and equation \eqref{eq:zerochi5} in terms of $\nu$ directly. We follow the same strategy used for Kerr BH and write the schematic expansion for the frequency and the composite monodromy parameter $\sigma$ as
\begin{equation}
M\omega = qQ + \beta_1 \nu + \beta_2 \nu^2+\ldots,\qquad
\sigma = 1 + \alpha_0+\alpha_1 \nu + \alpha_2\nu^2+\ldots,
\end{equation}
where the coefficients $\beta_i$ and $\alpha_i$ may encode non-analytic corrections in $\nu$. The values for $\alpha_i$ and $\beta_i$ can be found recursively from the expansions \eqref{eq:c5expansion} and \eqref{eq:zerochi5}. For instance, substitution of $\sigma$ into \eqref{eq:c5expansion} we arrive at
\begin{equation}
\sigma = 1+\alpha_0+ \frac{16qQ
	(2q^2Q^2-s(3+4s)-6{}_{s}\lambda_{\ell})}{\alpha_0(\alpha_0^2-1)}
\beta_1 \nu + \ldots,
\end{equation}
where
\begin{equation}
{}_{s}\lambda_{\ell} = (\ell-s)(s+\ell+1),\qquad
\alpha_0 =\pm\sqrt{(1 + 2\ell)^2-4q^2Q^2}.
\end{equation}
With the value of $\sigma$ calculated, we can substitute in \eqref{eq:zerotau5p} and find an equation for $\beta_1$. We note that
$\chi_V$ in the equation \eqref{eq:zerochi5} is analytic in $z_0$ and thus can be written as expansion in $\nu$. The non-analytic terms like $\log\nu$ arise from the expansion of the gamma functions in $\Theta_V$ \eqref{eq:theta5}. By considering the lowest-order
contribution in $\nu$, we find, after some algebra,
\begin{equation}
e^{-\frac{i\pi}{2}\alpha_0}\frac{\Xi(\alpha_0,\beta_1,qQ,s)}{\Xi(-\alpha_0,\beta_1,qQ,s)}(4\nu qQ)^{\alpha_0} =
1+\mathcal{O}(\nu,\nu\text{log}\nu),
\label{eq:zerotau5pnu}
\end{equation}
with 
\begin{equation}
\Xi(\phi,\beta_1,qQ,s) =\Gamma(1-\phi)^2 \Gamma(
\frac{1}{2}(1+\phi)-i\beta_1)\Gamma(\frac{1}{2}
(1+\phi)-iqQ-s) \Gamma(\frac{1}{2}(1+\phi)-iqQ+
s).
\end{equation}

Assuming $\alpha_0$ real and positive, the small $\nu$ regime of \eqref{eq:zerotau5pnu} simplifies considerably. The $\nu^{\alpha_0}$
term becomes small, and the equation can be solved provided the argument of one of the gamma functions becomes very
small, near its pole at vanishing argument. At any rate, this is the behavior we expect from the numerical analysis above the
critical value $qQ_c<qQ\leq 1$. Therefore, we find that the first correction to the non-damping frequency is
\begin{equation}
\beta_1=-\frac{1}{2}i(\alpha_0+1)+i\frac{(-i)^{\alpha_0} \Upsilon(\alpha_0,\ qQ,\ s)}{\Gamma(-\alpha_0)\Upsilon(-\alpha_0,\ qQ,\ s)} (4\nu q Q)^{\alpha_0}+\ldots
\end{equation}
with $\Upsilon(\phi,\ qQ,\ s)$ given in terms of Gamma functions,
\begin{equation}	
\Upsilon(\phi,\ qQ,\ s) =	\Gamma(1\mp\phi)^2 
\Gamma(\tfrac{1}{2}(1\pm\phi)-iqQ+s)
\Gamma(\tfrac{1}{2}(1\pm\phi)-iqQ-s).
\end{equation}
Again, for small $\nu$, we note that the second term becomes irrelevant. Dropping this term and summarizing the result, we find the
frequency to be
\begin{equation}
M\omega \simeq 
qQ -\frac{1}{2}i(\alpha_0+1)\nu = qQ
-\frac{i}{2}(1+\sqrt{(1+2\ell)^2-4q^{2}Q^{2}})\nu. 
\label{eq:omegaeq}
\end{equation}
These modes, illustrated in Fig. \ref{fig:criticalqQ}, are the non-damping $qQ>qQ_c$ QN frequencies with imaginary parts that
vanish linearly with $\nu$ as $\nu\rightarrow 0$. For $\alpha_0$ real, the near-extremal behavior of the real part of $M\omega$ is quadratic, which again corroborates the findings of last section. Similar remarks about the non-damping modes have been made by \cite{Richartz:2017qep} using the CF method, albeit with a slightly different expression for $\alpha_0$. 


\section{Confluence Limit and the Extremal RN Black Hole}
\label{sec:Conflimit_and_RNextremal}

For the modes with $\ell > |s|$ and $\ell = |s|$ with $qQ<qQ_c$, the QN frequencies approach the extremal limit still keeping a finite imaginary part. In such cases, the extremal limit coincides with the confluent limit of the \eqref{heuneq1}, written in terms of the parameters \eqref{parameters} as
\begin{equation}
\Lambda = \frac{1}{2}(\theta_{\mathrm{Rad},z_0}-\theta_{\mathrm{Rad},0}), \ \ \ \theta_{\circ}
=\theta_{\mathrm{Rad},z_0}+\theta_{\mathrm{Rad},0}, \ \ \ u_0 = \Lambda z_0 
, \ \ \ \Lambda \rightarrow \infty,
\label{eq:conflimit}
\end{equation}
with $\Lambda$ going to infinity, for $M\omega\neq qQ$, as
\begin{equation}
\Lambda = i\frac{M\omega-qQ}{\nu} - \frac{i}{6}(qQ-7M\omega)\nu+\mathcal{O}(\nu^3).
\label{eq:lambRN}
\end{equation}
Then, for $Q=M$, we have the following double-confluent Heun equation for the radial equation
\begin{equation}
\frac{d^2
	y}{dz^2}+\bigg[\frac{2-\theta_{\mathrm{Rad},\circ}}{z}-\frac{u_0}{z^2}\bigg]
\frac{dy}{dz}-\bigg[\frac{1}{4} 
+\frac{\theta_{\mathrm{Rad},\star}}{2z}+\frac{u_0k_{u_0}-u_0/2}{z^2}\bigg]y(z)=0, 
\label{doubheuneq}
\end{equation}
with two irregular singularities of rank 1 at $z=0$ and $z=\infty$ \cite{NIST:DLMF}. As listed in Table \ref{tab:DCHEDic}, the parameters of the equation are
\begin{equation}
\theta_{\mathrm{Rad},\circ} =  2s
+2i(-qQ+2Q\omega),\qquad
\theta_{\mathrm{Rad},\star} =-2s+2i(2Q\omega-qQ)
\label{eq:extremalparameters}
\end{equation}
with accessory parameter and modulus obtained by considering \eqref{modaccASnufunction} in the confluent limit
\eqref{eq:conflimit}
\begin{equation}
u_0k_{u_0} = {}_{s}\lambda_{l,m} +2s-i(1-2s)qQ +
2(qQ+i(1-2s)) M\omega
-8(M\omega)^2,
\ \
u_{0}= -4M\omega(M\omega-qQ). \\ 
\end{equation}

As seen in Chapter \ref{ChapIsoMethod}, the RH map for the double-confluent Heun equation can be cast in terms of the isomonodromic $\tau_{III}$-function, where the confluent limit \eqref{eq:conflimit} of the
$\tau_V$-function allows us to define the RHm for the radial equation \eqref{doubheuneq} as,
\begin{equation}
\tau_{III}(\{\tilde{\theta}\}_{\mathrm{Rad}};\sigma,\eta;u_0)=0, \ \ \ \,
u_0\frac{d}{du_0}\text{log}
\tau_{III}(\{\tilde{\theta}\}_{\mathrm{Rad}}^{-};\sigma-1,\eta;u_0)-\frac{(\theta_{\circ}-1)^2}{2}=
u_0 k_{u_0},
\label{eq:radialextremalsystemeqn}
\end{equation}
with $\{\tilde{\theta}\}_{\mathrm{Rad}} =
\{\theta_{\mathrm{Rad},\circ},\theta_{\mathrm{Rad},\star}\}$ and $\{\theta\}_{\mathrm{Rad}}^{-} =
\{\theta_{\mathrm{Rad},\circ}-1,\theta_{\mathrm{Rad},\star}-1\}$. From Appendix \ref{sec:tools}, equation \eqref{eq:fredholmIII}, also given in \cite{Gamayun:2013auu,daCunha:2021jkm}, we have again the first terms for the $\tau_{III}$-function
\begin{multline}
	\tau_{III}(\{\tilde{\theta}\};\sigma,\eta;u_0) = C_{III}(\{\tilde{\theta}\},\sigma)
	u_0^{\frac{1}{4}\sigma^2-\frac{1}{8}\theta_\circ^2}e^{\frac{1}{2}u_0}\times\\
	\left(
	1-\frac{\sigma-\theta_\circ\theta_\star}{2\sigma^2}u_0
	-\frac{(\sigma+\theta_\circ)(\sigma+\theta_\star)}{4\sigma^2
		(\sigma-1)^2}\kappa_{III}^{-1}u_0^{1-\sigma} -
	\frac{(\sigma-\theta_\circ)(\sigma-\theta_\star)}{4\sigma^2
		(\sigma+1)^2}\kappa_{III}u_0^{1+\sigma}+{ O}(u_0^2,u_0^{2\pm
		2\sigma})\right)
	\label{eq:tauIIIexpansion}
\end{multline}
where
\begin{equation}
\kappa_{III}=e^{i\pi\eta}
\frac{\Gamma(1-\sigma)^2}{\Gamma(1+\sigma)^2}
\frac{\Gamma(1+\tfrac{1}{2}(\theta_\star+\sigma))}{
	\Gamma(1+\tfrac{1}{2}(\theta_\star-\sigma))}
\frac{\Gamma(1+\tfrac{1}{2}(\theta_\circ+\sigma))}{
	\Gamma(1+\tfrac{1}{2}(\theta_\circ-\sigma))}.
\label{eq:kappaIII}
\end{equation}

Again, we list in Appendix \ref{sec:tools} expressions for $\tau_{III}$ in terms of Fredholm determinants, which we used in our numerical analysis. As discussed in the last part of the Chapter \ref{ChapIsoMethod}, the confluent limit of \eqref{eq:c5expansion} leads us to the expansion for $k_{u_0}$,
\begin{equation}
\begin{aligned}
u_0k_{u_0}=&\frac{(\sigma-1)^2-(\theta_\circ-1)^2}{4}
+\frac{\theta_\circ\theta_\star}{4}\bigg(\frac{1}{\sigma-2}-\frac{1}{\sigma}\bigg)u_0-
\\&-\bigg[\frac{\theta_\circ^2\theta_\star^2}{16}\bigg(\frac{1}{(\sigma-2)^3}-\frac{1}{2\sigma^3}\bigg)-\frac{\theta_\circ^2+\theta_\star^2-\theta_\circ^2\theta_\star^2}{
	8\sigma(\sigma-2)}
-\frac{(\theta_\circ^2-1)(\theta_\star^2-1)}{8(\sigma+1)(\sigma-3)}      
\bigg]{u_0}^2+{O}({u_0}^3).
\label{eq:c3expansion}
\end{aligned}
\end{equation}

Finally, following the same strategy applied to Kerr BH, we can take the confluent limit \eqref{eq:conflimit} of the condition
\eqref{eq:quantizationV}. We therefore have that the boundary conditions relating to the QNMs are written in terms of the monodromy
parameters as 
\begin{equation}
e^{i\pi\eta}=e^{-2\pi i\sigma}
\frac{\sin\tfrac{\pi}{2}(\theta_\star+\sigma)}{
	\sin\tfrac{\pi}{2}(\theta_\star-\sigma)}
\frac{\sin\tfrac{\pi}{2}(\theta_\circ+\sigma)}{
	\sin\tfrac{\pi}{2}(\theta_\circ-\sigma)}+\mathcal{O}(e^{2i\Lambda}),
\label{eq:quantizationIII}
\end{equation}
where the expression above is hold if $\Re M\omega > qQ$ in the extremal limit, as it can be seen in the expression for $\Lambda$ \eqref{eq:lambRN}. If $\Re M\omega < qQ$, then the argument of the exponential factor is replaced by $+2\pi i\sigma$. 

\begin{figure}[htb]
	\caption{Fundamental modes for $s=0, l=1,2$ (top) and $s=-\frac{1}{2}$, $\ell =\frac{3}{2}, \frac{5}{2}$ (bottom) in extremal RN black hole $(Q=M)$ as a function of $qQ$.}
	\begin{center}
		\includegraphics[width=0.95\textwidth]{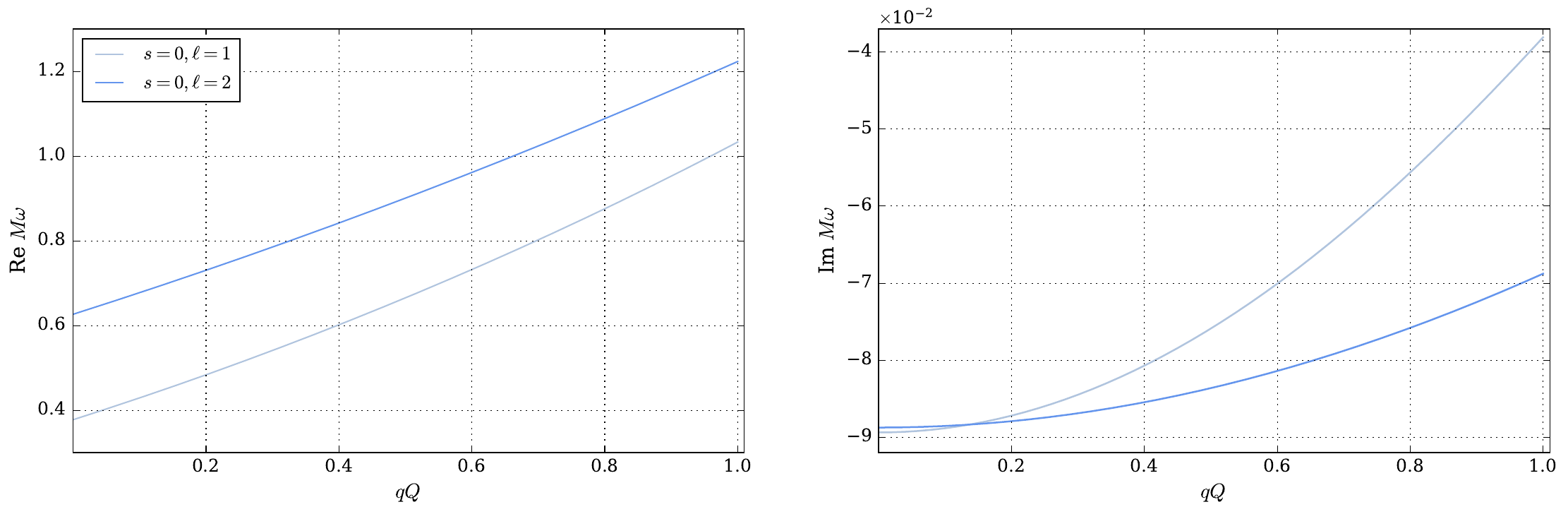}
		\includegraphics[width=0.95\textwidth]{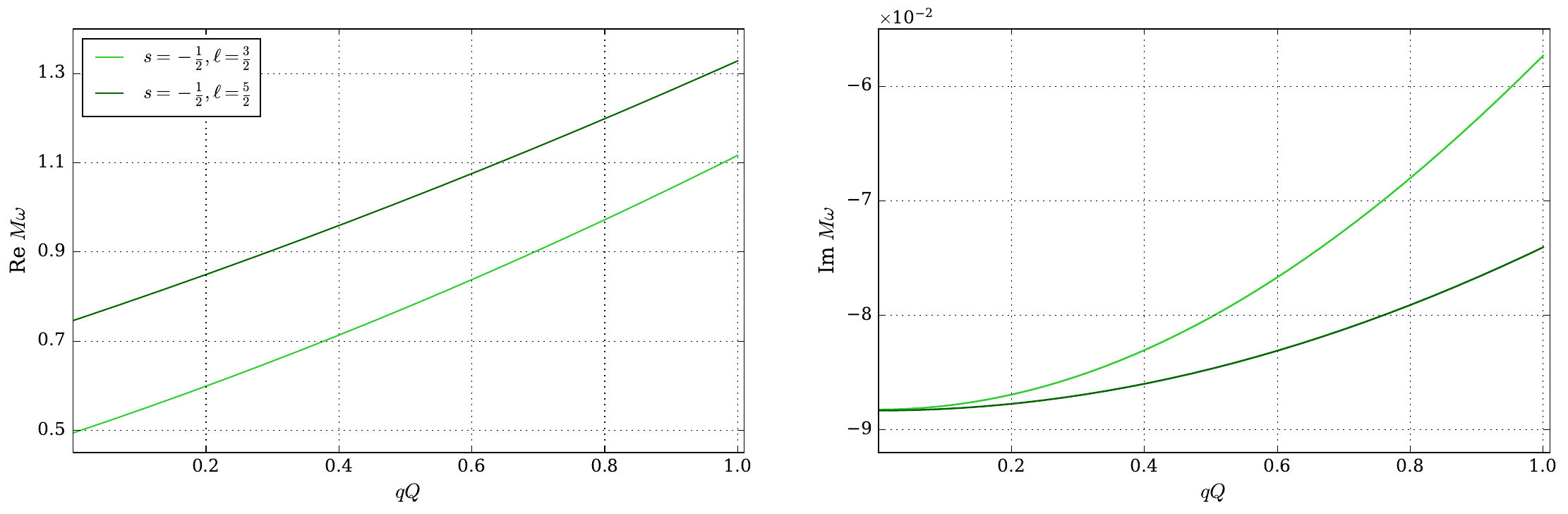}
	\end{center}
	\label{fig:s01over2_extremal}
\end{figure}

As in the non-extremal case, the resolution of the equations \eqref{eq:radialextremalsystemeqn} can be done directly from the
implementation of the $\tau_{III}$ function and accessory parameter $k_{u_0}$, where from the last one, one calculates the value of $\sigma$ via Muller's method. Then, $\sigma$ and $e^{i\pi\eta}$ are replaced in $\tau_{III}(\{\tilde{\theta}\};\sigma,\eta;u_0)=0$. The final step consists in obtaining the roots (QN frequencies) from the $\tau_{III}$ using again the finding roots algorithm. In Fig. \ref{fig:s01over2_extremal}, we plot the fundamental spin-$0$ and spin-$\frac{1}{2}$ perturbations for the extremal RN black hole as a
function of $qQ$, for the $\ell>|s|$ case. As expected, modes with higher values of $\ell$ tend to decay faster, and the difference
increases with $qQ$.

In Tables \ref{tab:scalar} and \ref{tab:dirac}, we give further support that the extremal limit of the $\ell>|s|$ modes is actually
smooth, by solving the non-extremal equations \eqref{eq:radialsystemeqn} at $Q/M$ very close to $1$ and the actual
extremal value computed numerically from \eqref{eq:radialextremalsystemeqn}. The values for $qQ=0$ in the scalar case show good accordance with \cite{PhysRevD.53.7033}, whereas the spinor case agrees with \cite{Richartz:2015saa} for $qQ<0.1$. Unlike the continued fraction method used in these references, the RH map allows us to compute the fundamental mode frequency for arbitrary values of $qQ$. 
\begin{figure}[htb]
	\caption{QNMs for $\ell=|s|$ in the scalar (blue) and spinorial (red) cases. For the interaction parameter less that the critical
		value of \eqref{eq:criticalqQ}, the values are calculable from the $\tau_{III}$ function. Above the critical value, the modes are
		non-damping and the imaginary values of the eigenfrequencies go abruptly to zero.}	
	\begin{center}
		\includegraphics[width=0.95\textwidth]{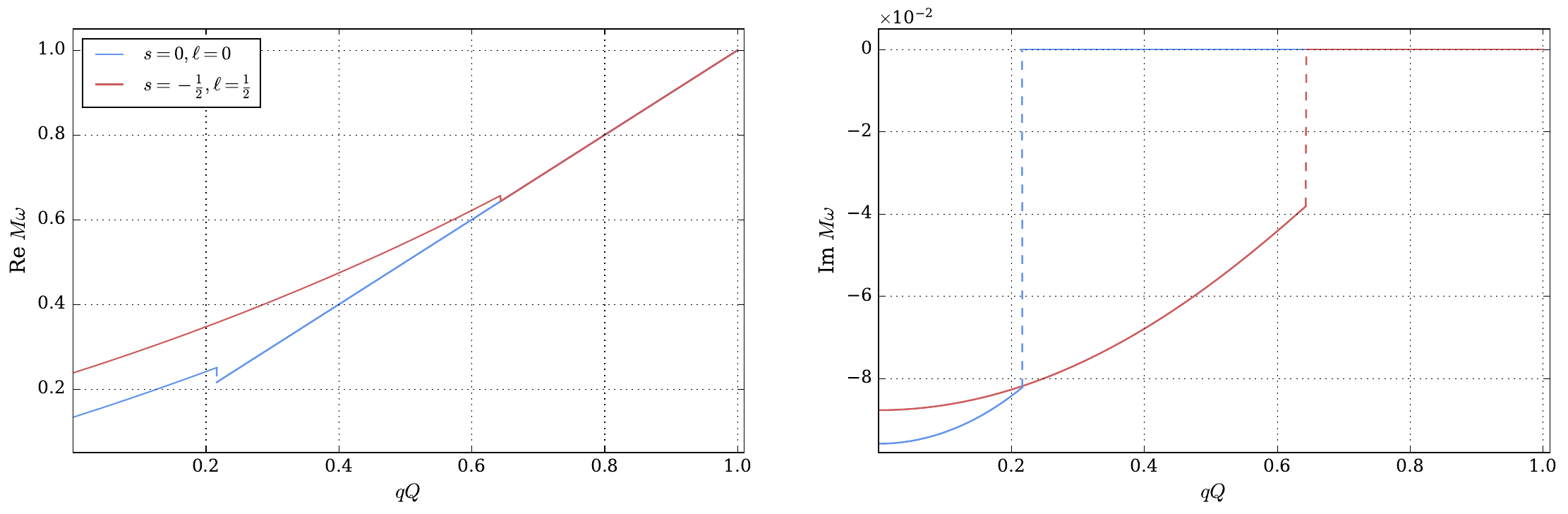}
	\end{center} 
	\label{fig:seql_extremal}
\end{figure}

For $\ell = |s|$ the QN frequencies will be divided into two different classes, listed in \ref{item:case}.  For the interaction
parameter $qQ$ smaller than the critical value given in \eqref{eq:criticalqQ}, the behavior of the frequencies is similar to the $\ell
>|s|$ case in which there will be a non-zero damping factor in the extremal limit. The values for the frequencies at extremality can also be computed from the equations \eqref{eq:radialextremalsystemeqn}. For $qQ$ above the critical value, however, the mode ``decouples'' from the double-confluent equations, and the frequency becomes non-damped with value $qQ/M$. The behavior is illustrated in Fig. \ref{fig:seql_extremal}. Because of the non-damping modes, the question of stability of the extremal RN black hole under scalar and spinorial perturbations can only be settled by considering higher order terms in the Maxwell-Einstein equations.

\begin{table}
	\centering
	\caption{A comparison between the fundamental modes for scalar $s=0$ perturbations of the RN black hole computed from \eqref{eq:radialsystemeqn} with $Q/M$ very close to $1$, and those computed in the extremal case using \eqref{eq:radialextremalsystemeqn} for various values of the interaction parameter $qQ$.}
	\scalebox{0.81}{
		\begin{tabular}{|c|c|c|c|c|}
			\hline
			qQ& \multicolumn{2}{c|}{$\ell=1$} & \multicolumn{2}{c|}{$\ell=2$}  \\ \hline
			&  $Q/M=0.999999$ &  $Q/M=1$  & $Q/M=0.999999$ & $Q/M=1$    \\
			\hline
			0.0 & $0.3776416 - 0.0893845i$  & $0.3776418 - 0.0893843i$  &
			$0.6265722
			-
			0.0887485i$
			&   $0.6265727 - 0.0887483i$    \\
			\hline
			0.1 & $0.4291343 - 0.0888385i$  & $0.4291346 - 0.0888382i$  &
			$0.6775330 - 0.0885422i$   &   $0.6775336 - 0.0885419i$   \\ \hline
			0.2 &  $0.4836193 - 0.0872035i$ & $0.4836196 - 0.0872031i$  &
			$0.7304163 - 0.0879239i$   &    $0.7304170 - 0.0879236i$       \\ \hline
			0.3 & $0.5411167 - 0.0844886i$  & $0.5411172 - 0.0844882i$  &
			$0.7852239
			-
			0.0868961i$
			&     $0.7852248 - 0.0868957i$      \\
			\hline
			0.4 & $0.6016610 - 0.0807101i$  & $0.6016616 - 0.0807095i$  &
			$0.8419591 - 0.0854624i$   &     $0.8419601 - 0.0854620i$      \\ \hline
			0.5 & $0.6653021 - 0.0758925i$  & $0.6653028 - 0.0758917i$  &
			$0.9006264
			-
			0.0836285i$
			&     $0.9006276 - 0.0836280i$      \\
			\hline
			0.6 & $0.7321070 - 0.0700711i$  & $0.7321078 - 0.0700701i$  &
			$0.9612317 - 0.0814015i$   &     $0.9612330 - 0.0814008i$      \\ \hline
			0.7 & $0.8021625 - 0.0632953i$  & $0.8021634 - 0.0632939i$  &
			$1.0237821 - 0.0787904i$   &     $1.0237836 - 0.0787896i$      \\ \hline
			0.8 & $0.8755790 - 0.0556340i$  & $0.8755799 - 0.0556320i$  &
			$1.0882862
			-
			0.0758061i$
			&    $1.0882879 - 0.0758052i$       \\
			\hline
			0.9 & $0.9524949 - 0.0471845i$  & $0.9524959 - 0.0471819i$  &
			$1.1547541
			-
			0.0724620i$
			&    $1.1547561 - 0.0724608i$       \\
			\hline
			1.0 & $1.0330846 - 0.0380869i$  & $1.0330854 - 0.0380833i$  &
			$1.2231973
			-
			0.0687733i$
			&     $1.2231995 - 0.0687720i$      \\
			\hline
		\end{tabular}
	}
	\label{tab:scalar}
\end{table}

\begin{table}
	\centering
	\caption{The comparison between fundamental modes for spinorial $s=-1/2$ perturbations of the RN black hole obtained from near-extremal \eqref{eq:radialsystemeqn} and extremal \eqref{eq:radialextremalsystemeqn} as a function of  $qQ$.}
	\scalebox{0.81}{
		\begin{tabular}{|c|c|c|c|c|}
			\hline
			qQ& \multicolumn{2}{c|}{$\ell=3/2$} &
			\multicolumn{2}{c|}{$\ell=5/2$}  \\ \hline
			&  $Q/M=0.999999$ & $Q/M=1$  & $Q/M=0.999999$    &
			$Q/M=1$    \\ \hline
			0.0 & $0.4941127 - 0.0882400i$  &  $0.4941131 - 0.0882398i$  &
			$0.7460834
			-
			0.0883267i$
			&     $0.7460841 - 0.0883265i$     \\ \hline
			0.1 & $0.5453239 - 0.0879134i$  & $0.5453244 - 0.0879132i$  &
			$0.7969051
			-
			0.0881804i$
			&    $0.7969059 - 0.0881802i$       \\ \hline
			0.2 & $0.5989590 - 0.0869355i$  & $0.5989596 - 0.0869352i$  &
			$0.8493706
			-
			0.0877419i$
			&     $0.8493715 - 0.0877416i$      \\ \hline
			0.3 &  $0.6550227 - 0.0853120i$  & $0.6550234 - 0.0853119i$  &
			$0.9034806
			-
			0.0870124i$
			&   $0.9034816 - 0.0870120i$        \\ \hline
			0.4 & $0.7135223 - 0.0830539i$  & $0.7135231 - 0.0830534i$  &
			$0.9592359
			-
			0.0859942i$
			&    $0.9592370 - 0.0859938i$       \\ \hline
			0.5 & $0.7744686 - 0.0801747i$  & $0.7744696 - 0.0801740i$  &
			$1.0166382
			-
			0.0846905i$
			&     $1.0166394 - 0.0846901i$      \\ \hline
			0.6 &  $0.8378753 - 0.0766939i$  & $0.8378764 - 0.0766930i$  &
			$1.0756891
			-
			0.0831055i$
			&    $1.0756906 - 0.0831049i$       \\ \hline
			0.7 & $0.9037594 - 0.0726358i$  & $0.9037607 - 0.0726347i$  &
			$1.1363911
			-
			0.0812441i$
			&    $1.1363927 - 0.0812434i$       \\ \hline
			0.8 & $0.9721413 - 0.0680306i$  & $0.9721429 - 0.0680293i$  &
			$1.1987468
			-
			0.0791125i$
			&    $1.1987486 - 0.0791117i$       \\ \hline
			0.9 & $1.0430450 - 0.0629152i$  & $1.0430468 - 0.0629135i$  &
			$1.2627593
			-
			0.0767178i$
			&     $1.2627613 - 0.0767169i$      \\ \hline
			1.0 & $1.1164980 - 0.0573342i$  & $1.1165001 - 0.0573321i$  &
			$1.3284321 - 0.0740684i$   &     $1.3284344 - 0.0740674i$      \\ \hline
		\end{tabular}
	}
	\label{tab:dirac}
\end{table}

\section{\texorpdfstring{Overtone Results: $\{\sigma,\eta\}$ and $\{\nu,\rho\}$ maps}%
{}}
\label{sec:RHmapslstRN}

Following the same procedure used for Kerr BH, where it was shown in Sec. \ref{sec:RHmapslst} the overtone frequencies for the case $a/M=0$ -- see Table \ref{tab:gravrad}. We can consider the expressions for $e^{i\pi\eta}$ and $X_{\pm}$, defined in \eqref{eq:nurhoparameter}, and the accessory parameter expansion for large $t$ \eqref{eq:contfracINF} to investigate the overtone frequencies for RN BH through the RHm that depends of the monodromy parameters $\{\nu,\rho\}$, as defined in \eqref{eq:tauVcondinf}. Where again we point out that, in this subsection, $\nu$ is the parameter of the $\tau_{V}$-function expanded for large $t$, as written in \eqref{eq:taufirstterm}. The previous "$\nu$" was defined only in the study of the extremal limit $Q\rightarrow M$.

As illustrated in Fig. \ref{fig:maps}, $\{\sigma,\eta\}$ and $\{\nu,\rho\}$ are related by the expression $X_{\pm}$ in \eqref{eq:nurhoparameter}. From this relation, we can show that the frequencies obtained from the RHm for \eqref{eq:radialsystemeqn} are exactly the same computed from the RHm for large $t$, where for Reissner-Nordström BH the map in \eqref{eq:tauVcondinf} relates $c_{z_0}$ and $z_0$ in equation \eqref{heuneq1} with the parameters $\{\nu,\rho\}$:
\begin{equation}
\frac{d}{dz_0}\text{log}(\tau_{V}(\{\theta\}_{\mathrm{Rad}}^{-};\nu-1, \rho;z_0)) =c_{z_0}+\frac{\theta_{\mathrm{Rad},0}(\theta_{\mathrm{Rad},z_0}-1)}{2z_0}, \quad \tau_{V}(\{\theta\}_{\mathrm{Rad}};\nu,\rho;z_0) =0,
\label{eq:tauVcondinfRN}
\end{equation}
$\{\theta\}_{\mathrm{Rad}}^{-}$ and $\{\theta\}_{\mathrm{Rad}}$ follow the same definition used in the map \eqref{eq:radialsystemeqn}, and depende of $qQ$, $Q/M$, $s$, $\ell$, and $M\omega$.

Again, to show the equivalence between the maps, we replace the expression for $e^{i\pi\eta}$ into the expression for $X_{\pm}$ and, then, compute a condition for $\nu$ from the equation $e^{-i\frac{\pi}{2} \nu} = X_-$. The first step is to use the expression that relates $\eta$ with $\{\theta\}(=\{\theta\}_{\mathrm{Rad}})$ and $\sigma$ written in \eqref{eq:quantizationV}, and the equation \eqref{eq:nurhoparameter} to show that the condition for $\nu$ is given by the same expression derived for Kerr BH, but now with the parameters $\theta_{{\mathrm{Rad}},z_0}$ and $\theta_{{\mathrm{Rad}},\star}$ associated with the parameters for the RN BH:
\begin{equation}
\nu_k = -\tfrac{1}{2}\theta_{\mathrm{Rad},z_0}-\tfrac{1}{4}(\theta_{\mathrm{Rad},\star}+1)+
k,\qquad k\in\mathbb{Z}
\label{eq:nuradquantcondRN}
\end{equation}
where depending on the overtone mode, one has $k>0$.

Using the condition for $\nu_{k}$ and fixing the values of $s$ and $\ell$, we can compute the overtone frequencies for different values of $Q/M$ and $qQ$. As example, we will only consider the case $qQ=0.001$ and $Q/M = 0.999$, which allows us to compare and complement the results already know in the literature, and obtained from the continued fraction method \cite{Richartz:2015saa}. Thus, from equation \eqref{heuneq1}, one has the following parameters for the problem:
\begin{equation}
\begin{gathered}
\theta_{\mathrm{Rad},0}=  s+\frac{i}{2\pi
	T_{-}}\bigg(\omega-\frac{qQ}{r_{-}}\bigg), \quad \theta_{\mathrm{Rad},z_0}=
s+ \frac{i}{2\pi
	T_{+}}\bigg(\omega-\frac{qQ}{r_{+}}\bigg), \\
\theta_{\mathrm{Rad},\star}=2i(2M\omega -qQ)-2s, \qquad
2\pi T_{\pm} = \frac{r_{\pm}-r_{\mp}}{2 r^{2}_{\pm}},   \qquad
r_{\pm} = M\pm\sqrt{M^2 -Q^2},
\label{RNpar}
\end{gathered}
\end{equation}
with modulus $z_0= 2i\omega(r_+ - r_-)$ and accessory parameter:
\begin{equation}
z_0c_{z_0} =   {}_{s}\lambda_{l,m}
+2s-i(1-2s)qQ+(2qQ+i(1-3s))\omega r_++i(1-s)\omega r_-
-2\omega^2 r_+^2
\label{RNacc}
\end{equation}
where the angular eigenvalue is given by ${}_{s}\lambda_{\ell m}=(\ell-s)(\ell+s+1)$. Substituting the parameters above and the conditition \eqref{eq:nuradquantcondRN} into equation \eqref{eq:contfracINF} and fixing the values of $s$, $\ell$, $qQ$, and $Q/M$, one obtains a function that depends only on $M\omega$. The final procedure consists in fixing the value of $N_c$ and computes the overtone frequencies. We have considered $N_c = 10^{4}$ in \eqref{eq:contfracINF} and computed the frequencies using the Muller's method. 

We remark that the expression above for $\nu_k$ simplifies the RHm \eqref{eq:tauVcondinfkerr}, where with $\nu_{k}$ calculated we arrive at the values for the overtones just searching for roots of the equation \eqref{eq:contfracINF} derived in Chapter \ref{ChapIsoMethod}. Eliminating the necessity of treating directly with the $\tau_V$-function expanded at $t=i\infty$, i.e. equation \eqref{eq:taufirstterm}.

\begin{table}[htb]
	\centering
	\caption{First overtones for $s=0,-1/2$ with $\ell=0$ (scalar case) and $\ell=1/2$ (spinorial case) in the Reissner-Nordström black hole. In the first column, we list the values for $M{}_s\omega_{\ell}$ found from \eqref{eq:contfracINF}, using the condition \eqref{eq:nuradquantcondRN}. We present, in the second column, the values obtained from the RHm \eqref{eq:radialsystemeqn} for the $\tau_{V}$-function expanded around $t=0$ with the condition \eqref{eq:quantizationV}.}
	\scalebox{0.84}{
		\begin{tabular}{|c|c|c|}
			\hline
			$n$ &  $M{}_{s}\omega_{\ell}$ -- large $t$
			& $M{}_{s}\omega_{\ell}$ -- small $t$ \\ \hline
			& $s=0$ and $\ell=0$ & $s=0$ and $\ell =0$ \\ \hline
			1 & $0.1339593261340725 - 0.0958430988307012i$ 
			& $0.1339593261340725 - 0.0958430988307012i$ \\ \hline
			2 & $0.0934644981823586 - 0.3306521443097391i$
			& $0.0934644981823586 - 0.3306521443097391i$ \\ \hline
			3 & $0.0755824097642539 - 0.5883276888253196i$
			& $0.0755824097642539 - 0.5883276888253196i$ \\ \hline
			4 & $0.0678510134676415 - 0.8443059932399761i$
			& $0.0678510134676415 - 0.8443059932399761i$ \\ \hline
			5 & $0.0635274710575977 - 1.0984317467320831i$
			& $0.0635274710575977 - 1.0984317467320831i$ \\ \hline
			6 & $0.0327213322000730 - 1.3464553946699416i$
			& $0.0327213322000730 - 1.3464553946699416i$ \\ \hline
			& $s=-1/2$ and $\ell=1/2$ & $s=-1/2$ and $\ell =1/2$ \\ \hline
			1 & $0.2385496381125137 - 0.0878110903057080i$ 
			& $0.2385496381125137 - 0.0878110903057080i$ \\ \hline
			2 & $0.1969774646281255 - 0.2813676377174057i$
			& $0.1969774646281255 - 0.2813676377174057i$ \\ \hline
			3 & $0.1462034288927990 - 0.5156262925493441i$
			& $0.1462034288927990 - 0.5156262925493441i$ \\ \hline
			4 & $0.1134460243851753 - 0.7696014318789649i$
			& $0.1134460243851753 - 0.7696014318789649i$ \\ \hline
			5 & $0.0933756448011380 - 1.0259812646166524i$
			& $0.0933756448011380 - 1.0259812646166524i$ \\ \hline
			6 & $0.0797388708490743 - 1.2815581659615172i$
			& $0.0797388708490743 - 1.2815581659615172i$ \\ \hline
			
		\end{tabular} 
		\label{tab:RNscalarSpin}
	}
\end{table}

\begin{figure}[htb]
	\caption{Contour plot of the equation \eqref{eq:contfracINF}, with $\nu_{k=0}$, for the scalar ($s=\ell=0$) and spinorial ($s=-1/2$, $\ell=1/2$) cases, with $qQ=0.001$ and $Q/M=0.999$. Note that the first three zeros, in both cases, corresponde to the first three overtones, listed in Table \ref{tab:RNscalarSpin} and in \cite{Richartz:2015saa}(Table VII).}
	\begin{subfigure}{.5\textwidth}
		\centering
		\includegraphics[width=0.95\textwidth]{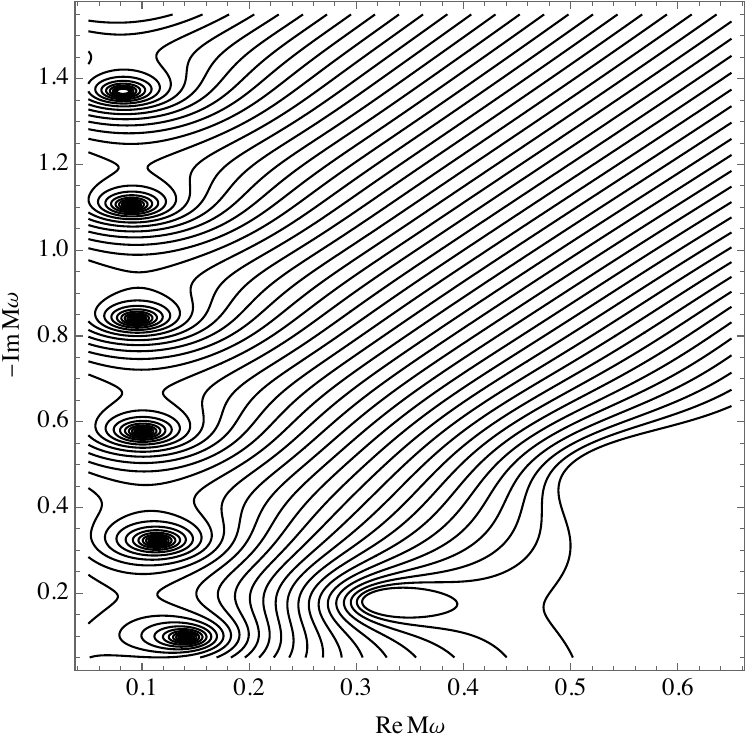}
		\caption{Scalar case.}
	\end{subfigure}%
	\begin{subfigure}{.5\textwidth}
		\centering
		\includegraphics[width=0.95\textwidth]{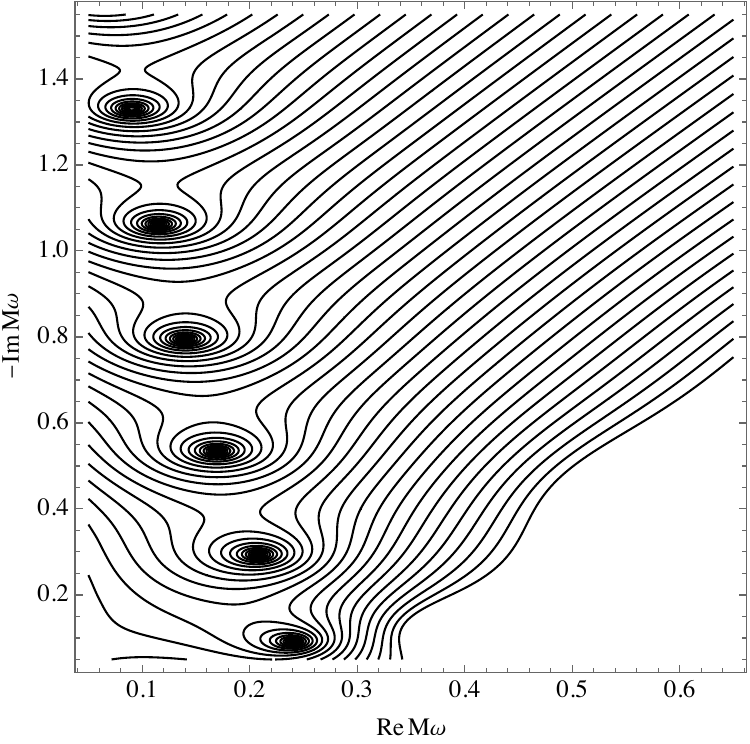}
		\caption{Spinorial case.}
	\end{subfigure}
	\label{fig:contourRN}
\end{figure}

The numerical results for the first six overtone frequencies, for the scalar $s=0$ and spinorial $s=-1/2$ cases, can be checked in Table \ref{tab:RNscalarSpin}, and match those obtained from the $\tau_{V}$-function in the map \eqref{eq:radialsystemeqn}. The first three values are already known in the literature \cite{Richartz:2015saa}. Additionally, we present the contour plot for \eqref{eq:contfracINF} in Fig. \ref{fig:contourRN}, where we can see that the zeros in the expression \eqref{eq:contfracINF} are the overtone frequencies. Note that, when the overtone number $n$ increases the real part of the frequency goes to zero, while the imaginary part becomes more negative.

\section{Conclusion of the Chapter}
\label{sec:discussion}

In this Chapter, we solved the Riemann-Hilbert maps introduced in Chapter \ref{ChapIsoMethod} to study quasinormal modes frequencies associated with scalar and spinorial perturbations in the Reissner-Nordström black hole. The maps consist in mapping the accessory parameter and modulus of the confluent and double-confluent Heun equations into monodromy parameters of the isomonodromic $\tau_{V}$ and $\tau_{III}$-functions expanded in their branch points. In turn, the expansions for the two isomonodromic $\tau$-funcions are given in terms of $c=1$ irregular conformal blocks \cite{Gamayun:2013auu,Lisovyy:2018mnj} or Fredholm determinant formulation \cite{daCunha:2021jkm}. 

For the Reissner-Nordström black hole, the analysis is comparatively simpler when compared with Kerr BH, since the angular problem is solved exactly in terms of spin-weighted spherical harmonics. We have therefore treated with the radial equation involved in the problem, where the analysis for values of $Q\in[0,M]$ and $qQ$ varying was made. Near the extremality, we have found two distinct behaviors for the QN frequencies for $\ell>|s|$ and $\ell=|s|$, with the last one developing a crucial rule in our analysis. There were verified the appearance of non-damping modes for $\ell=|s|$, with relaxation time diverging with the inverse temperature as one approaches $Q=M$. We have found that these non-damping modes appear above a critical value for the interaction parameter \eqref{eq:criticalqQ}. Below such a critical values the QN frequencies have a finite imaginary part in the limit $Q=M$.

This parameter $qQ$ plays a role for RN similar to the angular momentum to Kerr, and, unlike $m$, it can be varied continuously. We have found the critical value $qQ_c$ to be a bifurcation point in the sense that for small enough extremality parameter $\nu$, the behavior of the QN frequencies changes dramatically as the interaction parameter crosses $qQ_c$. The actual mechanism for this transition certainly needs clarification, and we hope to return to this issue in the future. Here we have kept our analysis to scalar and spinor modes, but, given the wide scope of the RH map, we have every reason to expect that the method will work for higher spin perturbations.

We finished by showing in Sec. \ref{sec:RHmapslstRN} that the overtone frequencies for $Q/M=0.999$ and $qQ=0.001$ can be obtained using two different maps defined in \eqref{eq:radialsystemeqn} and \eqref{eq:tauVcondinfRN}. In this way, one has two forms of computing the overtone frequencies for the Reissner-Nordström black hole. In addition, Table \ref{tab:RNscalarSpin} gives an insight into the relationship between the RH maps described in Fig. \ref{fig:maps}.

\chapter{Conclusion}
\thispagestyle{myheadings}

The analytical and numerical control of the computation of the QN frequencies is fundamental to understanding the physical properties of black holes, such as their mass, charge, and angular momentum. Additionally, given the last achievement reached by the LIGO collaboration, the theoretical results can be verified by the data obtained in the detectors. In this thesis, we presented and applied the isomonodromy method in the study of linear perturbations in the four-dimensional Kerr and Reissner-Nordström black holes. We made use of two conditions for the $\tau_V$ and $\tau_{III}$-functions together with the accessory parameter expansion, as described in Chapter \ref{ChapIsoMethod}, to numerically compute the quasinormal modes' frequencies. Furthermore, we analyzed with great precision the extremal limit in both black holes focusing on the fundamental QN frequencies.

We organized the thesis in the following form: In Chapter \ref{chap1}, it was presented an overview of black hole perturbations theory focusing on the theory applied in the Kerr and RN black holes. For the first case, $s$-spin field perturbations are described by the Teukolsky master equation, where, via variable changes, such an equation can be separated into ODEs, whose solutions are associated, for given boundary conditions, with the quasinormal mode solutions. For the Reissner-Nordström black hole, we considered linear perturbations of massless charged scalar and spinorial fields in this spacetime, whose radial differential equation involved has the same form as the radial Teukolsky master equation. Finally, since the treatment of linear perturbations for both BHs leads to the confluent Heun equation, for the non-extremal cases $a<M$ and $Q<M$ and double-confluent Heun equation, for the extremal cases $a=M$ and $Q=M$, we defined, in the final part of the chapter, two tables that list the parameters associated with the non-extremal and extremal cases, relevant to the QN frequency analysis.

We showed in Chapter \ref{ChapIsoMethod} that the study of isomonodromic deformations theory for both confluent and double-confluent Heun equations is given in terms of monodromy parameters and isomonodromic $\tau$-functions. From the approach, we have seen that the conditions for the $\tau$-functions are related to Riemann-Hilbert maps, which connect the accessory parameter and modulus of the differential equations involved (CHE or DCHE) to monodromy parameters, which, in turn, parametrize the expansions for the $\tau$-functions. 

Based on the theory introduced, we dealt with the isomonodromic deformations approach in the CHE, which led us to two conditions for the transcendental $\tau_{V}$-function. Given the branch points structure, we have seen that such a function is parametrized by two monodromy parameters, where for the expansion around $t=0$ and $t=i\infty$, one has $\{\sigma,\eta\}$ and $\{\nu,\rho\}$, respectively. 

Furthermore, it was presented that via the coalescence process, we can define a new Reimann-Hilbert map associated with the DCHE, whose map is written in terms of the $\tau_{III}$-function expanded at the branch point $u=0$. We concluded by showing that the accessory parameters for the CHE and DCHE can be computed using series representation of the solution of the differential equations, which led us to the numerical control of the Riemann-Hilbert maps defined for both differential equations.

The numerical results of our work were presented in Chapters 4 and 5 (plus Appendix). First, in Chapter 4, we applied the isomonodromic method to compute the QN frequencies associated with gravitational, scalar, and electromagnetic perturbations in the Kerr black hole, where all these perturbations are described by the Teukolsky Master equation. We demonstrated that the method works with a high precision for all values of the ratio $a/M$ in the interval $[0,1]$. One has, therefore, that the three Riemann- Hilbert maps defined in Chapter \ref{ChapIsoMethod} provide us numerical procedure alternative to the continued fraction method, given that, via the maps, one has precise control of the behavior for the QN frequencies at the extremal limit $r_-\rightarrow r_+$. In the method the eigenvalue problem for the radial TME equation is reduced to solving two transcendental equations for the non-extremal and extremal Kerr BH, which are numerically amenable given the analytical properties of the functions involved.

We found two distinct behaviors as $r_-\rightarrow r_+$; for the $\ell=|s|$ modes, we observed that the frequencies approach $\omega=m/(2M)$, in a manner linear with the temperature. For the modes with $\ell\neq|s|$, the extremal limit was seen to be described by the isomonodromic $\tau_{III}$-function by using the Fredholm determinant formulation for the $\tau$-function. By keeping the first near-extremal correction, we were able to assert that the extremal value for the frequencies approached the extremal value in a manner quadratic with the temperature, then in the final part of the chapter, we showed that the overtone frequencies for $a/M=0$ can be obtained using two different RH maps, leading to two forms of computing the overtone frequencies for the Schwarzschild black hole.

Next, in Chapter \ref{chap:RNBH}, we solved the Riemann-Hilbert maps to study quasinormal modes frequencies associated with scalar and spinorial perturbations in the Reissner-Nordström black hole. Again, the maps consist in mapping the accessory parameter and modulus of the confluent and double-confluent Heun equations into monodromy parameters of the isomonodromic $\tau_{V}$ and $\tau_{III}$-functions expanded in their branch points. In turn, the expansions for the two isomonodromic $\tau$-funcions are given in terms of $c=1$ irregular conformal blocks \cite{Gamayun:2013auu,Lisovyy:2018mnj} or Fredholm determinant \cite{daCunha:2021jkm}. 

For this case, the analysis is comparatively simpler when compared with the rotating Kerr BH, since the angular problem is solved exactly in terms of spin-weighted spherical harmonics. We therefore treated only with the radial equation involved in the problem, where the analysis for different values of $Q\in[0,M]$ and $qQ$ varying was made. Near the extremality, we found two distinct behaviors for the QN frequencies, for $\ell>|s|$ and $\ell=|s|$. There were verified the appearance of non-damping modes for $\ell=|s|$, with relaxation time diverging with the inverse temperature as one approach $Q=M$. We found that these non-damping modes appear above a critical value for the interaction parameter $qQ$. Below such a critical values the QN frequencies have a finite imaginary part in the limit $Q=M$. The actual mechanism for this transition certainly needs clarification, and we hope to return to this issue in the future. Here we kept our analysis to scalar and spinor modes, but given the wide scope of the RH maps, we have every reason to expect that the method will work for higher spin perturbations. In the final part of the chapter, we showed that the overtone frequencies for $Q/M=0.999$ and $qQ=0.001$ can be obtained using two RH maps. In this way, one has two forms of computing the overtone frequencies for the Reissner-Nordström black hole.

As was anticipated in \cite{CarneirodaCunha:2019tia}, the method does not converge faster than the standard continuous fraction method but, on the other hand, has analytical properties more transparent. Specifically, one can derive asymptotic formulas for the frequency in the extremal limit. We hope the results showed in this thesis provide promise for the application of the isomonodromy method to study problems related to physics and mathematics. For physics, we hope that the method developed in the papers \cite{CarneirodaCunha:2019tia,daCunha:2021jkm,Cavalcante:2021scq,daCunha:2022ewy} and presented in this thesis proves to be an invaluable tool in the calculation of black hole perturbations, both in the asymptotically flat case and otherwise \cite{Novaes:2018fry,Barragan-Amado:2018pxh}. With the comparatively nice properties of the isomonodromic $\tau$-functions involved, one can perhaps utilize them to study nonlinear perturbations and stability in black holes \cite{Green:2019nam}.  

The quasinormal mode frequencies for gravitational, electromagnetic, and scalar perturbations in the Kerr BH and scalar and Dirac perturbations in the RN BHs have been computed by many authors. Rather than listing numerical tables of well-known results, we set up a web page \cite{GithubIM} that provides tabulated values of the frequencies with more results obtained via the isomonodromy method. On this web page, we also provide the notebooks in $\texttt{MATHEMATICA}$ that compute the overtone frequencies in the contour maps \eqref{fig:contour_schw} and \eqref{fig:contourRN}. Finally, a script in \href{http://julialang.org}{Julia} language for the computation of QN frequencies is available.


\newpage
\thispagestyle{empty}
\begin{center}\centering
	\vspace*{\fill}
	\resizebox{!}{0,60cm}{\textbf{Appendices }}
	\vspace*{\fill}
\end{center}
\newpage

\begin{appendices}
	
	\chapter{Newman-Penrose formalism}
	\label{sec:NPformApp}
	
	We present in this appendix some useful formulas of the Newman-Penrose formalism.
	
	\section{Null tetrad}
	
	Any NP tetrad must satisfy the following orthogonality relations:
	\begin{equation}
	\ell^{\mu}n_{\mu} = 1, \qquad m^{\mu}{m^{*}}_{\mu} = -1
	\end{equation}
	satisfying
	\begin{equation}
	\begin{aligned}
	\ell^{\nu}\ell_{\nu} = n^{\nu}n_{\nu} = m^{\nu}m_{\nu} = {m^{*}}^{\nu}{m^{*}}_{\nu}=0,\\
	\ell^{\nu}m_{\nu} = \ell^{\nu}{m^{*}}_{\nu} = n^{\nu}m_{\nu} = n^{\nu}{m^{*}}_{\nu} =0.
	\end{aligned}
	\end{equation}
	With the metric expressed by 
	\begin{equation}
	\begin{aligned}
	g^{\mu\nu} &= \ell^{\mu}n^{\nu} + n^{\mu}\ell^{\nu} - m^{\mu}{m^{*}}^{\nu} - {m^{*}}^{\mu}m^{\nu} \\ g_{\mu\nu} &= \ell_{\mu}n_{\nu} + n_{\mu}\ell_{\nu} - m_{\mu}{m^{*}}_{\nu} - {m^{*}}_{\mu}m_{\nu}
	\end{aligned}
	\end{equation}
	\section{Directional Derivatives, Spin Coefficients, Ricci and Weyl Tensor}
	
	The covariant derivatives are expressed using four separate symbols $\nabla_{e_{a}} = \{D,\ \Delta,\ \delta,\ \delta^{*} \}$, with $e_{a} =\{\ell_{\mu},\ n_{\mu},\ m_{\mu},\ {m^{*}}_{\mu} \}$, $a=1,2,3,4$ , which name a directional covariant derivative operator for each tetrad direction. The operators are defined as 
	\begin{equation}
	D =l^{{\mu}}\nabla_{{\mu}},\quad \Delta =n^{{\mu}}\nabla_{{\mu}},\quad \delta = m^{\mu}\nabla_{\mu} , \quad \delta^{*} ={{m}^{*}}^{{\mu}}\nabla_{{\mu}}
	\end{equation}
	which reduce to $D=l^{{\mu}}\partial_{{\mu}},\ \Delta =n^{{\mu}}\partial_{{\mu}},\ \delta =m^{{\mu}}\partial_{{\mu}},\ \delta^{*} ={ m^{*}}^{{\mu}}\partial_{{\mu}}$ when acting on scalar functions. The operators satisfy the following commutation relations, as defined in \cite{Chandrasekhar:1976ap}:
	\begin{equation}
	\begin{aligned}
	[\Delta,D] &= \Delta D - D \Delta = (\gamma+\gamma^{*})D+(\varepsilon+\varepsilon^{*})\Delta - (\tau+\pi^{*})\delta^{*}-(\tau^{*}+\pi)\delta\\
	[\delta,D] &= \delta D - D \delta = (\alpha^{*}+\beta-\pi^{*})D + \kappa \Delta - \sigma \delta^{*}-(\rho^{*}+\varepsilon-\varepsilon^{*})\delta\\
	[\delta,\Delta] &= \delta \Delta - \Delta \delta = -\nu^{*}D + (\tau-\alpha^{*}-\beta)\Delta + \lambda^{*}\delta^{*}+(\mu-\gamma+\gamma^{*})\delta\\
	[\delta^{*},\delta] &= \delta^{*}\delta - \delta \delta^{*} = (\mu^{*}-\mu)D + (\rho^{*} - \rho)\Delta - (\alpha^{*}-\beta)\delta^{*}-(\beta^{*}-\alpha)\delta
	\end{aligned}
	\end{equation}

        \vspace{0.2cm}
	\textbf{Twelve spin coefficients}
        \vspace{0.2cm}
 
	Spin coefficients are the primary quantities in Newman-Penrose formalism, with which all other NP quantities could be calculated indirectly using the NP field equations, as defined in \eqref{sec:NPeq}. In the formalism, one has twelve complex spin coefficients associated with Ricci rotation coefficient labeled by $\gamma_{abc} = e_{c}\ .\ \nabla_{e_b}e_a = -\gamma_{acb}$,
	\begin{equation}
	\begin{aligned}
	\kappa &=\gamma_{311} = -m^{\mu}Dl_{\mu}=-m^{\mu} l^{\nu} \nabla_{\nu} l_{\mu}\,,\qquad \tau = \gamma_{312}= -m^{\mu}\Delta l_{\mu}=-m^{\mu} n^{\nu} \nabla_{\nu} l_{\mu},\\
	\sigma &=\gamma_{313} =  - m^{\mu} \delta l_{\mu} = - m^{\mu} m^{\nu} \nabla_{\nu} l_{\mu} , \qquad \rho =\gamma_{314} =  - m^{\mu} \delta^{*}l_{\mu} = - {m^{\mu} m^{*}}^{\nu} \nabla_{\nu} l_{\mu} ,\quad\\
	\pi &=\gamma_{421} =  {m^{*}}^{\mu} D n_{\mu} = {m^{*}}^{\nu} l^{\nu} \nabla_{\nu} n_{\mu}, \qquad \ \nu =\gamma_{422} = {m^{*}}^{\mu} \Delta n_{\mu} = {m^{*}}^{\mu} n^{\nu} \nabla_{\nu} n_{\mu},\\
	\mu &=\gamma_{423} ={m^{*}}^{\mu} \delta n_{\mu} = {m^{*}}^{\mu} m^{\nu} \nabla_{\nu} n_{\mu} ,\qquad \lambda =\gamma_{424} =  {m^{*}}^{\mu} {\delta^{*}}n_{\mu} = {m^{*}}^{\nu} {m^{*}}^{\nu} \nabla_{\nu} n_{\mu},\\ \varepsilon &=\frac{1}{2}(\gamma_{211}-\gamma_{431}) =-\frac{1}{2}(n^{{\mu}}D\ell_{{\mu}}-{m^{*}}^{{\mu}}Dm_{\mu}) = -\frac{1}{2}(n^{\mu}\ell^{\nu}\nabla_{\nu}\ell_{\mu} - {m^{*}}^{{\nu}}\ell^{{\nu}}\nabla_bm_{\mu}),\\\gamma &=\frac{1}{2}(\gamma_{342}-\gamma_{122})= -\frac{1}{2}(n^{{\mu}}\Delta\ell_{{\mu}}-{m^{*}}^{{\mu}}\Delta m_{\mu}) = -\frac{1}{2}(n^{\mu} n^{\nu}\nabla_{\nu}\ell_{\mu} - {m^{*}}^{{\mu}}n^{{\nu}}\nabla_{\nu} m_{\mu}),\\ \beta &=\frac{1}{2}(\gamma_{213}-\gamma_{433})= -\frac{1}{2}(n^{{\mu}}\delta\ell_{{\mu}}-{m^{*}}^{{\mu}}\delta m_{\mu}) = -\frac{1}{2}(n^{\mu} m^{\nu}\nabla_{\nu}\ell_{\mu} - {m^{*}}^{{\mu}}m^{{\nu}}\nabla_{\nu} m_{\mu}),  \\ \alpha &=\frac{1}{2}(\gamma_{344}-\gamma_{124})= -\frac{1}{2}(n^{{\mu}}\delta^{*}\ell_{{\mu}}-{m^{*}}^{{\mu}}\delta^{*} m_{\mu}) = -\frac{1}{2}(n^{\mu} {m^{*}}^{\nu}\nabla_{\nu}\ell_{\mu} - {m^{*}}^{{\mu}}{m^{*}}^{{\nu}}\nabla_{\nu} m_{\mu}).
	\end{aligned}
	\end{equation}
	\\
	\textbf{Weyl-NP and Ricci-NP scalars}
	\\
	The 10 independent components of the Weyl tensor can be encoded into 5 complex Weyl-NP scalars, 
	\begin{equation}
	\begin{aligned}
	\Psi_0 &=-C_{1313} =  -C_{\mu \nu \rho \sigma}\ell^{\mu} m^{\nu} \ell^{\rho} m^{\sigma}, \quad \Psi_1 = -C_{1213} = -C_{\mu \nu \rho \sigma}\ell^{\mu} n^{\nu} \ell^{\rho} m^{\sigma}, \\  \Psi_2 &=-C_{1342} =  -C_{\mu \nu \rho \sigma}\ell^{\mu} m^{\nu} {m^{*}}^{\rho} n^{\sigma},\quad \Psi_3 =-C_{1242} =-C_{\mu \nu \rho \sigma}\ell^{\mu} n^{\nu} {m^{*}}^{\rho} n^{\sigma}, \\ \Psi_4 &=-C_{2424}= -C_{\mu \nu \rho \sigma}n^{\mu} {m^{*}}^{\nu} n^{\rho} {m^{*}}^{\sigma},
	\end{aligned}
	\end{equation}
	with additional properties
	\begin{equation}
	\begin{aligned}
	C_{1334} &= C_{1231} = \Psi_{1},\qquad \quad C_{1241} = C_{1443} = \Psi_{1}^{*} \quad \qquad
	C_{1212} = C_{3434} = -(\Psi_{2}+\Psi_{2}^{*}),\\  C_{1234} &= C_{1234} = \Psi_{2}-\Psi_{2}^{*}\quad
	C_{2443} = -C_{1242} = \Psi_{3},\qquad C_{1232} = C_{2343} = -\Psi_{3}^{*}
	\end{aligned}
	\end{equation}
	and $C_{1314} = C_{2324}=C_{1332} = C_{1442}=0$.
	
	On the other hand, the 10 independent components of the Ricci tensor are encoded into 4 real scalars $\{\Phi_{00}$, $\Phi_{11}$, $\Phi_{22}$, $\Lambda\}$ and 3 complex scalars $\{\Phi_{10},\Phi_{20},\Phi_{21}\}$, with their complex conjugates, 
	\begin{equation}
	\begin{aligned}
	\Phi_{00} &= -\frac{1}{2}R_{\mu\nu}\ell^{\mu} \ell^{\nu}, \quad \Phi_{11} = -\frac{1}{4}R_{\mu\nu}(\ell^{\mu} n^{\nu}+m^{\mu} {m^{*}}^{\nu}), \\ \Phi_{22} &= -\frac{1}{2}R_{\mu\nu}n^{\mu} n^{\nu},\qquad \Lambda = \frac{R}{24} = \frac{1}{12}(R_{12}-R_{34}),\\
	\Phi_{01} &= -\frac{1}{2}R_{\mu\nu}\ell^{\mu} m^{\nu}, \quad \Phi_{10} = -\frac{1}{2}R_{\mu\nu}\ell^{\mu} {m^{*}}^{\nu} = -\Phi_{01}^{*},\\ \Phi_{02} &= -\frac{1}{2}R_{\mu\nu}m^{\mu} m^{\nu}, \quad \Phi_{20} = -\frac{1}{2}R_{\mu\nu}{m^{*}}^{\mu} {m^{*}}^{\nu} = -\Phi_{02}^{*}, \\ \Phi_{12} &=- \frac{1}{2}R_{\mu\nu}m^{\mu} n^{\nu},  \quad \Phi_{21} = -\frac{1}{2}R_{\mu\nu}{m^{*}}^{\nu} {n}^{\nu} =-\Phi_{12}^{*}.
	\end{aligned}
	\end{equation}
	
	\section{Newman-Penrose field Equations}
	\label{sec:NPeq}
	
	In a complex null tetrad, Ricci identities give rise to the following NP field equations connecting spin coefficients, Weyl-NP and Ricci-NP scalars. These equations in various notations can be found in several texts. The notation in Frolov and Novikov \cite{frolov1998black} is identical. 
	\begin{equation}
	\begin{aligned}
	&D\rho -\delta^{*}\kappa = (\rho^2 +\sigma \sigma^{*})+(\varepsilon+\varepsilon^{*})\rho - \kappa^{*}\tau-\kappa(3\alpha+\beta^{*}-\pi)+\Phi_{00}\\
	&D\sigma-\delta \kappa = (\rho+\rho^{*})\sigma+(3\varepsilon-\varepsilon^{*})\sigma - (\tau-\pi^{*}+\alpha^{*}+3\beta)\kappa+\Psi_{0}\\
	&D\tau - \Delta \kappa = (\tau+\pi^{*})\rho+(\tau^{*}+\pi)\sigma + (\varepsilon -\varepsilon^{*})\tau - (3\gamma+\gamma^{*})\kappa+\Psi_{1}+\Phi_{01}\\
	&D\alpha - \delta^{*}\varepsilon = (\rho +\varepsilon^{*}-2\varepsilon)\alpha+\beta\sigma^{*} - \beta^{*}\varepsilon - \kappa\lambda-\kappa^{*}\gamma+(\varepsilon+\rho)\pi+\Phi_{10}\\
	&D\beta - \delta \varepsilon = (\alpha+\pi)\sigma + (\rho^{*}-\varepsilon^{*})\beta - (\mu + \gamma)\kappa - (\alpha^{*}-\pi^{*})\varepsilon+\Psi_{1}\\
	&D\gamma - \Delta\varepsilon = (\tau+\pi^{*})\alpha + (\tau^{*}+\pi)\beta - (\varepsilon+\varepsilon^{*})\gamma - (\gamma+\gamma^{*})\varepsilon +\tau\pi-\nu\kappa+\Psi_2 + \Phi_{11}-\Lambda\\
	&D\lambda - \delta^{*}\pi = (\rho\lambda+\sigma^{*}\mu)+\pi^2 + (\alpha-\beta^{*})\pi - \nu\kappa^{*} - (3\varepsilon - \varepsilon^{*})\lambda+\Phi_{20}\\
	&D\mu - \delta \pi = (\rho^{*}\mu+\sigma\lambda)+\pi\pi^{*} - (\varepsilon + \varepsilon^{*})\mu - \pi(\alpha^{*}-\beta) - \nu\kappa+\Psi_{2}+2\Lambda\\
	&D\nu-\Delta\pi = (\pi+\tau^{*})\mu+(\pi^{*}+\tau)\lambda + (\gamma - \gamma^{*})\pi - (3\varepsilon + \varepsilon^{*})\nu + \Psi_{3} + \Phi_{21}\\
	&\Delta \lambda - \delta^{*}\nu = -(\mu+\mu^{*})\lambda - (3\gamma-\gamma^{*})\lambda+(3\alpha+\beta^{*}+\pi-\tau^{*})\nu - \Psi_{4}\\
	&\delta\rho - \delta^{*}\sigma = \rho(\alpha^{*}+\beta)-\sigma(3\alpha-\beta^{*})+(\rho-\rho^{*})\tau + (\mu-\mu^{*})\kappa - \Psi_1 + \Phi_{01}\\
	&\delta\alpha - \delta^{*}\beta = (\mu\rho-\lambda\sigma)+\alpha\alpha^{*}+\beta\beta^{*} - 2\alpha\beta+\gamma(\rho - \rho^{*})+\varepsilon(\mu-\mu^{*}) - \Psi_2 + \Phi_{11}+\Lambda\\
	&\delta\lambda - \delta^{*}\mu = (\rho-\rho^{*})\nu+ (\mu-\mu^{*})\pi+\mu(\alpha+\beta^{*})+\lambda(\alpha^{*}-3\beta)-\Psi_3 + \Phi_{21}\\
	&\delta\gamma - \Delta\beta = (\tau-\alpha^{*}-\beta)\gamma+\mu\tau-\sigma\nu-\varepsilon\nu^{*}-\beta(\gamma-\gamma^{*}-\mu)+\alpha\lambda^{*}+\Phi_{12}\\
	&\delta \tau - \Delta \sigma = (\mu \sigma+\lambda^{*}\rho)+(\tau+\beta-\alpha^{*})\tau - (3\gamma - \gamma^{*})\sigma -\kappa\nu^{*}+\Phi_{02}\\
	&\Delta\rho-\delta^{*}\tau = -(\rho\mu^{*}+\sigma\lambda)+(\beta^{*}-\alpha-\tau^{*})\tau+(\gamma+\gamma^{*})\rho +\nu\kappa-\Psi_2 - 2\Lambda\\
	&\Delta\alpha - \delta^{*}\gamma = (\rho+\varepsilon)\lambda + (\gamma^{*}-\mu^{*})\alpha + (\beta^{*}-\tau^{*})\gamma - \Psi_3.
	\label{eq:RicciId}
	\end{aligned}
	\end{equation}
	\section{Bianchi Identities}
	
	The Bianchi identities form the last of equations of the NP formalism, where the equation
	\begin{equation}
	\nabla_{[\lambda}R_{\mu\nu]\alpha\beta}= \nabla_{\lambda}R_{\mu\nu\alpha\beta}+\nabla_{\nu}R_{\lambda\mu\alpha\beta}+\nabla_{\mu}R_{\nu\lambda\alpha\beta}=0
	\end{equation}
	is rewritten as
	\begin{equation}
	\nabla_{[k}R_{mn]pq} = -2R_{pqs}{}_{[m}\gamma^{s}_{nk]}+\gamma^{s}_{p[m}R_{nk]sq}-\gamma^{s}_{q[m}R_{nk]sp}.
	\end{equation}
	For completeness, we list below the equations written in terms os spin coefficients,
	\begin{equation}
	\begin{aligned}
	\delta^{*}\Psi_0-& D\Psi_{1} + D\Phi_{01}-\delta \Phi_{00} = (4\alpha -\pi)\Psi_{0} - 2(2\rho+\varepsilon)\Psi_{1}+\\&+3\kappa\Psi_{2}+(\pi^{*}-2\alpha^{*}-2\beta)\Phi_{00}+2(\varepsilon+\rho^{*})\Phi_{01}+2\sigma\Phi_{10}-2\kappa\Phi_{11}-\kappa^{*}\Phi_{02}\\ 
	\Delta\Psi_0-& \delta\Psi_{1} + D\Phi_{02}-\delta \Phi_{01} = (4\gamma -\mu)\Psi_{0} - 2(2\tau+\beta)\Psi_{1}+\\&+3\sigma\Psi_{2}+(2\varepsilon-2\varepsilon^{*}+\rho^{*})\Phi_{02}+2(\pi^{*}-\beta)\Phi_{01}+2\sigma\Phi_{11}-2\kappa\Phi_{12}-\lambda^{*}\Phi_{00}\\
	\delta^{*}\Psi_3-& D\Psi_{4} + \delta^{*}\Phi_{21}-\Delta \Phi_{20} = (4\varepsilon -\rho)\Psi_{4} - 2(2\pi+\alpha)\Psi_{3}+\\&+3\lambda\Psi_{2}+(2\gamma-2\gamma^{*}+\mu^{*})\Phi_{20}+2(\tau^{*}-\alpha)\Phi_{21}+2\lambda\Phi_{11}-2\nu\Phi_{10}-\sigma^{*}\Phi_{22}\\
	\Delta\Psi_3-& \delta\Psi_{4} + \delta^{*}\Phi_{22}-\Delta \Phi_{21} = (4\beta -\tau)\Psi_{4} - 2(2\mu+\gamma)\Psi_{3}+\\&+3\nu\Psi_{2}+(\tau^{*}-2\beta^{*}-2\alpha)\Phi_{22}+2(\gamma+\mu^{*})\Phi_{21}+2\lambda\Phi_{12}-2\nu\Phi_{11}-\nu^{*}\Phi_{20}\\
	D\Psi_2-& \delta^{*}\Psi_{1} + \Delta\Phi_{00}-\delta^{*} \Phi_{01}+2D\Lambda = -\lambda\Psi_{0} + 2(\pi-\alpha)\Psi_{1}+\\&+3\rho\Psi_{2}-2\kappa\Psi_{3}+(2\gamma+2\gamma^{*}-\mu^{*})\Phi_{00}-2(\tau^{*}+\alpha)\Phi_{01}-2\tau\Phi_{10}+2\rho\Phi_{11}+\sigma^{*}\Phi_{02}\\
	\Delta\Psi_2-& \delta\Psi_{3} + D\Phi_{22}-\delta \Phi_{21}+2\Delta\Lambda = \sigma\Psi_{4} + 2(\beta-\tau)\Psi_{3}+\\&-3\mu\Psi_{2}+2\nu\Psi_{1}+(\rho^{*}-2\varepsilon-2\varepsilon^{*})\Phi_{22}+2(\pi^{*}+\beta)\Phi_{21}+2\pi\Phi_{12}-2\mu\Phi_{11}-\lambda^{*}\Phi_{20}\\
	D\Psi_3-& \delta^{*}\Psi_{2} - D\Phi_{21}+\delta \Phi_{20}-2\delta^{*}\Lambda = -\kappa\Psi_{4} + 2(\rho-\varepsilon)\Psi_{3}+\\&+3\pi\Psi_{2}-2\lambda\Psi_{1}+(2\alpha^{*}-2\beta-\pi^{*})\Phi_{20}-2(\rho^{*}-\varepsilon)\Phi_{21}-2\pi\Phi_{11}+2\mu\Phi_{10}+\kappa^{*}\Phi_{22}\\
	\Delta\Psi_1-& \delta\Psi_{2} - \Delta\Phi_{01}+\delta^{*} \Phi_{02}-2\delta\Lambda = \nu\Psi_{0} + 2(\gamma-\mu)\Psi_{1}-3\tau\Psi_{2}\\&+2\sigma\Psi_{3}+(\tau^{*}-2\beta^{*}+2\alpha)\Phi_{02}+2(\mu^{*}-\gamma)\Phi_{01}+2\tau\Phi_{11}\\
	D\Phi_{11}-& \delta\Phi_{10} - \delta^{*}\Phi_{01}+\Delta \Phi_{00}+3D\Lambda = (2\gamma-\mu+2\gamma^{*}-\mu^{*})\Phi_{00} + (\pi-2\alpha-2\tau^{*})\Phi_{01}+\\&+(\pi^{*}-2\alpha^{*}-2\tau)\Phi_{10}+2(\rho+\rho^{*} )\Phi_{11}+\sigma^{*}\Phi_{02}+\sigma\Phi_{20}-\kappa^{*}\Phi_{12}-\kappa\Phi_{21}\\
	D\Phi_{12}-& \delta\Phi_{11} - \delta^{*}\Phi_{02}+\Delta \Phi_{01}+3\delta\Lambda = (-2\alpha+2\beta^{*}+\pi-\tau)\Phi_{02} + (\rho^{*}+2\rho-2\varepsilon^{*})\Phi_{12}+\\&+ 2(\pi^{*}-\tau)\Phi_{11}+(2\gamma-2\mu^{*}-\mu)\Phi_{01}+\nu^{*}\Phi_{00}-\lambda^{*}\Phi_{10}+\sigma\Phi_{21}-\kappa\Phi_{22}\\
	D\Phi_{22}-& \delta\Phi_{21} - \delta^{*}\Phi_{12}+\Delta \Phi_{11}+3\Delta\Lambda = (-2\varepsilon-2\varepsilon^{*}+\rho+\rho^{*})\Phi_{22} + (2\beta^{*}+2\pi-\tau^{*})\Phi_{12}+\\&+ (2\beta+2\pi^{*}-\tau)\Phi_{21}-2(\mu+\mu^{*})\Phi_{11}+\nu\Phi_{01}+\nu^{*}\Phi_{10}-\lambda^{*}\Phi_{20}-\lambda\Phi_{02}\\
	\label{eq:BianchiID}
	\end{aligned}
	\end{equation}
	
	\section{Symmetries in NP equations}
	\label{sec:NPSym}
	
	The full set of NP equations is invariant under the interchange $\ell^{\mu} \leftrightarrow n^{\mu}$, $m^{\mu} \leftrightarrow {m^{*}}^{\mu}$.
 
	\vspace{0.2cm}
	\textbf{Directional derivatives and spin coefficients: }
	\begin{equation}
	\begin{aligned}
	& D=l^{\mu} \nabla_{\mu} \quad \stackrel{\ell^{\mu} \leftrightarrow n^{\mu}}{\longleftrightarrow}\quad  n^{\mu} \nabla_{\mu} \quad  \stackrel{m^{\mu} \leftrightarrow {m^{*}}^{\mu}}{\longleftrightarrow} \quad  n^{\mu} \nabla_{\mu}=\Delta, \\
	& \delta=m^{\mu} \nabla_{\mu} \quad \stackrel{\ell^{\mu} \leftrightarrow n^{\mu}}{\longleftrightarrow}\quad  m^{\mu} \nabla_{\mu} \quad  \stackrel{m^{\mu} \leftrightarrow {m^{*}}^{\mu}}{\longleftrightarrow} \quad  {m^{*}}^{\mu} \nabla_{\mu}=\delta^{*}, \\
	& \kappa=\gamma_{311} \quad  \stackrel{\ell^{\mu} \leftrightarrow n^{\mu}}{\longleftrightarrow}\quad  \gamma_{322}  \quad \stackrel{m^{\mu} \leftrightarrow {m^{*}}^{\mu}}{\longleftrightarrow} \quad  \gamma_{422}=-\gamma_{242}=-\nu, \\
	& \tau=\gamma_{312} \quad  \stackrel{\ell^{\mu} \leftrightarrow n^{\mu}}{\longleftrightarrow} \quad  \gamma_{321} \quad  \stackrel{m^{\mu} \leftrightarrow {m^{*}}^{\mu}}{\longleftrightarrow} \quad  \gamma_{421}=-\gamma_{241}=-\pi, \\
	& \sigma=\gamma_{313} \quad  \stackrel{\ell^{\mu} \leftrightarrow n^{\mu}}{\longleftrightarrow} \quad  \gamma_{323} \quad  \stackrel{m^{\mu} \leftrightarrow {m^{*}}^{\mu}}{\longleftrightarrow} \quad  \gamma_{424}=-\gamma_{244}=-\lambda, \\
	& \rho=\gamma_{314} \quad  \stackrel{\ell^{\mu} \leftrightarrow n^{\mu}}{\longleftrightarrow} \quad  \gamma_{324} \quad \stackrel{m^{\mu} \leftrightarrow {m^{*}}^{\mu}}{\longleftrightarrow} \quad  \gamma_{423}=-\gamma_{243}=-\mu, \\
	& \varepsilon=\frac{1}{2}\left(\gamma_{211}+\gamma_{341}\right) \quad \stackrel{\ell^{\mu} \leftrightarrow n^{\mu}}{\longleftrightarrow} \quad \frac{1}{2}\left(\gamma_{122}+\gamma_{342}\right) \quad \stackrel{m^{\mu} \leftrightarrow {m^{*}}^{\mu}}{\longleftrightarrow} \quad \frac{1}{2}\left(\gamma_{122}+\gamma_{432}\right)=-\gamma \\
	& \alpha=\frac{1}{2}\left(\gamma_{214}+\gamma_{344}\right) \quad \stackrel{\ell^{\mu} \leftrightarrow n^{\mu}}{\longleftrightarrow} \quad \frac{1}{2}\left(\gamma_{124}+\gamma_{344}\right) \quad \stackrel{m^{\mu} \leftrightarrow {m^{*}}^{\mu}}{\longleftrightarrow} \quad \frac{1}{2}\left(\gamma_{123}+\gamma_{433}\right)=-\beta.
	\end{aligned}
	\end{equation}
	\\
	\textbf{Weyl-NP scalars: }
	\begin{equation}
	\begin{aligned}
	& \Psi_{0}=-C_{1313} \quad \stackrel{\ell^{\mu} \leftrightarrow n^{\mu}}{\longleftrightarrow}\quad -C_{2323}\quad \stackrel{m^{\mu} \leftrightarrow {m^{*}}^{\mu}}{\longleftrightarrow}\quad-C_{2424}=\Psi_{4}, \\
	& \Psi_{1}=-C_{1213} \quad \stackrel{\ell^{\mu} \leftrightarrow n^{\mu}}{\longleftrightarrow}\quad-C_{2123}\quad \stackrel{m^{\mu} \leftrightarrow {m^{*}}^{\mu}}{\longleftrightarrow}\quad-C_{2124}=-C_{1242}=\Psi_{3}, \\
	& \Psi_{2}=-C_{1342} \quad \stackrel{\ell^{\mu} \leftrightarrow n^{\mu}}{\longleftrightarrow}\quad-C_{2341}\quad \stackrel{m^{\mu} \leftrightarrow {m^{*}}^{\mu}}{\longleftrightarrow}\quad-C_{2431}=-C_{1342}=\Psi_{2} 
	\end{aligned}
	\end{equation}
	\\
	\textbf{Maxwell tensor:}
	\begin{equation}
	\begin{aligned}
	\Phi_0=F_{\mu\nu}\ell^{\mu}m^{\nu} \quad \stackrel{\ell^{\mu} \leftrightarrow n^{\mu}}{\longleftrightarrow}& \quad  F_{\mu\nu}n^{\mu}m^{\nu} \quad  \stackrel{m^{\nu} \leftrightarrow {m^{*}}^{\nu}}{\longleftrightarrow} \quad  F_{\mu\nu}n^{\mu}{m^{*}}^{\nu}=-\Phi_2, \\
	\Phi_1=\frac{1}{2}F_{\mu\nu}(\ell^{\mu}n^{\nu}+{m^*}^{\mu}m^{\nu}) \quad &\stackrel{\ell^{\mu} \leftrightarrow n^{\mu}}{\longleftrightarrow}\quad \frac{1}{2}F_{\mu\nu}(n^{\mu}\ell^{\nu}+{m^*}^{\mu}m^{\nu}) \quad \\ &\stackrel{m^{\mu} \leftrightarrow {m^{*}}^{\mu}}{\longleftrightarrow}  \frac{1}{2}F_{\mu\nu}(n^{\mu}\ell^{\nu}+m^{\mu}{m^*}^{\nu})=-\Phi_1.
	\end{aligned}
	\end{equation}
	with
	\begin{equation}
	F_{\mu\nu} = 2[\Phi_1 (n_{[\mu}\ell_{\nu]}+m_{[\mu}{m^{*}}_{\nu]})+\Phi_2 \ell_{[\mu}m_{\nu]}+ \Phi_{0}{m^{*}}_{[\mu}n_{\nu]}]+ c.c.
	\label{eq:Ftensor}
	\end{equation}
	where "c.c" denotes "complex conjugate of the preceding terms".
	
	\chapter{\texorpdfstring{Conditions for the $\tau_V$-function}%
 {}}
	\label{chap:AppB}
	
	In this appendix, we derive the expression \eqref{eq:3.39} for the $\tau_{V}$-function. These conditions comes from Toda equation, which relates the logarithm derivative of the $\tau_{V}$-function with two different $\tau_{V}$-functions. The proof of \eqref{eq:3.39} is a little laborious and straightforward: Initially, we consider a basis of solutions where $A_t$ is diagonal,
	
	\begin{equation}
	\frac{\partial \Phi(z)}{\partial z}= \bigg[A_\infty +\frac{A_0}{z}+\frac{1}{z-t}\begin{pmatrix}
	\hat{\alpha}_t&0\\
	0&\hat{\beta}_t
	\end{pmatrix}\bigg]\Phi(z) .
	\label{eq:system}
	\end{equation}
	From the fundamental solution $\Phi(z)$, we define the derived solutions that are represented by
	\begin{equation}
	\Phi^{+}(z) =L^{+}(z)\Phi(z) = \begin{pmatrix}
	1  &  0 \\
	p^{+} & 1
	\end{pmatrix}\begin{pmatrix}
	z-t  &  0 \\
	0 & 1
	\end{pmatrix}\begin{pmatrix}
	1  &  q^{+} \\
	0 & 1
	\end{pmatrix}\Phi(z),
	\label{eq:A1}
	\end{equation}
	\begin{equation}
	\Phi^{-}(z) =L^{-}(z)\Phi(z) = \begin{pmatrix}
	1  &  p^{-} \\
	0 & 1
	\end{pmatrix}\begin{pmatrix}
	(z-t)^{-1}  &  0 \\
	0 & 1
	\end{pmatrix}\begin{pmatrix}
	1  &  0 \\
	q^{-} & 1
	\end{pmatrix}\Phi(z),
	\label{eq:A2}
	\end{equation}
	here we dropped the $t$ dependence. From \eqref{eq:system}, $\Phi^{\pm}(z)$ satisfy
	\begin{equation}
	\frac{\partial\Phi^{\pm}}{\partial z}[\Phi^{\pm}(z)]^{-1} = A_{\infty}^{\pm}+\frac{A^{\pm}_0}{z}+\frac{1}{z-t}\begin{pmatrix}
	\hat{\alpha}_t \pm 1 & 0\\
	0 & \hat{\beta}_t
	\end{pmatrix}. 
	\label{eq:A3}
	\end{equation}
	Note that, $\Phi^{\pm}(z)$ is related to the  possible shifts represented in \eqref{eq:3.40}. In terms of the set $\{\theta\}$, it is clear that the set $\{\theta\}^{\pm}$ of $\Phi^\pm(z)$ are related to that of $\Phi(z)$ by \eqref{eq:A1} and \eqref{eq:A2}. Given $\Phi^\pm(z)$, one can establish the \textit{Toda equation} \eqref{eq:3.39} by comparing the corresponding expressions for each $\tau_V$-function \eqref{eq:3.39} with the shifts $\{\theta\}^{\pm}$ taken account, with $\{\sigma, \eta\}$ kept invariant.
	
	The parameters $p^{\pm}$ and $q^{\pm}$ are determined from requirement that the transformation of $A_{\infty}$ does not include term proportional to $z$ and that the transformation of $A_t$ is still diagonal. 
	
	Let us start by considering $\Phi^{+}(z)$, where replacing \eqref{eq:A1} in \eqref{eq:A3}, we find
	\begin{equation}
	\frac{\partial\Phi^{+}}{\partial z}[\Phi^{+}(z)]^{-1}= \frac{\partial L(z)^{+}}{\partial z}(L^{+}(z))^{-1}+ L^{+}(z)A(z)(L^{+}(z))^{-1}.
	\label{eq:A4}
	\end{equation}
	Then, defining the matrices $A^{+}_\infty$ and $A^{+}_{0}$ as function of $A_\infty$ and $A_{0}$ respectively,
	\begin{equation}
	A^{+}_{i} = \begin{pmatrix}
	1&q^{+}\\0&1
	\end{pmatrix}A_{i}\begin{pmatrix}
	1&-q^{+}\\0&1
	\end{pmatrix} = \begin{pmatrix}
	\tilde{a}_i&\tilde{b}_i\\\tilde{c}_i&\tilde{d}_i
	\end{pmatrix}, \quad A_i =\begin{pmatrix}
	a_i&b_i\\c_i&d_i
	\end{pmatrix} \quad i =\infty, 0,
	\label{eq:original}
	\end{equation}
	with
	\begin{equation}
	\begin{pmatrix}
	\tilde{a}_i&\tilde{b}_i\\\tilde{c}_i&\tilde{d}_i
	\end{pmatrix}=\begin{pmatrix}
	a_i +q^{+}c_i&b_i-(a_i-d_i)q^{+}-c_i(q^{+})^{2}\\c_i&d_i-q^{+}c_i
	\end{pmatrix}
	\label{eq:def}.
	\end{equation}
	After some algebraic manipulation in \eqref{eq:A4}, it is possible to prove that the equations \eqref{eq:A3} and \eqref{eq:A4} lead to the following expression for $A^{+}_{\infty}$, $A^{+}_{0}$, and $A^{+}_{t}$:
	\begin{equation}
	A^{+}_\infty= \begin{pmatrix}
	1&0\\p^{+}&1
	\end{pmatrix}\begin{pmatrix}
	\tilde{a}_\infty&0\\0&\tilde{d}_\infty
	\end{pmatrix}\begin{pmatrix}
	1&0\\-p^{+}&1
	\end{pmatrix}\newline -
	(z-t)\tilde{b}_{\infty}\begin{pmatrix}
	p^{+}&-1\\(p^{+})^{2}&-p^{+}
	\end{pmatrix} +\frac{1}{z-t}\begin{pmatrix}
	0&0\\\tilde{c}_{\infty}&0
	\end{pmatrix},
	\label{eq:3.42}
	\end{equation}
	\begin{equation}
	\frac{1}{z}A^{+}_0= \frac{1}{z}\begin{pmatrix}
	1&0\\p^{+}&1
	\end{pmatrix}\begin{pmatrix}
	\tilde{a}_0&-\tilde{b}_0 t\\-\frac{\tilde{c}_0}{t}&\tilde{d}_0
	\end{pmatrix}\begin{pmatrix}
	1&0\\-p^{+}&1
	\end{pmatrix}\newline	-\tilde{b}_{0}\begin{pmatrix}
	p^{+}&-1\\(p^{+})^{2}&-p^{+}
	\end{pmatrix} +\frac{1}{z-t}\begin{pmatrix}
	0&0\\\frac{\tilde{c}_{0}}{t}&0
	\end{pmatrix},
	\label{eq:3.421}
	\end{equation}
	\begin{equation}
	\frac{1}{z-t}A^{+}_t= \frac{1}{z-t}\begin{pmatrix}
	\hat{\alpha}_t&0\\0&\hat{\beta}_t
	\end{pmatrix}\newline+q^{+}(\hat{\alpha}_t-\hat{\beta}_t)\begin{pmatrix}
	p^{+}&-1\\(p^{+})^{2}&-p^{+}
	\end{pmatrix}	+\frac{1}{z-t}\begin{pmatrix}
	0&0\\p^{+}(\hat{\alpha}_t -\hat{\beta}_t)&0
	\end{pmatrix}.
	\label{eq:3.422}
	\end{equation}
	Adding these terms in \eqref{eq:A3} and taking
	\begin{equation}
	\frac{\partial L^{+}(z) }{\partial z}L^{+}(z) =\frac{1}{z-t} \begin{pmatrix}
	1&0\\0&0
	\end{pmatrix}+\frac{1}{z-t} \begin{pmatrix}
	0&0\\p^{+}&0
	\end{pmatrix}
	\label{eq:extra},
	\end{equation}
	we find the constraint that the extra terms must satisfy to recover the system: from $A^{+}_\infty$, the term proportional to $z$ must vanish, which implies $\tilde{b}_\infty =0$. The extra terms proportional to $(z-t)^{-1}$ also must be canceled, then with $A^{+}_{\infty}$, $A^{+}_{0}$, $A^{+}_{t}$, and equation \eqref{eq:extra} we have the following constraint, $p^{+}(\hat{\alpha}_t-\hat{\beta}_t+1)=-\tilde{c}_0/t-\tilde{c}_\infty$. 
	Finally, using the first contraint, we find the explicit form of $q^{+}$:
	\begin{equation}
	q^{+} = \frac{\hat{\alpha}_\infty - a_\infty}{c_\infty}, \quad \quad q^{+} = -	\frac{\hat{\beta}_\infty - d_\infty}{c_\infty},
	\end{equation}
	where $\hat{\alpha}_\infty$ and $\hat{\beta}_\infty$ are eingevalues of $A_\infty$. Replacing $q^{+}$ in the definition \eqref{eq:def} we have $\tilde{a}_\infty =  \hat{\alpha}_\infty$, $\tilde{d}_\infty = \hat{\beta}_\infty$, and $\tilde{c}_\infty = c_\infty$.
	Therefore, $A_t$ will keep the form in \eqref{eq:A3} with the matrices $A^{+}_0$ and $A^{+}_t$ written as 
	\begin{equation}
	A^{+}_0 =\begin{pmatrix}
	1&0\\p^{+}&1
	\end{pmatrix}\begin{pmatrix}
	\tilde{a}_0&-\tilde{b}_0 t\\-\tilde{c}_0/t&\tilde{d}_0
	\end{pmatrix}\begin{pmatrix}
	1&0\\-p^{+}&1
	\end{pmatrix},\newline
	\end{equation}
	\begin{equation}
	A^{+}_\infty=
	\begin{pmatrix}1&0\\p^{+}&1\end{pmatrix}\begin{pmatrix}\hat{\alpha}_\infty&\tilde{b}_0-q^{+}(\hat{\alpha}_t-\hat{\beta}_t)\\0&\hat{\beta}_\infty\end{pmatrix}\begin{pmatrix}1&0\\-p^{+}&1\end{pmatrix}.
	\end{equation}
	The calculation for  $\hat{\alpha_t}-1$ is entirely analogous. Starting from the original linear system, we have
	\begin{equation}
	\Phi^{-}(z) =L^{-}(z)\Phi(z) = \begin{pmatrix}
	1  &  p^{-} \\
	0 & 1
	\end{pmatrix}\begin{pmatrix}
	(z-t)^{-1}  &  0 \\
	0 & 1
	\end{pmatrix}\begin{pmatrix}
	1  &  0 \\
	q^{-} & 1
	\end{pmatrix}\Phi(z),
	\label{eq:3.423}
	\end{equation}
	that satisfies,
	\begin{equation}
	\frac{\partial\Phi^{-}}{\partial z}[\Phi^{-}(z)]^{-1} = A^{-}+\frac{A^{-}_0}{z}+\frac{1}{z-t}\begin{pmatrix}
	\hat{\alpha}_t - 1 & 0\\
	0 & \hat{\beta}_t
	\end{pmatrix}. 
	\label{eq:3.43}
	\end{equation}
	Again we define the matrices $A^{-}_\infty$ and $A^{-}_{0}$ by
	\begin{equation}
	A^{-}_{i} = \begin{pmatrix}
	1&0\\q^{-}&1
	\end{pmatrix}A_{i}\begin{pmatrix}
	1&0\\-q^{-}&1
	\end{pmatrix} = \begin{pmatrix}
	\tilde{a}_i&\tilde{b}_i\\\tilde{c}_i&\tilde{d}_i
	\end{pmatrix}, \quad \quad i =\infty, 0.
	\end{equation}
	\begin{equation}
	\begin{pmatrix}
	\tilde{a}_i&\tilde{b}_i\\\tilde{c}_i&\tilde{d}_i
	\end{pmatrix}=\begin{pmatrix}
	a_i -q^{-}c_i&b_i\\c_i+(a_i-d_i)q^{-}-c_i(q^{-})^{2}&d_i+q^{-}c_i
	\end{pmatrix}.
	\end{equation}
	In this case, the matrices $A^{-}_\infty$, $A^{-}_t$, $A^{-}_0$, and the extra term are given directly by
	\begin{equation}
	A^{-}_\infty= \begin{pmatrix}
	1&p^{-}\\0&1
	\end{pmatrix}\begin{pmatrix}
	\tilde{a}_\infty&0\\0&\tilde{d}_\infty
	\end{pmatrix}\begin{pmatrix}
	1&-p^{-}\\0&1
	\end{pmatrix}\newline -
	(z-t)\tilde{c}_{\infty}\begin{pmatrix}
	-p^{-}&(p^{-})^{2}\\-1&p^{-}
	\end{pmatrix} +\frac{\tilde{b}_{\infty}}{z-t}\begin{pmatrix}
	0&1\\0&0
	\end{pmatrix},
	\label{eq:3.424}
	\end{equation}
	\begin{equation}
	\frac{A^{-}_t}{z-t}=\frac{1}{z-t} \begin{pmatrix}
	\hat{\alpha}_t&0\\0&\hat{\beta}_t
	\end{pmatrix}\newline -
	q^{-}(\hat{\alpha}_t-\hat{\beta}_t)\begin{pmatrix}
	-p^{-}&(p^{-})^{2}\\-1&p^{-}
	\end{pmatrix}+\frac{1}{z-t}\begin{pmatrix}
	0&-p^{-}(\hat{\alpha}_t-\hat{\beta}_t)\\0&0
	\end{pmatrix}
	\label{eq:3.425},
	\end{equation}
	\begin{equation}
	\frac{A^{-}_0}{z}=\frac{1}{z} \begin{pmatrix}
	1&p^{-}\\0&1
	\end{pmatrix}\begin{pmatrix}
	\tilde{a}_0&-\tilde{b}_0/t\\-\tilde{c}_0 t&\tilde{b}_0
	\end{pmatrix}\begin{pmatrix}
	1&-p^{-}\\0&1
	\end{pmatrix}\newline- \tilde{c}_0\begin{pmatrix}
	-p^{-}&(p^{-})^{2}\\-1&p^{-}
	\end{pmatrix}
	\newline +\frac{\tilde{b}_0/t}{z-t}\begin{pmatrix}
	0&1\\0&0
	\end{pmatrix},
	\label{eq:3.426}
	\end{equation}
	\begin{equation}
	\frac{\partial L^{-}}{\partial z}[L^{-}]^{-1} = \frac{1}{z-t}\begin{pmatrix}
	-1&0\\0&0
	\end{pmatrix} + \frac{1}{z-t}\begin{pmatrix}
	0&p^{-}\\0&0
	\end{pmatrix}.
	\end{equation} 
	Since the term $(z-t)$ must vanish to keep the form of the system, we must take $\tilde{c}_\infty =0$, which results in
	\begin{equation}
	q^{-} = -\frac{\hat{\alpha}_\infty - a_\infty}{b_\infty}, \quad \quad q^{-} = 	\frac{\hat{\beta}_\infty - d_\infty}{b_\infty},
	\end{equation}
	as well as the constraint $p^{-}(\hat{\alpha}_t-\hat{\beta}_t-1)= -\tilde{b}_0/t - \tilde{b}_\infty$, that cancels the terms $(z-t)^{-1}$. Thus, the matrix $A^{-}_{0}$ and $A^{-}_t$ are expressed by
	\begin{equation}
	A^{-}_0 = \begin{pmatrix}
	1&p^{-}\\0&1
	\end{pmatrix}\begin{pmatrix}
	\tilde{a}_0&-\tilde{b}_0/t\\-\tilde{c}_0 t&\tilde{b}_0
	\end{pmatrix}\begin{pmatrix}
	1&-p^{-}\\0&1
	\end{pmatrix},
	\end{equation}
	\begin{equation}	
	A^{-}_\infty
	\begin{pmatrix}1&p^{-}\\0&1\end{pmatrix}\begin{pmatrix}\hat{\alpha}_\infty&0\\ \tilde{c}_0+q^{-}(\hat{\alpha}_t-\hat{\beta}_t)&\hat{\beta}_\infty\end{pmatrix}\begin{pmatrix}1&-p^{-}\\0&1\end{pmatrix}.
	\end{equation}
	
	Now we are ready to write the Toda equation. Using the definition of isomonodromy flow \eqref{eq:2.33}, we write the Hamiltonians of this new system as
	\begin{equation}
	H^{\pm} =\frac{d}{dt}\text{log}\tau^{\pm}_V(t)= \text{Tr}A^{\pm}_{\infty}A^{\pm}_t +\frac{1}{t}\text{Tr}A^{\pm}_0A^{\pm}_t.
	\label{eq:hpm}
	\end{equation}
	Here, replacing the matrices to the system $\Phi^{\pm}(z)$ and using $q^{\pm}$, we find the following Hamiltonians:
	\begin{equation}
	H^{+} = \hat{\alpha}_\infty+\frac{1}{t}a_0+\frac{1}{t}c_0q^{+}+\hat{\alpha}_t\bigg(a_\infty+\frac{a_0}{t} \bigg)+\hat{\beta}_t\bigg(d_\infty+\frac{d_0}{t} \bigg),
	\end{equation}
	\begin{equation}
	H^{-} = -\hat{\alpha}_\infty-\frac{1}{t}a_0+\frac{1}{t}b_0q^{-}+\hat{\alpha}_t\bigg(a_\infty+\frac{a_0}{t} \bigg)+\hat{\beta}_t\bigg(d_\infty+\frac{d_0}{t} \bigg). 
	\end{equation}
	It is not difficult to prove that the last two terms in the equations above is related to the isomonodromy flow of the original system. Thus, we rewrite the equations as
	\begin{equation}
	\begin{aligned}
	H^{+}-H &=  \hat{\alpha}_\infty+\frac{1}{t}a_0+\frac{1}{t}c_0q^{+},\\
	H^{+}-H &=-  \hat{\alpha}_\infty-\frac{1}{t}a_0+\frac{1}{t}b_0q^{-}.
	\label{eq:hamils}
	\end{aligned}
	\end{equation}
	Let us take the derivate of the Hamiltonian of the original system, which is expressed as
	\begin{equation}
	H = \frac{d}{dt}\text{log}\tau_V(\{\theta\};\sigma, \eta;t) = \text{Tr}A_\infty A_t+\frac{1}{t}\text{Tr}A_0A_t.
	\label{eq:HamilAppendix}
	\end{equation}
	where $\{\theta\}$ is the set of parameters of the deformed confluent Heun equation, $\{\theta\}=\{\theta_{ 0},\theta_{t},\theta_{\star}\}$, and encodes the behavior of the solution around each singularity. $\{\sigma,\eta\}$ paramatrizes the $\tau_{V}$-function expansion at $t=0$. We have directly
	\begin{equation}
	\frac{d}{dt}t\frac{d}{dt}\text{log}\tau_{V}(\{\theta\};\sigma, \eta;t) =\text{Tr}A_\infty A_t= a_\infty\hat{\alpha}_t+d_\infty\hat{\beta}_t.
	\label{eq:derivate}
	\end{equation}
	Taking a second derivate in \eqref{eq:derivate} and then using the Schlesinger equations defined in \eqref{eq:2.36}, we arrive at,
	\begin{equation}
	\frac{d^{2}}{dt^{2}}t\frac{d}{dt}\text{log}\tau_{V}(\{\theta\};\sigma, \eta;t)= \frac{1}{t}\text{Tr}(A_\infty[A_0,A_t])=-\frac{1}{t}(\hat{\alpha}_t-\hat{\beta}_t)(b_0c_\infty-c_0b_\infty).
	\end{equation} 
	Using the equations for $q^{+}$, $q^{-}$, \eqref{eq:hpm}, and \eqref{eq:hamils} we find the relations between the second and first derivate on $t$,
	\begin{equation}
	\frac{d^{2}}{dt^{2}}t\frac{d}{dt}\text{log}\tau_{V}(\{\theta\};\sigma, \eta;t)=(\hat{\alpha}_t-\hat{\beta}_t)\frac{b_\infty c_\infty}{\hat{\alpha}_\infty -a_\infty} \frac{d}{dt}\text{log}\frac{\tau^{+}_V(\{\theta\}^{+};\sigma, \eta;t) \tau^{-}_V(\{\theta\}^{-};\sigma, \eta;t)}{\tau_{V}(\{\theta\};\sigma, \eta;t)}.
	\label{eq:Toda}
	\end{equation}
	Then, after some algebra, we can prove the useful relations 
	\begin{equation}
	\begin{aligned}
	(\hat{\alpha}_t-\hat{\beta}_t)\frac{b_\infty c_\infty}{\hat{\alpha}_\infty -a_\infty} &= (\hat{\alpha}_t-\hat{\beta}_t)(a_\infty-\hat{\beta}_\infty),\\
	(\hat{\alpha}_t-\hat{\beta}_t)(a_\infty-\hat{\beta}_\infty) = \frac{d}{dt}t\frac{d}{dt}\text{log}\tau_{V}(&\{\theta\};\sigma,\eta;t) - (\hat{\alpha}_\infty+\hat{\beta}_\infty)\hat{\beta}_t-\hat{\beta}_\infty(\hat{\alpha}_t-\hat{\beta}_t).
	\end{aligned}
	\end{equation}
	We finally stablish the Toda equation for the isomonodromic $\tau_V$-function by replacing the equations above in \eqref{eq:Toda}and then integrating on t,
	\begin{equation}
	\frac{d}{dt}t\frac{d}{dt}\text{log}\tau_{V}(\{\theta\};\sigma,\eta;t)-\hat{\alpha}_\infty\hat{\beta}_t-b_\infty\hat{\alpha}_t= K_V \frac{\tau^{+}_V(\{\theta\}^{+};\sigma,\eta;t) \tau^{-}_V(\{\theta\}^{-};\sigma,\eta;t)}{\tau^{2}_{V}(\{\theta\};\sigma,\eta;t)}
	\label{eq:tauPlusMinus}
	\end{equation}
	where $K_V$ is just a constant independent of $t$.
	
	\subsection{Relevant parametrization for the linear system}
	In the previous section, it was convenient to parametrize the linear system in such a way that $A_t$ was diagonal. However, this is not the parametrization used in the Chapter \eqref{ChapIsoMethod}. Nevertheless, we can use the results above for the system where $A_{\infty}$ is diagonal. Thus, let consider now the parametrization where $A_{\infty}$ is diagonal:
	\begin{equation}
	A_\infty=\frac{1}{2} \begin{pmatrix}
	1&0 \\
	0&-1
	\end{pmatrix}  \quad \bar{A}_i= \begin{pmatrix}
	\bar{a}_i&\bar{b}_i\\
	\bar{c}_i&\bar{d}_i
	\end{pmatrix}= \begin{pmatrix}
	p_i+\theta_{i}&-q_ip_i\\
	\frac{1}{q_i}(p_i+\theta_i)&-p_i
	\end{pmatrix}, \quad i=0,t,
	\end{equation}
	where the bar reminds us that $\bar{A}_t$ is not diagonal. In order to follow the same strategy used in the previous section, we diagonalize $A_t$ by using a inversible matric, $G_t$
	\begin{equation}
	A_t =G_t\begin{pmatrix}
	\theta_t&0\\
	0 & 0 \\
	\end{pmatrix}G^{-1}_t, \quad G_t = \begin{pmatrix}
	q_t& 1\\
	1 & \frac{p_t+\theta_t}{q_tp_t	}
	\end{pmatrix},
	\end{equation}
	which satisfy the constraint \eqref{eq:2.48}. From the last constraint in \eqref{eq:2.48} one obtains
	\begin{equation}
	p_0+p_t =-\frac{\theta_0+\theta_t+\theta_\infty}{2},
	\end{equation}
	where we observe that such constraint reduces the number of free parameters to three. Conjugating with $G_t$, we have the same matrix $A_\infty$ used in the last section,
	\begin{equation}
	A_\infty =\frac{1}{2\theta_t}\begin{pmatrix}
	\bar{a}_t-\bar{d}_t& 2\bar{c}_t\\
	2\bar{b}_t&-\bar{a}_t+\bar{d}_t
	\end{pmatrix}.
	\end{equation}
	Using the same $G_t$, one has that the matrix $A_0$ is also computed straightforwardly. Morover, we have that $q^{+}$ and $q^{-}$ are given by
	\begin{equation}
	q^{+}=\frac{\hat{\alpha}_\infty-a_\infty}{c_\infty} =\frac{\bar{d}_t}{\bar{b}_t}, \quad q^{-}=-\frac{\hat{\alpha}_\infty-a_\infty}{b_\infty} =-\frac{\bar{d}_t}{\bar{c}_t},
	\end{equation}
	which for the boundary condition $\lambda(t)=t$ or $\bar{b}_t=0$ means that $q^{+}$ will diverge.
	In terms of $H^{\pm}$, we have 
	\begin{equation}
	H^{+}-H =\frac{1}{2}+\frac{1}{t}\bigg(\bar{a}_0+\frac{\bar{b}_0}{\bar{b}_t}\bar{d}_t \bigg), \quad \quad 	H^{-}-H =-\frac{1}{2}-\frac{1}{t}\bigg(\bar{a}_0+\frac{\bar{c}_0}{\bar{c}_t}\bar{d}_t \bigg).
	\label{eq:buu}
	\end{equation}
	We now introduce the position of the apparent singularity $\lambda$ and the canonically conjugate parameter $\mu$, 
	\begin{equation}
	\lambda =\frac{t\bar{b}_0}{\bar{b}_0+\bar{b}_t}, \quad\quad \mu =\frac{1}{2}+\frac{\bar{a}_0}{\lambda}+\frac{\bar{a}_t}{\lambda-t}.
	\end{equation}
	Therefore, by replacing $\lambda$ and $\mu$
	\begin{equation}
	H^{+}-H =\frac{1}{2}+\frac{\lambda}{t}\bigg(\mu-\frac{1}{2}\bigg)-\frac{\lambda}{t(\lambda-t)}\theta_{t}
	\end{equation} 
	\begin{equation}
	H^{-}-H = -\frac{1}{2}+\frac{(\lambda-t)\big(\mu-\frac{1}{2}\big)-\frac{1}{2}(\theta_{0}+\theta_{t}-\theta_{\star})}{\lambda\big(\mu-\frac{1}{2}\big)-\frac{1}{2}(\theta_{ 0}+\theta_{t}-\theta_{\star})}\bigg(\frac{\lambda}{t}\big(\mu-\frac{1}{2}\big)-\frac{\theta_0}{t}  \bigg).
	\end{equation}
	After some algebra, we use the isomonodromy flow definition \eqref{eq:HamilAppendix} to that 
	\begin{gather}
		\frac{d}{dt}\log\frac{\tau^{+}_V(\{\theta\}^+;\sigma,\eta;t)}{\tau_V(\{\theta\};\sigma,\eta;t)} =
		-\frac{1}{2}-\frac{\lambda}{t}\left(\mu-\frac{1}{2}\right)
		+\frac{\lambda}{t(\lambda-t)}\theta_t \\
		\frac{d}{dt}\log\frac{\tau^{-}_V(\{\theta\}^-;\sigma,\eta;t)}{\tau_V(\{\theta\};\sigma,\eta;t)} =
		\frac{1}{2}-\frac{(\lambda-t)\left(\mu-\frac{1}{2}\right)-
			\frac{1}{2}(\theta_0+\theta_t-\theta_\star)}{
			\lambda\left(\mu-\frac{1}{2}\right)-\frac{1}{2}(\theta_0
			+\theta_t-\theta_\star)}\left(
		\frac{\lambda}{t}\left(\mu-\frac{1}{2}\right)-\frac{\theta_0}{t}
		\right).
		\label{eq:3.427}
	\end{gather}
	Note that, the first line has a divergent limit $\lambda\rightarrow t$, we conclude, by taking the integration on $t$ and then taking the limit, that the second condition in \eqref{eq:3.38} can be substitued by the simpler one
	\begin{equation}
	\tau_V(\{\theta\};\sigma,\eta;t)=0,
	\label{eq:secondcond}
	\end{equation}
	where $\{\theta\}$ is defined by the parameters of the \eqref{eq:2.53} and denoted by
	\begin{equation}
	\{\theta\}=\{\theta_0,\theta_{t},\theta_\star\}.
	\label{eq:3.44}
	\end{equation}
	Thus, in terms of monodromy data, the first condition in \eqref{eq:3.38} is given by
	\begin{equation}
	c_{t}=\frac{d}{dt}\log\tau_V(\{\theta\}^-;\sigma,\eta;t)-\frac{\theta_0(\theta_{t}-1)}{2t}
	\label{eq:3.45}
	\end{equation}
	with the shifted monodromy data $\{\theta\}^-=\{\theta_{ 0},\theta_{t}-1,\theta_{\star}+1\}$.

	\chapter{\texorpdfstring{Fredholm Determinant representation: $\tau_V$ and $\tau_{III}$-functions}%
 {}}
	\label{sec:tools}
	
	The Fredholm determinant representation for the PV $\tau$-function uses the Riemann-Hilbert problem formulation in terms of Plamelj (projection) operators and matrices. The idea is to introduce projection operators which act on the space of (pair of) functions in the complex plane to give analytic functions with prescribed monodromy (Cauchy-Riemann operators). Details can be found in \cite{Gavrylenko:2016zlf}. 
		
	\section{\texorpdfstring{Fredholm representation for the $\tau_{V}$ function}%
{}}
	\label{subsec:tauV}
	
	The expression for the $\tau_V$ which allows us to compute the QNMs efficiently is given by
	\begin{equation}
	\tau_V(\{\theta\};\sigma,\eta;t)=
	t^{\tfrac{1}{4}(\sigma^2-\theta_0^2-\theta_t^2)}
	e^{\tfrac{1}{2}\theta_t t}
	\det(\mathbbold{1}-\mathsf{A}
	\kappa_V^{\tfrac{1}{2}\sigma_3}t^{\tfrac{1}{2}\sigma\sigma_3}
	\mathsf{D}_c(t)
	\kappa_V^{-\tfrac{1}{2}\sigma_3}t^{-\tfrac{1}{2}\sigma\sigma_3})
	\label{eq:fredholmV}
	\end{equation}
	\begin{equation}
	(\mathsf{A}g)(z)=\oint_{ C} \frac{dz'}{2\pi i}A(z,z')g(z'),\qquad
	(\mathsf{D}_cg)(z)=\oint_{ C} \frac{dz'}{2\pi i}D_c(z,z')g(z'),\qquad
	g(z')=\begin{pmatrix}
	f_+(z) \\
	f_-(z)
	\end{pmatrix}
	\label{eq:fredholmad}
	\end{equation}
	with ${ C}$ a circle of radius $R<1$ and kernels given
	explicitly for $|t|<R$, by
	\begin{equation}
	\begin{gathered}
	A(z,z')=\frac{\Psi^{-1}(\sigma,\theta_t,\theta_0;z')
		\Psi(\sigma,\theta_t,\theta_0;
		z)-\mathbbold{1}}{z-z'},\\ 
	D_c(z,z')=\frac{\mathbbold{1}-\Psi_c^{-1}(-\sigma,\theta_\star;t/z')
		\Psi_c(-\sigma,\theta_\star;t/z)}{z-z'},
	\end{gathered}
	\end{equation}
	where the parametrices $\Psi$ and $\Psi_c$ are matrices whose entries
	are given by
	\begin{equation}
	\displaystyle
	\Psi(\sigma,\theta_t,\theta_0;z) = \begin{pmatrix}
	\phi(\sigma,\theta_t,\theta_0;z) &
	\chi(-\sigma,\theta_t,\theta_0;z) \\
	\chi(\sigma,\theta_t,\theta_0;z) &
	\phi(-\sigma,\theta_t,\theta_0,z)
	\end{pmatrix},
	\end{equation}
	with $\phi$ and $\chi$ in terms of Gauss' hypergeometric series
	\begin{equation}
	\displaystyle 
	\begin{gathered}
	\phi(\sigma,\theta_t,\theta_0;z) = {_2F_1}(
	\tfrac{1}{2}(\sigma-\theta_t+\theta_0),\tfrac{1}{2}(\sigma-\theta_t-\theta_0);
	\sigma;z) \\
	\chi(\sigma,\theta_t,\theta_0;z) =
	\frac{\theta_0^2-(\sigma-\theta_t)^2}{4\sigma(1+\sigma)}
	z\,{_2F_1}(
	1+\tfrac{1}{2}(\sigma-\theta_t+\theta_0),
	1+\tfrac{1}{2}(\sigma-\theta_t-\theta_0);
	2+\sigma;z)
	\end{gathered}
	\end{equation}
	and 
	\begin{gather}
		\Psi_c(-\sigma,\theta_\star;t/z)=
		\begin{pmatrix}
			\phi_c(-\sigma,\theta_\star;t/z) &
			\chi_c(-\sigma,\theta_\star; t/z) \\
			\chi_c(\sigma,\theta_\star; t/z) &
			\phi_c(\sigma,\theta_\star; t/z)
		\end{pmatrix},
		\nonumber \\
		\phi_c(\pm\sigma,\theta_\star;t/z) = 
		{_1F_1}(\tfrac{-\theta_\star\pm\sigma}{2};\pm\sigma;-t/z), \\
		\chi_c(\pm\sigma,\theta_\star;t/z) =
		\pm\frac{-\theta_\star\pm\sigma}{2\sigma(1\pm\sigma)}\,
		\frac{t}{z}
		\,{_1F_1}(1+\tfrac{-\theta_\star\pm\sigma}{2},2\pm\sigma;-t/z),
		\label{eq:confluentparametrix}
	\end{gather}
	where  ${_1F_1}$ the confluent (Kummer's) hypergeometric series. In turn, $\kappa_V$ in \eqref{eq:fredholmV} is expressed in terms of ratio of Gamma functions as
	\begin{equation}
	\kappa_V =e^{i\pi\eta}\Pi_{V}=
	e^{i\pi\eta}\frac{\Gamma(1-\sigma)^2}{\Gamma(1+\sigma)^2}
	\frac{\Gamma(1+\tfrac{1}{2}(\theta_\star+\sigma))}{
		\Gamma(1+\tfrac{1}{2}(\theta_\star-\sigma))}
	\frac{\Gamma(1+\tfrac{1}{2}(\theta_t+\theta_0+\sigma))
		\Gamma(1+\tfrac{1}{2}(\theta_t-\theta_0+\sigma))}{
		\Gamma(1+\tfrac{1}{2}(\theta_t+\theta_0-\sigma))
		\Gamma(1+\tfrac{1}{2}(\theta_t-\theta_0-\sigma))}.
	\label{eq:kappaV}
	\end{equation}
	
	From the definition \eqref{eq:fredholmV}, one can expand the determinant and work out the small-$t$ expansion of the $\tau$-function for Painlevé V, recovering the results found in the literature \cite{Jimbo:1982aa,Lisovyy:2018mnj,Andreev:2000aa,CarneirodaCunha:2019tia}
	\begin{equation}
	\tau_V(\{\theta\};\sigma,\eta;t)=C_{V}(\{\theta\};\sigma)
	t^{\frac{1}{4}(\sigma^2-\theta_0^2-\theta_t^2)}
	e^{\frac{1}{2}\theta_tt}\hat{\tau}_V(\{\theta\};\sigma,\eta;t),
	\end{equation}
	where $\hat{\tau}_V$ comprises the expansion of the Fredholm determinant in \eqref{eq:fredholmV}. 
	
	Essentially, to recover the Nekrasov expansion in \cite{Gamayun:2013auu}, we must use a different basis for
	truncation. First we expand the parametrices 
	\begin{equation}
	\Psi(\sigma,\theta_t,\theta_0;z)=\mathbbold{1}+
	\sum_{n=1}^{\infty}{ G}_n(\sigma,\theta_t,\theta_0)z^n,\qquad
	\Psi_c(-\sigma,\theta_\star;t/z)=\mathbbold{1}+
	\sum_{n=1}^{\infty}{ G}_{c,n}(-\sigma,\theta_\star)(t/z)^n
	\end{equation}
	and compute the matrix elements associated to $\mathsf{A}$ and $\mathsf{D}_c$ in the Fourier basis $g(z')=\sum_n g_n (z')^n$. The matrix-valued coefficients ${ G}_n(\sigma,\theta_t,\theta_0)$ and ${ G}_{c,n}(-\sigma,\theta_\star)$ can be computed from the expansion of the Gauss hypergeometric series:
	\begin{equation}
	{ G}_n(\sigma,\theta_t,\theta_0) =
	\begin{pmatrix}
	\frac{(\frac{1}{2}(\sigma-\theta_t+\theta_0))_n
		(\frac{1}{2}(\sigma-\theta_t-\theta_0))_n}{
		(\sigma)_nn!} &
	\frac{(\frac{1}{2}(\sigma-\theta_t+\theta_0))_n
		(\frac{1}{2}(\sigma-\theta_t-\theta_0))_n}{
		(-\sigma)_{n+1}(n-1)!} \\
	-\frac{(\frac{1}{2}(-\sigma-\theta_t+\theta_0))_n
		(\frac{1}{2}(-\sigma-\theta_t-\theta_0))_n}{
		(\sigma)_{n+1}(n-1)!} &
	\frac{(\frac{1}{2}(-\sigma-\theta_t+\theta_0))_n
		(\frac{1}{2}(-\sigma-\theta_t-\theta_0))_n}{
		(-\sigma)_nn!}
	\end{pmatrix},
	\qquad n\ge 1;
	\label{eq:parametrixexp}
	\end{equation}
	and
	\begin{equation}
	{ G}_{c,n}(-\sigma,\theta_\star) =
	\begin{pmatrix}
	\frac{(\frac{1}{2}(-\sigma-\theta_\star))_n}{
		(-\sigma)_nn!} &
	\frac{(\frac{1}{2}(-\sigma-\theta_\star))_n}{
		(\sigma)_{n+1}(n-1)!} \\
	\frac{(\frac{1}{2}(\sigma-\theta_\star))_n}{
		(-\sigma)_{n+1}(n-1)!} &
	\frac{(\frac{1}{2}(\sigma-\theta_\star))_n}{
		(\sigma)_nn!}
	\end{pmatrix},
	\qquad n\ge 1;
	\label{eq:confparametrixexp}
	\end{equation}
	where $(z)_n=\Gamma(z+n)/\Gamma(z)$ is the Pochhammer symbol. The
	kernels $A(z,z')$ and $D_c(z,z')$ can be suitably expanded:
	\begin{equation}
	\begin{gathered}
	A(z,z')={ G}_{1}+{ G}_{2}z+({ G}_2-{ G}_{1}^2)z'+\ldots, \\
	D_c(z,z')=t{ G}_{c,1}\frac{1}{zz'}+t^2{
		G}_{c,2}\frac{1}{z^2z'}+t^2({ G}_{c,2}-({
		G}_{c,1})^2)\frac{1}{z(z')^2}+\ldots
	\end{gathered}
	\label{eq:kernelexp}
	\end{equation}
	The resulting matrices are semi-infinite, and truncation to order $N$ gives an approximation to the isomonodromic $\tau_V$-function of order ${O}(t^N,|t|^{(1\pm\Re\tilde{\sigma})N})$. We refer to \cite{Gavrylenko:2016zlf} for the corresponding calculation for the isomonodromic $\tau_{VI}$-function\footnote{An arbitrary-precision implementation of the Painlevé VI, V, and III $\tau$-functions in
		\href{http://julialang.org}{Julia} programming language can be obtained in \href{https://github.com/strings-ufpe/painleve}{\texttt{https://github.com/strings-ufpe/painleve}}.}. From the
	computational point of view, this expansion is costlier, but has the bonus of not requiring pre-existing implementations of special
	functions, except the gamma function.

	When interpreted in terms of two
	variables $\mu=\kappa_V t^\sigma$ and $t$, the series obtained from the
	small $t$ expansion is analytic in $t$ and meromorphic in $\mu$:
	\begin{multline}
		\hat{\tau}_V(\{\theta\};\sigma,\eta;t)=
		1-\left(\frac{\theta_t}{2}-\frac{\theta_\star}{4}
		+\frac{\theta_\star(\theta_0^2-\theta_t^2)}{4\sigma^2}\right)t
		\\ -
		\frac{(\theta_\star+\sigma)((\sigma+\theta_t)^2-
			\theta_0^2)}{8\sigma^2(\sigma-1)^2}\kappa_V^{-1}
		t^{1-\sigma}-
		\frac{(\theta_\star-\sigma)((\sigma-\theta_t)^2-
			\theta_0^2)}{8\sigma^2(\sigma+1)^2}
		\kappa_V\, t^{1+\sigma}+{\mathcal O}(t^2,|t|^{2\pm
			2\Re\sigma}).
		\label{eq:expansiontauV}
	\end{multline}
	The structure of the expansion \eqref{eq:expansiontauV} imports a
	great deal of structure from the conformal block expansion it is
	derived from \cite{Gamayun:2013auu}. One notes from \eqref{eq:expansiontauV} that $\tau_V$ is meromorphic in $\kappa_V t^{\sigma}$ so the equation $\tau_{V}(\{\theta\};\sigma,\eta;t)=0$ can be inverted to define a series for $\kappa_V t^{\sigma}$, or $e^{i\pi\eta}$ in terms of
	$t$. Assuming that $\sigma$ satisfies $0<\Re(\sigma)<1$, we arrive at the expression 
	\begin{equation}
	\Theta_{V}(\{\theta\};\sigma)e^{i\pi\eta}t^{\sigma-1} =
	\chi_{V}(\{\theta\};\sigma;t),
	\label{eq:zerotau5p}
	\end{equation} 
	with $\Theta_{V}(\{\theta\};\sigma)$ expressed in terms of ratios of
	gamma functions 
	\begin{equation}
	\Theta_V(\{\theta\};\tilde{\sigma})=
	\frac{\Gamma^2(2-\sigma)}{\Gamma^2(\sigma)}
	\frac{\Gamma(\tfrac{1}{2}(\theta_\star+\sigma))}{
		\Gamma(1+\tfrac{1}{2}(\theta_\star-\sigma))}
	\frac{\Gamma(\tfrac{1}{2}(\theta_t+\theta_0+\sigma))}{
		\Gamma(1+\tfrac{1}{2}(\theta_t+\theta_0-\sigma))}
	\frac{\Gamma(\tfrac{1}{2}(\theta_t-\theta_0+\sigma))}{
		\Gamma(1+\tfrac{1}{2}(\theta_t-\theta_0-\sigma))}
	\label{eq:theta5}
	\end{equation}
	and the function $\chi_{V}(\{\theta\};\sigma;t)$ analytic for
	$t_0$ small, with expansion 
	\begin{equation}
	\begin{aligned}
	\chi_V(\{\theta\};\tilde{\sigma};t)
	=&1+(\tilde{\sigma}-1)\frac{\theta_\star
		(\theta_t^2-\theta_0^2)}{\tilde{\sigma}^2(\tilde{\sigma}-2)^2}t+
	\bigg[\frac{(\theta_t^2-\theta_{0}^2)^2+2\theta_\star^2
		(\theta_t^2+\theta_{0}^2)}{64}\left(
	\frac{1}{(\tilde{\sigma}-2)^2}-\frac{1}{\tilde{\sigma}^2}\right)+
	\\ + &\frac{\theta_\star^2(\theta_t^2-\theta_{0}^2)^2}{64}
	\bigg(\frac{5}{\tilde{\sigma}^4}-\frac{1}{(\tilde{\sigma}-2)^4}
	-\frac{2}{(\tilde{\sigma}-2)^2}+\frac{2}{\tilde{\sigma}(\tilde{\sigma}-2)}\bigg)+
	\\
	+&\frac{(1-\theta_\star^2)(\theta_t^2-(\theta_{0}-1)^2)(\theta_t^2
		-(\theta_{0}+1)^2)}{128}\left(\frac{1}{(\tilde{\sigma}+1)^2}-
	\frac{1}{(\tilde{\sigma}-3)^2}\right)\bigg]t^2+{ O}(t^3).
	\label{eq:chi5}
	\end{aligned}
	\end{equation}
	For $\Re(\sigma)<0$, we also have an expression similar for
	\eqref{eq:zerotau5p}, where one simply changes to $\sigma \rightarrow
	-\sigma$ and $e^{i\pi\eta} \rightarrow e^{-i\pi\eta}$. Although it
	will not be necessary for our purposes, one can in principle compute
	the expansion of $\chi_V$ to arbitrary order in $t$.
	
	Finally, we can incorporate the condition \eqref{eq:quantizationV}
	into the expression \eqref{eq:zerotau5p}. Using the definition of
	$\Theta_V$, and the identity $\Gamma(z)\Gamma(1-z) =\pi/\sin(\pi z)$,
	one obtains, when $\eta$ satisfies \eqref{eq:quantizationV}, 
	\begin{equation}
	\Theta_V(\{\theta\},\sigma) e^{i\pi\eta} =
	-e^{-i\pi\sigma}\Theta_V(-\{\theta\},\sigma),
	\end{equation}
	Substituting back into
	\eqref{eq:zerotau5p}, we find
	\begin{equation}
	-e^{-i\pi\sigma}\Theta_V(-\{\theta\},\sigma)
	t^{\sigma-1}=\chi_V(\{\theta\};\sigma;t),
	\label{eq:zerochi5}
	\end{equation}
	again assuming $\Re(\sigma)>0$. For $\Re(\sigma)<0$, one finds the
	right expansion by simply sending $\sigma\rightarrow -\sigma$ in
	\eqref{eq:zerochi5}. With the expression \eqref{eq:zerochi5}, one can
	then consider for the QNM problem the overdetermined system of
	equations consisted by \eqref{eq:c5expansion} and \eqref{eq:zerochi5},
	with $\sigma$ the free parameter.

	\section{\texorpdfstring{Fredholm representation for the $\tau_{III}$ function}%
 {}}

	\begin{equation}
	(\mathsf{A}_cg)(z)=\oint_{ C} \frac{dz'}{2\pi i}A_c(z,z')g(z'),\quad
	(\mathsf{D}_cg)(z)=\oint_{ C} \frac{dz'}{2\pi i}D_c(z,z')g(z'),\quad
	g(z')=\begin{pmatrix}
	f_+(z) \\
	f_-(z)
	\end{pmatrix}
	\label{eq:fredholmad1}
	\end{equation}
	with ${ C}$ a circle of radius $R<1$ and kernels given explicitly for $|t|<R$, by
	\begin{equation}
	\begin{gathered}
	A_c(z,z')=\frac{\Psi^{-1}_c(\sigma,\theta_t,\theta_0;tz')
		\Psi_c(\sigma,\theta_t,\theta_0;
		tz)-\mathbbold{1}}{z-z'},\\ 
	D_c(z,z')=\frac{\mathbbold{1}-\Psi_c^{-1}(-\sigma,\theta_\star;1/z')
		\Psi_c(-\sigma,\theta_\star;1/z)}{z-z'},
	\end{gathered}
	\end{equation}

	As the calculation is relatively short and to our knowledge not present in the literature we include it here. This derivation will complement the discussion of the confluent limit of the Riemann-Hilbert map for the $\tau_{V}$-function, presented in Chapter 3~. We start by considering \eqref{eq:fredholmV} and deform the circle ${C}$ multiplying its radius by $t$. This has the effect of shifting the $t$ dependence in the argument from the kernel of $\mathsf{D}$ to $\mathsf{A}$, so that
	\begin{equation}
	\begin{gathered}
	\tilde{A}(z,z')=\frac{\Psi^{-1}(\sigma,\theta_t,\theta_0;tz')
		\Psi(\sigma,\theta_t,\theta_0;
		tz)-\mathbbold{1}}{z-z'},\\ 
	\tilde{D}(z,z')=\frac{\mathbbold{1}-\Psi_c^{-1}(-\sigma,\theta_\star;1/z')
		\Psi_c(-\sigma,\theta_\star;1/z)}{z-z'}.
	\end{gathered}
	\end{equation}
	With this provision, we can now implement the confluent limit on the parametrix $\Psi$:
	\begin{equation}
	\lim_{\Lambda\rightarrow \infty}
	\Psi(\sigma,\Lambda+\tfrac{1}{2}
	\theta_\circ,-\Lambda+\tfrac{1}{2}\theta_\circ;uz/\Lambda)
	= \Psi_c(\sigma,\theta_\circ;uz)+
	\frac{uz}{\Lambda}\Psi^{(1)}_c(\sigma;\theta_\circ,uz)+\ldots,
	\end{equation}
	where $\theta_\circ = \theta_t+\theta_0$ is fixed and $\Psi_c(\sigma,\theta_\circ;uz)$ is the same confluent parametrix as above \eqref{eq:confluentparametrix} and the first
	$\Lambda^{-1}$ correction
	\begin{equation}
	\Psi^{(1)}_c(\sigma;\theta_\circ;uz)=
	\begin{pmatrix}
	\phi^{(1)}_c(\sigma;\theta_\circ;uz) &
	\chi^{(1)}_c(-\sigma;\theta_\circ;uz) \\
	\chi^{(1)}_c(\sigma;\theta_\circ;uz)
	& \phi^{(1)}_c(-\sigma;\theta_\circ;uz)
	\end{pmatrix},
	\end{equation}
	is also given in terms of confluent hypergeometric functions
	\begin{equation}
	\begin{aligned}
	\phi^{(1)}_c(\pm\sigma;\theta_\circ;uz&)
	=\frac{\pm\sigma-\theta_\circ}{2}
	{_1F_1}(\tfrac{1}{2}(\pm\sigma-\theta_\circ);\pm\sigma;-uz),\\
	\chi^{(1)}_c(\pm\sigma;\theta_\circ;uz)
	=&-\frac{\pm\sigma-\theta_\circ}{2(1\pm\sigma)}\bigg[
	{_1F_1}(1+\tfrac{(\pm\sigma-\theta_\circ)}{2};2\pm\sigma;-uz)\\
	&\mp\frac{uz}{\sigma}
	\bigg(1+\frac{\pm\sigma-\theta_\circ}{2}\bigg)
	\,{_1F_1}(2+\tfrac{(\pm\sigma-\theta_\circ)}{2};2\pm\sigma;-uz)\bigg].
	\end{aligned}
	\end{equation}
	Given the expansion of the parametrix, the expansion of the kernel $\tilde{A}(z,z')$ then follows 
	\begin{equation}
	\tilde{A}(z,z')=\tilde{A}_c(z,z')+\frac{u}{\Lambda}\tilde{A}_c^{(1)}(z,z')
	+{ O}(\Lambda^{-2}),
	\label{eq:akernelexpansion}
	\end{equation}
	where
	\begin{gather}
		\tilde{A}_c(z,z')=\frac{\Psi_c^{-1}(\sigma,\theta_\circ;uz')
			\Psi_c(\sigma,\theta_\circ;uz)-\mathbbold{1}}{z-z'},\\
		\tilde{A}_c^{(1)}(z,z') = \frac{z\Psi_c(\sigma,\theta_\circ;uz')
			\Psi^{(1)}_c(\sigma,\theta_\circ;uz)-z'
			\Psi_c^{-1}\Psi_c^{(1)}\Psi_c^{-1}(\sigma,\theta_\circ;uz')
			\Psi_c(\sigma,\theta_\circ;uz)}{z-z'}.
	\end{gather}
	
	We now turn into the confluent limit of the monodromy parameters. We will assume that $\eta$ has a well-defined limit, which is certainly the case in our application. The parameter $\kappa$ has a well-defined function in terms of $\Lambda$, provided $\Lambda$ is not close to the negative real axis. Expanding the gamma functions in \eqref{eq:kappaV}, we find
	\begin{equation}
	\begin{aligned}
	\kappa t^\sigma =
	e^{i\pi\eta} u^{\sigma}\Pi_{III}\left(1+
	\frac{\sigma}{2\Lambda}+{ O}(\Lambda^{-2})\right),
	\end{aligned}
	\end{equation}
	with
	$$\Pi_{III}=\frac{\Gamma(1-\sigma)^2}{
		\Gamma(1+\sigma)^2}
	\frac{\Gamma(1+\tfrac{1}{2}(\theta_\star+\sigma))}{
		\Gamma(1+\tfrac{1}{2}(\theta_\star-\sigma))}
	\frac{\Gamma(1+\tfrac{1}{2}(\theta_\circ+\sigma))}{
		\Gamma(1+\tfrac{1}{2}(\theta_\circ-\sigma))}.$$
	
	For convenience, we will refer to $\kappa_{III}=\Pi_{III}e^{i\pi\eta}$ and $\mu=\kappa_{III}u^{\sigma}$ in the following expressions.
	
	The first term in the $\mathsf{A}$ kernel expansion \eqref{eq:akernelexpansion} gives the Painlevé III $\tau$-function
	\begin{equation}
	\tau_{III}(\{\tilde{\theta}\};\sigma,\eta;u) =
	u^{\frac{1}{4}\sigma^2-\frac{1}{8}\theta_\circ^2}
	e^{\frac{1}{2}u}\det(
	\mathbbold{1}-\mathsf{A}_c\kappa_{III}^{\frac{1}{2}\sigma_3}
	u^{\frac{1}{2}\sigma\sigma_3}\mathsf{D}_c(u)
	\kappa_{III}^{-\frac{1}{2}\sigma_3}u^{-\frac{1}{2}\sigma\sigma_3}).
	\label{eq:fredholmIII}
	\end{equation}
	with $\{\tilde{\theta}\}=\{\theta_\star,\theta_\circ\}$ and $\sigma_{3} = \diag(1,-1)$. This definition can be compared to the expansion given in \cite{Gamayun:2013auu} by comparing the first terms, see
	\eqref{eq:tauIIIexpansion} below. 
	
	It will be interesting to consider the first order term in $\Lambda^{-1}$ of the expansion of \eqref{eq:fredholmV}. Using
	well-known properties of the determinant, we have
	\begin{equation}
	\begin{aligned}
	\det(
	\mathbbold{1}-\mathsf{A}\kappa_{V}^{\frac{1}{2}\sigma_3}
	t^{\frac{1}{2}\sigma\sigma_3}\mathsf{D}_c(t)
	\kappa_{V}^{-\frac{1}{2}\sigma_3}t^{-\frac{1}{2}\sigma\sigma_3})=
	\det(
	\mathbbold{1}-\mathsf{A}_c
	\kappa_{III}^{\frac{1}{2}\sigma_3}u^{\frac{1}{2}\sigma\sigma_3}
	\mathsf{D}_c(u)
	\kappa_{III}^{-\frac{1}{2}\sigma_3}u^{-\frac{1}{2}\sigma\sigma_3}) \times
	\\\ \bigg[ 1-\frac{1}{\Lambda}Tr((\mathbbold{1}-
	\mathsf{A}_c\mu^{\frac{1}{2}\sigma_3}\mathsf{D}_c(u)
	\mu^{-\frac{1}{2}\sigma_3})^{-1}
	((\mathsf{A}^{(1)}_c \mu^{\frac{1}{2}\sigma_3}\mathsf{D}_c(u)
	\mu^{-\frac{1}{2}\sigma_3}
	+\tfrac{1}{4}\sigma\mathsf{A}_c \mu^{\frac{1}{2}\sigma_3}
	[\sigma_3,\mathsf{D}_c(u)]
	\mu^{-\frac{1}{2}\sigma_3}))\\+{O}(\Lambda^{-2}))\bigg],
	\end{aligned}
	\end{equation}
	where, again, $\mu=e^{i\pi\eta}\Pi_{III}u^{\sigma}$. We note that the correction is well-defined even when the determinant
	vanishes. Generically, for finite-dimensional matrices,
	\begin{equation}
	(\det \mathsf{M}) \mathsf{M}^{-1} = \text{adj}(\mathsf{M}),
	\end{equation}
	is the \textit{adjugate} to $\mathsf{M}$, which is the transpose of the cofactor matrix.
	
	The calculation of the $\tau_{III}$ from \eqref{eq:fredholmIII} follows the same strategy of Sec. \ref{subsec:tauV}, expanding the parametrices
	\begin{equation}
	\Psi_c(\sigma,\theta_\circ;z) = \sum_{n=0}^{\infty}{
		G}_{c,n}(\sigma,\theta_\circ) z^{n},\qquad
	\Psi_c(-\sigma,\theta_\star;u/z)=\sum_{n=0}^{\infty}{
		G}_{c,n}(-\sigma,\theta_\star)(u/z)^n,
	\end{equation}
	where the coefficients ${ G}_{c,n}$ are the same as \eqref{eq:confparametrixexp}. The expansion of the confluent kernels
	$A_c(z,z')$ and $D_c(z,z')$ is analogous to \eqref{eq:kernelexp}, and the expansion of the Painlevé III $\tau$-function \eqref{eq:fredholmIII}
	gives 
	\begin{multline}
		\tau_{III}(\{\tilde{\theta}\};\sigma,\eta;u) =
		u^{\frac{1}{4}\sigma^2-\frac{1}{8}\theta_\circ^2}e^{\frac{1}{2}u}\times\\
		\left(
		1-\frac{\sigma-\theta_\circ\theta_\star}{2\sigma^2}u
		-\frac{(\sigma+\theta_\circ)(\sigma+\theta_\star)}{4\sigma^2
			(\sigma-1)^2}\kappa_{III}^{-1}u^{1-\sigma} -
		\frac{(\sigma-\theta_\circ)(\sigma-\theta_\star)}{4\sigma^2
			(\sigma+1)^2}\kappa_{III}u^{1+\sigma}+{ O}(u^2,u^{2\pm
			2\sigma})\right).
		\label{eq:tauIIIexpansion1}
	\end{multline}
	For small $u$, the first correction in order $\Lambda^{-1}$ is surprisingly simple 
	\begin{equation}
	\tau_V( \{\theta\};
	\sigma,\eta;\tfrac{1}{\Lambda}u)
	=
	\left(1+\frac{\theta_\star-2\theta_\circ}{4\Lambda}u\right)
	\tau_{III}(\{\tilde{\theta}\};\sigma,\eta;u)
	+{O}(\Lambda^{-2},u^2,u^{2\pm 2\sigma}),
	\end{equation}
	so, to first order in $\Lambda^{-1}$, the zero of the $\tau$-function does not change from the extremal value. Note that, in the confluent limit $\{\theta\}=\{\theta_0, \theta_t, \theta_{\star}\}$ changes to $\{\tilde{\theta}\} = \{\theta_{\circ},\theta_{\star}\}$, with
	\begin{equation}
	\theta_0 = \Lambda-\tfrac{1}{2}\theta_\circ \qquad	\theta_{t}=\Lambda+\tfrac{1}{2}\theta_\circ, \qquad \Lambda \rightarrow \infty.
	\end{equation}

\end{appendices}

\phantompart
\postextual
\phantompart
\printindex

\bibliography{referencesC2}

\begin{thebibliography}{170}%
\makeatletter
\providecommand \@ifxundefined [1]{%
 \@ifx{#1\undefined}
}%
\providecommand \@ifnum [1]{%
 \ifnum #1\expandafter \@firstoftwo
 \else \expandafter \@secondoftwo
 \fi
}%
\providecommand \@ifx [1]{%
 \ifx #1\expandafter \@firstoftwo
 \else \expandafter \@secondoftwo
 \fi
}%
\providecommand \natexlab [1]{#1}%
\providecommand \enquote  [1]{``#1''}%
\providecommand \bibnamefont  [1]{#1}%
\providecommand \bibfnamefont [1]{#1}%
\providecommand \citenamefont [1]{#1}%
\providecommand \href@noop [0]{\@secondoftwo}%
\providecommand \href [0]{\begingroup \@sanitize@url \@href}%
\providecommand \@href[1]{\@@startlink{#1}\@@href}%
\providecommand \@@href[1]{\endgroup#1\@@endlink}%
\providecommand \@sanitize@url [0]{\catcode `\\12\catcode `\$12\catcode
  `\&12\catcode `\#12\catcode `\^12\catcode `\_12\catcode `\%12\relax}%
\providecommand \@@startlink[1]{}%
\providecommand \@@endlink[0]{}%
\providecommand \url  [0]{\begingroup\@sanitize@url \@url }%
\providecommand \@url [1]{\endgroup\@href {#1}{\urlprefix }}%
\providecommand \urlprefix  [0]{URL }%
\providecommand \Eprint [0]{\href }%
\providecommand \doibase [0]{http://dx.doi.org/}%
\providecommand \selectlanguage [0]{\@gobble}%
\providecommand \bibinfo  [0]{\@secondoftwo}%
\providecommand \bibfield  [0]{\@secondoftwo}%
\providecommand \translation [1]{[#1]}%
\providecommand \BibitemOpen [0]{}%
\providecommand \bibitemStop [0]{}%
\providecommand \bibitemNoStop [0]{.\EOS\space}%
\providecommand \EOS [0]{\spacefactor3000\relax}%
\providecommand \BibitemShut  [1]{\csname bibitem#1\endcsname}%
\let\auto@bib@innerbib\@empty
\bibitem [{\citenamefont {e~Poesia.}(2013)}]{encanto}%
  \BibitemOpen
  \bibfield  {author} {\bibinfo {author} {\bibfnamefont {E.~C.}\ \bibnamefont
  {e~Poesia.}},\ }\href {https://www.youtube.com/watch?v=XGrr8dZc-bI} {\enquote
  {\bibinfo {title} {Com quem vim},}\ } (\bibinfo {year} {2013})\BibitemShut
  {NoStop}%
\bibitem [{\citenamefont {Davis}\ \emph {et~al.}(1971)\citenamefont {Davis},
  \citenamefont {Ruffini}, \citenamefont {Press},\ and\ \citenamefont
  {Price}}]{PhysRevLett.27.1466}%
  \BibitemOpen
  \bibfield  {author} {\bibinfo {author} {\bibfnamefont {M.}~\bibnamefont
  {Davis}}, \bibinfo {author} {\bibfnamefont {R.}~\bibnamefont {Ruffini}},
  \bibinfo {author} {\bibfnamefont {W.~H.}\ \bibnamefont {Press}}, \ and\
  \bibinfo {author} {\bibfnamefont {R.~H.}\ \bibnamefont {Price}},\ }\href
  {\doibase 10.1103/PhysRevLett.27.1466} {\bibfield  {journal} {\bibinfo
  {journal} {Phys. Rev. Lett.}\ }\textbf {\bibinfo {volume} {27}},\ \bibinfo
  {pages} {1466} (\bibinfo {year} {1971})}\BibitemShut {NoStop}%
\bibitem [{\citenamefont
  {Chandrasekhar}(1976{\natexlab{a}})}]{Chandrasekhar:1976ap}%
  \BibitemOpen
  \bibfield  {author} {\bibinfo {author} {\bibfnamefont {S.}~\bibnamefont
  {Chandrasekhar}},\ }\href {\doibase 10.1098/rspa.1976.0090} {\bibfield
  {journal} {\bibinfo  {journal} {Proc. Roy. Soc. Lond.}\ }\textbf {\bibinfo
  {volume} {A349}},\ \bibinfo {pages} {571} (\bibinfo {year}
  {1976}{\natexlab{a}})}\BibitemShut {NoStop}%
\bibitem [{\citenamefont {Chandrasekhar}(1976{\natexlab{b}})}]{10.2307/79115}%
  \BibitemOpen
  \bibfield  {author} {\bibinfo {author} {\bibfnamefont {S.}~\bibnamefont
  {Chandrasekhar}},\ }\href {http://www.jstor.org/stable/79115} {\bibfield
  {journal} {\bibinfo  {journal} {Proceedings of the Royal Society of London.
  Series A, Mathematical and Physical Sciences}\ }\textbf {\bibinfo {volume}
  {348}},\ \bibinfo {pages} {39} (\bibinfo {year}
  {1976}{\natexlab{b}})}\BibitemShut {NoStop}%
\bibitem [{\citenamefont {Conte}(2011)}]{conte}%
  \BibitemOpen
  \bibfield  {author} {\bibinfo {author} {\bibfnamefont {R.}~\bibnamefont
  {Conte}},\ }\href {https://books.google.com.br/books?id=VvH1vQAACAAJ} {\emph
  {\bibinfo {title} {The Painlev\'e Property: One Century Later}}},\ CRM Series
  in Mathematical Physics\ (\bibinfo  {publisher} {Springer New York},\
  \bibinfo {year} {2011})\BibitemShut {NoStop}%
\bibitem [{\citenamefont {Cardoso}(2004)}]{Cardoso:2003pj}%
  \BibitemOpen
  \bibfield  {author} {\bibinfo {author} {\bibfnamefont {V.}~\bibnamefont
  {Cardoso}},\ }\href {https://api.semanticscholar.org/CorpusID:117105085}
  {\bibfield  {journal} {\bibinfo  {journal} {arXiv: General Relativity and
  Quantum Cosmology}\ } (\bibinfo {year} {2004})}\BibitemShut {NoStop}%
\bibitem [{\citenamefont {Richartz}(2016)}]{Richartz:2015saa}%
  \BibitemOpen
  \bibfield  {author} {\bibinfo {author} {\bibfnamefont {M.}~\bibnamefont
  {Richartz}},\ }\href {\doibase 10.1103/PhysRevD.93.064062} {\bibfield
  {journal} {\bibinfo  {journal} {Phys. Rev. D}\ }\textbf {\bibinfo {volume}
  {93}},\ \bibinfo {pages} {064062} (\bibinfo {year} {2016})},\ \Eprint
  {http://arxiv.org/abs/1509.04260} {arXiv:1509.04260 [gr-qc]} \BibitemShut
  {NoStop}%
\bibitem [{\citenamefont {Berti}\ \emph
  {et~al.}(2006{\natexlab{a}})\citenamefont {Berti}, \citenamefont {Cardoso},\
  and\ \citenamefont {Casals}}]{Berti:2005gp}%
  \BibitemOpen
  \bibfield  {author} {\bibinfo {author} {\bibfnamefont {E.}~\bibnamefont
  {Berti}}, \bibinfo {author} {\bibfnamefont {V.}~\bibnamefont {Cardoso}}, \
  and\ \bibinfo {author} {\bibfnamefont {M.}~\bibnamefont {Casals}},\ }\href
  {\doibase 10.1103/PhysRevD.73.109902, 10.1103/PhysRevD.73.024013} {\bibfield
  {journal} {\bibinfo  {journal} {Phys.Rev.}\ }\textbf {\bibinfo {volume}
  {D73}},\ \bibinfo {pages} {024013} (\bibinfo {year} {2006}{\natexlab{a}})},\
  \Eprint {http://arxiv.org/abs/gr-qc/0511111} {arXiv:gr-qc/0511111 [gr-qc]}
  \BibitemShut {NoStop}%
\bibitem [{\citenamefont {{Press}}(1971)}]{1971ApJ...170L.105P}%
  \BibitemOpen
  \bibfield  {author} {\bibinfo {author} {\bibfnamefont {W.~H.}\ \bibnamefont
  {{Press}}},\ }\href {\doibase 10.1086/180849} {\bibfield  {journal} {\bibinfo
   {journal} {Astrophys.J.Lett.}\ }\textbf {\bibinfo {volume} {170}},\ \bibinfo
  {pages} {L105} (\bibinfo {year} {1971})}\BibitemShut {NoStop}%
\bibitem [{\citenamefont
  {Nollert}(1999{\natexlab{a}})}]{Hans-Peter_Nollert_1999}%
  \BibitemOpen
  \bibfield  {author} {\bibinfo {author} {\bibfnamefont {H.-P.}\ \bibnamefont
  {Nollert}},\ }\href {\doibase 10.1088/0264-9381/16/12/201} {\bibfield
  {journal} {\bibinfo  {journal} {Classical and Quantum Gravity}\ }\textbf
  {\bibinfo {volume} {16}},\ \bibinfo {pages} {R159} (\bibinfo {year}
  {1999}{\natexlab{a}})}\BibitemShut {NoStop}%
\bibitem [{\citenamefont {Kokkotas}\ and\ \citenamefont
  {Schmidt}(1999)}]{Kokkotas:1999bd}%
  \BibitemOpen
  \bibfield  {author} {\bibinfo {author} {\bibfnamefont {K.~D.}\ \bibnamefont
  {Kokkotas}}\ and\ \bibinfo {author} {\bibfnamefont {B.~G.}\ \bibnamefont
  {Schmidt}},\ }\href {\doibase 10.12942/lrr-1999-2} {\bibfield  {journal}
  {\bibinfo  {journal} {Living Rev. Rel.}\ }\textbf {\bibinfo {volume} {2}},\
  \bibinfo {pages} {2} (\bibinfo {year} {1999})},\ \Eprint
  {http://arxiv.org/abs/gr-qc/9909058} {arXiv:gr-qc/9909058} \BibitemShut
  {NoStop}%
\bibitem [{\citenamefont {Shutterstock}(2023)}]{berimbau}%
  \BibitemOpen
  \bibfield  {author} {\bibinfo {author} {\bibnamefont {Shutterstock}},\ }\href
  {https://www.shutterstock.com/pt/search/berimbau} {\enquote {\bibinfo {title}
  {single-string percussion berimbau},}\ } (\bibinfo {year} {2023})\BibitemShut
  {NoStop}%
\bibitem [{\citenamefont {Vishveshwara}(1970)}]{Vishveshwara:1970zz}%
  \BibitemOpen
  \bibfield  {author} {\bibinfo {author} {\bibfnamefont {C.~V.}\ \bibnamefont
  {Vishveshwara}},\ }\href {\doibase 10.1038/227936a0} {\bibfield  {journal}
  {\bibinfo  {journal} {Nature}\ }\textbf {\bibinfo {volume} {227}},\ \bibinfo
  {pages} {936} (\bibinfo {year} {1970})}\BibitemShut {NoStop}%
\bibitem [{\citenamefont {CERN}(2023)}]{infallingmass}%
  \BibitemOpen
  \bibfield  {author} {\bibinfo {author} {\bibnamefont {CERN}},\ }\href
  {https://cds.cern.ch/record/2198943/plots} {\enquote {\bibinfo {title}
  {Document server:testing the black hole "no-hair" hypothesis/plots},}\ }
  (\bibinfo {year} {2023})\BibitemShut {NoStop}%
\bibitem [{\citenamefont {Chandrasekhar}\ and\ \citenamefont
  {Detweiler}(1975)}]{1975RSPSA.344..441C}%
  \BibitemOpen
  \bibfield  {author} {\bibinfo {author} {\bibfnamefont {S.}~\bibnamefont
  {Chandrasekhar}}\ and\ \bibinfo {author} {\bibfnamefont {S.}~\bibnamefont
  {Detweiler}},\ }\href {\doibase 10.1098/rspa.1975.0112} {\bibfield  {journal}
  {\bibinfo  {journal} {Proceedings of the Royal Society of London Series A}\
  }\textbf {\bibinfo {volume} {344}},\ \bibinfo {pages} {441} (\bibinfo {year}
  {1975})}\BibitemShut {NoStop}%
\bibitem [{\citenamefont {Detweiler}(1980)}]{Detweiler1980Black}%
  \BibitemOpen
  \bibfield  {author} {\bibinfo {author} {\bibfnamefont {S.}~\bibnamefont
  {Detweiler}},\ }\href {\doibase 10.1086/158109} {\bibfield  {journal}
  {\bibinfo  {journal} {The Astrophysical Journal}\ }\textbf {\bibinfo {volume}
  {239}},\ \bibinfo {pages} {292+} (\bibinfo {year} {1980})}\BibitemShut
  {NoStop}%
\bibitem [{\citenamefont {{Schutz}}\ and\ \citenamefont
  {{Will}}(1985)}]{osti_6061112}%
  \BibitemOpen
  \bibfield  {author} {\bibinfo {author} {\bibfnamefont {B.~F.}\ \bibnamefont
  {{Schutz}}}\ and\ \bibinfo {author} {\bibfnamefont {C.~M.}\ \bibnamefont
  {{Will}}},\ }\href {\doibase 10.1086/184453} {\bibfield  {journal} {\bibinfo
  {journal} {Astrophys. J.}\ }\textbf {\bibinfo {volume} {291}},\ \bibinfo
  {pages} {L33} (\bibinfo {year} {1985})}\BibitemShut {NoStop}%
\bibitem [{\citenamefont {Iyer}\ and\ \citenamefont
  {Will}(1987)}]{PhysRevD.35.3621}%
  \BibitemOpen
  \bibfield  {author} {\bibinfo {author} {\bibfnamefont {S.}~\bibnamefont
  {Iyer}}\ and\ \bibinfo {author} {\bibfnamefont {C.~M.}\ \bibnamefont
  {Will}},\ }\href {\doibase 10.1103/PhysRevD.35.3621} {\bibfield  {journal}
  {\bibinfo  {journal} {Phys. Rev. D}\ }\textbf {\bibinfo {volume} {35}},\
  \bibinfo {pages} {3621} (\bibinfo {year} {1987})}\BibitemShut {NoStop}%
\bibitem [{\citenamefont {Iyer}(1987)}]{PhysRevD.35.3632}%
  \BibitemOpen
  \bibfield  {author} {\bibinfo {author} {\bibfnamefont {S.}~\bibnamefont
  {Iyer}},\ }\href {\doibase 10.1103/PhysRevD.35.3632} {\bibfield  {journal}
  {\bibinfo  {journal} {Phys. Rev. D}\ }\textbf {\bibinfo {volume} {35}},\
  \bibinfo {pages} {3632} (\bibinfo {year} {1987})}\BibitemShut {NoStop}%
\bibitem [{\citenamefont {Konoplya}(2004)}]{Konoplya:2004ip}%
  \BibitemOpen
  \bibfield  {author} {\bibinfo {author} {\bibfnamefont {R.~A.}\ \bibnamefont
  {Konoplya}},\ }\href@noop {} {\bibfield  {journal} {\bibinfo  {journal} {J.
  Phys. Stud.}\ }\textbf {\bibinfo {volume} {8}},\ \bibinfo {pages} {93}
  (\bibinfo {year} {2004})}\BibitemShut {NoStop}%
\bibitem [{\citenamefont {Kokkotas}\ and\ \citenamefont
  {Schutz}(1988{\natexlab{a}})}]{PhysRevD.37.3378}%
  \BibitemOpen
  \bibfield  {author} {\bibinfo {author} {\bibfnamefont {K.~D.}\ \bibnamefont
  {Kokkotas}}\ and\ \bibinfo {author} {\bibfnamefont {B.~F.}\ \bibnamefont
  {Schutz}},\ }\href {\doibase 10.1103/PhysRevD.37.3378} {\bibfield  {journal}
  {\bibinfo  {journal} {Phys. Rev. D}\ }\textbf {\bibinfo {volume} {37}},\
  \bibinfo {pages} {3378} (\bibinfo {year} {1988}{\natexlab{a}})}\BibitemShut
  {NoStop}%
\bibitem [{\citenamefont {Seidel}\ and\ \citenamefont
  {Iyer}(1990)}]{PhysRevD.41.374}%
  \BibitemOpen
  \bibfield  {author} {\bibinfo {author} {\bibfnamefont {E.}~\bibnamefont
  {Seidel}}\ and\ \bibinfo {author} {\bibfnamefont {S.}~\bibnamefont {Iyer}},\
  }\href {\doibase 10.1103/PhysRevD.41.374} {\bibfield  {journal} {\bibinfo
  {journal} {Phys. Rev. D}\ }\textbf {\bibinfo {volume} {41}},\ \bibinfo
  {pages} {374} (\bibinfo {year} {1990})}\BibitemShut {NoStop}%
\bibitem [{\citenamefont {Kokkotas}(1991{\natexlab{a}})}]{KDKokkotas1991}%
  \BibitemOpen
  \bibfield  {author} {\bibinfo {author} {\bibfnamefont {K.~D.}\ \bibnamefont
  {Kokkotas}},\ }\href {\doibase 10.1088/0264-9381/8/12/006} {\bibfield
  {journal} {\bibinfo  {journal} {Classical and Quantum Gravity}\ }\textbf
  {\bibinfo {volume} {8}},\ \bibinfo {pages} {2217} (\bibinfo {year}
  {1991}{\natexlab{a}})}\BibitemShut {NoStop}%
\bibitem [{\citenamefont {{Kokkotas}}(1993)}]{1993NCimB.108..991K}%
  \BibitemOpen
  \bibfield  {author} {\bibinfo {author} {\bibfnamefont {K.~D.}\ \bibnamefont
  {{Kokkotas}}},\ }\href {\doibase 10.1007/BF02822861} {\bibfield  {journal}
  {\bibinfo  {journal} {Nuovo Cimento B Serie}\ }\textbf {\bibinfo {volume}
  {108}},\ \bibinfo {pages} {991} (\bibinfo {year} {1993})}\BibitemShut
  {NoStop}%
\bibitem [{\citenamefont {{P{\"o}schl}}\ and\ \citenamefont
  {{Teller}}(1933)}]{1933ZPhy...83..143P}%
  \BibitemOpen
  \bibfield  {author} {\bibinfo {author} {\bibfnamefont {G.}~\bibnamefont
  {{P{\"o}schl}}}\ and\ \bibinfo {author} {\bibfnamefont {E.}~\bibnamefont
  {{Teller}}},\ }\href {\doibase 10.1007/BF01331132} {\bibfield  {journal}
  {\bibinfo  {journal} {Zeitschrift fur Physik}\ }\textbf {\bibinfo {volume}
  {83}},\ \bibinfo {pages} {143} (\bibinfo {year} {1933})}\BibitemShut
  {NoStop}%
\bibitem [{\citenamefont {Ferrari}\ and\ \citenamefont
  {Mashhoon}(1984)}]{PhysRevD.30.295}%
  \BibitemOpen
  \bibfield  {author} {\bibinfo {author} {\bibfnamefont {V.}~\bibnamefont
  {Ferrari}}\ and\ \bibinfo {author} {\bibfnamefont {B.}~\bibnamefont
  {Mashhoon}},\ }\href {\doibase 10.1103/PhysRevD.30.295} {\bibfield  {journal}
  {\bibinfo  {journal} {Phys. Rev. D}\ }\textbf {\bibinfo {volume} {30}},\
  \bibinfo {pages} {295} (\bibinfo {year} {1984})}\BibitemShut {NoStop}%
\bibitem [{\citenamefont {Leaver}(1985)}]{Leaver:1985ax}%
  \BibitemOpen
  \bibfield  {author} {\bibinfo {author} {\bibfnamefont {E.}~\bibnamefont
  {Leaver}},\ }\href {\doibase 10.1098/rspa.1985.0119} {\bibfield  {journal}
  {\bibinfo  {journal} {Proc.Roy.Soc.Lond.}\ }\textbf {\bibinfo {volume}
  {A402}},\ \bibinfo {pages} {285} (\bibinfo {year} {1985})}\BibitemShut
  {NoStop}%
\bibitem [{\citenamefont {Onozawa}(1997)}]{PhysRevD.55.3593}%
  \BibitemOpen
  \bibfield  {author} {\bibinfo {author} {\bibfnamefont {H.}~\bibnamefont
  {Onozawa}},\ }\href {\doibase 10.1103/PhysRevD.55.3593} {\bibfield  {journal}
  {\bibinfo  {journal} {Phys. Rev. D}\ }\textbf {\bibinfo {volume} {55}},\
  \bibinfo {pages} {3593} (\bibinfo {year} {1997})}\BibitemShut {NoStop}%
\bibitem [{\citenamefont {Berti}\ \emph {et~al.}(2003)\citenamefont {Berti},
  \citenamefont {Cardoso}, \citenamefont {Kokkotas},\ and\ \citenamefont
  {Onozawa}}]{PhysRevD.68.124018}%
  \BibitemOpen
  \bibfield  {author} {\bibinfo {author} {\bibfnamefont {E.}~\bibnamefont
  {Berti}}, \bibinfo {author} {\bibfnamefont {V.}~\bibnamefont {Cardoso}},
  \bibinfo {author} {\bibfnamefont {K.~D.}\ \bibnamefont {Kokkotas}}, \ and\
  \bibinfo {author} {\bibfnamefont {H.}~\bibnamefont {Onozawa}},\ }\href
  {\doibase 10.1103/PhysRevD.68.124018} {\bibfield  {journal} {\bibinfo
  {journal} {Phys. Rev. D}\ }\textbf {\bibinfo {volume} {68}},\ \bibinfo
  {pages} {124018} (\bibinfo {year} {2003})}\BibitemShut {NoStop}%
\bibitem [{\citenamefont {Leaver}(1990)}]{PhysRevD.41.2986}%
  \BibitemOpen
  \bibfield  {author} {\bibinfo {author} {\bibfnamefont {E.~W.}\ \bibnamefont
  {Leaver}},\ }\href {\doibase 10.1103/PhysRevD.41.2986} {\bibfield  {journal}
  {\bibinfo  {journal} {Phys. Rev. D}\ }\textbf {\bibinfo {volume} {41}},\
  \bibinfo {pages} {2986} (\bibinfo {year} {1990})}\BibitemShut {NoStop}%
\bibitem [{\citenamefont {Abbott}\ \emph {et~al.}(2016)\citenamefont {Abbott}
  \emph {et~al.}}]{Abbott:2016blz}%
  \BibitemOpen
  \bibfield  {author} {\bibinfo {author} {\bibfnamefont {B.~P.}\ \bibnamefont
  {Abbott}} \emph {et~al.} (\bibinfo {collaboration} {LIGO Scientific,
  Virgo}),\ }\href {\doibase 10.1103/PhysRevLett.116.061102} {\bibfield
  {journal} {\bibinfo  {journal} {Phys. Rev. Lett.}\ }\textbf {\bibinfo
  {volume} {116}},\ \bibinfo {pages} {061102} (\bibinfo {year} {2016})},\
  \Eprint {http://arxiv.org/abs/1602.03837} {arXiv:1602.03837 [gr-qc]}
  \BibitemShut {NoStop}%
\bibitem [{\citenamefont {Riemann}(1857)}]{Riemann1857}%
  \BibitemOpen
  \bibfield  {author} {\bibinfo {author} {\bibfnamefont {B.}~\bibnamefont
  {Riemann}},\ }\href {http://eudml.org/doc/147699} {\bibfield  {journal}
  {\bibinfo  {journal} {Journal für die reine und angewandte Mathematik}\
  }\textbf {\bibinfo {volume} {54}},\ \bibinfo {pages} {115} (\bibinfo {year}
  {1857})}\BibitemShut {NoStop}%
\bibitem [{\citenamefont {Schlesinger}(1912)}]{Schlesinger1912}%
  \BibitemOpen
  \bibfield  {author} {\bibinfo {author} {\bibfnamefont {L.}~\bibnamefont
  {Schlesinger}},\ }\href {http://eudml.org/doc/149373} {\bibfield  {journal}
  {\bibinfo  {journal} {Journal für die reine und angewandte Mathematik}\
  }\textbf {\bibinfo {volume} {141}},\ \bibinfo {pages} {96} (\bibinfo {year}
  {1912})}\BibitemShut {NoStop}%
\bibitem [{\citenamefont {Garnier}(1912)}]{Garnier}%
  \BibitemOpen
  \bibfield  {author} {\bibinfo {author} {\bibfnamefont {R.}~\bibnamefont
  {Garnier}},\ }\href {\doibase 10.24033/asens.644} {\bibfield  {journal}
  {\bibinfo  {journal} {Annales scientifiques de l'\'Ecole Normale
  Sup\'erieure}\ }\textbf {\bibinfo {volume} {3e s{\'e}rie, 29}},\ \bibinfo
  {pages} {1} (\bibinfo {year} {1912})}\BibitemShut {NoStop}%
\bibitem [{\citenamefont {Ablowitz}\ and\ \citenamefont
  {Segur}(1981)}]{doi:10.1137/1.9781611970883}%
  \BibitemOpen
  \bibfield  {author} {\bibinfo {author} {\bibfnamefont {M.~J.}\ \bibnamefont
  {Ablowitz}}\ and\ \bibinfo {author} {\bibfnamefont {H.}~\bibnamefont
  {Segur}},\ }\href {\doibase 10.1137/1.9781611970883} {\emph {\bibinfo {title}
  {Solitons and the Inverse Scattering Transform}}}\ (\bibinfo  {publisher}
  {Society for Industrial and Applied Mathematics},\ \bibinfo {year} {1981})\
  \Eprint
  {http://arxiv.org/abs/https://epubs.siam.org/doi/pdf/10.1137/1.9781611970883}
  {https://epubs.siam.org/doi/pdf/10.1137/1.9781611970883} \BibitemShut
  {NoStop}%
\bibitem [{\citenamefont {Its}\ and\ \citenamefont
  {Novokshenov}(2006)}]{its2006isomonodromic}%
  \BibitemOpen
  \bibfield  {author} {\bibinfo {author} {\bibfnamefont {A.}~\bibnamefont
  {Its}}\ and\ \bibinfo {author} {\bibfnamefont {V.}~\bibnamefont
  {Novokshenov}},\ }\href {https://books.google.com.br/books?id=beh7CwAAQBAJ}
  {\emph {\bibinfo {title} {The Isomonodromic Deformation Method in the Theory
  of Painleve Equations}}},\ Lecture Notes in Mathematics\ (\bibinfo
  {publisher} {Springer Berlin Heidelberg},\ \bibinfo {year}
  {2006})\BibitemShut {NoStop}%
\bibitem [{\citenamefont {Jimbo}\ and\ \citenamefont
  {Miwa}(1981{\natexlab{a}})}]{JIMBO198126}%
  \BibitemOpen
  \bibfield  {author} {\bibinfo {author} {\bibfnamefont {M.}~\bibnamefont
  {Jimbo}}\ and\ \bibinfo {author} {\bibfnamefont {T.}~\bibnamefont {Miwa}},\
  }\href {\doibase https://doi.org/10.1016/0167-2789(81)90003-8} {\bibfield
  {journal} {\bibinfo  {journal} {Physica D: Nonlinear Phenomena}\ }\textbf
  {\bibinfo {volume} {4}},\ \bibinfo {pages} {26 } (\bibinfo {year}
  {1981}{\natexlab{a}})}\BibitemShut {NoStop}%
\bibitem [{\citenamefont {Jimbo}\ \emph
  {et~al.}(1981{\natexlab{a}})\citenamefont {Jimbo}, \citenamefont {Miwa},\
  and\ \citenamefont {Ueno}}]{JIMBO1981306}%
  \BibitemOpen
  \bibfield  {author} {\bibinfo {author} {\bibfnamefont {M.}~\bibnamefont
  {Jimbo}}, \bibinfo {author} {\bibfnamefont {T.}~\bibnamefont {Miwa}}, \ and\
  \bibinfo {author} {\bibfnamefont {K.}~\bibnamefont {Ueno}},\ }\href {\doibase
  https://doi.org/10.1016/0167-2789(81)90013-0} {\bibfield  {journal} {\bibinfo
   {journal} {Physica D: Nonlinear Phenomena}\ }\textbf {\bibinfo {volume}
  {2}},\ \bibinfo {pages} {306 } (\bibinfo {year}
  {1981}{\natexlab{a}})}\BibitemShut {NoStop}%
\bibitem [{\citenamefont {Jimbo}\ and\ \citenamefont
  {Miwa}(1981{\natexlab{b}})}]{JIMBO1981407}%
  \BibitemOpen
  \bibfield  {author} {\bibinfo {author} {\bibfnamefont {M.}~\bibnamefont
  {Jimbo}}\ and\ \bibinfo {author} {\bibfnamefont {T.}~\bibnamefont {Miwa}},\
  }\href {\doibase https://doi.org/10.1016/0167-2789(81)90021-X} {\bibfield
  {journal} {\bibinfo  {journal} {Physica D: Nonlinear Phenomena}\ }\textbf
  {\bibinfo {volume} {2}},\ \bibinfo {pages} {407 } (\bibinfo {year}
  {1981}{\natexlab{b}})}\BibitemShut {NoStop}%
\bibitem [{\citenamefont {Motl}\ and\ \citenamefont
  {Neitzke}(2003)}]{Motl:2003cd}%
  \BibitemOpen
  \bibfield  {author} {\bibinfo {author} {\bibfnamefont {L.}~\bibnamefont
  {Motl}}\ and\ \bibinfo {author} {\bibfnamefont {A.}~\bibnamefont {Neitzke}},\
  }\href {\doibase 10.4310/ATMP.2003.v7.n2.a4} {\bibfield  {journal} {\bibinfo
  {journal} {Adv. Theor. Math. Phys.}\ }\textbf {\bibinfo {volume} {7}},\
  \bibinfo {pages} {307} (\bibinfo {year} {2003})},\ \Eprint
  {http://arxiv.org/abs/hep-th/0301173} {arXiv:hep-th/0301173} \BibitemShut
  {NoStop}%
\bibitem [{\citenamefont {Neitzke}(2003)}]{Neitzke:2003mz}%
  \BibitemOpen
  \bibfield  {author} {\bibinfo {author} {\bibfnamefont {A.}~\bibnamefont
  {Neitzke}},\ }\href {https://api.semanticscholar.org/CorpusID:15786058}
  {\bibfield  {journal} {\bibinfo  {journal} {arXiv: High Energy Physics -
  Theory}\ } (\bibinfo {year} {2003})}\BibitemShut {NoStop}%
\bibitem [{\citenamefont {Castro}\ \emph
  {et~al.}(2013{\natexlab{a}})\citenamefont {Castro}, \citenamefont {Lapan},
  \citenamefont {Maloney},\ and\ \citenamefont
  {Rodriguez}}]{PhysRevD.88.044003}%
  \BibitemOpen
  \bibfield  {author} {\bibinfo {author} {\bibfnamefont {A.}~\bibnamefont
  {Castro}}, \bibinfo {author} {\bibfnamefont {J.~M.}\ \bibnamefont {Lapan}},
  \bibinfo {author} {\bibfnamefont {A.}~\bibnamefont {Maloney}}, \ and\
  \bibinfo {author} {\bibfnamefont {M.~J.}\ \bibnamefont {Rodriguez}},\ }\href
  {\doibase 10.1103/PhysRevD.88.044003} {\bibfield  {journal} {\bibinfo
  {journal} {Phys. Rev. D}\ }\textbf {\bibinfo {volume} {88}},\ \bibinfo
  {pages} {044003} (\bibinfo {year} {2013}{\natexlab{a}})}\BibitemShut
  {NoStop}%
\bibitem [{\citenamefont {Castro}\ \emph
  {et~al.}(2013{\natexlab{b}})\citenamefont {Castro}, \citenamefont {Lapan},
  \citenamefont {Maloney},\ and\ \citenamefont {Rodriguez}}]{Castro2013b}%
  \BibitemOpen
  \bibfield  {author} {\bibinfo {author} {\bibfnamefont {A.}~\bibnamefont
  {Castro}}, \bibinfo {author} {\bibfnamefont {J.~M.}\ \bibnamefont {Lapan}},
  \bibinfo {author} {\bibfnamefont {A.}~\bibnamefont {Maloney}}, \ and\
  \bibinfo {author} {\bibfnamefont {M.~J.}\ \bibnamefont {Rodriguez}},\ }\href
  {\doibase 10.1088/0264-9381/30/16/165005} {\bibfield  {journal} {\bibinfo
  {journal} {Class.Quant.Grav.}\ }\textbf {\bibinfo {volume} {30}},\ \bibinfo
  {pages} {165005} (\bibinfo {year} {2013}{\natexlab{b}})},\ \Eprint
  {http://arxiv.org/abs/1304.3781} {arXiv:1304.3781 [hep-th]} \BibitemShut
  {NoStop}%
\bibitem [{\citenamefont {Novaes}\ and\ \citenamefont {Carneiro~da
  Cunha}(2014)}]{Novaes:2014lha}%
  \BibitemOpen
  \bibfield  {author} {\bibinfo {author} {\bibfnamefont {F.}~\bibnamefont
  {Novaes}}\ and\ \bibinfo {author} {\bibfnamefont {B.}~\bibnamefont
  {Carneiro~da Cunha}},\ }\href {\doibase 10.1007/JHEP07(2014)132} {\bibfield
  {journal} {\bibinfo  {journal} {JHEP}\ }\textbf {\bibinfo {volume} {07}},\
  \bibinfo {pages} {132} (\bibinfo {year} {2014})},\ \Eprint
  {http://arxiv.org/abs/1404.5188} {arXiv:1404.5188 [hep-th]} \BibitemShut
  {NoStop}%
\bibitem [{\citenamefont {Carneiro~da Cunha}\ and\ \citenamefont
  {Novaes}(2015{\natexlab{a}})}]{daCunha:2015ana}%
  \BibitemOpen
  \bibfield  {author} {\bibinfo {author} {\bibfnamefont {B.}~\bibnamefont
  {Carneiro~da Cunha}}\ and\ \bibinfo {author} {\bibfnamefont {F.}~\bibnamefont
  {Novaes}},\ }\href {\doibase 10.1007/JHEP11(2015)144} {\bibfield  {journal}
  {\bibinfo  {journal} {JHEP}\ }\textbf {\bibinfo {volume} {11}},\ \bibinfo
  {pages} {144} (\bibinfo {year} {2015}{\natexlab{a}})},\ \Eprint
  {http://arxiv.org/abs/1506.06588} {arXiv:1506.06588 [hep-th]} \BibitemShut
  {NoStop}%
\bibitem [{\citenamefont {Novaes}\ \emph {et~al.}(2019)\citenamefont {Novaes},
  \citenamefont {Marinho}, \citenamefont {Lencs\'es},\ and\ \citenamefont
  {Casals}}]{Novaes:2018fry}%
  \BibitemOpen
  \bibfield  {author} {\bibinfo {author} {\bibfnamefont {F.}~\bibnamefont
  {Novaes}}, \bibinfo {author} {\bibfnamefont {C.}~\bibnamefont {Marinho}},
  \bibinfo {author} {\bibfnamefont {M.}~\bibnamefont {Lencs\'es}}, \ and\
  \bibinfo {author} {\bibfnamefont {M.}~\bibnamefont {Casals}},\ }\href
  {\doibase 10.1007/JHEP05(2019)033} {\bibfield  {journal} {\bibinfo  {journal}
  {JHEP}\ }\textbf {\bibinfo {volume} {05}},\ \bibinfo {pages} {033} (\bibinfo
  {year} {2019})},\ \Eprint {http://arxiv.org/abs/1811.11912} {arXiv:1811.11912
  [gr-qc]} \BibitemShut {NoStop}%
\bibitem [{\citenamefont {Barrag\'an~Amado}\ \emph {et~al.}(2019)\citenamefont
  {Barrag\'an~Amado}, \citenamefont {Carneiro Da~Cunha},\ and\ \citenamefont
  {Pallante}}]{Barragan-Amado:2018pxh}%
  \BibitemOpen
  \bibfield  {author} {\bibinfo {author} {\bibfnamefont {J.}~\bibnamefont
  {Barrag\'an~Amado}}, \bibinfo {author} {\bibfnamefont {B.}~\bibnamefont
  {Carneiro Da~Cunha}}, \ and\ \bibinfo {author} {\bibfnamefont
  {E.}~\bibnamefont {Pallante}},\ }\href {\doibase 10.1103/PhysRevD.99.105006}
  {\bibfield  {journal} {\bibinfo  {journal} {Phys. Rev. D}\ }\textbf {\bibinfo
  {volume} {99}},\ \bibinfo {pages} {105006} (\bibinfo {year} {2019})},\
  \Eprint {http://arxiv.org/abs/1812.08921} {arXiv:1812.08921 [hep-th]}
  \BibitemShut {NoStop}%
\bibitem [{\citenamefont {Amado}\ \emph {et~al.}(2020)\citenamefont {Amado},
  \citenamefont {Carneiro~da Cunha},\ and\ \citenamefont
  {Pallante}}]{Amado:2020zsr}%
  \BibitemOpen
  \bibfield  {author} {\bibinfo {author} {\bibfnamefont {J.~B.}\ \bibnamefont
  {Amado}}, \bibinfo {author} {\bibfnamefont {B.}~\bibnamefont {Carneiro~da
  Cunha}}, \ and\ \bibinfo {author} {\bibfnamefont {E.}~\bibnamefont
  {Pallante}},\ }\href {\doibase 10.1007/JHEP04(2020)155} {\bibfield  {journal}
  {\bibinfo  {journal} {JHEP}\ }\textbf {\bibinfo {volume} {04}},\ \bibinfo
  {pages} {155} (\bibinfo {year} {2020})},\ \Eprint
  {http://arxiv.org/abs/2002.06108} {arXiv:2002.06108 [hep-th]} \BibitemShut
  {NoStop}%
\bibitem [{\citenamefont {Carneiro~da Cunha}\ and\ \citenamefont
  {Cavalcante}(2020)}]{CarneirodaCunha:2019tia}%
  \BibitemOpen
  \bibfield  {author} {\bibinfo {author} {\bibfnamefont {B.}~\bibnamefont
  {Carneiro~da Cunha}}\ and\ \bibinfo {author} {\bibfnamefont {J.~P.}\
  \bibnamefont {Cavalcante}},\ }\href {\doibase 10.1103/PhysRevD.102.105013}
  {\bibfield  {journal} {\bibinfo  {journal} {Phys. Rev. D}\ }\textbf {\bibinfo
  {volume} {102}},\ \bibinfo {pages} {105013} (\bibinfo {year} {2020})},\
  \Eprint {http://arxiv.org/abs/1906.10638} {arXiv:1906.10638 [hep-th]}
  \BibitemShut {NoStop}%
\bibitem [{\citenamefont {Cavalcante}\ and\ \citenamefont
  {da~Cunha}(2021)}]{Cavalcante:2021scq}%
  \BibitemOpen
  \bibfield  {author} {\bibinfo {author} {\bibfnamefont {J.~P.}\ \bibnamefont
  {Cavalcante}}\ and\ \bibinfo {author} {\bibfnamefont {B.~C.}\ \bibnamefont
  {da~Cunha}},\ }\href {\doibase 10.1103/PhysRevD.104.124040} {\bibfield
  {journal} {\bibinfo  {journal} {Phys. Rev. D}\ }\textbf {\bibinfo {volume}
  {104}},\ \bibinfo {pages} {124040} (\bibinfo {year} {2021})},\ \Eprint
  {http://arxiv.org/abs/2109.06929} {arXiv:2109.06929 [gr-qc]} \BibitemShut
  {NoStop}%
\bibitem [{\citenamefont {Gamayun}\ \emph {et~al.}(2013)\citenamefont
  {Gamayun}, \citenamefont {Iorgov},\ and\ \citenamefont
  {Lisovyy}}]{Gamayun:2013auu}%
  \BibitemOpen
  \bibfield  {author} {\bibinfo {author} {\bibfnamefont {O.}~\bibnamefont
  {Gamayun}}, \bibinfo {author} {\bibfnamefont {N.}~\bibnamefont {Iorgov}}, \
  and\ \bibinfo {author} {\bibfnamefont {O.}~\bibnamefont {Lisovyy}},\ }\href
  {\doibase 10.1088/1751-8113/46/33/335203} {\bibfield  {journal} {\bibinfo
  {journal} {J. Phys.}\ }\textbf {\bibinfo {volume} {A46}},\ \bibinfo {pages}
  {335203} (\bibinfo {year} {2013})},\ \Eprint {http://arxiv.org/abs/1302.1832}
  {arXiv:1302.1832 [hep-th]} \BibitemShut {NoStop}%
\bibitem [{\citenamefont {Lisovyy}\ \emph {et~al.}(2018)\citenamefont
  {Lisovyy}, \citenamefont {Nagoya},\ and\ \citenamefont
  {Roussillon}}]{Lisovyy:2018mnj}%
  \BibitemOpen
  \bibfield  {author} {\bibinfo {author} {\bibfnamefont {O.}~\bibnamefont
  {Lisovyy}}, \bibinfo {author} {\bibfnamefont {H.}~\bibnamefont {Nagoya}}, \
  and\ \bibinfo {author} {\bibfnamefont {J.}~\bibnamefont {Roussillon}},\
  }\href {\doibase 10.1063/1.5031841} {\bibfield  {journal} {\bibinfo
  {journal} {J. Math. Phys.}\ }\textbf {\bibinfo {volume} {59}},\ \bibinfo
  {pages} {091409} (\bibinfo {year} {2018})},\ \Eprint
  {http://arxiv.org/abs/1806.08344} {arXiv:1806.08344 [math-ph]} \BibitemShut
  {NoStop}%
\bibitem [{\citenamefont {Carneiro~da Cunha}\ and\ \citenamefont
  {Novaes}(2015{\natexlab{b}})}]{CarneirodaCunha:2015hzd}%
  \BibitemOpen
  \bibfield  {author} {\bibinfo {author} {\bibfnamefont {B.}~\bibnamefont
  {Carneiro~da Cunha}}\ and\ \bibinfo {author} {\bibfnamefont {F.}~\bibnamefont
  {Novaes}},\ }\href {\doibase 10.1007/JHEP11(2015)144} {\bibfield  {journal}
  {\bibinfo  {journal} {JHEP}\ }\textbf {\bibinfo {volume} {11}},\ \bibinfo
  {pages} {144} (\bibinfo {year} {2015}{\natexlab{b}})},\ \Eprint
  {http://arxiv.org/abs/1506.06588} {arXiv:1506.06588 [hep-th]} \BibitemShut
  {NoStop}%
\bibitem [{\citenamefont {da~Cunha}\ and\ \citenamefont
  {Cavalcante}(2021)}]{daCunha:2021jkm}%
  \BibitemOpen
  \bibfield  {author} {\bibinfo {author} {\bibfnamefont {B.~C.}\ \bibnamefont
  {da~Cunha}}\ and\ \bibinfo {author} {\bibfnamefont {J.~P.}\ \bibnamefont
  {Cavalcante}},\ }\href {\doibase 10.1103/PhysRevD.104.084051} {\bibfield
  {journal} {\bibinfo  {journal} {Phys. Rev. D}\ }\textbf {\bibinfo {volume}
  {104}},\ \bibinfo {pages} {084051} (\bibinfo {year} {2021})},\ \Eprint
  {http://arxiv.org/abs/2105.08790} {arXiv:2105.08790 [hep-th]} \BibitemShut
  {NoStop}%
\bibitem [{\citenamefont {da~Cunha}\ and\ \citenamefont
  {Cavalcante}(2022)}]{daCunha:2022ewy}%
  \BibitemOpen
  \bibfield  {author} {\bibinfo {author} {\bibfnamefont {B.~C.}\ \bibnamefont
  {da~Cunha}}\ and\ \bibinfo {author} {\bibfnamefont {J.~a.~P.}\ \bibnamefont
  {Cavalcante}},\ }\href@noop {} {\bibfield  {journal} {\bibinfo  {journal}
  {(submitted-JHEP)}\ } (\bibinfo {year} {2022})},\ \Eprint
  {http://arxiv.org/abs/2211.03551} {arXiv 2211.03551:2211.03551 [hep-th]}
  \BibitemShut {NoStop}%
\bibitem [{\citenamefont {Regge}\ and\ \citenamefont
  {Wheeler}(1957)}]{PhysRev.108.1063}%
  \BibitemOpen
  \bibfield  {author} {\bibinfo {author} {\bibfnamefont {T.}~\bibnamefont
  {Regge}}\ and\ \bibinfo {author} {\bibfnamefont {J.~A.}\ \bibnamefont
  {Wheeler}},\ }\href {\doibase 10.1103/PhysRev.108.1063} {\bibfield  {journal}
  {\bibinfo  {journal} {Phys. Rev.}\ }\textbf {\bibinfo {volume} {108}},\
  \bibinfo {pages} {1063} (\bibinfo {year} {1957})}\BibitemShut {NoStop}%
\bibitem [{\citenamefont {Zerilli}(1970{\natexlab{a}})}]{PhysRevLett.24.737}%
  \BibitemOpen
  \bibfield  {author} {\bibinfo {author} {\bibfnamefont {F.~J.}\ \bibnamefont
  {Zerilli}},\ }\href {\doibase 10.1103/PhysRevLett.24.737} {\bibfield
  {journal} {\bibinfo  {journal} {Phys. Rev. Lett.}\ }\textbf {\bibinfo
  {volume} {24}},\ \bibinfo {pages} {737} (\bibinfo {year}
  {1970}{\natexlab{a}})}\BibitemShut {NoStop}%
\bibitem [{\citenamefont {Zerilli}(1970{\natexlab{b}})}]{PhysRevD.2.2141}%
  \BibitemOpen
  \bibfield  {author} {\bibinfo {author} {\bibfnamefont {F.~J.}\ \bibnamefont
  {Zerilli}},\ }\href {\doibase 10.1103/PhysRevD.2.2141} {\bibfield  {journal}
  {\bibinfo  {journal} {Phys. Rev. D}\ }\textbf {\bibinfo {volume} {2}},\
  \bibinfo {pages} {2141} (\bibinfo {year} {1970}{\natexlab{b}})}\BibitemShut
  {NoStop}%
\bibitem [{\citenamefont {Teukolsky}(2015)}]{Teukolsky:2014vca}%
  \BibitemOpen
  \bibfield  {author} {\bibinfo {author} {\bibfnamefont {S.~A.}\ \bibnamefont
  {Teukolsky}},\ }\href {\doibase 10.1088/0264-9381/32/12/124006} {\bibfield
  {journal} {\bibinfo  {journal} {Class. Quant. Grav.}\ }\textbf {\bibinfo
  {volume} {32}},\ \bibinfo {pages} {124006} (\bibinfo {year} {2015})},\
  \Eprint {http://arxiv.org/abs/1410.2130} {arXiv:1410.2130 [gr-qc]}
  \BibitemShut {NoStop}%
\bibitem [{\citenamefont {Brill}\ \emph {et~al.}(1972)\citenamefont {Brill},
  \citenamefont {Chrzanowski}, \citenamefont {Pereira}, \citenamefont
  {Fackerell},\ and\ \citenamefont {Ipser}}]{PhysRevD.5.1913}%
  \BibitemOpen
  \bibfield  {author} {\bibinfo {author} {\bibfnamefont {D.~R.}\ \bibnamefont
  {Brill}}, \bibinfo {author} {\bibfnamefont {P.~L.}\ \bibnamefont
  {Chrzanowski}}, \bibinfo {author} {\bibfnamefont {C.~M.}\ \bibnamefont
  {Pereira}}, \bibinfo {author} {\bibfnamefont {E.~D.}\ \bibnamefont
  {Fackerell}}, \ and\ \bibinfo {author} {\bibfnamefont {J.~R.}\ \bibnamefont
  {Ipser}},\ }\href {\doibase 10.1103/PhysRevD.5.1913} {\bibfield  {journal}
  {\bibinfo  {journal} {Phys. Rev. D}\ }\textbf {\bibinfo {volume} {5}},\
  \bibinfo {pages} {1913} (\bibinfo {year} {1972})}\BibitemShut {NoStop}%
\bibitem [{\citenamefont {Fackerell}\ and\ \citenamefont
  {Ipser}(1972)}]{PhysRevD.5.2455}%
  \BibitemOpen
  \bibfield  {author} {\bibinfo {author} {\bibfnamefont {E.~D.}\ \bibnamefont
  {Fackerell}}\ and\ \bibinfo {author} {\bibfnamefont {J.~R.}\ \bibnamefont
  {Ipser}},\ }\href {\doibase 10.1103/PhysRevD.5.2455} {\bibfield  {journal}
  {\bibinfo  {journal} {Phys. Rev. D}\ }\textbf {\bibinfo {volume} {5}},\
  \bibinfo {pages} {2455} (\bibinfo {year} {1972})}\BibitemShut {NoStop}%
\bibitem [{\citenamefont {Newman}\ and\ \citenamefont
  {Penrose}(1962)}]{doi:10.1063/1.1724257}%
  \BibitemOpen
  \bibfield  {author} {\bibinfo {author} {\bibfnamefont {E.}~\bibnamefont
  {Newman}}\ and\ \bibinfo {author} {\bibfnamefont {R.}~\bibnamefont
  {Penrose}},\ }\href {\doibase 10.1063/1.1724257} {\bibfield  {journal}
  {\bibinfo  {journal} {Journal of Mathematical Physics}\ }\textbf {\bibinfo
  {volume} {3}},\ \bibinfo {pages} {566} (\bibinfo {year} {1962})},\ \Eprint
  {http://arxiv.org/abs/https://doi.org/10.1063/1.1724257}
  {https://doi.org/10.1063/1.1724257} \BibitemShut {NoStop}%
\bibitem [{\citenamefont {Teukolsky}(1972)}]{PhysRevLett.29.1114}%
  \BibitemOpen
  \bibfield  {author} {\bibinfo {author} {\bibfnamefont {S.~A.}\ \bibnamefont
  {Teukolsky}},\ }\href {\doibase 10.1103/PhysRevLett.29.1114} {\bibfield
  {journal} {\bibinfo  {journal} {Phys. Rev. Lett.}\ }\textbf {\bibinfo
  {volume} {29}},\ \bibinfo {pages} {1114} (\bibinfo {year}
  {1972})}\BibitemShut {NoStop}%
\bibitem [{\citenamefont {Teukolsky}(1973)}]{Teukolsky:1973ha}%
  \BibitemOpen
  \bibfield  {author} {\bibinfo {author} {\bibfnamefont {S.~A.}\ \bibnamefont
  {Teukolsky}},\ }\href {\doibase 10.1086/152444} {\bibfield  {journal}
  {\bibinfo  {journal} {Astrophys. J.}\ }\textbf {\bibinfo {volume} {185}},\
  \bibinfo {pages} {635} (\bibinfo {year} {1973})}\BibitemShut {NoStop}%
\bibitem [{\citenamefont
  {Chandrasekhar}(1976{\natexlab{c}})}]{doi:10.1098/rspa.1976.0090}%
  \BibitemOpen
  \bibfield  {author} {\bibinfo {author} {\bibfnamefont {S.}~\bibnamefont
  {Chandrasekhar}},\ }\href {\doibase 10.1098/rspa.1976.0090} {\bibfield
  {journal} {\bibinfo  {journal} {Proceedings of the Royal Society of London.
  A. Mathematical and Physical Sciences}\ }\textbf {\bibinfo {volume} {349}},\
  \bibinfo {pages} {571} (\bibinfo {year} {1976}{\natexlab{c}})},\ \Eprint
  {http://arxiv.org/abs/https://royalsocietypublishing.org/doi/pdf/10.1098/rspa.1976.0090}
  {https://royalsocietypublishing.org/doi/pdf/10.1098/rspa.1976.0090}
  \BibitemShut {NoStop}%
\bibitem [{\citenamefont
  {Chandrasekhar}(1998)}]{chandrasekhar1998mathematical}%
  \BibitemOpen
  \bibfield  {author} {\bibinfo {author} {\bibfnamefont {S.}~\bibnamefont
  {Chandrasekhar}},\ }\href {https://books.google.com.br/books?id=LBOVcrzFfhsC}
  {\emph {\bibinfo {title} {The Mathematical Theory of Black Holes}}},\ Oxford
  classic texts in the physical sciences\ (\bibinfo  {publisher} {Clarendon
  Press},\ \bibinfo {year} {1998})\BibitemShut {NoStop}%
\bibitem [{\citenamefont {Zerilli}(1974)}]{PhysRevD.9.860}%
  \BibitemOpen
  \bibfield  {author} {\bibinfo {author} {\bibfnamefont {F.~J.}\ \bibnamefont
  {Zerilli}},\ }\href {\doibase 10.1103/PhysRevD.9.860} {\bibfield  {journal}
  {\bibinfo  {journal} {Phys. Rev. D}\ }\textbf {\bibinfo {volume} {9}},\
  \bibinfo {pages} {860} (\bibinfo {year} {1974})}\BibitemShut {NoStop}%
\bibitem [{\citenamefont {Moncrief}(1974{\natexlab{a}})}]{PhysRevD.10.1057}%
  \BibitemOpen
  \bibfield  {author} {\bibinfo {author} {\bibfnamefont {V.}~\bibnamefont
  {Moncrief}},\ }\href {\doibase 10.1103/PhysRevD.10.1057} {\bibfield
  {journal} {\bibinfo  {journal} {Phys. Rev. D}\ }\textbf {\bibinfo {volume}
  {10}},\ \bibinfo {pages} {1057} (\bibinfo {year}
  {1974}{\natexlab{a}})}\BibitemShut {NoStop}%
\bibitem [{\citenamefont {Moncrief}(1974{\natexlab{b}})}]{PhysRevD.9.2707}%
  \BibitemOpen
  \bibfield  {author} {\bibinfo {author} {\bibfnamefont {V.}~\bibnamefont
  {Moncrief}},\ }\href {\doibase 10.1103/PhysRevD.9.2707} {\bibfield  {journal}
  {\bibinfo  {journal} {Phys. Rev. D}\ }\textbf {\bibinfo {volume} {9}},\
  \bibinfo {pages} {2707} (\bibinfo {year} {1974}{\natexlab{b}})}\BibitemShut
  {NoStop}%
\bibitem [{\citenamefont {Moncrief}(1975)}]{PhysRevD.12.1526}%
  \BibitemOpen
  \bibfield  {author} {\bibinfo {author} {\bibfnamefont {V.}~\bibnamefont
  {Moncrief}},\ }\href {\doibase 10.1103/PhysRevD.12.1526} {\bibfield
  {journal} {\bibinfo  {journal} {Phys. Rev. D}\ }\textbf {\bibinfo {volume}
  {12}},\ \bibinfo {pages} {1526} (\bibinfo {year} {1975})}\BibitemShut
  {NoStop}%
\bibitem [{\citenamefont {Chandrasekhar}(1979)}]{doi:10.1098/rspa.1979.0028}%
  \BibitemOpen
  \bibfield  {author} {\bibinfo {author} {\bibfnamefont {S.}~\bibnamefont
  {Chandrasekhar}},\ }\href {\doibase 10.1098/rspa.1979.0028} {\bibfield
  {journal} {\bibinfo  {journal} {Proceedings of the Royal Society of London.
  A. Mathematical and Physical Sciences}\ }\textbf {\bibinfo {volume} {365}},\
  \bibinfo {pages} {453} (\bibinfo {year} {1979})},\ \Eprint
  {http://arxiv.org/abs/https://royalsocietypublishing.org/doi/pdf/10.1098/rspa.1979.0028}
  {https://royalsocietypublishing.org/doi/pdf/10.1098/rspa.1979.0028}
  \BibitemShut {NoStop}%
\bibitem [{\citenamefont {Stephani}\ \emph {et~al.}(2003)\citenamefont
  {Stephani}, \citenamefont {Kramer}, \citenamefont {MacCallum}, \citenamefont
  {Hoenselaers},\ and\ \citenamefont
  {Herlt}}]{stephani_kramer_maccallum_hoenselaers_herlt_2003}%
  \BibitemOpen
  \bibfield  {author} {\bibinfo {author} {\bibfnamefont {H.}~\bibnamefont
  {Stephani}}, \bibinfo {author} {\bibfnamefont {D.}~\bibnamefont {Kramer}},
  \bibinfo {author} {\bibfnamefont {M.}~\bibnamefont {MacCallum}}, \bibinfo
  {author} {\bibfnamefont {C.}~\bibnamefont {Hoenselaers}}, \ and\ \bibinfo
  {author} {\bibfnamefont {E.}~\bibnamefont {Herlt}},\ }\href {\doibase
  10.1017/CBO9780511535185} {\emph {\bibinfo {title} {Exact Solutions of
  Einstein's Field Equations}}},\ \bibinfo {edition} {2nd}\ ed.,\ Cambridge
  Monographs on Mathematical Physics\ (\bibinfo  {publisher} {Cambridge
  University Press},\ \bibinfo {year} {2003})\BibitemShut {NoStop}%
\bibitem [{\citenamefont {Frolov}\ and\ \citenamefont
  {Novikov}(1998)}]{frolov1998black}%
  \BibitemOpen
  \bibfield  {author} {\bibinfo {author} {\bibfnamefont {V.}~\bibnamefont
  {Frolov}}\ and\ \bibinfo {author} {\bibfnamefont {I.}~\bibnamefont
  {Novikov}},\ }\href {https://books.google.com.br/books?id=n0kHI6CVWZUC}
  {\emph {\bibinfo {title} {Black Hole Physics: Basic Concepts and New
  Developments}}},\ Fundamental Theories of Physics\ (\bibinfo  {publisher}
  {Springer Netherlands},\ \bibinfo {year} {1998})\BibitemShut {NoStop}%
\bibitem [{\citenamefont {Schwarzschild}(1916)}]{Schwarzschild:1916uq}%
  \BibitemOpen
  \bibfield  {author} {\bibinfo {author} {\bibfnamefont {K.}~\bibnamefont
  {Schwarzschild}},\ }\href@noop {} {\bibfield  {journal} {\bibinfo  {journal}
  {Sitzungsber. Preuss. Akad. Wiss. Berlin (Math. Phys. )}\ }\textbf {\bibinfo
  {volume} {1916}},\ \bibinfo {pages} {189} (\bibinfo {year} {1916})},\ \Eprint
  {http://arxiv.org/abs/physics/9905030} {arXiv:physics/9905030} \BibitemShut
  {NoStop}%
\bibitem [{\citenamefont {Wald}(1974)}]{WALD1974548}%
  \BibitemOpen
  \bibfield  {author} {\bibinfo {author} {\bibfnamefont {R.}~\bibnamefont
  {Wald}},\ }\href {\doibase https://doi.org/10.1016/0003-4916(74)90125-0}
  {\bibfield  {journal} {\bibinfo  {journal} {Annals of Physics}\ }\textbf
  {\bibinfo {volume} {82}},\ \bibinfo {pages} {548} (\bibinfo {year}
  {1974})}\BibitemShut {NoStop}%
\bibitem [{\citenamefont {Kinnersley}(1968)}]{Kinnersley:1968zz}%
  \BibitemOpen
  \bibfield  {author} {\bibinfo {author} {\bibfnamefont {W.~M.}\ \bibnamefont
  {Kinnersley}},\ }\emph {\bibinfo {title} {{Type D gravitational fields}}},\
  \href@noop {} {Ph.D. thesis},\ \bibinfo  {school} {Caltech} (\bibinfo {year}
  {1968})\BibitemShut {NoStop}%
\bibitem [{\citenamefont {Wald}(1973)}]{doi:10.1063/1.1666203}%
  \BibitemOpen
  \bibfield  {author} {\bibinfo {author} {\bibfnamefont {R.~M.}\ \bibnamefont
  {Wald}},\ }\href {\doibase 10.1063/1.1666203} {\bibfield  {journal} {\bibinfo
   {journal} {Journal of Mathematical Physics}\ }\textbf {\bibinfo {volume}
  {14}},\ \bibinfo {pages} {1453} (\bibinfo {year} {1973})},\ \Eprint
  {http://arxiv.org/abs/https://doi.org/10.1063/1.1666203}
  {https://doi.org/10.1063/1.1666203} \BibitemShut {NoStop}%
\bibitem [{\citenamefont {O'Toole}\ \emph {et~al.}(2022)\citenamefont
  {O'Toole}, \citenamefont {Macedo}, \citenamefont {Stratton},\ and\
  \citenamefont {Wardell}}]{BHPToolkit}%
  \BibitemOpen
  \bibfield  {author} {\bibinfo {author} {\bibfnamefont {C.}~\bibnamefont
  {O'Toole}}, \bibinfo {author} {\bibfnamefont {R.}~\bibnamefont {Macedo}},
  \bibinfo {author} {\bibfnamefont {T.}~\bibnamefont {Stratton}}, \ and\
  \bibinfo {author} {\bibfnamefont {B.}~\bibnamefont {Wardell}},\ }\href@noop
  {} {\enquote {\bibinfo {title} {{Black Hole Perturbation Toolkit}},}\
  }\bibinfo {howpublished} {(\href{http://bhptoolkit.org/}{bhptoolkit.org})}
  (\bibinfo {year} {2022})\BibitemShut {NoStop}%
\bibitem [{\citenamefont {Teukolsky}(1974)}]{Teukolsky1974PerturbationsOA}%
  \BibitemOpen
  \bibfield  {author} {\bibinfo {author} {\bibfnamefont {S.~A.}\ \bibnamefont
  {Teukolsky}},\ }in\ \href@noop {} {\emph {\bibinfo {booktitle} {Astrophysical
  Journal}}},\ Vol.\ \bibinfo {volume} {185}\ (\bibinfo {year} {1974})\ pp.\
  \bibinfo {pages} {635--648}\BibitemShut {NoStop}%
\bibitem [{\citenamefont {{Press}}\ and\ \citenamefont
  {{Teukolsky}}(1973)}]{1973ApJ...185..649P}%
  \BibitemOpen
  \bibfield  {author} {\bibinfo {author} {\bibfnamefont {W.~H.}\ \bibnamefont
  {{Press}}}\ and\ \bibinfo {author} {\bibfnamefont {S.~A.}\ \bibnamefont
  {{Teukolsky}}},\ }\href {\doibase 10.1086/152445} {\bibfield  {journal}
  {\bibinfo  {journal} {Astrophysical Journal}\ }\textbf {\bibinfo {volume}
  {185}},\ \bibinfo {pages} {649} (\bibinfo {year} {1973})}\BibitemShut
  {NoStop}%
\bibitem [{\citenamefont {{Teukolsky}}\ and\ \citenamefont
  {{Press}}(1974)}]{1974ApJ...193..443T}%
  \BibitemOpen
  \bibfield  {author} {\bibinfo {author} {\bibfnamefont {S.~A.}\ \bibnamefont
  {{Teukolsky}}}\ and\ \bibinfo {author} {\bibfnamefont {W.~H.}\ \bibnamefont
  {{Press}}},\ }\href {\doibase 10.1086/153180} {\bibfield  {journal} {\bibinfo
   {journal} {Astrophysical Journal}\ }\textbf {\bibinfo {volume} {193}},\
  \bibinfo {pages} {443} (\bibinfo {year} {1974})}\BibitemShut {NoStop}%
\bibitem [{\citenamefont {Starobinskil}\ and\ \citenamefont
  {Churilov}(1974)}]{Starobinskil:1974nkd}%
  \BibitemOpen
  \bibfield  {author} {\bibinfo {author} {\bibfnamefont {A.~A.}\ \bibnamefont
  {Starobinskil}}\ and\ \bibinfo {author} {\bibfnamefont {S.~M.}\ \bibnamefont
  {Churilov}},\ }\href@noop {} {\bibfield  {journal} {\bibinfo  {journal} {Sov.
  Phys. JETP}\ }\textbf {\bibinfo {volume} {65}},\ \bibinfo {pages} {1}
  (\bibinfo {year} {1974})}\BibitemShut {NoStop}%
\bibitem [{\citenamefont
  {Chandrasekhar}(1976{\natexlab{d}})}]{doi:10.1098/rspa.1976.0022}%
  \BibitemOpen
  \bibfield  {author} {\bibinfo {author} {\bibfnamefont {S.}~\bibnamefont
  {Chandrasekhar}},\ }\href {\doibase 10.1098/rspa.1976.0022} {\bibfield
  {journal} {\bibinfo  {journal} {Proceedings of the Royal Society of London.
  A. Mathematical and Physical Sciences}\ }\textbf {\bibinfo {volume} {348}},\
  \bibinfo {pages} {39} (\bibinfo {year} {1976}{\natexlab{d}})},\ \Eprint
  {http://arxiv.org/abs/https://royalsocietypublishing.org/doi/pdf/10.1098/rspa.1976.0022}
  {https://royalsocietypublishing.org/doi/pdf/10.1098/rspa.1976.0022}
  \BibitemShut {NoStop}%
\bibitem [{\citenamefont {Starobinsky}(1973)}]{Starobinsky:1973aij}%
  \BibitemOpen
  \bibfield  {author} {\bibinfo {author} {\bibfnamefont {A.~A.}\ \bibnamefont
  {Starobinsky}},\ }\href@noop {} {\bibfield  {journal} {\bibinfo  {journal}
  {Sov. Phys. JETP}\ }\textbf {\bibinfo {volume} {37}},\ \bibinfo {pages} {28}
  (\bibinfo {year} {1973})}\BibitemShut {NoStop}%
\bibitem [{\citenamefont {G\"uven}(1980)}]{PhysRevD.22.2327}%
  \BibitemOpen
  \bibfield  {author} {\bibinfo {author} {\bibfnamefont {R.}~\bibnamefont
  {G\"uven}},\ }\href {\doibase 10.1103/PhysRevD.22.2327} {\bibfield  {journal}
  {\bibinfo  {journal} {Phys. Rev. D}\ }\textbf {\bibinfo {volume} {22}},\
  \bibinfo {pages} {2327} (\bibinfo {year} {1980})}\BibitemShut {NoStop}%
\bibitem [{\citenamefont {Teixeira~da Costa}(2020)}]{TeixeiradaCosta:2019skg}%
  \BibitemOpen
  \bibfield  {author} {\bibinfo {author} {\bibfnamefont {R.}~\bibnamefont
  {Teixeira~da Costa}},\ }\href {\doibase 10.1007/s00220-020-03796-z}
  {\bibfield  {journal} {\bibinfo  {journal} {Commun. Math. Phys.}\ }\textbf
  {\bibinfo {volume} {378}},\ \bibinfo {pages} {705} (\bibinfo {year}
  {2020})},\ \Eprint {http://arxiv.org/abs/1910.02854} {arXiv:1910.02854
  [gr-qc]} \BibitemShut {NoStop}%
\bibitem [{\citenamefont {Breuer}\ \emph {et~al.}(1977)\citenamefont {Breuer},
  \citenamefont {Ryan}, \citenamefont {Waller},\ and\ \citenamefont
  {Chandrasekhar}}]{doi:10.1098/rspa.1977.0187}%
  \BibitemOpen
  \bibfield  {author} {\bibinfo {author} {\bibfnamefont {R.~A.}\ \bibnamefont
  {Breuer}}, \bibinfo {author} {\bibfnamefont {M.~P.}\ \bibnamefont {Ryan}},
  \bibinfo {author} {\bibfnamefont {S.}~\bibnamefont {Waller}}, \ and\ \bibinfo
  {author} {\bibfnamefont {S.}~\bibnamefont {Chandrasekhar}},\ }\href {\doibase
  10.1098/rspa.1977.0187} {\bibfield  {journal} {\bibinfo  {journal}
  {Proceedings of the Royal Society of London. A. Mathematical and Physical
  Sciences}\ }\textbf {\bibinfo {volume} {358}},\ \bibinfo {pages} {71}
  (\bibinfo {year} {1977})},\ \Eprint
  {http://arxiv.org/abs/https://royalsocietypublishing.org/doi/pdf/10.1098/rspa.1977.0187}
  {https://royalsocietypublishing.org/doi/pdf/10.1098/rspa.1977.0187}
  \BibitemShut {NoStop}%
\bibitem [{\citenamefont {Casals}\ \emph {et~al.}(2019)\citenamefont {Casals},
  \citenamefont {Ottewill},\ and\ \citenamefont {Warburton}}]{Casals:2018cgx}%
  \BibitemOpen
  \bibfield  {author} {\bibinfo {author} {\bibfnamefont {M.}~\bibnamefont
  {Casals}}, \bibinfo {author} {\bibfnamefont {A.~C.}\ \bibnamefont
  {Ottewill}}, \ and\ \bibinfo {author} {\bibfnamefont {N.}~\bibnamefont
  {Warburton}},\ }\href {\doibase 10.1098/rspa.2018.0701} {\bibfield  {journal}
  {\bibinfo  {journal} {Proc. Roy. Soc. Lond. A}\ }\textbf {\bibinfo {volume}
  {475}},\ \bibinfo {pages} {20180701} (\bibinfo {year} {2019})},\ \Eprint
  {http://arxiv.org/abs/1810.00432} {arXiv:1810.00432 [gr-qc]} \BibitemShut
  {NoStop}%
\bibitem [{\citenamefont {{Goldberg}}\ \emph {et~al.}(1967)\citenamefont
  {{Goldberg}}, \citenamefont {{Macfarlane}}, \citenamefont {{Newman}},
  \citenamefont {{Rohrlich}},\ and\ \citenamefont
  {{Sudarshan}}}]{1967JMP.....8.2155G}%
  \BibitemOpen
  \bibfield  {author} {\bibinfo {author} {\bibfnamefont {J.~N.}\ \bibnamefont
  {{Goldberg}}}, \bibinfo {author} {\bibfnamefont {A.~J.}\ \bibnamefont
  {{Macfarlane}}}, \bibinfo {author} {\bibfnamefont {E.~T.}\ \bibnamefont
  {{Newman}}}, \bibinfo {author} {\bibfnamefont {F.}~\bibnamefont
  {{Rohrlich}}}, \ and\ \bibinfo {author} {\bibfnamefont {E.~C.~G.}\
  \bibnamefont {{Sudarshan}}},\ }\href {\doibase 10.1063/1.1705135} {\bibfield
  {journal} {\bibinfo  {journal} {Journal of Mathematical Physics}\ }\textbf
  {\bibinfo {volume} {8}},\ \bibinfo {pages} {2155} (\bibinfo {year}
  {1967})}\BibitemShut {NoStop}%
\bibitem [{\citenamefont {Berti}\ \emph
  {et~al.}(2006{\natexlab{b}})\citenamefont {Berti}, \citenamefont {Cardoso},\
  and\ \citenamefont {Casals}}]{PhysRevD.73.024013}%
  \BibitemOpen
  \bibfield  {author} {\bibinfo {author} {\bibfnamefont {E.}~\bibnamefont
  {Berti}}, \bibinfo {author} {\bibfnamefont {V.}~\bibnamefont {Cardoso}}, \
  and\ \bibinfo {author} {\bibfnamefont {M.}~\bibnamefont {Casals}},\ }\href
  {\doibase 10.1103/PhysRevD.73.024013} {\bibfield  {journal} {\bibinfo
  {journal} {Phys. Rev. D}\ }\textbf {\bibinfo {volume} {73}},\ \bibinfo
  {pages} {024013} (\bibinfo {year} {2006}{\natexlab{b}})}\BibitemShut
  {NoStop}%
\bibitem [{\citenamefont {Casals}\ and\ \citenamefont
  {Ottewill}(2005)}]{PhysRevD.71.064025}%
  \BibitemOpen
  \bibfield  {author} {\bibinfo {author} {\bibfnamefont {M.}~\bibnamefont
  {Casals}}\ and\ \bibinfo {author} {\bibfnamefont {A.~C.}\ \bibnamefont
  {Ottewill}},\ }\href {\doibase 10.1103/PhysRevD.71.064025} {\bibfield
  {journal} {\bibinfo  {journal} {Phys. Rev. D}\ }\textbf {\bibinfo {volume}
  {71}},\ \bibinfo {pages} {064025} (\bibinfo {year} {2005})}\BibitemShut
  {NoStop}%
\bibitem [{\citenamefont
  {Nollert}(1999{\natexlab{b}})}]{Nollert1999QuasinormalMT}%
  \BibitemOpen
  \bibfield  {author} {\bibinfo {author} {\bibfnamefont {H.-P.}\ \bibnamefont
  {Nollert}},\ }\href@noop {} {\bibfield  {journal} {\bibinfo  {journal}
  {Classical and Quantum Gravity}\ }\textbf {\bibinfo {volume} {16}} (\bibinfo
  {year} {1999}{\natexlab{b}})}\BibitemShut {NoStop}%
\bibitem [{\citenamefont {Casals}\ and\ \citenamefont
  {da~Costa}(2022)}]{Casals:2021ugr}%
  \BibitemOpen
  \bibfield  {author} {\bibinfo {author} {\bibfnamefont {M.}~\bibnamefont
  {Casals}}\ and\ \bibinfo {author} {\bibfnamefont {R.~T.}\ \bibnamefont
  {da~Costa}},\ }\href {\doibase 10.1007/s00220-022-04410-0} {\bibfield
  {journal} {\bibinfo  {journal} {Commun. Math. Phys.}\ }\textbf {\bibinfo
  {volume} {394}},\ \bibinfo {pages} {797} (\bibinfo {year} {2022})},\ \Eprint
  {http://arxiv.org/abs/2105.13329} {arXiv:2105.13329 [gr-qc]} \BibitemShut
  {NoStop}%
\bibitem [{\citenamefont {Whiting}(1989)}]{doi:10.1063/1.528308}%
  \BibitemOpen
  \bibfield  {author} {\bibinfo {author} {\bibfnamefont {B.~F.}\ \bibnamefont
  {Whiting}},\ }\href {\doibase 10.1063/1.528308} {\bibfield  {journal}
  {\bibinfo  {journal} {Journal of Mathematical Physics}\ }\textbf {\bibinfo
  {volume} {30}},\ \bibinfo {pages} {1301} (\bibinfo {year} {1989})},\ \Eprint
  {http://arxiv.org/abs/https://doi.org/10.1063/1.528308}
  {https://doi.org/10.1063/1.528308} \BibitemShut {NoStop}%
\bibitem [{\citenamefont {Berti}\ \emph {et~al.}(2009)\citenamefont {Berti},
  \citenamefont {Cardoso},\ and\ \citenamefont {Starinets}}]{Berti:2009kk}%
  \BibitemOpen
  \bibfield  {author} {\bibinfo {author} {\bibfnamefont {E.}~\bibnamefont
  {Berti}}, \bibinfo {author} {\bibfnamefont {V.}~\bibnamefont {Cardoso}}, \
  and\ \bibinfo {author} {\bibfnamefont {A.~O.}\ \bibnamefont {Starinets}},\
  }\href {\doibase 10.1088/0264-9381/26/16/163001} {\bibfield  {journal}
  {\bibinfo  {journal} {Class. Quant. Grav.}\ }\textbf {\bibinfo {volume}
  {26}},\ \bibinfo {pages} {163001} (\bibinfo {year} {2009})},\ \Eprint
  {http://arxiv.org/abs/0905.2975} {arXiv:0905.2975 [gr-qc]} \BibitemShut
  {NoStop}%
\bibitem [{\citenamefont {Berti}(2016)}]{BertiBangalore}%
  \BibitemOpen
  \bibfield  {author} {\bibinfo {author} {\bibfnamefont {E.}~\bibnamefont
  {Berti}},\ }\href {https://www.icts.res.in/event/page/3071} {\enquote
  {\bibinfo {title} {Black hole perturbation theory},}\ } (\bibinfo {year}
  {2016})\BibitemShut {NoStop}%
\bibitem [{\citenamefont {Kokkotas}(1991{\natexlab{b}})}]{Kokkotas1991}%
  \BibitemOpen
  \bibfield  {author} {\bibinfo {author} {\bibfnamefont {K.~D.}\ \bibnamefont
  {Kokkotas}},\ }\href {\doibase 10.1088/0264-9381/8/12/006} {\bibfield
  {journal} {\bibinfo  {journal} {Classical and Quantum Gravity}\ }\textbf
  {\bibinfo {volume} {8}},\ \bibinfo {pages} {2217} (\bibinfo {year}
  {1991}{\natexlab{b}})}\BibitemShut {NoStop}%
\bibitem [{\citenamefont {Berti}\ and\ \citenamefont
  {Kokkotas}(2003)}]{Berti:2003zu}%
  \BibitemOpen
  \bibfield  {author} {\bibinfo {author} {\bibfnamefont {E.}~\bibnamefont
  {Berti}}\ and\ \bibinfo {author} {\bibfnamefont {K.~D.}\ \bibnamefont
  {Kokkotas}},\ }\href {\doibase 10.1103/PhysRevD.68.044027} {\bibfield
  {journal} {\bibinfo  {journal} {Phys. Rev. D}\ }\textbf {\bibinfo {volume}
  {68}},\ \bibinfo {pages} {044027} (\bibinfo {year} {2003})},\ \Eprint
  {http://arxiv.org/abs/hep-th/0303029} {arXiv:hep-th/0303029} \BibitemShut
  {NoStop}%
\bibitem [{\citenamefont {Nollert}(1993)}]{PhysRevD.47.5253}%
  \BibitemOpen
  \bibfield  {author} {\bibinfo {author} {\bibfnamefont {H.-P.}\ \bibnamefont
  {Nollert}},\ }\href {\doibase 10.1103/PhysRevD.47.5253} {\bibfield  {journal}
  {\bibinfo  {journal} {Phys. Rev. D}\ }\textbf {\bibinfo {volume} {47}},\
  \bibinfo {pages} {5253} (\bibinfo {year} {1993})}\BibitemShut {NoStop}%
\bibitem [{\citenamefont {Onozawa}\ \emph
  {et~al.}(1996{\natexlab{a}})\citenamefont {Onozawa}, \citenamefont {Mishima},
  \citenamefont {Okamura},\ and\ \citenamefont {Ishihara}}]{Onozawa:1995vu}%
  \BibitemOpen
  \bibfield  {author} {\bibinfo {author} {\bibfnamefont {H.}~\bibnamefont
  {Onozawa}}, \bibinfo {author} {\bibfnamefont {T.}~\bibnamefont {Mishima}},
  \bibinfo {author} {\bibfnamefont {T.}~\bibnamefont {Okamura}}, \ and\
  \bibinfo {author} {\bibfnamefont {H.}~\bibnamefont {Ishihara}},\ }\href
  {\doibase 10.1103/PhysRevD.53.7033} {\bibfield  {journal} {\bibinfo
  {journal} {Phys. Rev. D}\ }\textbf {\bibinfo {volume} {53}},\ \bibinfo
  {pages} {7033} (\bibinfo {year} {1996}{\natexlab{a}})},\ \Eprint
  {http://arxiv.org/abs/gr-qc/9603021} {arXiv:gr-qc/9603021} \BibitemShut
  {NoStop}%
\bibitem [{\citenamefont {Leaver}(1992)}]{PhysRevD.45.4713}%
  \BibitemOpen
  \bibfield  {author} {\bibinfo {author} {\bibfnamefont {E.~W.}\ \bibnamefont
  {Leaver}},\ }\href {\doibase 10.1103/PhysRevD.45.4713} {\bibfield  {journal}
  {\bibinfo  {journal} {Phys. Rev. D}\ }\textbf {\bibinfo {volume} {45}},\
  \bibinfo {pages} {4713} (\bibinfo {year} {1992})}\BibitemShut {NoStop}%
\bibitem [{\citenamefont {Wald}(1984)}]{Wald:1984}%
  \BibitemOpen
  \bibfield  {author} {\bibinfo {author} {\bibfnamefont {R.~M.}\ \bibnamefont
  {Wald}},\ }\href@noop {} {\emph {\bibinfo {title} {{General Relativity}}}}\
  (\bibinfo  {publisher} {The University of Chicago Press},\ \bibinfo {year}
  {1984})\BibitemShut {NoStop}%
\bibitem [{\citenamefont {Wu}\ and\ \citenamefont {Zhao}(2004)}]{Wu:2004vb}%
  \BibitemOpen
  \bibfield  {author} {\bibinfo {author} {\bibfnamefont {Y.-J.}\ \bibnamefont
  {Wu}}\ and\ \bibinfo {author} {\bibfnamefont {Z.}~\bibnamefont {Zhao}},\
  }\href {\doibase 10.1103/PhysRevD.69.084015} {\bibfield  {journal} {\bibinfo
  {journal} {Phys. Rev. D}\ }\textbf {\bibinfo {volume} {69}},\ \bibinfo
  {pages} {084015} (\bibinfo {year} {2004})}\BibitemShut {NoStop}%
\bibitem [{\citenamefont {Chang}\ and\ \citenamefont
  {Shen}(2007)}]{Chang2007MassiveCQ}%
  \BibitemOpen
  \bibfield  {author} {\bibinfo {author} {\bibfnamefont {J.}~\bibnamefont
  {Chang}}\ and\ \bibinfo {author} {\bibfnamefont {Y.}~\bibnamefont {Shen}},\
  }\href@noop {} {\bibfield  {journal} {\bibinfo  {journal} {International
  Journal of Theoretical Physics}\ }\textbf {\bibinfo {volume} {46}},\ \bibinfo
  {pages} {1570} (\bibinfo {year} {2007})}\BibitemShut {NoStop}%
\bibitem [{\citenamefont {Richartz}\ and\ \citenamefont
  {Giugno}(2014)}]{Richartz:2014jla}%
  \BibitemOpen
  \bibfield  {author} {\bibinfo {author} {\bibfnamefont {M.}~\bibnamefont
  {Richartz}}\ and\ \bibinfo {author} {\bibfnamefont {D.}~\bibnamefont
  {Giugno}},\ }\href {\doibase 10.1103/PhysRevD.90.124011} {\bibfield
  {journal} {\bibinfo  {journal} {Phys. Rev. D}\ }\textbf {\bibinfo {volume}
  {90}},\ \bibinfo {pages} {124011} (\bibinfo {year} {2014})},\ \Eprint
  {http://arxiv.org/abs/1409.7440} {arXiv:1409.7440 [gr-qc]} \BibitemShut
  {NoStop}%
\bibitem [{\citenamefont {Nakamura}\ and\ \citenamefont
  {Sato}(1976)}]{NAKAMURA1976371}%
  \BibitemOpen
  \bibfield  {author} {\bibinfo {author} {\bibfnamefont {T.}~\bibnamefont
  {Nakamura}}\ and\ \bibinfo {author} {\bibfnamefont {H.}~\bibnamefont
  {Sato}},\ }\href {\doibase https://doi.org/10.1016/0370-2693(76)90591-8}
  {\bibfield  {journal} {\bibinfo  {journal} {Physics Letters B}\ }\textbf
  {\bibinfo {volume} {61}},\ \bibinfo {pages} {371} (\bibinfo {year}
  {1976})}\BibitemShut {NoStop}%
\bibitem [{\citenamefont {Stewart}\ and\ \citenamefont
  {Stewart}(1993)}]{stewart1993advanced}%
  \BibitemOpen
  \bibfield  {author} {\bibinfo {author} {\bibfnamefont {J.}~\bibnamefont
  {Stewart}}\ and\ \bibinfo {author} {\bibfnamefont {J.~M.}\ \bibnamefont
  {Stewart}},\ }\href@noop {} {\emph {\bibinfo {title} {Advanced general
  relativity}}}\ (\bibinfo  {publisher} {Cambridge university press},\ \bibinfo
  {year} {1993})\BibitemShut {NoStop}%
\bibitem [{\citenamefont {Goldberg}\ \emph {et~al.}(1967)\citenamefont
  {Goldberg}, \citenamefont {Macfarlane}, \citenamefont {Newman}, \citenamefont
  {Rohrlich},\ and\ \citenamefont {Sudarshan}}]{doi:10.1063/1.1705135}%
  \BibitemOpen
  \bibfield  {author} {\bibinfo {author} {\bibfnamefont {J.~N.}\ \bibnamefont
  {Goldberg}}, \bibinfo {author} {\bibfnamefont {A.~J.}\ \bibnamefont
  {Macfarlane}}, \bibinfo {author} {\bibfnamefont {E.~T.}\ \bibnamefont
  {Newman}}, \bibinfo {author} {\bibfnamefont {F.}~\bibnamefont {Rohrlich}}, \
  and\ \bibinfo {author} {\bibfnamefont {E.~C.~G.}\ \bibnamefont {Sudarshan}},\
  }\href {\doibase 10.1063/1.1705135} {\bibfield  {journal} {\bibinfo
  {journal} {Journal of Mathematical Physics}\ }\textbf {\bibinfo {volume}
  {8}},\ \bibinfo {pages} {2155} (\bibinfo {year} {1967})},\ \Eprint
  {http://arxiv.org/abs/https://doi.org/10.1063/1.1705135}
  {https://doi.org/10.1063/1.1705135} \BibitemShut {NoStop}%
\bibitem [{\citenamefont {JING}\ \emph {et~al.}(2007)\citenamefont {JING},
  \citenamefont {PAN},\ and\ \citenamefont
  {HE}}]{doi:10.1142/S0218271807009322}%
  \BibitemOpen
  \bibfield  {author} {\bibinfo {author} {\bibfnamefont {J.}~\bibnamefont
  {JING}}, \bibinfo {author} {\bibfnamefont {Q.}~\bibnamefont {PAN}}, \ and\
  \bibinfo {author} {\bibfnamefont {X.}~\bibnamefont {HE}},\ }\href {\doibase
  10.1142/S0218271807009322} {\bibfield  {journal} {\bibinfo  {journal}
  {International Journal of Modern Physics D}\ }\textbf {\bibinfo {volume}
  {16}},\ \bibinfo {pages} {81} (\bibinfo {year} {2007})},\ \Eprint
  {http://arxiv.org/abs/https://doi.org/10.1142/S0218271807009322}
  {https://doi.org/10.1142/S0218271807009322} \BibitemShut {NoStop}%
\bibitem [{\citenamefont {Richartz}\ and\ \citenamefont
  {Saa}(2011)}]{PhysRevD.84.104021}%
  \BibitemOpen
  \bibfield  {author} {\bibinfo {author} {\bibfnamefont {M.}~\bibnamefont
  {Richartz}}\ and\ \bibinfo {author} {\bibfnamefont {A.}~\bibnamefont {Saa}},\
  }\href {\doibase 10.1103/PhysRevD.84.104021} {\bibfield  {journal} {\bibinfo
  {journal} {Phys. Rev. D}\ }\textbf {\bibinfo {volume} {84}},\ \bibinfo
  {pages} {104021} (\bibinfo {year} {2011})}\BibitemShut {NoStop}%
\bibitem [{\citenamefont {Kokkotas}\ and\ \citenamefont
  {Schutz}(1988{\natexlab{b}})}]{Kokkotas:1988fm}%
  \BibitemOpen
  \bibfield  {author} {\bibinfo {author} {\bibfnamefont {K.~D.}\ \bibnamefont
  {Kokkotas}}\ and\ \bibinfo {author} {\bibfnamefont {B.~F.}\ \bibnamefont
  {Schutz}},\ }\href {\doibase 10.1103/PhysRevD.37.3378} {\bibfield  {journal}
  {\bibinfo  {journal} {Phys. Rev. D}\ }\textbf {\bibinfo {volume} {37}},\
  \bibinfo {pages} {3378} (\bibinfo {year} {1988}{\natexlab{b}})}\BibitemShut
  {NoStop}%
\bibitem [{{\relax DLMF}()}]{NIST:DLMF}%
  \BibitemOpen
  {\relax DLMF},\ \href {http://dlmf.nist.gov/} {\enquote {\bibinfo {title}
  {{NIST Digital Library of Mathematical Functions}},}\ }\bibinfo
  {howpublished} {http://dlmf.nist.gov/, Release 1.1.2 of 2021-06-15} (\bibinfo
  {year} {2023}),\ \bibinfo {note} {f.~W.~J. Olver, A.~B. {Olde Daalhuis},
  D.~W. Lozier, B.~I. Schneider, R.~F. Boisvert, C.~W. Clark, B.~R. Miller,
  B.~V. Saunders, H.~S. Cohl, and M.~A. McClain, eds.}\BibitemShut {Stop}%
\bibitem [{\citenamefont {Haraoka}(2020)}]{haraoka2020linear}%
  \BibitemOpen
  \bibfield  {author} {\bibinfo {author} {\bibfnamefont {Y.}~\bibnamefont
  {Haraoka}},\ }\href {https://books.google.com.br/books?id=mxGXzQEACAAJ}
  {\emph {\bibinfo {title} {Linear Differential Equations in the Complex
  Domain: From Classical Theory to Forefront}}},\ Lecture Notes in Mathematics\
  (\bibinfo  {publisher} {Springer International Publishing},\ \bibinfo {year}
  {2020})\BibitemShut {NoStop}%
\bibitem [{\citenamefont {Somasundaram}(2001)}]{somasundaram2001ordinary}%
  \BibitemOpen
  \bibfield  {author} {\bibinfo {author} {\bibfnamefont {D.}~\bibnamefont
  {Somasundaram}},\ }\href {https://books.google.com.br/books?id=yOUPngEACAAJ}
  {\emph {\bibinfo {title} {Ordinary Differential Equations: A First Course}}}\
  (\bibinfo  {publisher} {CRC Press},\ \bibinfo {year} {2001})\BibitemShut
  {NoStop}%
\bibitem [{\citenamefont {Jimbo}\ \emph
  {et~al.}(1981{\natexlab{b}})\citenamefont {Jimbo}, \citenamefont {Miwa},\
  and\ \citenamefont {Ueno}}]{Jimbo:1981aa}%
  \BibitemOpen
  \bibfield  {author} {\bibinfo {author} {\bibfnamefont {M.}~\bibnamefont
  {Jimbo}}, \bibinfo {author} {\bibfnamefont {T.}~\bibnamefont {Miwa}}, \ and\
  \bibinfo {author} {\bibfnamefont {A.~K.}\ \bibnamefont {Ueno}},\ }\href@noop
  {} {\bibfield  {journal} {\bibinfo  {journal} {Physica}\ }\textbf {\bibinfo
  {volume} {D2}},\ \bibinfo {pages} {306} (\bibinfo {year}
  {1981}{\natexlab{b}})}\BibitemShut {NoStop}%
\bibitem [{\citenamefont {Hille}(1997)}]{hille1997ordinary}%
  \BibitemOpen
  \bibfield  {author} {\bibinfo {author} {\bibfnamefont {E.}~\bibnamefont
  {Hille}},\ }\href {https://books.google.com.br/books?id=I1OR4t8UZCgC} {\emph
  {\bibinfo {title} {Ordinary Differential Equations in the Complex Domain}}},\
  Dover books on mathematics\ (\bibinfo  {publisher} {Dover Publications},\
  \bibinfo {year} {1997})\BibitemShut {NoStop}%
\bibitem [{\citenamefont {Iwasaki}(1991)}]{iwasaki1991gauss}%
  \BibitemOpen
  \bibfield  {author} {\bibinfo {author} {\bibfnamefont {K.}~\bibnamefont
  {Iwasaki}},\ }\href {https://books.google.com.br/books?id=LSrvAAAAMAAJ}
  {\emph {\bibinfo {title} {From Gauss to Painlev{\'e}: a modern theory of
  special functions}}},\ Aspects of mathematics\ (\bibinfo  {publisher}
  {Vieweg},\ \bibinfo {year} {1991})\BibitemShut {NoStop}%
\bibitem [{\citenamefont {Meyer}(1989)}]{10.2307/2031404}%
  \BibitemOpen
  \bibfield  {author} {\bibinfo {author} {\bibfnamefont {R.~E.}\ \bibnamefont
  {Meyer}},\ }\href {http://www.jstor.org/stable/2031404} {\bibfield  {journal}
  {\bibinfo  {journal} {SIAM Review}\ }\textbf {\bibinfo {volume} {31}},\
  \bibinfo {pages} {435} (\bibinfo {year} {1989})}\BibitemShut {NoStop}%
\bibitem [{\citenamefont {Berry}(1988)}]{Berry1988}%
  \BibitemOpen
  \bibfield  {author} {\bibinfo {author} {\bibfnamefont {M.~V.}\ \bibnamefont
  {Berry}},\ }\href {\doibase 10.1007/BF02698550} {\bibfield  {journal}
  {\bibinfo  {journal} {Publications Math{\'e}matiques de l'Institut des Hautes
  {\'E}tudes Scientifiques}\ }\textbf {\bibinfo {volume} {68}},\ \bibinfo
  {pages} {211} (\bibinfo {year} {1988})}\BibitemShut {NoStop}%
\bibitem [{\citenamefont {Gilbert}\ and\ \citenamefont
  {Wood}(2004)}]{GILBERT2004247}%
  \BibitemOpen
  \bibfield  {author} {\bibinfo {author} {\bibfnamefont {D.}~\bibnamefont
  {Gilbert}}\ and\ \bibinfo {author} {\bibfnamefont {A.}~\bibnamefont {Wood}},\
  }\href {\doibase https://doi.org/10.1016/j.cam.2004.01.012} {\bibfield
  {journal} {\bibinfo  {journal} {Journal of Computational and Applied
  Mathematics}\ }\textbf {\bibinfo {volume} {171}},\ \bibinfo {pages} {247}
  (\bibinfo {year} {2004})},\ \bibinfo {note} {special issue on the occasion of
  the eightieth birthday of Prof. W.M. Everitt}\BibitemShut {NoStop}%
\bibitem [{\citenamefont {Sibuya}(2008)}]{sibuya2008linear}%
  \BibitemOpen
  \bibfield  {author} {\bibinfo {author} {\bibfnamefont {Y.}~\bibnamefont
  {Sibuya}},\ }\href {https://books.google.com.br/books?id=FPqcAwAAQBAJ} {\emph
  {\bibinfo {title} {Linear Differential Equations in the Complex Domain:
  Problems of Analytic Continuation}}},\ Translations of Mathematical
  Monographs\ (\bibinfo  {publisher} {American Mathematical Society},\ \bibinfo
  {year} {2008})\BibitemShut {NoStop}%
\bibitem [{\citenamefont {Andreev}\ and\ \citenamefont
  {Kitaev}(2000{\natexlab{a}})}]{Andreev:1995in}%
  \BibitemOpen
  \bibfield  {author} {\bibinfo {author} {\bibfnamefont {F.~V.}\ \bibnamefont
  {Andreev}}\ and\ \bibinfo {author} {\bibfnamefont {A.~V.}\ \bibnamefont
  {Kitaev}},\ }\href {\doibase 10.1088/0951-7715/13/5/319} {\bibfield
  {journal} {\bibinfo  {journal} {Nonlinearity}\ }\textbf {\bibinfo {volume}
  {13}},\ \bibinfo {pages} {1801} (\bibinfo {year}
  {2000}{\natexlab{a}})}\BibitemShut {NoStop}%
\bibitem [{\citenamefont
  {Jean-Pierre}(1989)}]{oai:repository.dl.itc.u-tokyo.ac.jp:00039422}%
  \BibitemOpen
  \bibfield  {author} {\bibinfo {author} {\bibfnamefont {R.}~\bibnamefont
  {Jean-Pierre}},\ }\href@noop {} {\bibfield  {journal} {\bibinfo  {journal}
  {Journal of the Faculty of Science, the University of Tokyo. Sect. 1 A,
  Mathematics}\ }\textbf {\bibinfo {volume} {36}},\ \bibinfo {pages} {703}
  (\bibinfo {year} {1989})},\ \bibinfo {note} {application/pdf}\BibitemShut
  {NoStop}%
\bibitem [{\citenamefont {Jimbo}\ and\ \citenamefont
  {Miwa}(1981{\natexlab{c}})}]{Jimbo:1981ab}%
  \BibitemOpen
  \bibfield  {author} {\bibinfo {author} {\bibfnamefont {M.}~\bibnamefont
  {Jimbo}}\ and\ \bibinfo {author} {\bibfnamefont {T.}~\bibnamefont {Miwa}},\
  }\href@noop {} {\bibfield  {journal} {\bibinfo  {journal} {Physica}\ }\textbf
  {\bibinfo {volume} {D2}},\ \bibinfo {pages} {407} (\bibinfo {year}
  {1981}{\natexlab{c}})}\BibitemShut {NoStop}%
\bibitem [{\citenamefont {Jimbo}\ and\ \citenamefont
  {Miwa}(1981{\natexlab{d}})}]{Jimbo:1981ac}%
  \BibitemOpen
  \bibfield  {author} {\bibinfo {author} {\bibfnamefont {M.}~\bibnamefont
  {Jimbo}}\ and\ \bibinfo {author} {\bibfnamefont {T.}~\bibnamefont {Miwa}},\
  }\href@noop {} {\bibfield  {journal} {\bibinfo  {journal} {Physica}\ }\textbf
  {\bibinfo {volume} {D4}},\ \bibinfo {pages} {26} (\bibinfo {year}
  {1981}{\natexlab{d}})}\BibitemShut {NoStop}%
\bibitem [{\citenamefont {Fuchs}(1907)}]{Fuchs1907}%
  \BibitemOpen
  \bibfield  {author} {\bibinfo {author} {\bibfnamefont {R.}~\bibnamefont
  {Fuchs}},\ }\href {http://eudml.org/doc/158289} {\bibfield  {journal}
  {\bibinfo  {journal} {Mathematische Annalen}\ }\textbf {\bibinfo {volume}
  {63}},\ \bibinfo {pages} {301} (\bibinfo {year} {1907})}\BibitemShut
  {NoStop}%
\bibitem [{\citenamefont {Clarkson}(2003)}]{CLARKSON2003127}%
  \BibitemOpen
  \bibfield  {author} {\bibinfo {author} {\bibfnamefont {P.~A.}\ \bibnamefont
  {Clarkson}},\ }\href {\doibase https://doi.org/10.1016/S0377-0427(02)00589-7}
  {\bibfield  {journal} {\bibinfo  {journal} {Journal of Computational and
  Applied Mathematics}\ }\textbf {\bibinfo {volume} {153}},\ \bibinfo {pages}
  {127} (\bibinfo {year} {2003})},\ \bibinfo {note} {proceedings of the 6th
  International Symposium on Orthogonal Poly nomials, Special Functions and
  their Applications, Rome, Italy, 18-22 June 2001}\BibitemShut {NoStop}%
\bibitem [{\citenamefont {Okamoto}(1980{\natexlab{a}})}]{10.3792/pjaa.56.264}%
  \BibitemOpen
  \bibfield  {author} {\bibinfo {author} {\bibfnamefont {K.}~\bibnamefont
  {Okamoto}},\ }\href {\doibase 10.3792/pjaa.56.264} {\bibfield  {journal}
  {\bibinfo  {journal} {Proceedings of the Japan Academy, Series A,
  Mathematical Sciences}\ }\textbf {\bibinfo {volume} {56}},\ \bibinfo {pages}
  {264 } (\bibinfo {year} {1980}{\natexlab{a}})}\BibitemShut {NoStop}%
\bibitem [{\citenamefont {Okamoto}(1981)}]{okamoto1981isomonodromic}%
  \BibitemOpen
  \bibfield  {author} {\bibinfo {author} {\bibfnamefont {K.}~\bibnamefont
  {Okamoto}},\ }\href {https://books.google.com.br/books?id=1tHvOgAACAAJ}
  {\emph {\bibinfo {title} {Isomonodromic Deformation and Painlev{\'e}
  Equations, and the Garnier System}}},\ Institut de Recherche Math{\'e}matique
  Avanc{\'e}e Strasbourg: Publication\ (\bibinfo  {publisher} {Universit{\'e}
  Louis Pasteur, Institut de Recherche Math{\'e}matique Avanc{\'e}e},\ \bibinfo
  {year} {1981})\BibitemShut {NoStop}%
\bibitem [{\citenamefont {Mahoux}(1999)}]{Mahoux1999}%
  \BibitemOpen
  \bibfield  {author} {\bibinfo {author} {\bibfnamefont {G.}~\bibnamefont
  {Mahoux}},\ }\enquote {\bibinfo {title} {Introduction to the theory of
  isomonodromic deformations of linear ordinary differential equations with
  rational coefficients},}\ in\ \href {\doibase 10.1007/978-1-4612-1532-5_2}
  {\emph {\bibinfo {booktitle} {The Painlev{\'e} Property: One Century
  Later}}}\ (\bibinfo  {publisher} {Springer New York},\ \bibinfo {address}
  {New York, NY},\ \bibinfo {year} {1999})\ pp.\ \bibinfo {pages}
  {35--76}\BibitemShut {NoStop}%
\bibitem [{\citenamefont {Cosgrove}\ and\ \citenamefont
  {Scoufis}(1993)}]{cosgrove}%
  \BibitemOpen
  \bibfield  {author} {\bibinfo {author} {\bibfnamefont {C.~M.}\ \bibnamefont
  {Cosgrove}}\ and\ \bibinfo {author} {\bibfnamefont {G.}~\bibnamefont
  {Scoufis}},\ }\href {\doibase https://doi.org/10.1002/sapm199388125}
  {\bibfield  {journal} {\bibinfo  {journal} {Studies in Applied Mathematics}\
  }\textbf {\bibinfo {volume} {88}},\ \bibinfo {pages} {25} (\bibinfo {year}
  {1993})},\ \Eprint
  {http://arxiv.org/abs/https://onlinelibrary.wiley.com/doi/pdf/10.1002/sapm199388125}
  {https://onlinelibrary.wiley.com/doi/pdf/10.1002/sapm199388125} \BibitemShut
  {NoStop}%
\bibitem [{\citenamefont {Jimbo}(1982)}]{Jimbo:1982aa}%
  \BibitemOpen
  \bibfield  {author} {\bibinfo {author} {\bibfnamefont {M.}~\bibnamefont
  {Jimbo}},\ }\href@noop {} {\bibfield  {journal} {\bibinfo  {journal} {Publ.
  Res. Inst. Math. Sci.}\ }\textbf {\bibinfo {volume} {18}},\ \bibinfo {pages}
  {1137} (\bibinfo {year} {1982})}\BibitemShut {NoStop}%
\bibitem [{\citenamefont {McCoy}\ and\ \citenamefont
  {Tang}(1986{\natexlab{a}})}]{MCCOY198642}%
  \BibitemOpen
  \bibfield  {author} {\bibinfo {author} {\bibfnamefont {B.~M.}\ \bibnamefont
  {McCoy}}\ and\ \bibinfo {author} {\bibfnamefont {S.}~\bibnamefont {Tang}},\
  }\href {\doibase https://doi.org/10.1016/0167-2789(86)90053-9} {\bibfield
  {journal} {\bibinfo  {journal} {Physica D: Nonlinear Phenomena}\ }\textbf
  {\bibinfo {volume} {19}},\ \bibinfo {pages} {42} (\bibinfo {year}
  {1986}{\natexlab{a}})}\BibitemShut {NoStop}%
\bibitem [{\citenamefont {McCoy}\ and\ \citenamefont
  {Tang}(1986{\natexlab{b}})}]{MCCOY1986190}%
  \BibitemOpen
  \bibfield  {author} {\bibinfo {author} {\bibfnamefont {B.~M.}\ \bibnamefont
  {McCoy}}\ and\ \bibinfo {author} {\bibfnamefont {S.}~\bibnamefont {Tang}},\
  }\href {\doibase https://doi.org/10.1016/0167-2789(86)90176-4} {\bibfield
  {journal} {\bibinfo  {journal} {Physica D: Nonlinear Phenomena}\ }\textbf
  {\bibinfo {volume} {18}},\ \bibinfo {pages} {190} (\bibinfo {year}
  {1986}{\natexlab{b}})}\BibitemShut {NoStop}%
\bibitem [{\citenamefont {Jimbo}\ \emph {et~al.}(1980)\citenamefont {Jimbo},
  \citenamefont {Miwa}, \citenamefont {M\^ori},\ and\ \citenamefont
  {Sato}}]{JIMBO198080}%
  \BibitemOpen
  \bibfield  {author} {\bibinfo {author} {\bibfnamefont {M.}~\bibnamefont
  {Jimbo}}, \bibinfo {author} {\bibfnamefont {T.}~\bibnamefont {Miwa}},
  \bibinfo {author} {\bibfnamefont {Y.}~\bibnamefont {M\^ori}}, \ and\ \bibinfo
  {author} {\bibfnamefont {M.}~\bibnamefont {Sato}},\ }\href {\doibase
  https://doi.org/10.1016/0167-2789(80)90006-8} {\bibfield  {journal} {\bibinfo
   {journal} {Physica D: Nonlinear Phenomena}\ }\textbf {\bibinfo {volume}
  {1}},\ \bibinfo {pages} {80} (\bibinfo {year} {1980})}\BibitemShut {NoStop}%
\bibitem [{\citenamefont {McCoy}\ and\ \citenamefont
  {Tang}(1986{\natexlab{c}})}]{MCCOY1986187}%
  \BibitemOpen
  \bibfield  {author} {\bibinfo {author} {\bibfnamefont {B.~M.}\ \bibnamefont
  {McCoy}}\ and\ \bibinfo {author} {\bibfnamefont {S.}~\bibnamefont {Tang}},\
  }\href {\doibase https://doi.org/10.1016/0167-2789(86)90030-8} {\bibfield
  {journal} {\bibinfo  {journal} {Physica D: Nonlinear Phenomena}\ }\textbf
  {\bibinfo {volume} {20}},\ \bibinfo {pages} {187} (\bibinfo {year}
  {1986}{\natexlab{c}})}\BibitemShut {NoStop}%
\bibitem [{\citenamefont {Nagoya}(2015)}]{Nagoya:2015cja}%
  \BibitemOpen
  \bibfield  {author} {\bibinfo {author} {\bibfnamefont {H.}~\bibnamefont
  {Nagoya}},\ }\href {\doibase 10.1063/1.4937760} {\bibfield  {journal}
  {\bibinfo  {journal} {J. Math. Phys.}\ }\textbf {\bibinfo {volume} {56}},\
  \bibinfo {pages} {123505} (\bibinfo {year} {2015})},\ \Eprint
  {http://arxiv.org/abs/1505.02398} {arXiv:1505.02398 [math-ph]} \BibitemShut
  {NoStop}%
\bibitem [{\citenamefont {Nagoya}(2016)}]{nag2}%
  \BibitemOpen
  \bibfield  {author} {\bibinfo {author} {\bibfnamefont {H.}~\bibnamefont
  {Nagoya}},\ }\href {\doibase 10.48550/ARXIV.1611.08971} {\enquote {\bibinfo
  {title} {Conformal blocks and painlev{\'e} functions},}\ } (\bibinfo {year}
  {2016})\BibitemShut {NoStop}%
\bibitem [{\citenamefont {Gavrylenko}\ and\ \citenamefont
  {Lisovyy}(2018)}]{Gavrylenko:2016zlf}%
  \BibitemOpen
  \bibfield  {author} {\bibinfo {author} {\bibfnamefont {P.}~\bibnamefont
  {Gavrylenko}}\ and\ \bibinfo {author} {\bibfnamefont {O.}~\bibnamefont
  {Lisovyy}},\ }\href {\doibase 10.1007/s00220-018-3224-7} {\bibfield
  {journal} {\bibinfo  {journal} {Commun. Math. Phys.}\ }\textbf {\bibinfo
  {volume} {363}},\ \bibinfo {pages} {1} (\bibinfo {year} {2018})},\ \Eprint
  {http://arxiv.org/abs/1608.00958} {arXiv:1608.00958 [math-ph]} \BibitemShut
  {NoStop}%
\bibitem [{\citenamefont {Its}\ \emph {et~al.}(2018)\citenamefont {Its},
  \citenamefont {Lisovyy},\ and\ \citenamefont {Prokhorov}}]{its:hal-01797601}%
  \BibitemOpen
  \bibfield  {author} {\bibinfo {author} {\bibfnamefont {A.}~\bibnamefont
  {Its}}, \bibinfo {author} {\bibfnamefont {O.}~\bibnamefont {Lisovyy}}, \ and\
  \bibinfo {author} {\bibfnamefont {A.}~\bibnamefont {Prokhorov}},\ }\href
  {\doibase 10.1215/00127094-2017-0055} {\bibfield  {journal} {\bibinfo
  {journal} {{Duke Math.J.}}\ }\textbf {\bibinfo {volume} {167}},\ \bibinfo
  {pages} {1347} (\bibinfo {year} {2018})}\BibitemShut {NoStop}%
\bibitem [{\citenamefont {Andreev}\ and\ \citenamefont
  {Kitaev}(2000{\natexlab{b}})}]{Andreev:2000aa}%
  \BibitemOpen
  \bibfield  {author} {\bibinfo {author} {\bibfnamefont {F.~V.}\ \bibnamefont
  {Andreev}}\ and\ \bibinfo {author} {\bibfnamefont {A.~V.}\ \bibnamefont
  {Kitaev}},\ }\href@noop {} {\bibfield  {journal} {\bibinfo  {journal}
  {Nonlinearity}\ }\textbf {\bibinfo {volume} {13}},\ \bibinfo {pages} {1801}
  (\bibinfo {year} {2000}{\natexlab{b}})}\BibitemShut {NoStop}%
\bibitem [{\citenamefont {Okamoto}(1987)}]{198747}%
  \BibitemOpen
  \bibfield  {author} {\bibinfo {author} {\bibfnamefont {K.}~\bibnamefont
  {Okamoto}},\ }\href {\doibase 10.4099/math1924.13.47} {\bibfield  {journal}
  {\bibinfo  {journal} {Japanese journal of mathematics. New series}\ }\textbf
  {\bibinfo {volume} {13}},\ \bibinfo {pages} {47} (\bibinfo {year}
  {1987})}\BibitemShut {NoStop}%
\bibitem [{\citenamefont {Miwa}(1981{\natexlab{a}})}]{Miwa:1980yj}%
  \BibitemOpen
  \bibfield  {author} {\bibinfo {author} {\bibfnamefont {T.}~\bibnamefont
  {Miwa}},\ }\href@noop {} {\bibfield  {journal} {\bibinfo  {journal}
  {Publications of The Research Institute for Mathematical Sciences}\ }\textbf
  {\bibinfo {volume} {17}},\ \bibinfo {pages} {703} (\bibinfo {year}
  {1981}{\natexlab{a}})}\BibitemShut {NoStop}%
\bibitem [{\citenamefont {Bothner}(2021)}]{Bothner_2021}%
  \BibitemOpen
  \bibfield  {author} {\bibinfo {author} {\bibfnamefont {T.}~\bibnamefont
  {Bothner}},\ }\href {\doibase 10.1088/1361-6544/abb543} {\bibfield  {journal}
  {\bibinfo  {journal} {Nonlinearity}\ }\textbf {\bibinfo {volume} {34}},\
  \bibinfo {pages} {R1} (\bibinfo {year} {2021})}\BibitemShut {NoStop}%
\bibitem [{\citenamefont {Its}(2003)}]{Its2003TheRP}%
  \BibitemOpen
  \bibfield  {author} {\bibinfo {author} {\bibfnamefont {A.}~\bibnamefont
  {Its}},\ }in\ \href@noop {} {\emph {\bibinfo {booktitle} {The Riemann-Hilbert
  Problem and Integrable Systems}}}\ (\bibinfo {year} {2003})\BibitemShut
  {NoStop}%
\bibitem [{\citenamefont {Mebkhout}(1980)}]{10.1007/3-540-09996-4_31}%
  \BibitemOpen
  \bibfield  {author} {\bibinfo {author} {\bibfnamefont {Z.}~\bibnamefont
  {Mebkhout}},\ }in\ \href@noop {} {\emph {\bibinfo {booktitle} {Complex
  Analysis, Microlocal Calculus and Relativistic Quantum Theory}}},\ \bibinfo
  {editor} {edited by\ \bibinfo {editor} {\bibfnamefont {D.}~\bibnamefont
  {Iagolnitzer}}}\ (\bibinfo  {publisher} {Springer Berlin Heidelberg},\
  \bibinfo {address} {Berlin, Heidelberg},\ \bibinfo {year} {1980})\ pp.\
  \bibinfo {pages} {90--110}\BibitemShut {NoStop}%
\bibitem [{\citenamefont {Anselmo}\ \emph {et~al.}(2020)\citenamefont
  {Anselmo}, \citenamefont {Carneiro~da Cunha}, \citenamefont {Nelson},\ and\
  \citenamefont {Crowdy}}]{Anselmo:2020bmt}%
  \BibitemOpen
  \bibfield  {author} {\bibinfo {author} {\bibfnamefont {T.}~\bibnamefont
  {Anselmo}}, \bibinfo {author} {\bibfnamefont {B.}~\bibnamefont {Carneiro~da
  Cunha}}, \bibinfo {author} {\bibfnamefont {R.}~\bibnamefont {Nelson}}, \ and\
  \bibinfo {author} {\bibfnamefont {D.~G.}\ \bibnamefont {Crowdy}},\ }\href
  {\doibase 10.1088/1751-8121/ab9f71} {\bibfield  {journal} {\bibinfo
  {journal} {J. Phys. A}\ }\textbf {\bibinfo {volume} {53}},\ \bibinfo {pages}
  {355201} (\bibinfo {year} {2020})}\BibitemShut {NoStop}%
\bibitem [{\citenamefont {Anselmo}\ \emph {et~al.}(2018)\citenamefont
  {Anselmo}, \citenamefont {Nelson}, \citenamefont {Carneiro~da Cunha},\ and\
  \citenamefont {Crowdy}}]{Anselmo:2018zre}%
  \BibitemOpen
  \bibfield  {author} {\bibinfo {author} {\bibfnamefont {T.}~\bibnamefont
  {Anselmo}}, \bibinfo {author} {\bibfnamefont {R.}~\bibnamefont {Nelson}},
  \bibinfo {author} {\bibfnamefont {B.}~\bibnamefont {Carneiro~da Cunha}}, \
  and\ \bibinfo {author} {\bibfnamefont {D.~G.}\ \bibnamefont {Crowdy}},\
  }\href {\doibase 10.1098/rspa.2018.0080} {\bibfield  {journal} {\bibinfo
  {journal} {Proc. Roy. Soc. Lond. A}\ }\textbf {\bibinfo {volume} {474}},\
  \bibinfo {pages} {20180080} (\bibinfo {year} {2018})}\BibitemShut {NoStop}%
\bibitem [{\citenamefont {Amado}\ \emph {et~al.}(2017)\citenamefont {Amado},
  \citenamefont {Carneiro~da Cunha},\ and\ \citenamefont
  {Pallante}}]{Amado:2017kao}%
  \BibitemOpen
  \bibfield  {author} {\bibinfo {author} {\bibfnamefont {J.~B.}\ \bibnamefont
  {Amado}}, \bibinfo {author} {\bibfnamefont {B.}~\bibnamefont {Carneiro~da
  Cunha}}, \ and\ \bibinfo {author} {\bibfnamefont {E.}~\bibnamefont
  {Pallante}},\ }\href {\doibase 10.1007/JHEP08(2017)094} {\bibfield  {journal}
  {\bibinfo  {journal} {JHEP}\ }\textbf {\bibinfo {volume} {08}},\ \bibinfo
  {pages} {094} (\bibinfo {year} {2017})},\ \Eprint
  {http://arxiv.org/abs/1702.01016} {arXiv:1702.01016 [hep-th]} \BibitemShut
  {NoStop}%
\bibitem [{\citenamefont {Filipuk}\ \emph {et~al.}(2019)\citenamefont
  {Filipuk}, \citenamefont {Ishkhanyan},\ and\ \citenamefont
  {Derezi'nski}}]{Filipuk2019OnTD}%
  \BibitemOpen
  \bibfield  {author} {\bibinfo {author} {\bibfnamefont {G.}~\bibnamefont
  {Filipuk}}, \bibinfo {author} {\bibfnamefont {A.~M.}\ \bibnamefont
  {Ishkhanyan}}, \ and\ \bibinfo {author} {\bibfnamefont {J.}~\bibnamefont
  {Derezi'nski}},\ }\href@noop {} {\bibfield  {journal} {\bibinfo  {journal}
  {Journal of Contemporary Mathematical Analysis (Armenian Academy of
  Sciences)}\ }\textbf {\bibinfo {volume} {55}},\ \bibinfo {pages} {200}
  (\bibinfo {year} {2019})}\BibitemShut {NoStop}%
\bibitem [{\citenamefont {Ohyama}\ \emph {et~al.}(2006)\citenamefont {Ohyama},
  \citenamefont {Kawamuko}, \citenamefont {Sakai},\ and\ \citenamefont
  {Okamoto}}]{Ohyama2006StudiesOT}%
  \BibitemOpen
  \bibfield  {author} {\bibinfo {author} {\bibfnamefont {Y.}~\bibnamefont
  {Ohyama}}, \bibinfo {author} {\bibfnamefont {H.}~\bibnamefont {Kawamuko}},
  \bibinfo {author} {\bibfnamefont {H.}~\bibnamefont {Sakai}}, \ and\ \bibinfo
  {author} {\bibfnamefont {K.}~\bibnamefont {Okamoto}},\ }in\ \href@noop {}
  {\emph {\bibinfo {booktitle} {Studies on the Painlev\'e Equations, V, Third
  Painlev\'e Equations of Special Type PIII(D7) and PIII(D8)}}}\ (\bibinfo
  {year} {2006})\BibitemShut {NoStop}%
\bibitem [{\citenamefont {Okamoto}(1986)}]{Okamoto1986StudiesOT}%
  \BibitemOpen
  \bibfield  {author} {\bibinfo {author} {\bibfnamefont {K.}~\bibnamefont
  {Okamoto}},\ }\href@noop {} {\bibfield  {journal} {\bibinfo  {journal}
  {Annali di Matematica Pura ed Applicata}\ }\textbf {\bibinfo {volume}
  {146}},\ \bibinfo {pages} {337} (\bibinfo {year} {1986})}\BibitemShut
  {NoStop}%
\bibitem [{\citenamefont {Derezi'nski}\ \emph {et~al.}(2020)\citenamefont
  {Derezi'nski}, \citenamefont {Ishkhanyan},\ and\ \citenamefont
  {Latosi'nski}}]{Derezinski2020FromHC}%
  \BibitemOpen
  \bibfield  {author} {\bibinfo {author} {\bibfnamefont {J.}~\bibnamefont
  {Derezi'nski}}, \bibinfo {author} {\bibfnamefont {A.~M.}\ \bibnamefont
  {Ishkhanyan}}, \ and\ \bibinfo {author} {\bibfnamefont {A.}~\bibnamefont
  {Latosi'nski}},\ }\href@noop {} {\bibfield  {journal} {\bibinfo  {journal}
  {arXiv: Classical Analysis and ODEs}\ } (\bibinfo {year} {2020})}\BibitemShut
  {NoStop}%
\bibitem [{\citenamefont {Okamoto}(1980{\natexlab{b}})}]{10.3792/pjaa.56.367}%
  \BibitemOpen
  \bibfield  {author} {\bibinfo {author} {\bibfnamefont {K.}~\bibnamefont
  {Okamoto}},\ }\href {\doibase 10.3792/pjaa.56.367} {\bibfield  {journal}
  {\bibinfo  {journal} {Proceedings of the Japan Academy, Series A,
  Mathematical Sciences}\ }\textbf {\bibinfo {volume} {56}},\ \bibinfo {pages}
  {367 } (\bibinfo {year} {1980}{\natexlab{b}})}\BibitemShut {NoStop}%
\bibitem [{\citenamefont {Okamoto}(1999)}]{Okamoto1999}%
  \BibitemOpen
  \bibfield  {author} {\bibinfo {author} {\bibfnamefont {K.}~\bibnamefont
  {Okamoto}},\ }\enquote {\bibinfo {title} {The hamiltonians associated to the
  painlev{\'e} equations},}\ in\ \href {\doibase 10.1007/978-1-4612-1532-5_13}
  {\emph {\bibinfo {booktitle} {The Painlev{\'e} Property: One Century
  Later}}}\ (\bibinfo  {publisher} {Springer New York},\ \bibinfo {address}
  {New York, NY},\ \bibinfo {year} {1999})\ pp.\ \bibinfo {pages}
  {735--787}\BibitemShut {NoStop}%
\bibitem [{\citenamefont {Chen}\ and\ \citenamefont {Its}(2010)}]{CHEN2010270}%
  \BibitemOpen
  \bibfield  {author} {\bibinfo {author} {\bibfnamefont {Y.}~\bibnamefont
  {Chen}}\ and\ \bibinfo {author} {\bibfnamefont {A.}~\bibnamefont {Its}},\
  }\href {\doibase https://doi.org/10.1016/j.jat.2009.05.005} {\bibfield
  {journal} {\bibinfo  {journal} {Journal of Approximation Theory}\ }\textbf
  {\bibinfo {volume} {162}},\ \bibinfo {pages} {270} (\bibinfo {year}
  {2010})}\BibitemShut {NoStop}%
\bibitem [{\citenamefont {Its}\ \emph {et~al.}(2014)\citenamefont {Its},
  \citenamefont {Lisovyy},\ and\ \citenamefont {Tykhyy}}]{Its:2014lga}%
  \BibitemOpen
  \bibfield  {author} {\bibinfo {author} {\bibfnamefont {A.}~\bibnamefont
  {Its}}, \bibinfo {author} {\bibfnamefont {O.}~\bibnamefont {Lisovyy}}, \ and\
  \bibinfo {author} {\bibfnamefont {Y.}~\bibnamefont {Tykhyy}},\ }\href
  {\doibase 10.1093/imrn/rnu209} {\bibfield  {journal} {\bibinfo  {journal}
  {International Mathematics Research Notices}\ }\textbf {\bibinfo {volume}
  {2015}},\ \bibinfo {pages} {8903} (\bibinfo {year} {2014})},\ \Eprint
  {http://arxiv.org/abs/https://academic.oup.com/imrn/article-pdf/2015/18/8903/5183221/rnu209.pdf}
  {https://academic.oup.com/imrn/article-pdf/2015/18/8903/5183221/rnu209.pdf}
  \BibitemShut {NoStop}%
\bibitem [{\citenamefont {Lisovyy}\ and\ \citenamefont
  {Naidiuk}(2021)}]{Lisovyy:2021bkm}%
  \BibitemOpen
  \bibfield  {author} {\bibinfo {author} {\bibfnamefont {O.}~\bibnamefont
  {Lisovyy}}\ and\ \bibinfo {author} {\bibfnamefont {A.}~\bibnamefont
  {Naidiuk}},\ }\href {\doibase 10.1007/s11005-021-01400-6} {\bibfield
  {journal} {\bibinfo  {journal} {Lett. Math. Phys.}\ }\textbf {\bibinfo
  {volume} {111}},\ \bibinfo {pages} {137} (\bibinfo {year} {2021})},\ \Eprint
  {http://arxiv.org/abs/2101.05715} {arXiv:2101.05715 [math-ph]} \BibitemShut
  {NoStop}%
\bibitem [{\citenamefont {Jones}\ \emph {et~al.}(1980)\citenamefont {Jones},
  \citenamefont {Thron},\ and\ \citenamefont {Henrici}}]{jones1980continued}%
  \BibitemOpen
  \bibfield  {author} {\bibinfo {author} {\bibfnamefont {W.}~\bibnamefont
  {Jones}}, \bibinfo {author} {\bibfnamefont {W.}~\bibnamefont {Thron}}, \ and\
  \bibinfo {author} {\bibfnamefont {P.}~\bibnamefont {Henrici}},\ }\href
  {https://books.google.com.br/books?id=Okr1-6PjWPYC} {\emph {\bibinfo {title}
  {Continued Fractions: Analytic Theory and Applications}}},\ Encyclopedia of
  mathematics and its applications\ (\bibinfo  {publisher} {Addison-Wesley
  Publishing Company},\ \bibinfo {year} {1980})\BibitemShut {NoStop}%
\bibitem [{\citenamefont {Seidel}(1989)}]{Seidel:1988ue}%
  \BibitemOpen
  \bibfield  {author} {\bibinfo {author} {\bibfnamefont {E.}~\bibnamefont
  {Seidel}},\ }\href {\doibase 10.1088/0264-9381/6/7/012} {\bibfield  {journal}
  {\bibinfo  {journal} {Class. Quant. Grav.}\ }\textbf {\bibinfo {volume}
  {6}},\ \bibinfo {pages} {1057} (\bibinfo {year} {1989})}\BibitemShut
  {NoStop}%
\bibitem [{\citenamefont {Berti}\ \emph
  {et~al.}(2006{\natexlab{c}})\citenamefont {Berti}, \citenamefont {Cardoso},\
  and\ \citenamefont {Will}}]{Berti:2005ys}%
  \BibitemOpen
  \bibfield  {author} {\bibinfo {author} {\bibfnamefont {E.}~\bibnamefont
  {Berti}}, \bibinfo {author} {\bibfnamefont {V.}~\bibnamefont {Cardoso}}, \
  and\ \bibinfo {author} {\bibfnamefont {C.~M.}\ \bibnamefont {Will}},\ }\href
  {\doibase 10.1103/PhysRevD.73.064030} {\bibfield  {journal} {\bibinfo
  {journal} {Phys. Rev.}\ }\textbf {\bibinfo {volume} {D73}},\ \bibinfo {pages}
  {064030} (\bibinfo {year} {2006}{\natexlab{c}})},\ \Eprint
  {http://arxiv.org/abs/gr-qc/0512160} {arXiv:gr-qc/0512160 [gr-qc]}
  \BibitemShut {NoStop}%
\bibitem [{Git({\natexlab{a}})}]{GithubIM}%
  \BibitemOpen
  \href@noop {} {}\bibinfo {howpublished}
  {\url{https://github.com/JoaoPCavalcante/isomonodromic-method}}
  ({\natexlab{a}})\BibitemShut {NoStop}%
\bibitem [{\citenamefont {Miwa}(1981{\natexlab{b}})}]{Miwa:1981aa}%
  \BibitemOpen
  \bibfield  {author} {\bibinfo {author} {\bibfnamefont {T.}~\bibnamefont
  {Miwa}},\ }\href@noop {} {\bibfield  {journal} {\bibinfo  {journal}
  {Publications of the Research Institute for Mathematical Sciences}\ }\textbf
  {\bibinfo {volume} {17}},\ \bibinfo {pages} {703} (\bibinfo {year}
  {1981}{\natexlab{b}})}\BibitemShut {NoStop}%
\bibitem [{\citenamefont {Cohen-Tannoudji}\ \emph {et~al.}(2019)\citenamefont
  {Cohen-Tannoudji}, \citenamefont {Diu},\ and\ \citenamefont
  {Lalo{\"e}}}]{cohen2019quantum}%
  \BibitemOpen
  \bibfield  {author} {\bibinfo {author} {\bibfnamefont {C.}~\bibnamefont
  {Cohen-Tannoudji}}, \bibinfo {author} {\bibfnamefont {B.}~\bibnamefont
  {Diu}}, \ and\ \bibinfo {author} {\bibfnamefont {F.}~\bibnamefont
  {Lalo{\"e}}},\ }\href {https://books.google.com.br/books?id=tVI\_EAAAQBAJ}
  {\emph {\bibinfo {title} {Quantum Mechanics, Volume 1: Basic Concepts, Tools,
  and Applications}}}\ (\bibinfo  {publisher} {Wiley},\ \bibinfo {year}
  {2019})\BibitemShut {NoStop}%
\bibitem [{\citenamefont {Richartz}\ \emph {et~al.}(2017)\citenamefont
  {Richartz}, \citenamefont {Herdeiro},\ and\ \citenamefont
  {Berti}}]{Richartz:2017qep}%
  \BibitemOpen
  \bibfield  {author} {\bibinfo {author} {\bibfnamefont {M.}~\bibnamefont
  {Richartz}}, \bibinfo {author} {\bibfnamefont {C.~A.~R.}\ \bibnamefont
  {Herdeiro}}, \ and\ \bibinfo {author} {\bibfnamefont {E.}~\bibnamefont
  {Berti}},\ }\href {\doibase 10.1103/PhysRevD.96.044034} {\bibfield  {journal}
  {\bibinfo  {journal} {Phys. Rev. D}\ }\textbf {\bibinfo {volume} {96}},\
  \bibinfo {pages} {044034} (\bibinfo {year} {2017})},\ \Eprint
  {http://arxiv.org/abs/1706.01112} {arXiv:1706.01112 [gr-qc]} \BibitemShut
  {NoStop}%
\bibitem [{\citenamefont {Casals}\ and\ \citenamefont
  {Longo~Micchi}(2019)}]{Casals:2019vdb}%
  \BibitemOpen
  \bibfield  {author} {\bibinfo {author} {\bibfnamefont {M.}~\bibnamefont
  {Casals}}\ and\ \bibinfo {author} {\bibfnamefont {L.~F.}\ \bibnamefont
  {Longo~Micchi}},\ }\href {\doibase 10.1103/PhysRevD.99.084047} {\bibfield
  {journal} {\bibinfo  {journal} {Phys. Rev. D}\ }\textbf {\bibinfo {volume}
  {99}},\ \bibinfo {pages} {084047} (\bibinfo {year} {2019})},\ \Eprint
  {http://arxiv.org/abs/1901.04586} {arXiv:1901.04586 [gr-qc]} \BibitemShut
  {NoStop}%
\bibitem [{Git({\natexlab{b}})}]{Github}%
  \BibitemOpen
  \href@noop {} {}\bibinfo {howpublished}
  {\url{https://github.com/strings-ufpe/painleve}} ({\natexlab{b}})\BibitemShut
  {NoStop}%
\bibitem [{\citenamefont {Cho}(2003)}]{Cho:2003qe}%
  \BibitemOpen
  \bibfield  {author} {\bibinfo {author} {\bibfnamefont {H.~T.}\ \bibnamefont
  {Cho}},\ }\href {\doibase 10.1103/PhysRevD.68.024003} {\bibfield  {journal}
  {\bibinfo  {journal} {Phys. Rev. D}\ }\textbf {\bibinfo {volume} {68}},\
  \bibinfo {pages} {024003} (\bibinfo {year} {2003})},\ \Eprint
  {http://arxiv.org/abs/gr-qc/0303078} {arXiv:gr-qc/0303078} \BibitemShut
  {NoStop}%
\bibitem [{\citenamefont {Onozawa}\ \emph
  {et~al.}(1996{\natexlab{b}})\citenamefont {Onozawa}, \citenamefont {Mishima},
  \citenamefont {Okamura},\ and\ \citenamefont {Ishihara}}]{PhysRevD.53.7033}%
  \BibitemOpen
  \bibfield  {author} {\bibinfo {author} {\bibfnamefont {H.}~\bibnamefont
  {Onozawa}}, \bibinfo {author} {\bibfnamefont {T.}~\bibnamefont {Mishima}},
  \bibinfo {author} {\bibfnamefont {T.}~\bibnamefont {Okamura}}, \ and\
  \bibinfo {author} {\bibfnamefont {H.}~\bibnamefont {Ishihara}},\ }\href
  {\doibase 10.1103/PhysRevD.53.7033} {\bibfield  {journal} {\bibinfo
  {journal} {Phys. Rev. D}\ }\textbf {\bibinfo {volume} {53}},\ \bibinfo
  {pages} {7033} (\bibinfo {year} {1996}{\natexlab{b}})}\BibitemShut {NoStop}%
\bibitem [{\citenamefont {Green}\ \emph {et~al.}(2020)\citenamefont {Green},
  \citenamefont {Hollands},\ and\ \citenamefont {Zimmerman}}]{Green:2019nam}%
  \BibitemOpen
  \bibfield  {author} {\bibinfo {author} {\bibfnamefont {S.~R.}\ \bibnamefont
  {Green}}, \bibinfo {author} {\bibfnamefont {S.}~\bibnamefont {Hollands}}, \
  and\ \bibinfo {author} {\bibfnamefont {P.}~\bibnamefont {Zimmerman}},\ }\href
  {\doibase 10.1088/1361-6382/ab7075} {\bibfield  {journal} {\bibinfo
  {journal} {Class. Quant. Grav.}\ }\textbf {\bibinfo {volume} {37}},\ \bibinfo
  {pages} {075001} (\bibinfo {year} {2020})},\ \Eprint
  {http://arxiv.org/abs/1908.09095} {arXiv:1908.09095 [gr-qc]} \BibitemShut
  {NoStop}%
\end{thebibliography}%



\end{document}